\documentclass{aa}  
\usepackage{amsmath}
\usepackage{graphicx}
\usepackage{txfonts}
\usepackage{natbib} 
\usepackage{subfig}
\usepackage[T1]{fontenc}
\usepackage{subfig}

\begin{document}

\def\mean#1{\left< #1 \right>}

   \title{The evolutionary nature of RV Tauri stars in the SMC and LMC} 
   \author{Rajeev Manick\inst{1}
          \and
          Hans Van Winckel\inst{1}
          \and
          Devika Kamath\inst{1,3,4,5}
          \and
          Sanjay Sekaran \inst{1}
          \and
          Katrien Kolenberg \inst{1, 2}
          }

      \institute{Instituut voor Sterrenkunde (IvS), KU Leuven, Celestijnenlaan 200D, B-3001 Leuven, Belgium
              \and
              Science, Engineering and Technology Group, Celestijnenlaan 200I, B-2201 Leuven, Belgium
              \and
              Department of Physics and Astronomy, Macquarie University, Sydney, NSW 2109, Australia
              \and
              Astronomy, Astrophysics and Astrophotonics Research Centre, Macquarie University, Sydney, NSW 2109, Australia
              \and
              Australian Astronomical Observatory, PO Box 915, North Ryde, NSW 1670, Australia \\
              \email{rajeev.manick@kuleuven.be}
             }
   \authorrunning{Manick et al. 2018}
   \titlerunning{Evolutionary nature of RV Tauri stars}

   \abstract
   {Based on their stellar parameters and the presence of a mid-IR excess due to circumstellar dust, RV Tauri stars have been classified as post-AGB stars.
   Our recent studies, however, reveal diverse spectral energy distributions (SEDs) among RV\,Tauri stars, suggesting they may occupy other evolutionary channels as well.}
   {The aim of this paper is to present the diverse SED characteristics of RV Tauri stars and investigate their evolutionary nature as a function of their SEDs.}
   {We carried out a systematic study of RV Tauri stars in the SMC and LMC because of their known distances and hence luminosities. 
   Their SEDs were classified into three groups: dusty (disc-type), non-dusty (non-IR), and uncertain.
   A period-luminosity-colour (PLC) relation was calibrated. The luminosities
   from the PLC were complemented with those found using their SEDs and the stars were 
   placed on a Hertzsprung Russell diagram (HRD). I-band time series were used to search for period changes via ($O-C$) analyses to identify period changes.}
   {
The four main results from this study are: 1) RV Tauri stars with a clear IR excess have disc-type SEDs, which indicates that the dust is trapped in a stable disc. 
Given the strong link between disc-type SEDs and binarity in the Galaxy, we postulate that these are binaries as well. These  
cover a range of luminosities and we argue that the more luminous binaries are post-AGB stars while the lower luminosity binaries are likely post-red giant branch (post-RGB) stars. 2) Two of these objects
have variable mean brightness with periods of 916 and 850 days, respectively, caused by variable extinction during orbital motion. 3) Non-dusty RV Tauri stars and objects 
with an uncertain SED evolve such that the circumstellar dust has dispersed. If they are single stars, they are post-AGB objects of low initial mass (< 1.25 M$_{\odot}$), while
if they are binaries, the low-luminosity portion of the sample are likely post-RGB stars. 4) We find that RV Tauri stars with dust are on 
average more luminous than the rest of the sample.
    }

   \keywords{Stars: AGB and post-AGB -- 
             Stars: binaries: general --
             Stars: chemically peculiar --
             Stars: evolution --
             Stars: Population II --
             Techniques: photometric
               }

   \maketitle
%

\section{Introduction}
RV Tauri stars are the most luminous stars among the type II Cepheids and are mostly of F to K spectral types. 
A defining characteristic of RV Tauri stars is their light curves, 
which alternate between deep and shallow minima. In most cases, the light curves can be decomposed
into two periodic signals: one with a formal period (the time between two adjacent deep minima) and 
the other with a fundamental period (the time between consecutive deep and shallow minima).
The fundamental period typically ranges between 20 days and 75 days and the formal period has a range that is
twice that of the fundamental period \citep{preston63,wallerstein02}. 

In addition to the pulsational variability, a number of RV Tauri stars display a long-term periodic variability in the mean flux of the order of 600 to 2600 days.
Stars exhibiting this long-term variability have been classified photometrically as RVb types \citep{pollard96}, while those with a constant mean flux in their 
time series are classified as RVa photometric types \citep[][]{evans85,pollard96}.

The mechanism giving rise to RV Tauri-like pulsations is not well understood \citep{takeuti83,percy07}. 
However, there are two leading hypotheses explaining their nature. The first explanation is that the stars are experiencing 
double-mode pulsators in which the periods are in ratio 2:1 \citep{takeuti83,fokin94} and 2) the stars are undergoing 
low-dimensional chaos \citep{buchler87}.

RV Tauri stars have been linked to post-AGB stars. Their high luminosities and the presence of 
circumstellar dust around many of these stars were the main arguments to classify the whole sample of RV Tauri
stars as post-AGB stars \citep[e.g.][]{lloydevans85,jura86}.
Later, with the advent of a period-luminosity (PL) relation for the type II 
Cepheids in the LMC, calibrated by \citet{alcock98}, and the resulting high luminosities from their work, the post-AGB nature of RV Tauri stars was corroborated. 
The observational picture is, however, more complicated.

Colour-colour plots based on surveys such as the 
Infra-Red Astronomical Satellite (IRAS), Wide-field Infrared Survey Explorer (WISE), \textit{Spitzer} (IRAC and MIPS) have led to 
a systematic characterisation of the IR excesses \citep[e.g.][]{lloydevans85,gezer2015,kamath2014,kamath2015}. Detailed  spectral energy distribution
(SED) studies of optically bright post-AGB candidates have shown that they can be subdivided into three main 
categories based on the types of their IR excess \citep{trams91,vanderveen94,bogaert94,vanwinckel03,deruyter06}:

\begin{itemize}
 \item Disc-type: a broad IR excess with a hot dust component starting at the dust sublimation temperature. 
 \item Shell-type: a double peaked SED with the first peak representing the photospheric component 
 and the second peak characterizing a detached and expanding shell of cold dust - a remnant of the AGB dust shell - peaking in the mid-IR \citep{volk89}. 
 \item Uncertain: the type of IR excess in their SEDs is not clear owing to the scarcity of data at long wavelengths \citep{kamath2014,kamath2015}.
\end{itemize}

In our systematic SED study of the $\sim$128 known RV Tauri stars in the Galaxy \citep{gezer2015,gezer2015err}, we showed that RV Tauri stars display three main types of IR properties in the SEDs: disc-type (23\%), 
uncertain (27\%), and non-dusty (50\%). The first two of these SED types (disc-type and uncertain) are similar to what is observed in post-AGB stars (as mentioned above). The latter class of objects, i.e. the non-dusty stars
show no IR excess in their SEDs and were hence classified as non-IR. Furthermore, the results of our study also demonstrated that when dust is present in RV Tauri stars, it is always in the form of a disc and not in the form of a clear detached shell.

It is now well established that post-AGB stars with a disc-type SED are likely all binaries \citep{vanwinckel09}.
The disc-binary correlation is bona fide among Galactic RV Tauri stars as well. 
At present, there are 12 spectroscopically confirmed binaries among Galactic RV Tauri stars with a disc, six of which are 
U\,Mon \citep{pollard95,pollard97}, IW\,Car \citep{pollard97}, AC\,Her \citep{vanwinckel98}, EN\,Tra 
\citep{vanwinckel99}, SX\,Cen \citep{maas02}, and RU\,Cen \citep{maas02}. The spectroscopic orbits of the additional six stars (DY\,Ori, EP\,Lyr, HP\,Lyr, IRAS\, 
17038$-$4815, IRAS\,09144$-$4933, and TW\,Cam) were presented in our previous paper (Manick et al. 2017). 
We also used the PLC relation to quantify their luminosities. We found that while the most luminous binaries comply with a post-AGB track, some objects are less luminous 
than the tip of the red giant branch (RGB) of a low-mass star. We argued that these objects evolve likely on tracks in which the binary interaction already took place on the 
RGB and these objects can be seen as examples of post-RGB stars. This class was introduced by \citet{kamath2016} to account for dusty objects at low 
luminosity that were discovered in their systematic large-scale survey of the SMC and LMC, respectively \citep{kamath2014,kamath2015}.

Chemically, most RV Tauri stars do not display signs of the third dredge-up, i.e. no enrichment in C and s-process elemental abundances in their photospheres \citep{giridhar94,gonzalez97b,gonzalez97a,vanwinckel98,giridhar00}. 
There are, however, three noticeable exceptions in which either one or both of these elements are seen: HD 158616 \citep{vanwinckel97}, MACHO 47.2496.8 \citep{reyniers07a}, and OGLE-SMC-T2CEP-018 (SMC J005107.19-734133.3) \citep{kamath2014}.
Instead, the photospheric composition of most RV Tauri stars with a disc-type SED displays
a phenomenon called depletion. Depleted photospheres are selectively under-abundant of refractory relative to non-refractory elements.
A seminal study by \citet{waters92} introduced the following explanation for the depletion process: the circumbinary disc is composed of dust and gas;
there is selective re-accretion of gas onto the stellar photosphere from the disc, while refractory elements that stick to the dust particles 
remain in the dust component. This idea is supported observationally by the fact that many disc sources are highly depleted \citep{maas02,giridhar05,gezer2015}.

While some observational aspects of a subset of RV Tauri stars indicate that they are likely 
post-AGB stars in binary systems, a few of their characteristics may suggest they are likely single post-AGB stars. One of these features is that 
none of the known optically bright RV Tauri stars display a clear shell-type SED, which is characteristic of an expanding detached shell around
a single star after an event of high mass loss on the AGB \citep{fujii02,vanwinckel03}. In addition, there is a significant percentage ($\sim$\,45\%) of the aforementioned non-dusty RV Tauri stars \citep{gezer2015}. 
These stars do not seem to comply with post-AGB evolution mainly because typically the AGB phase is terminated either by a superwind driven mass loss or binary interaction that sheds the outer envelope of the star. 
The ejected circumstellar material is either in the form of a detached shell (for single stars) or a hot dusty disc (for binaries). 
Thus a main characteristic of post-AGB stars is the large IR excess indicative of the circumstellar dust.

The different SED characteristics and the observed chemical patterns amongst RV Tauri stars suggest that
their evolutionary nature is more complex than their currently hypothesised post-AGB nature. Central to this research is the evolutionary nature of RV Tauri stars with different SED characteristics. 
We focus on a systematic study of RV Tauri stars in the SMC and LMC because the well-known distances to the Magellanic clouds allows for accurate luminosity determination.

In section \ref{section:sample}, we describe our sample selection in the SMC and LMC. We illustrate our time series and  
SED data in section \ref{section:data}. The analysis and results from the time series data are presented in section \ref{section:timeseriesanalysis}. 
We outline the IR classification of their SED characteristics 
in section \ref{section:characterisation}. The SED fitting method is described in section \ref{section:sedfitting}. The derivation of the luminosities based on the PLC relation and SEDs is presented 
in section \ref{section:luminosities}. We describe in section \ref{section:O_C}, the computation of the ($O-C$) for stars that show period variation.
In section \ref{section:discussion} we discuss our results and we present our conclusion in section \ref{section:conclusion}.

\section{Sample selection} \label{section:sample}
Our sample consists of RV Tauri stars identified in both the SMC and LMC by the OGLE-III survey.
We give a brief account of the selection process and classification of the RV Tauri variables. 
For a full description, see \citet{soszynski2008a} and \citet{soszynski10}. These authors classify a star as an RV Tauri pulsator if the star has a pulsational period greater than $\sim$\,20 days
and has the proper location on the PL diagram. A description of our sample is given in Tables~\ref{table:sampleLMC} and \ref{table:sampleSMC}.

\subsection{LMC Sample}
The initial selection process by \citet{soszynski2008a} consisted of a massive period search among 32 million objects in the OGLE-III time series. Stars that showed periodic 
variability (pulsating, eclipsing, and other variables) were selected. The pulsating variables were then plotted on a PL diagram and the type-II Cepheids were identified 
based on their position on the PL diagram. The outcome of this process was the identification of 42 RV Tauri variables, which we use as our LMC sample. Thirteen of the LMC stars were already part of the
MACHO database compiled by \citet{alcock98}. The MACHO names are included in the second column of Table \ref{table:sampleLMC}.

\subsection{SMC Sample}
The selection process of the SMC RV Tauri stars is similar to that adopted for the LMC targets, but with a period search among 6 million stars.
Based on this analysis, \citet{soszynski10} discovered 43 type II Cepheids in the SMC, of which 9 were classified as RV Tauri pulsators. These 9 stars
are part of our SMC sample.

\begin{table*}
\tiny
\centering                        
\begin{tabular}{lllllllll}
\hline 
Star & MACHO ID  & RA & DEC & I (mag) & V (mag) & P$_0$ (days) & N$_{epochs}$ & Ref. \\
\hline\hline    
OGLE-LMC-T2CEP-003 & &04:43:05.54 & -66:49:36.1 & 14.166 & 14.953 & 35.659 & 383  &   a   \\
OGLE-LMC-T2CEP-005 & &04:48:08.33 & -69:51:14.2 & 14.739 & 15.796 & 33.185 & 364  &   a   \\
OGLE-LMC-T2CEP-011 & &04:54:38.69 & -67:30:12.2 & 14.089 & 14.789 & 39.256 & 378  &   a,f   \\
OGLE-LMC-T2CEP-014 & &04:55:35.40 & -69:54:04.2 & 14.312 & 15.103 & 61.875 & 370  &   a   \\
OGLE-LMC-T2CEP-015$^{*}$ & MACHO 47.2496.8 & 04:55:43.25 & -67:51:09.9 & 14.061 & 15.243 & 56.521 & 386  &   a,b,e,f   \\
OGLE-LMC-T2CEP-016 & &04:55:45.99 & -69:07:46.3 & 15.458 & 15.891 & 20.295 & 370  &   a   \\
OGLE-LMC-T2CEP-025 & &05:01:38.78 & -68:35:19.0 & 14.042 & 15.102 & 67.965 & 370  &   a   \\
OGLE-LMC-T2CEP-029$^{*}$ & MACHO 19.3694.19 & 05:03:04.97 & -68:40:24.7 & 14.642 & 15.446 & 31.245 & 658  &   a,c,e,f   \\
OGLE-LMC-T2CEP-032 & &05:03:56.31 & -67:27:24.6 & 14.011 & 14.992 & 44.561 & 610  &   a   \\
OGLE-LMC-T2CEP-045 & &05:06:34.06 & -69:30:03.7 & 13.729 & 14.787 & 63.386 & 363  &   a   \\
OGLE-LMC-T2CEP-050 & &05:09:26.15 & -68:50:05.0 & 14.964 & 15.661 & 34.748 & 667  &   a   \\
OGLE-LMC-T2CEP-051 & &05:09:41.93 & -68:51:25.0 & 14.569 & 15.440 & 40.606 & 658  &   a   \\
OGLE-LMC-T2CEP-055 & &05:11:15.17 & -67:56:00.4 & 14.262 & 15.078 & 41.005 & 387  &   a   \\
OGLE-LMC-T2CEP-058 & MACHO 2.5026.30 & 05:11:33.52 & -68:35:53.7 & 15.511 & 16.594 & 21.482 & 384  &   a,e   \\
OGLE-LMC-T2CEP-065 & &05:14:00.76 & -68:57:56.8 & 14.699 & 15.611 & 35.054 & 783  &   a   \\
OGLE-LMC-T2CEP-067$^{*}$ & MACHO 79.5501.13 & 05:14:18.11 & -69:12:35.0 & 13.825 & 14.627 & 48.231 & 760  &   a,c,d,e,f   \\
OGLE-LMC-T2CEP-075 & &05:16:16.06 & -69:43:36.9 & 14.568 & 15.728 & 50.186 & 781  &   a   \\
OGLE-LMC-T2CEP-080 & MACHO 78.5856.2363 & 05:16:47.43 & -69:44:15.1 & 14.341 & 15.175 & 40.916 & 735  &   a,e   \\
OGLE-LMC-T2CEP-082 & &05:16:55.33 & -71:41:41.5 & 14.982 & 16.065 & 35.124 & 364  &   a   \\
OGLE-LMC-T2CEP-091 & &05:18:45.47 & -69:03:21.6 & 14.203 & 14.899 & 35.749 & 899  &   a,c,f   \\
OGLE-LMC-T2CEP-104 & MACHO 78.6698.38 & 05:21:49.10 & -70:04:34.3 & 14.937 & 15.830 & 24.879 & 867  &   a,e   \\
OGLE-LMC-T2CEP-108 & &05:22:11.27 & -68:11:31.3 & 14.746 & 15.477 & 30.010 & 738  &   a   \\
OGLE-LMC-T2CEP-112 & &05:22:58.36 & -69:26:20.9 & 14.065 & 14.749 & 39.397 & 885  &   a   \\
OGLE-LMC-T2CEP-115 & MACHO 77.7069.213 & 05:23:43.53 & -69:32:06.8 & 15.593 & 16.651 & 24.966 & 413  &   a,e  \\
OGLE-LMC-T2CEP-119$^{*}$ & &05:25:19.48 & -70:54:10.0 & 14.391 & 15.225 & 33.825 & 366  &   a,c,f   \\
OGLE-LMC-T2CEP-125 & &05:27:39.18 & -71:47:57.0 & 14.934 & 15.924 & 33.033 & 365  &   a   \\
OGLE-LMC-T2CEP-129 & &05:28:54.60 & -69:52:41.1 & 14.096 & 14.813 & 62.508 & 1069  &   a   \\
OGLE-LMC-T2CEP-135 & MACHO 82.8041.17 & 05:29:38.50 & -69:15:12.2 & 15.194 & 16.162 & 26.522 & 446  &   a,e   \\
OGLE-LMC-T2CEP-147$^{*}$ & MACHO 82.8405.15 & 05:31:51.00 & -69:11:46.3 & 13.678 & 14.391 & 46.795 & 524  &   a,d,e,f   \\
OGLE-LMC-T2CEP-149$^{*}$ & MACHO 81.8520.15 & 05:32:54.46 & -69:35:13.2 & 14.151 & 14.868 & 42.480 & 399  &   a,d,e,f   \\
OGLE-LMC-T2CEP-162 & MACHO 81.9362.25 & 05:37:44.96 & -69:54:16.5 & 15.112 & 16.200 & 30.394 & 639  &   a,e   \\
OGLE-LMC-T2CEP-169 & MACHO 14.9582.9  & 05:39:32.80 & -71:21:54.3 & 14.699 & 15.651 & 30.955 & 352  &   a,e   \\
OGLE-LMC-T2CEP-174$^{*}$ & MACHO 81.9728.14 & 05:40:00.50 & -69:42:14.7 & 13.693 & 14.457 & 46.818 & 372  &   a,c,d,e,f   \\
OGLE-LMC-T2CEP-180$^{*}$ & &05:43:12.87 & -68:33:57.1 & 14.502 & 15.303 & 30.996 & 351  &   a,c,f   \\
OGLE-LMC-T2CEP-190 & &05:50:20.46 & -69:30:54.6 & 14.307 & 15.147 & 38.361 & 355  &   a   \\
OGLE-LMC-T2CEP-191$^{*}$ & &05:51:22.54 & -69:53:51.4 & 14.220 & 15.017 & 34.344 & 359  &   a,c,f   \\
OGLE-LMC-T2CEP-192 & &05:53:55.69 & -70:17:11.4 & 15.233 & 16.148 & 26.194 & 327  &   a  \\
OGLE-LMC-T2CEP-198 & &04:39:35.27 & -69:33:28.9 & 15.271 & 16.474 & 38.274 & 436  &   a  \\
OGLE-LMC-T2CEP-199 & &05:04:44.83 & -68:58:31.8 & 13.658 & 13.806 & 37.203 & 756  &   a  \\
OGLE-LMC-T2CEP-200 & &05:13:56.44 & -69:31:58.3 & 15.092 & 16.124 & 34.916 & 917  &   a  \\
OGLE-LMC-T2CEP-202 & &05:21:49.09 & -70:46:01.4 & 15.167 & 16.359 & 38.135 & 738  &   a  \\
OGLE-LMC-T2CEP-203 & &05:22:33.79 & -69:38:08.5 & 15.395 & 16.723 & 37.126 & 1480  &   a  \\
\hline
    \textbf{References:} \\
    \multicolumn{7}{l}{\text{a:} \footnotesize{\citet{soszynski10}}} \\
    \multicolumn{7}{l}{\text{b:} \footnotesize{\citet{reyniers07a}}}  \\
    \multicolumn{7}{l}{\text{c:} \footnotesize{\citet{kamath2015}}} \\
    \multicolumn{7}{l}{\text{d:} \footnotesize{\citet{gielen09b}}} \\
    \multicolumn{7}{l}{\text{e:} \footnotesize{\citet{alcock98}}} \\
    \multicolumn{7}{l}{\text{f:} \footnotesize{\citet{groenewegen2017a}}} \\
\end{tabular}
\caption{Summary of our sample of RV Tauri stars in the LMC. Columns 3 and 4 give the position of the stars in the LMC. Columns 5 and 6 display the mean I and V magnitude of each star, respectively. 
In column 7 the fundamental period of each star is shown and column 8 represents the number of I-band data points available for each star. The stars marked by an asterisk (*) indicate those that have already been studied spectroscopically.}   
\label{table:sampleLMC}   
\end{table*}

\begin{table*}
\tiny
\centering                        
\begin{tabular}{llllllll}
\hline 
Star & RA & DEC & I (mag) & V (mag) & P$_0$ (days) & N$_{epochs}$ &Ref. \\
\hline\hline    
OGLE-SMC-T2CEP-07 & 00:42:57.49 & -73:07:14.8 & 13.572 & 14.440 & 30.961  & 1015  & a  \\
OGLE-SMC-T2CEP-12 & 00:45:19.69 & -73:20:13.7 & 15.415 & 16.369 & 29.219  & 1020  & a  \\
OGLE-SMC-T2CEP-18$^{*}$ & 00:51:07.23 & -73:41:33.4 & 14.813 & 15.627 & 39.519  & 710  & a,b,c  \\
OGLE-SMC-T2CEP-19 & 00:53:27.69 & -73:38:09.5 & 14.481 & 15.064 & 40.912  & 710  & a  \\
OGLE-SMC-T2CEP-20 & 00:53:35.98 & -72:34:21.8 & 14.894 & 15.890 & 50.623  & 1033  & a  \\
OGLE-SMC-T2CEP-24 & 00:55:20.55 & -73:21:37.9 & 14.511 & 15.308 & 43.961  & 870  & a  \\
OGLE-SMC-T2CEP-29 & 00:57:38.09 & -72:18:12.2 & 13.648 & 14.556 & 33.676  & 846  & a,c  \\
OGLE-SMC-T2CEP-41 & 01:13:19.05 & -73:18:07.8 & 15.353 & 16.068 & 29.118  & 653  & a  \\
OGLE-SMC-T2CEP-43 & 01:23:53.06 & -72:16:45.9 & 15.398 & 16.342 & 23.743  & 595  & a  \\
\hline
    \textbf{References:} \\
    \multicolumn{7}{l}{\text{a:} \footnotesize{\citet{soszynski2008a}}} \\
    \multicolumn{7}{l}{\text{b:} \footnotesize{\citet{kamath2014}}}  \\
    \multicolumn{7}{l}{\text{c:} \footnotesize{\citet{groenewegen2017a}}} \\
\end{tabular}
\caption{A summary of our sample of RV Tauri stars in the SMC. Columns 2 and 3 give the position of the stars in the SMC. Columns 4 and 5 display the mean I and V magnitude of each star, respectively. 
In column 6 the fundamental period of each star is shown and column 7 represents the number of I-band datapoints available for each star. The stars marked by an asterisk (*) indicate the ones which have already been studied spectroscopically.}   
\label{table:sampleSMC}   
\end{table*}

\section{Data} \label{section:data}
\subsection{Time series} \label{section:timeseries} 
We used time series data from the OGLE-III survey carried out using the 1.3 $\rm m$ Warsaw Telescope at Las Campanas Observatory, Chile \citep{soszynski2008a,soszynski10}.
The telescope is equipped with a large field CCD mosaic camera, consisting of eight 2048\,$\times$\,4096 pixel detectors. Each detector has 
a pixel size of 15 $\mu$m at a resolution of 0.26 arcsec per pixel and a field of view of 35 $\times$ 35.5 arcmins$^{2}$. For more details, see \citet{Udalski2003}.

The observations were carried out in two bands, I and V, with better sampling in the I band.
For both the SMC and LMC targets, the span of the time series data is $\sim$\,2400 days. The number of I-band data points for each star is 
included in column 8 of Table \ref{table:sampleLMC} and column 7 of Table \ref{table:sampleSMC}.

\subsection{Spectral energy distribution} \label{section:SEDs}
To construct the SEDs for the individual stars, we obtained various sets of broadband photometry available in the literature. The search was carried 
out homogeneously within a radius of 2 arcseconds of the 2MASS coordinates.
We excluded any photometric point that was marked in the catalogue as having an unreliable flux determination. 
There were a few objects that had the WISE [12] micron and (or) [22] micron fluxes flagged as upper limits (S/N $<$ 2) in the WISE catalogue. These points were also 
excluded in our analyses.

The final dataset consists of UBVRI photometry from the Johnson (UBVR) \citep{johnson1953}, and Sloan Digital Sky Survey (SDSS) I-band filters \citep{york2000}. The near-IR region is covered by J-, H- and K-band 
photometry from the Two Micron All Sky Survey (2\,MASS) \citep{skrutskie06} and JOHNSON J- and H-band photometry from the IRSF Magellanic Clouds Point Source Catalogue \citep{kato07}. 
The mid-IR data is composed of data from the WISE \citep{wright10} and 
\textit{Spitzer} (IRAC) survey \citep{houck04,fazio04}. The WISE data is comprised of four bands, centred at [3.4], [4.6], [12], and [22] $\mu$m (W1, W2, W3, W4) at angular resolutions 
of 6.1'', 6.4'', 6.5'', and 12.0'' in the four bands, respectively. The full width half maximum (FWHM) of the point spread function (PSF) of the four IRAC bands at [3.6], [4.5], [5.8] and [8.0] $\mu$m are
1.6'', 1.6'', 1.8'', and 1.9'', respectively. At longer wavelengths, the far-IR region is characterised by the \textit{Spitzer} MIPS 24 $\mu$m data \citep{rieke04}.
The MIPS 24 $\mu$m filter has a PSF of 6''.

\section{Time series analysis} \label{section:timeseriesanalysis}
\subsection{LMC} \label{section:timeserieslmc}
We pre-whitened the I-band photometric time series to find the dominant periods in the data. This was achieved using the Lomb-Scargle method for unequally spaced data \citep{lomb76,scargle82}. 
The obtained fundamental pulsation periods (P$_0$) and the formal pulsation periods (P$_1$) were consistently checked with those provided by \citet{soszynski2008a} and are the same for all stars. 

The time series, phase-folded on the formal pulsation periods (P$_1$), are shown in Appendix \ref{appendix:AppendixC} for the LMC targets. 
We note that the plots are grouped in terms of their SED types. Most of the phase-folded light curves show well-behaved RV Tauri-like pulsations 
with the exception of a few cases in which the alternating deep and shallow minima are not easily discernible. 
An example of such a case is OGLE-LMC-T2CEP-108 (see Figure \ref{figure:LMC108pulsation} in Appendix \ref{appendix:AppendixC}).

In some objects, we see evidence of modulation in the minima that is indicative of period variation \citep[e.g.][]{percy97}. 
An example of such a case is OGLE-LMC-T2CEP-191 (see Figure \ref{subfig:modulation} in Appendix \ref{appendix:AppendixC}). 
Stars that showed such variability were further analysed using an ($O-C$) diagram to determine the nature of the period change. 
The ($O-C$) variation is discussed further in Section \ref{section:O_C}.

In addition, we see long-term variability in the light curves of OGLE-LMC-T2CEP-032 and OGLE-LMC-T2CEP-200 
at periods of 850 days and 916 days, respectively. This variation is typical of the RVb nature and this is discussed further in Section \ref{section:RVb_bin}. The rest of 
the stars do not show any prominent long-term variability in the time series and are therefore RVa types.

\subsection{SMC} \label{section:timeseriessmc}
The time series data analysis of the SMC targets was carried out in the same way as the LMC ones. A few stars showed significant amplitude variations in the light curves, namely
OGLE-SMC-T2CEP-19, OGLE-SMC-T2CEP-20,OGLE-SMC-T2CEP-24, and OGLE-SMC-T2CEP-29. The amplitude variations in OGLE-SMC-T2CEP-24 was also reported by \citet{groenewegen2017a}.
Evidence of period-changes are seen in OGLE-SMC-T2CEP-19, OGLE-SMC-T2CEP-20, and OGLE-SMC-T2CEP-24, but are not significant enough to derive any conclusions (see section \ref{section:O_C}).

Apart from the pulsation periods, we detect significant long-term periodic variability in the periodograms of OGLE-SMC-T2CEP-07 and OGLE-SMC-T2CEP-29.
The period search in the time series of these two stars reveals a period of 196.7 days for OGLE-SMC-T2CEP-07 and 304.04 days for OGLE-SMC-T2CEP-29.
Long periodic variability in these two stars were already reported by \citet{soszynski10} and \citet{groenewegen2017a}. The periods were linked to their 
binary nature: OGLE-SMC-T2CEP-07 is an ellipsoidal variable and OGLE-SMC-T2CEP-29 is an eclipsing binary.

Ellipsoidal variables are non-eclipsing close binaries in which one or both components is or are distorted \citep{wood99,pawlak2014}. This results in a distinct light curve shape with periodic light variations due to the 
changing cross-sectional areas of at least one of the components. The brightness variations occur with a period that is equal to half of the orbital 
period of the binary star \citep{morris85}. The long-term period of 196.7 days that we detect in the light curve of OGLE-SMC-T2CEP-07 is therefore half of the orbital period of $\sim$ 393 days
given by \citet{soszynski10}.

\citet{soszynski10} found that OGLE-SMC-T2CEP-29 is an eclipsing binary with a period of 608.6 days. The light curve phased on the orbital period
shows occultations in the form of dips in the I-band flux. The significant long-term periodicity of 304.04 days that we detect in the time series 
for this object is half of the orbital period.

We believe that the long-term variability in these two stars is not due to the RVb phenomenon because the SEDs do not show the presence of dust in the 
IR region (see Figures \ref{figure:nonirsmc1} and \ref{figure:nonirsmc2} in Appendix \ref{appendix:AppendixB}).

\section{Infrared colour classification} \label{section:characterisation}
\begin{figure}
   \centering
   \includegraphics[width=9cm,height=7cm]{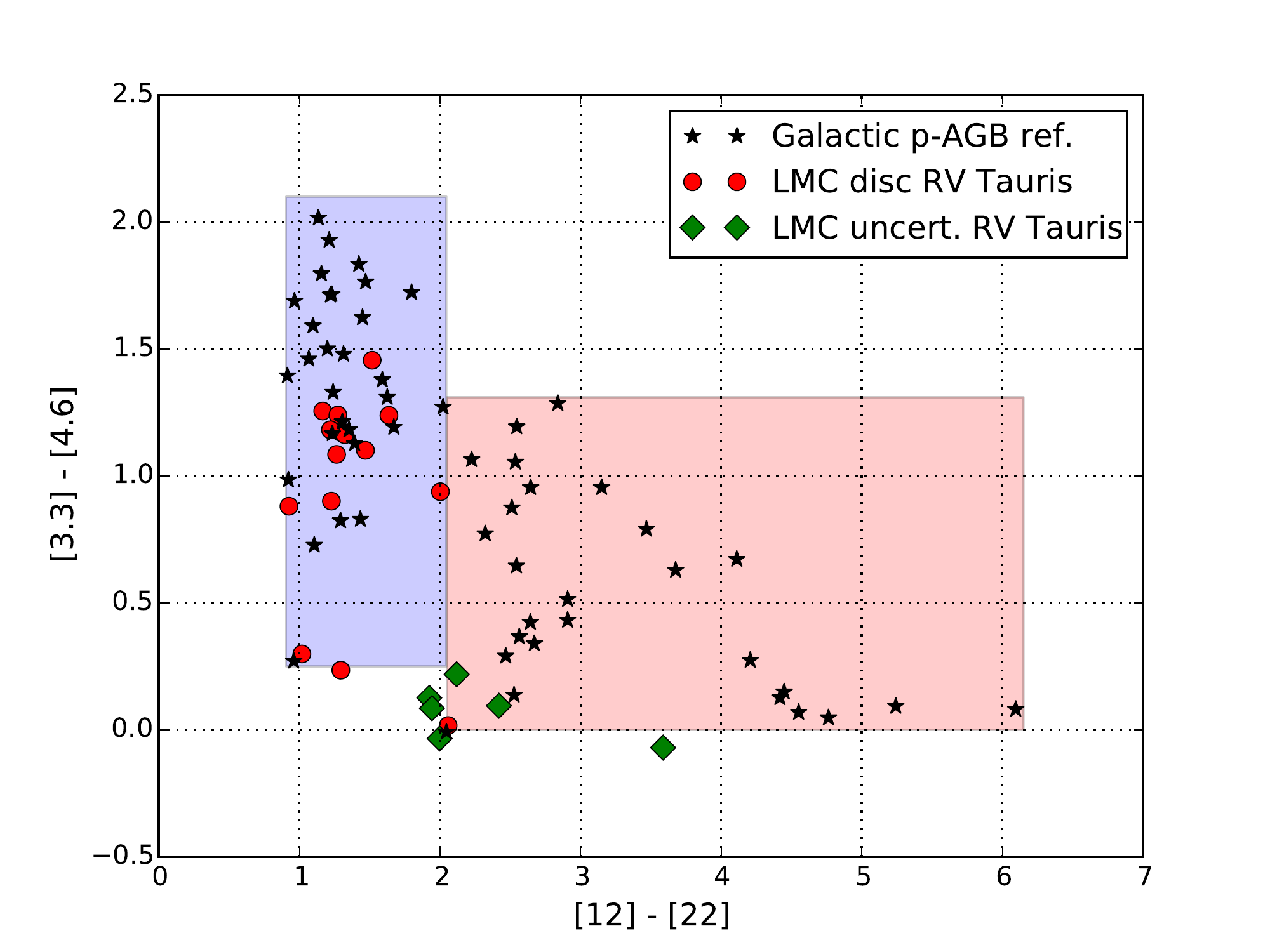} 
   \caption{WISE colour-colour plot of the sample RV Tauri stars in the LMC. The reference Galactic sample is shown by the black star markers.
   The blue shaded region corresponds to stars in the reference sample with disc-type characteristics and the red shaded 
   region displays stars in the reference sample with shell-type characteristics. The markers are colour-coded according to their IR characteristics.}
     \label{figure:color-color_LMC}
\end{figure}

The infrared colour classification of our sample stars was performed based on two colour-colour classification schemes: 1) 
the \textit{Spitzer} (IRAC and MIPS) colour-colour diagram by \citet{kamath2015}, which uses the [8]$-$[24] and [3.6]$-$[4.5] micron fluxes 
and 2) the WISE colour-colour plot by \citet{gezer2015}, which is based on the [12]$-$[22] and [3.3]$-$[4.6] micron fluxes.
A sample of Galactic post-AGB stars (shown by the black star markers in Figures \ref{figure:color-color_LMC} and \ref{figure:color-color_SMC}), 
which are already classified as disc or shell types, were used as reference in our WISE classification.

In the reference sample, the disc objects display a broad IR excess starting at $\sim$\,[2] microns, which is an observational signature of a disc.
These stars are typically located in the blue shaded region in Figures \ref{figure:color-color_LMC} and \ref{figure:color-color_SMC}. 
The SEDs of post-AGB stars representing shell types are double peaked with a mild [3.3]$-$[4.6] micron excess. These stars lie in the 
red shaded region of the WISE colour-colour plot; see Figures \ref{figure:color-color_LMC} and \ref{figure:color-color_SMC}. 

\subsection{LMC IR classification} \label{section:lmcsedclassification}
Most of the LMC targets had IR data from the WISE catalogue, while only few had \textit{Spitzer} data. There were a few cases in which either the WISE [22] micron or the MIPS [24] 
micron flux was not available and these stars were not included in the respective colour-colour plots. We note that the non-IRs are not plotted in any of the colour-colour diagrams. 
A summary of our LMC IR classification is shown in column 10 of Table \ref{table:lumsLMC}.

\subsubsection{Disc-type LMC stars}
The disc-type stars are identified by the red dots in Figure \ref{figure:color-color_LMC}. Among the disc-types, there are 14 stars for which both WISE [3.4]$-$[4.6] and [12]$-$[22] colour were available. 
Not included in this figure are two other objects namely, OGLE-LMC-T2CEP-149 and OGLE-LMC-T2CEP-200, which show clear disc-type SEDs (see Figures \ref{figure:seddisc12} and \ref{figure:seddisc16}), 
but no WISE data were available.

\begin{figure}
   \centering
   \includegraphics[width=9cm,height=7cm]{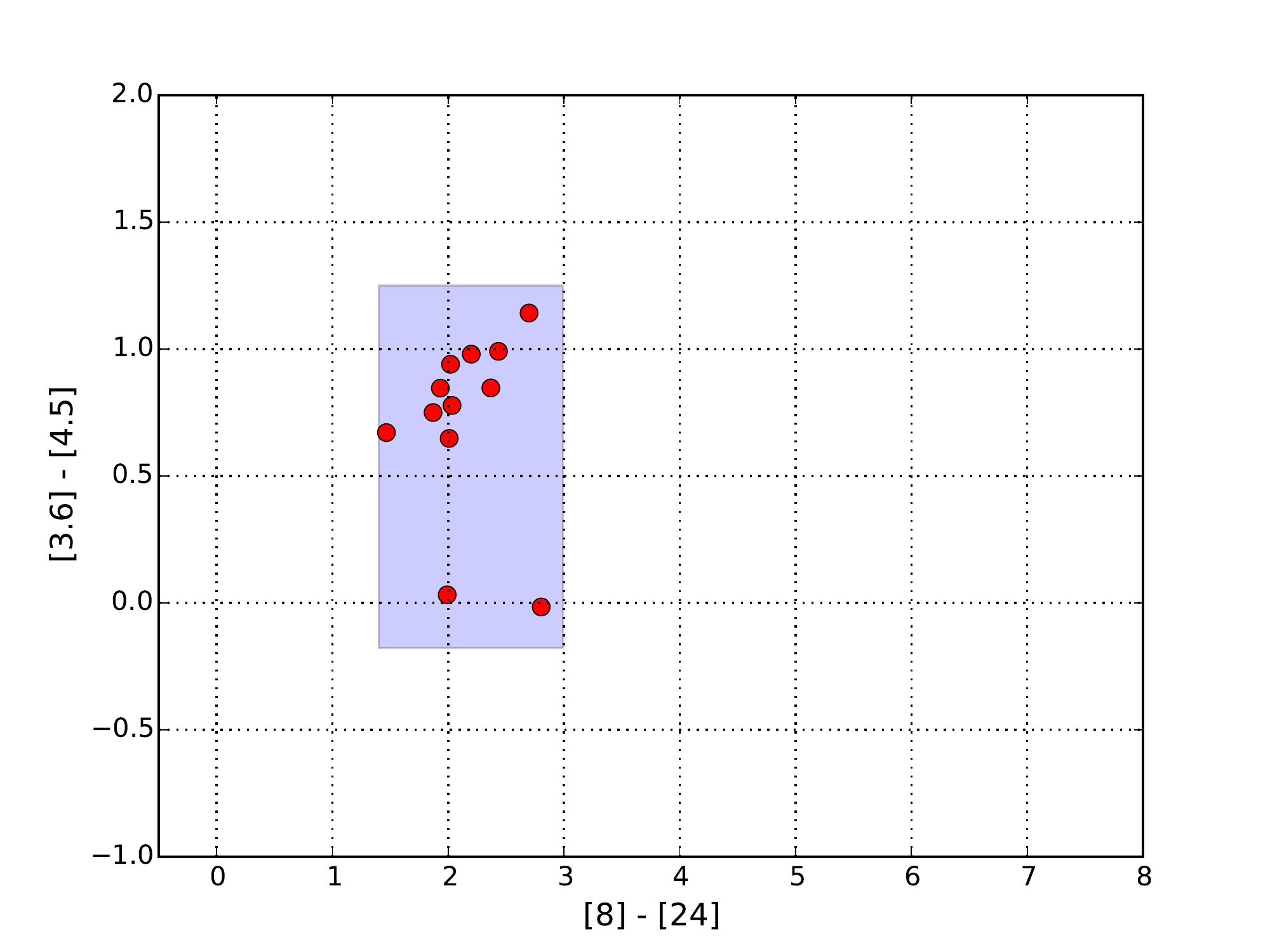} 
   \caption{Spitzer colour-colour plot of the sample RV Tauri stars in the LMC.
   The red markers show the 12 sample stars for which IRAC and MIPS fluxes were available. The stars 
   lie in the blue shaded region, which is defined as the region occupied by disc-type objects \citep{kamath2015}.}
     \label{figure:color-color_LMC_spitzer}
\end{figure}

We complemented the WISE classification with the \textit{Spitzer} classification for 12 stars that had both IRAC and MIPS data. 
These stars are indicated with red dots in Figure \ref{figure:color-color_LMC_spitzer}. 
These objects are located in the disc region of \citet{kamath2015}, shown by the blue shaded area, which contains post-AGB and post-RGB with a clear near-IR excess in their SEDs. 
The \textit{Spitzer} classification of our sample stars matches well with the WISE classification scheme. 
The total number of objects in the LMC we classified as disc types using these two classication schemes is 16. They are listed as ``disc'' in column 10 of Table \ref{table:lumsLMC}. 
Their SEDs are shown in Figures \ref{figure:seddisc1} to \ref{figure:seddisc16} in Appendix \ref{appendix:AppendixA}.

\subsubsection{Non-IR LMC stars}
We identified nine non-IR stars in the LMC. These stars show no clear IR excess at longer wavelengths (see Figures \ref{figure:nonirsed1} to \ref{figure:nonirsed9} in Appendix \ref{appendix:AppendixA}). 
In some cases there was a mild WISE [12] and/or [22] micron excess,
but closer inspection revealed that they were flagged as upper limits in the catalogue and were therefore removed from our analyses.
These stars are outlined as ``non-IR'' in column 10 of Table \ref{table:lumsLMC}.

\subsubsection{Uncertain LMC stars}
The rest of the stars has either one or a few points in excess in the reddest wavelength region, the characteristics of which are not clear. These constitute of 17 stars and are 
classified as uncertain (see Figures \ref{figure:uncertsed1} to \ref{figure:uncertsed17} in Appendix \ref{appendix:AppendixA}). Notably, among these 17 stars, there are a few in which 
there are indications of an IR excess. Examples of such stars are OGLE-LMC-T2CEP-011 (Figure \ref{figure:uncertsed1}), OGLE-LMC-T2CEP-025 (Figure \ref{figure:uncertsed3}), OGLE-LMC-T2CEP-055 (Figure \ref{figure:uncertsed6}), and
OGLE-LMC-T2CEP-112 (Figure \ref{figure:uncertsed11}). For these stars we clearly see an excess at around 2 microns but the WISE [12] micron fluxes are not in excess, thus introducing an ambiguity in their classification. 

There are other cases for which the high-amplitude pulsations induce a scatter in the IR region and the star appears as having an excess. 
Examples of such cases are OGLE-LMC-T2CEP-050 (Figure \ref{figure:uncertsed5}) and OGLE-LMC-T2CEP-058 (Figure \ref{figure:uncertsed7}).
These stars are still classified as uncertain because we do not know the nature of the IR excess. All these 
stars are listed as ``uncertain'' in column 10 of Table \ref{table:lumsLMC}.

\subsection{SMC IR classification}
We adopted the same criterion as in the LMC to classify the SMC IR colours. None of the SMC targets had \textit{Spitzer} data to be classified according to the \textit{Spitzer} colour-colour plot.
So, only a WISE classification was carried out. A summary of our SMC IR classification is shown in column 10 of Table \ref{table:lumsLMC}.

\subsubsection{Disc-type SMC star}
Based on the WISE classification, we found one disc-type object (OGLE-SMC-T2CEP-18) among the nine RV Tauri stars in the SMC. 
This star is plotted as a red dot in the SMC WISE colour-colour plot (see Figure \ref{figure:color-color_SMC}). 
The SED of this object is shown in Figure \ref{figure:sedsmcdisc1} in Appendix \ref{appendix:AppendixB}.

\subsubsection{Non-IR SMC stars}
Three stars, OGLE-SMC-T2CEP-07,OGLE-SMC-T2CEP-29, and OGLE-SMC-T2CEP-41, showed signs of excesses in the WISE data points, but closer inspection revealed that
the [12] and [22] micron fluxes were flagged as upper limits in the WISE catalogue. We therefore removed these flux points in our analyses, identifying them as non-IRs.
The SEDs of the non-IRs are shown in Figures \ref{figure:nonirsmc1} to \ref{figure:nonirsmc3} in Appendix \ref{appendix:AppendixB}.

\subsubsection{Uncertain SMC stars}
The rest of the stars in the SMC are classified as uncertain for the same reasons as described for the LMC objects 
(see Figures \ref{figure:uncertsmc1} to \ref{figure:uncertsmc5} in Appendix \ref{appendix:AppendixB}).
In the case of OGLE-SMC-T2CEP-20 (see Figure \ref{figure:uncertsmc3}), the high amplitude pulsations introduces a large scatter and it becomes difficult to infer the correct SED type.

\begin{figure}
   \centering
   \includegraphics[width=9cm,height=7cm]{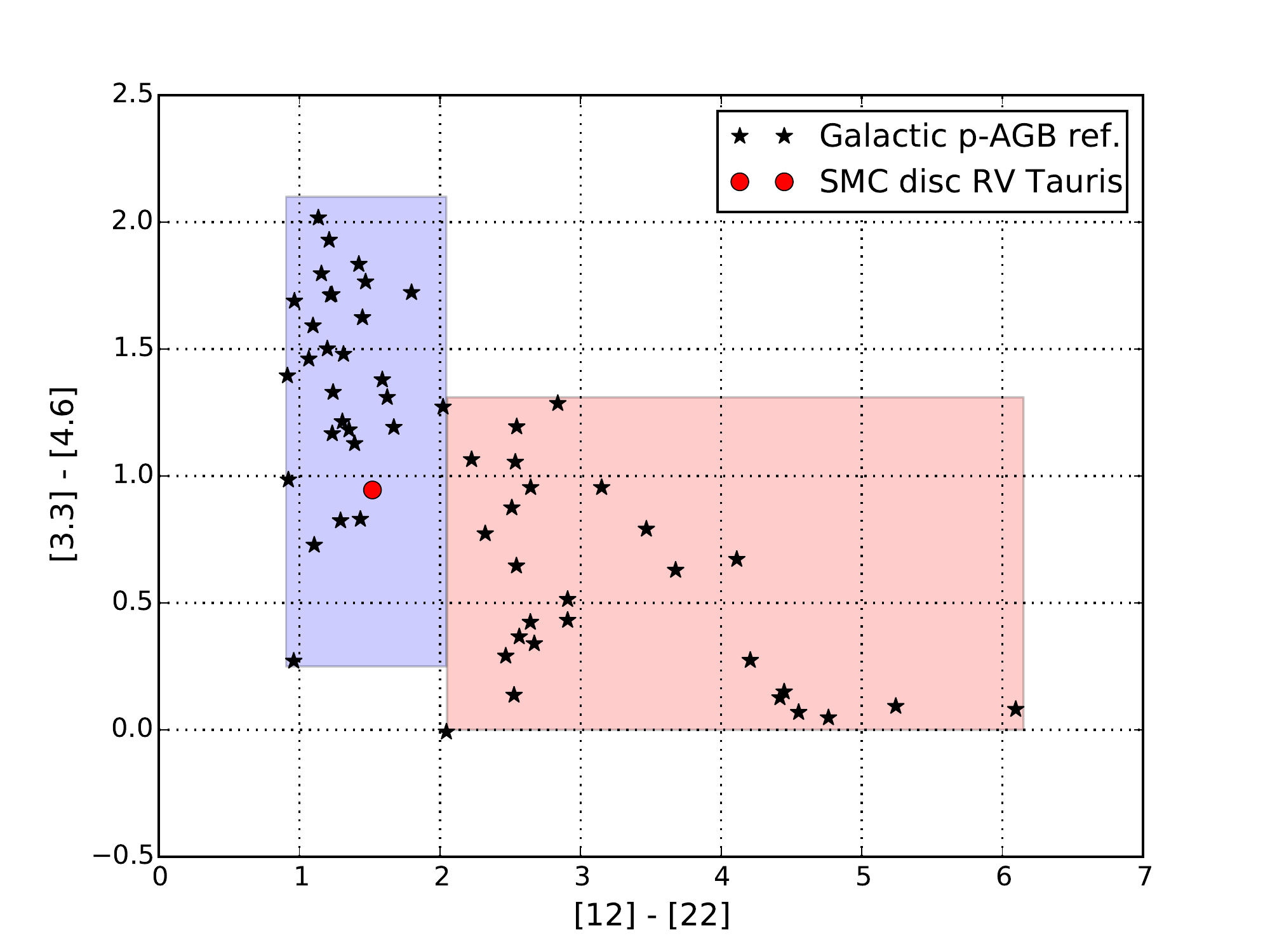} 
   \caption{WISE colour-colour plot of the disc-type RV Tauri star in the SMC (shown in red dot). The reference Galactic sample is shown by the black stars.
   The blue shaded region corresponds to stars in the reference sample with disc-type characteristics and the red shaded 
   region represents stars in the reference sample with shell-type characteristics.}
     \label{figure:color-color_SMC}
\end{figure}

Based on these aspects, we identified five stars in the SMC as uncertain. These stars could not be included in the WISE colour-colour plot 
(Figure \ref{figure:color-color_SMC}) because either the WISE [12] or WISE [22] micron flux was flagged as the upper limit in the catalogue.

\section{SED fitting} \label{section:sedfitting}
One of the challenges of fitting SEDs to large amplitude pulsators is that the pulsations induce a 
scatter in the photometric data points. To circumvent this problem, photometric data at specific phases of the pulsations are required. 
Since we do not have such equally phased data, we used all available photometric data points to construct the SEDs.

We analysed and fitted the SEDs of the SMC and LMC targets in a homogeneous and systematic way.
We carried out the SED fitting with a parameter-grid search to reach the optimised Kurucz model \citep[see][for details]{degroote2011}. 
The input parameters of the fit are effective temperature (T$_{\rm eff}$), metallicity ([Fe/H]), the extinction ($E(B-V)$), and $\log\,g$. 
The extinction, $E(B-V)$, was left as a free parameter in our SED fitting.
We assumed that the line-of-sight extinction follows the extinction law given by A$_\nu$ = R$_\nu$ $\times$ $E(B-V)$
\citep{fitzpatrick99}. We used a value of R$_\nu$ = 3.1 for the SMC and LMC, which roughly has the same value in the Milky Way \citep{fitzpatrick98}. 

The high reddening values for stars without circumstellar dust
excess indicates that the interstellar medium line-of-sight extinction is high
for these stars (see Table \ref{table:lumsLMC}). We notice that the reddening is high
for some of the non-dusty stars (e.g. OGLE-LMC-T2CEP-
005, OGLE-LMC-T2CEP-192, and OGLE-LMC-T2CEP-
198), while it is low for a few sources with circumstellar
dust (e.g. OGLE-LMC-T2CEP-003 and OGLE-LMCT2CEP-
067). The low reddening values in stars with dust is
only a contradiction when the dust is spherically symmetric.
In the case of a dusty disc, however, the viewing angle determines the line-of-sight extinction. All objects are optically
bright, so we are biased towards objects in which we do not
look into the plane of the disc; hence in these objects
our line of sight towards the star does not encounter circumstellar
material, yet these objects show an IR excess. The
high reddening values for stars without circumstellar dust
excess suggests that the ISM line-of-sight extinction is high
for these stars.

\subsection{Stars with known spectroscopic stellar parameters}
We obtained spectroscopic T$_{\rm eff}$, [Fe/H] and $\log$\,$g$ for ten stars in our LMC sample from the previous work of \citet{reyniers07a,gielen09b}, and \citet{kamath2015}.
These stars are indicated by an asterisk (*) in Tables \ref{table:sampleLMC} and \ref{table:sampleSMC}.
For these stars, the SED modelling was carried out using metallicity fixed to the obtained literature values. However, the T$_{\rm eff}$ and $\log$\,$g$ were let to 
vary by a small amount (5\%) in the fitting to account for the scatter in the data points due to pulsations.

The only RV Tauri star studied spectroscopically in the SMC is OGLE-SMC-T2CEP-18.
The SED fit for this star was carried out using the spectroscopic stellar parameters obtained by \citet{kamath2014}.
We emphasise that for this star the pulsational amplitude is large, which introduces a large scatter in the SED data points, 
and the SED fit is not optimal (see Figure \ref{figure:sedsmcdisc1}).

\subsection{Stars without known spectroscopic stellar parameters}
For the rest of the stars in the SMC and LMC, we assumed an effective temperature of 5500 K, which is the approximate mean T$_{\rm eff}$ of type II Cepheids in
the theoretical instability strip \citep{demers74,kiss07}.
This temperature was let to vary in the fitting by an even larger amount of $\sim$\,20\%. For these stars we used the mean metallicity of the LMC and SMC in the fitting, which have values of
$-$0.3 dex and $-$0.65 dex, respectively \citep{westerlund97,larsen2000}. In the fitting, [Fe/H] was let to vary within a range corresponding to the spread in metallicity for a sample of G and K giants, which have values of
0.45 and 0.15 for the SMC and LMC, respectively \citep{larsen2000}.
The SED data and the modelled SEDs are shown in Appendices \ref{appendix:AppendixA} and \ref{appendix:AppendixB} for the LMC and SMC, respectively.

\section{Luminosities} \label{section:luminosities}
This study is focussed on understanding the evolutionary nature of RV Tauri stars. Hence, it is crucial to constrain their luminosities as accurately as possible. We used the known mean distance to the LMC and SMC to obtain the luminosities.
 These luminosities combined with their effective temperatures  allow us to place 
our sample stars on the Hertzsprung Russell diagram (HRD) and therefore in an evolutionary context. We computed the luminosities using two different methods, i.e. 
the PLC relation ($L_\mathrm{PLC}$) and the SED fit ($L_\mathrm{SED}$).

\subsection{PLC relation and luminosity determination} \label{section:PLC_lum}
One of the main advantages of the type II Cepheids is that their pulsations can be used to determine their luminosities due to the correlation between their
pulsational periods and luminosities. Type II Cepheids in the Magellanic Clouds have been particularly useful in calibrating the P-L relation
because of their known distances. Some of the P-L relation studies of the type
II Cepheids in the Galactic globular clusters, SMC, and LMC are included in the work of \citet{nemec94,alcock98,ripepi2015}, and \citet{groenewegen2017b}.

\begin{figure}
   \centering
   \includegraphics[width=9cm,height=11cm]{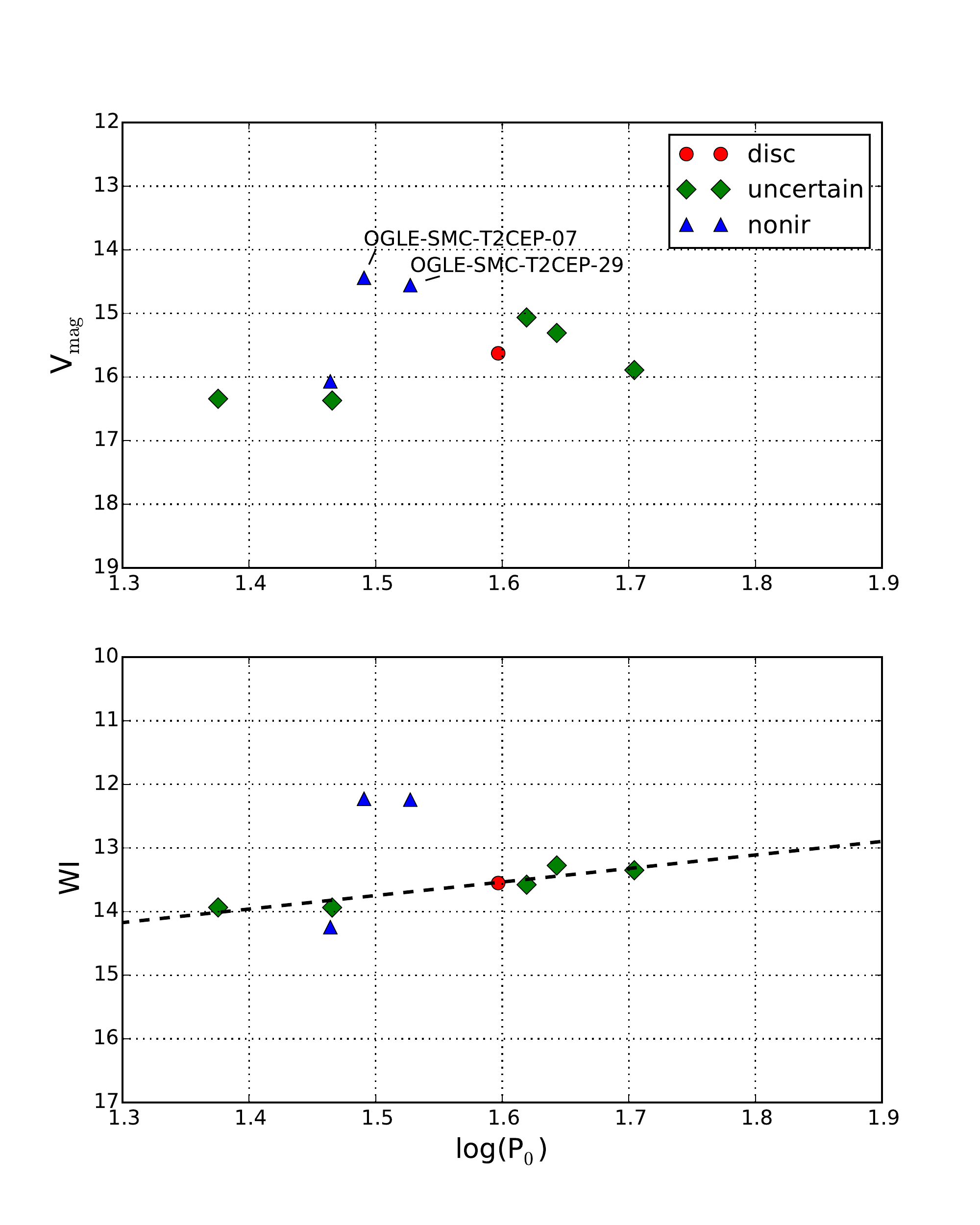} 
   \caption{Period-luminosity-colour relation for the RV Tauri stars in the SMC. The top panel shows the 
   V-band magnitudes plotted with the Log (P$_0$) of the sample stars. The two noticeable outliers are OGLE-SMC-T2CEP-07 
   (an ellipsoidal variable) and OGLE-SMC-T2CEP-29 (an eclipsing binary), see Section \ref{section:specialcasessmc}.
   The bottom panel shows the reddening-free Wesenheit index plotted with the Log (P$_0$) of the sample stars.}
     \label{figure:PLCsmc}
\end{figure}

\begin{figure}
   \centering
   \includegraphics[width=9cm,height=11cm]{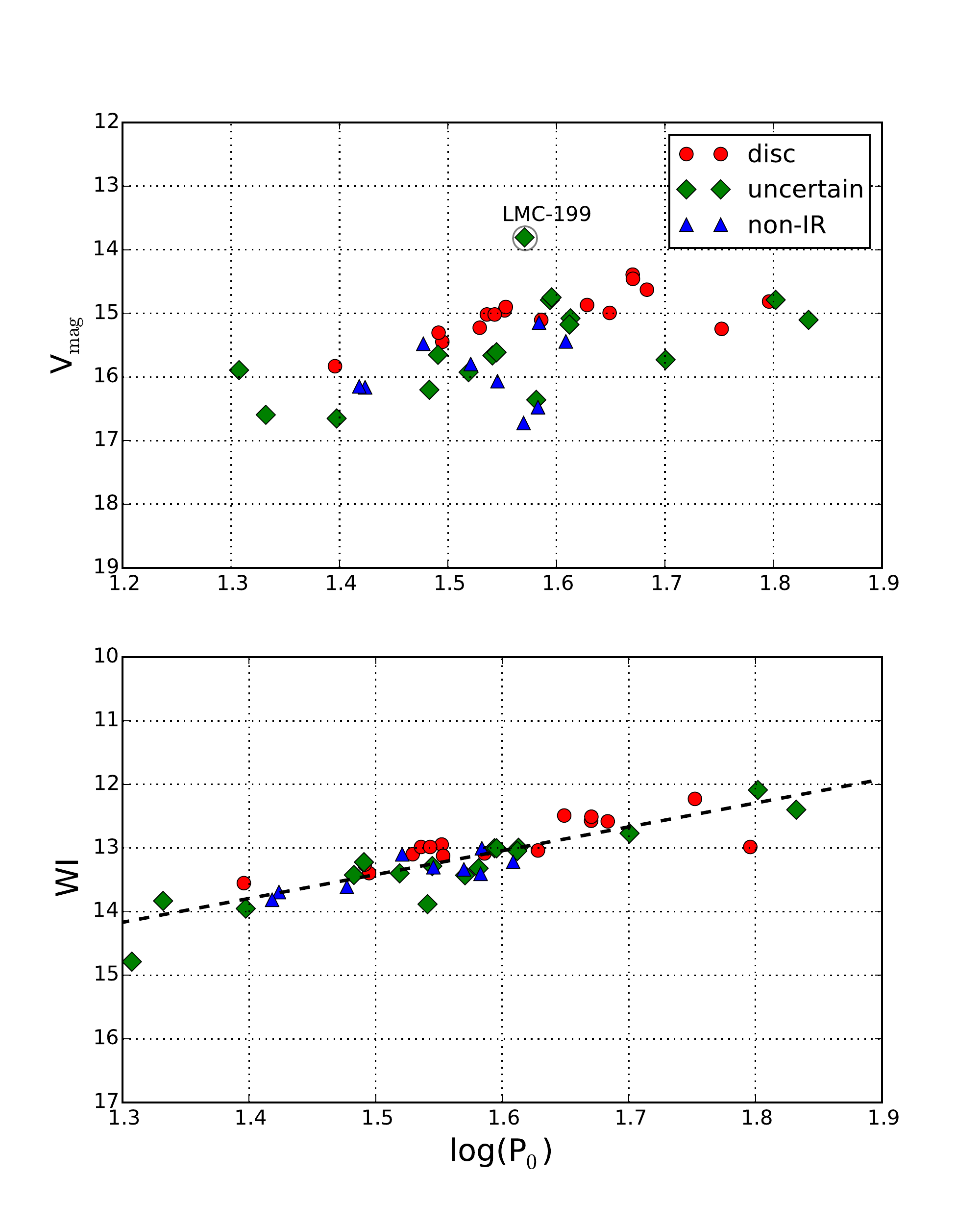}
   \caption{Period-luminosity-colour relation for the RV Tauri stars in the LMC. The top panel shows the 
   V-band magnitudes plotted with the Log (P$_0$) of the sample stars. The bottom panel represents the reddening-free Wesenheit index plotted with the Log (P$_0$) of the sample stars.
   In the top panel, OGLE-LMC-T2CEP-199 has the brightest V-mag among all the RV Tauri stars. This star is discussed in Section \ref{section:specialcasessmc}.}
     \label{figure:PLClmc}
\end{figure}

In the top panel of Figures \ref{figure:PLCsmc} and \ref{figure:PLClmc}, we show the fundamental period, P$_0$, plotted with the apparent V-band magnitude of the sample stars. The luminosity dependance of the 
periods is clear in the LMC (see top panel of Figure \ref{figure:PLClmc}), but the scatter is still significant. Given that there is a comparably lower number of RV Tauri stars in the SMC sample, 
the correlation is not clear (see top panel of Figure \ref{figure:PLCsmc}).

Given that the luminosity of a star is dependent on its effective temperature and radius, we expect a PLC relation to be that where the 
scatter is significantly reduced due to a colour correction term. The colour-corrected V-band magnitude is termed as the Wesenheit index, $WI$ \citep{ngeow05}, given by the equation 
\begin{equation} \label{equation:eq1}
 WI = V - 2.55\,(V - I)
.\end{equation}
The PLC relation plotted on the Wesenheit plane indeed shows a reduction in scatter and better correlation between the pulsation period and Wesenheit indices 
(see bottom panels of Figures \ref{figure:PLCsmc} and \ref{figure:PLClmc}). There are two noticeable exceptions in the SMC, namely OGLE-SMC-T2CEP-07 (an ellipsoidal variable) 
and OGLE-SMC-T2CEP-29 (an eclipsing binary). These stars are discussed in Section \ref{section:specialcasessmc}.

In principle, equation \ref{equation:eq1} can be rewritten in terms of the intrinsic properties of the star as the following equation \citep{vandenbergh68,madore76,madore82}:
\begin{equation} \label{equation:eq2}
WI = V_0 - 2.55\,(V - I)_0 
.\end{equation}
The Wesenheit index can be transformed into the absolute magnitude of each star, provided that 
the intrinsic $(V-I)_0$ colour correction terms are known for each individual star \citep[e.g.][]{alcock98}. 

The amount of extinction towards each star is caused mainly by two sources: interstellar and circumstellar dust. 
We accounted for the interstellar extinction using a mean reddening of 0.056\,mag to the SMC \citep{deb2017} and 
0.085\,mag for the LMC \citep{haschke2011}. These values were combined with the circumstellar extinction obtained from 
the dereddened SED model (see Section \ref{section:sedluminosities} and column 8 of Table \ref{table:lumsLMC}).

We computed $E(V-I)$ for the individual stars using the conversion relation by \citet{tammann03} and \citet{haschke2011}, i.e.
\begin{equation} \label{equation:eq3}
E(V-I) = 1.38\,E(B-V)
,\end{equation}
where $E(B-V)$ is the total reddening towards each star (interstellar and circumstellar).  
The $E(V-I)$ and $(V-I)$ values were then used to calculate the intrinsic $(V-I)_0$ for each star in the SMC and LMC. 

A linear regression in $\log\,(P_0)$ against the quantity $V_0$ $-$ 2.55\, $(V-I)_0$ yields the following PLC 
relation for the intrinsic apparent magnitude ($V_0$) of the targets: 
\begin{equation} \label{equation:eq4}
V_0 = m \times \log\,(P_0) + c + 2.55\,{(V - I)_{0}}
,\end{equation}
where $m$ is the slope and $c$ the intercept of the linear regression in the Wesenheit plane. We obtained $m = -1.33$ and $c = 15.42$ for SMC and $m = -3.75$ and $c = 19.04$ for the LMC.

\begin{figure}
   \centering
   \includegraphics[width=8cm,height=6cm]{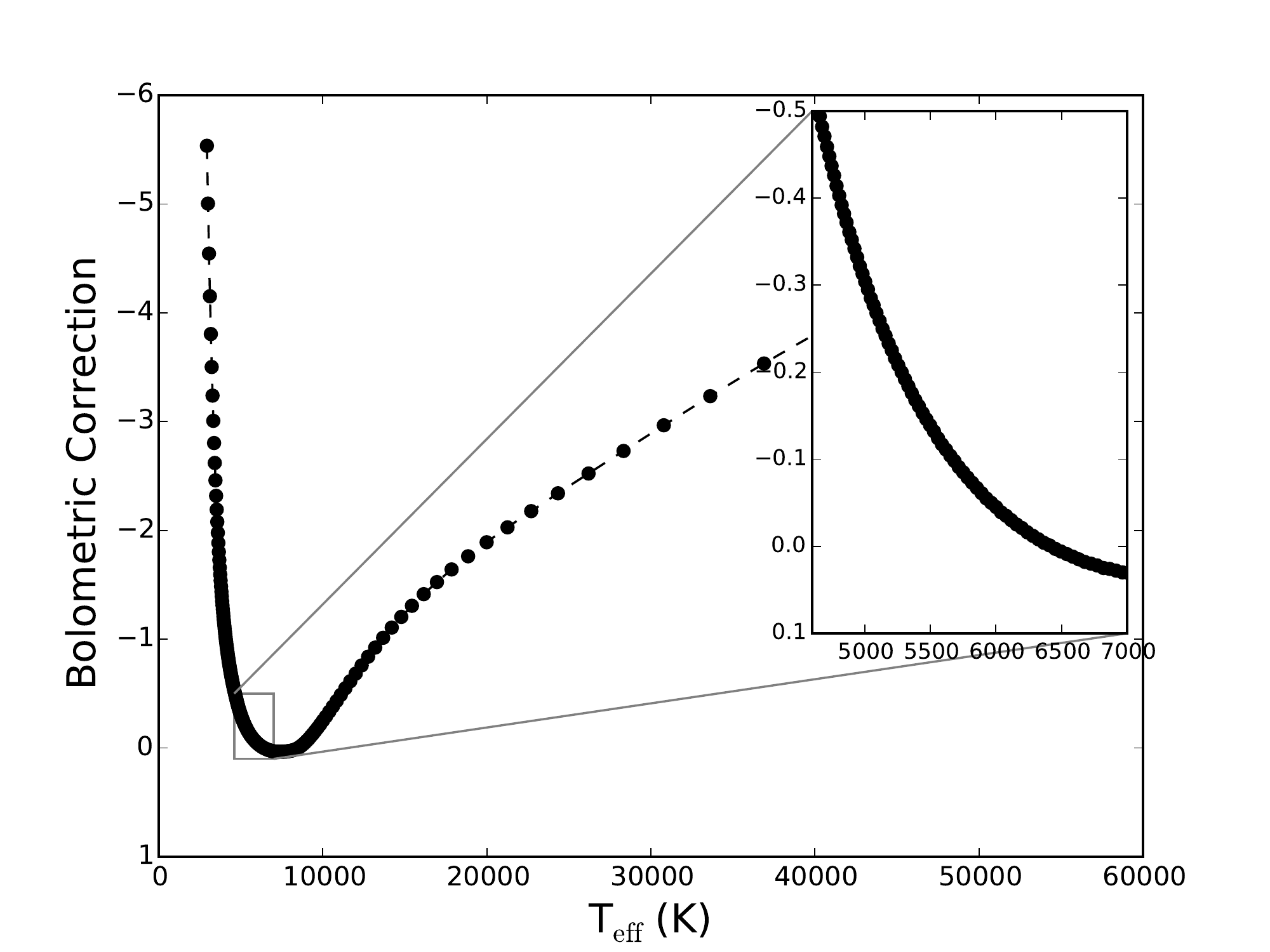} 
   \caption{Bolometric correction factor we used to obtain the bolometric magnitudes of our stars.}.
     \label{figure:BC_flower96}
\end{figure}

The parameters of the regression and the fundamental pulsation periods were then used to compute the bolometric absolute magnitude given by
\begin{equation} \label{equation:eq5}
M_{\rm bol,{WI}} = m \times \log(P_0) + c - \mu + BC + 2.55\,{(V - I)_0}
,\end{equation}
where $\mu$ is the distance modulus for the SMC and LMC and have values of 18.965 \citep{graczyk2013} and 18.49 \citep{walker2012, pietrzynski2013}, respectively. The value $BC$ is the bolometric correction for each star computed using 
the relation between $BC$ and effective temperature provided by \citet{flower96}. This relation is shown in Figure \ref{figure:BC_flower96}, which is reproduced using their data. A zoom-in is 
shown in the same plot to represent the approximate range of temperatures occupied by our sample stars. 
We used the absolute bolometric magnitude, $M_\mathrm{bol,{WI}}$ to compute luminosities for each star using the PLC, denoted by $L_{\rm PLC}$ (see column 3 of Table \ref{table:lumsLMC}).

\subsection{LMC luminosities from SED} \label{section:sedluminosities}
\begin{figure}
   \centering
   \includegraphics[width=8cm,height=6cm]{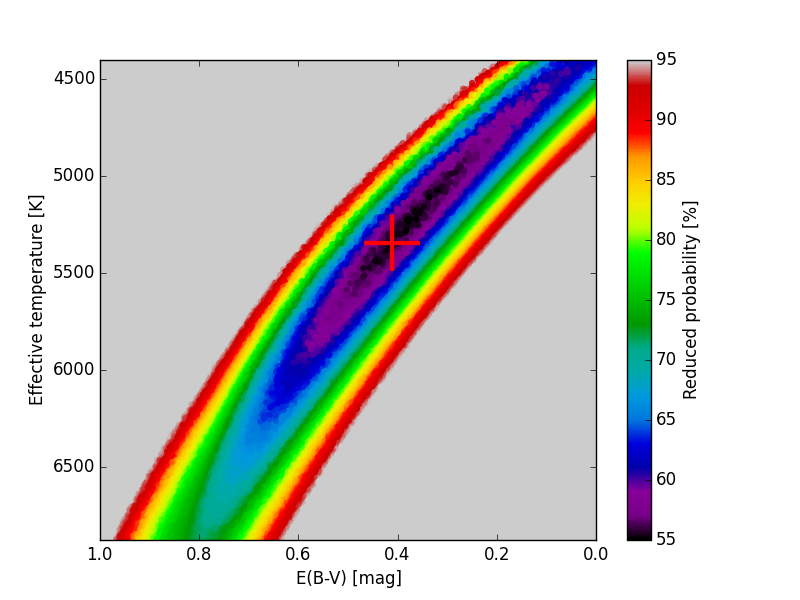} 
   \caption{Example of the $\chi^2$ plot of OGLE-LMC-T2CEP-169 to obtain the reddening parameter ($E(B-V)$) after the parameter grid search.}
     \label{figure:chi2}
\end{figure}

Kurucz model atmospheres were fit to the original data points and a dereddened SED 
model was computed using the extinction parameter, ($E(B-V)$), corresponding to the lowest $\chi^2$ value 
of the parameter grid search (see Figure \ref{figure:chi2} for an example). The bolometric luminosities, $L_\mathrm{SED}$, were then computed 
using the integrated flux below the dereddened photospheric SED model and assuming an average distance of 49.97 $\pm$ 1 kpc to the LMC \citep{walker2012, pietrzynski2013}. 
In this process we assumed that the flux from the star is radiated isotropically. The derived SED luminosities for the LMC sample are shown in column 5 of Table \ref{table:lumsLMC}. 

\subsection{SMC luminosities from SED}
The SED luminosities of the SMC sample stars were computed in the same way as described for the LMC targets. We assumed a mean distance of 62.1 $\pm$ 1.9 kpc to the SMC \citep{graczyk2013}.
The computed SMC luminosities from the SED ($L_\mathrm{SED}$) are shown in column 3 of Table \ref{table:lumsLMC}. 

\subsection{Luminosity correlation: $L_\mathrm{SED}$ vs $L_\mathrm{PLC}$}
\begin{figure}
   \centering
   \includegraphics[width=9cm,height=7cm]{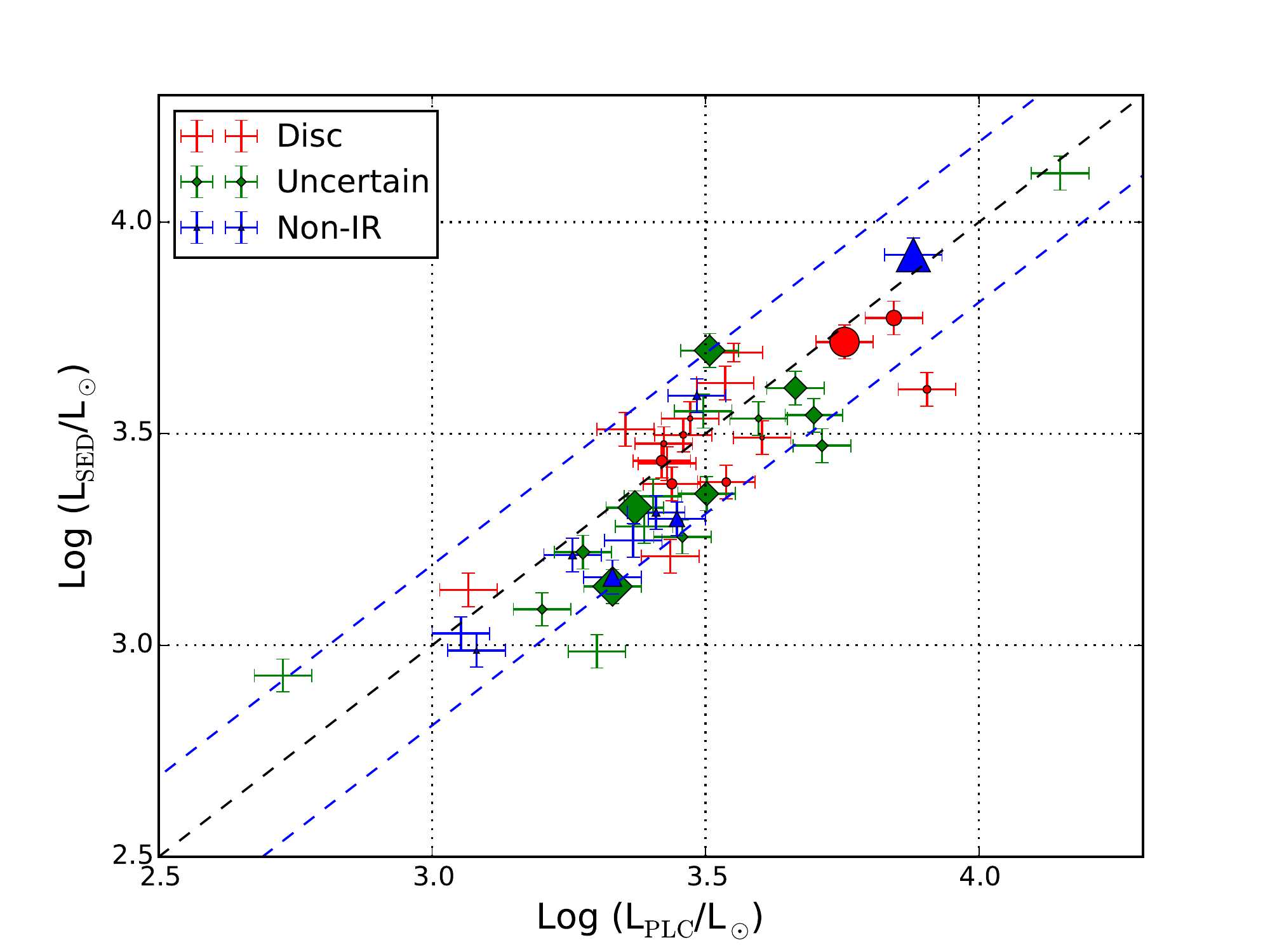}\\
   \includegraphics[width=9cm,height=7cm]{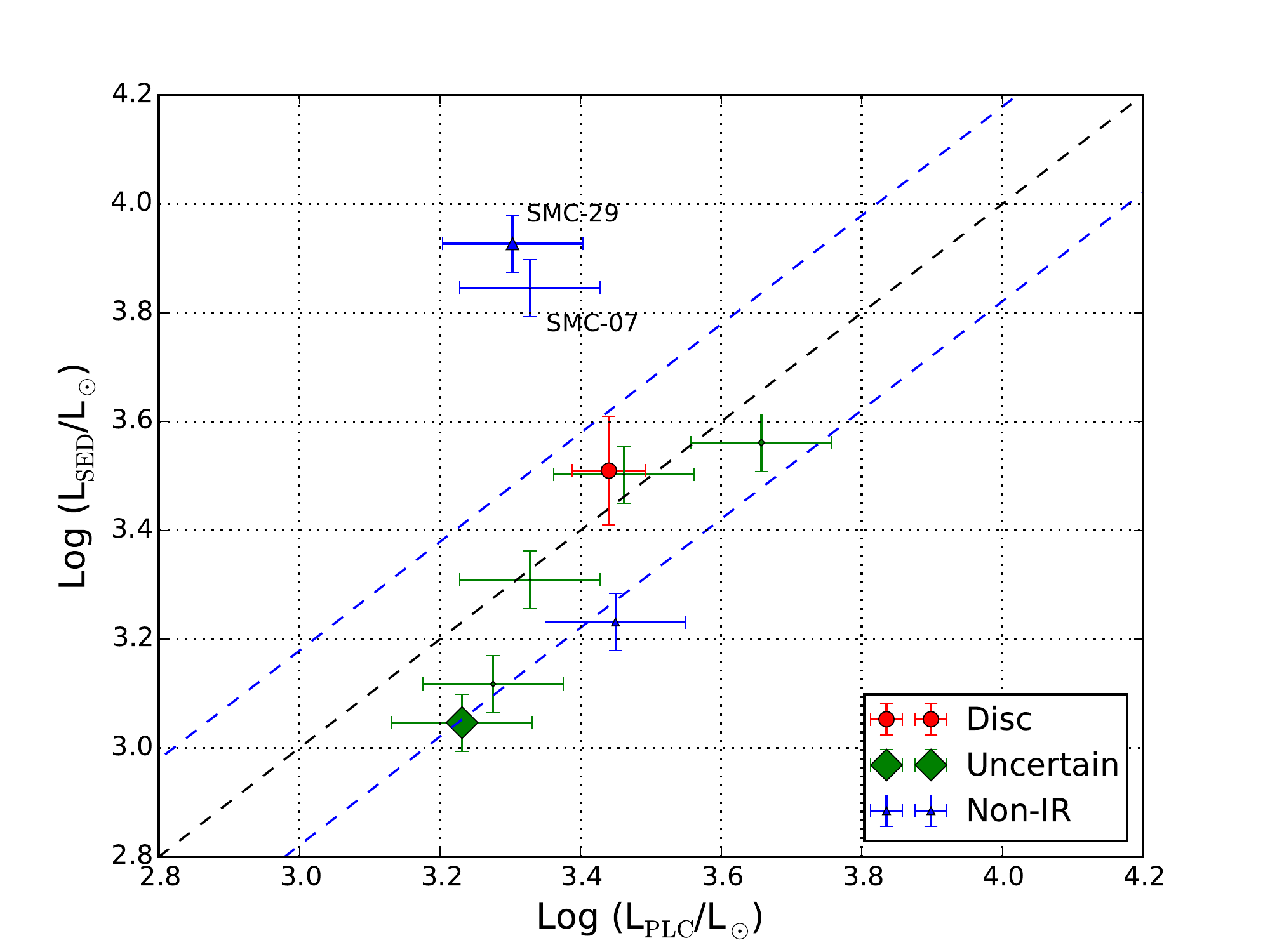}
   \caption{Comparison of the luminosities derived from the deredenned SED models and the PLC relation. The top panel indicates the LMC and the bottom panel indicates the SMC.
   The blue dashed lines show the 1$\sigma$ from the mean. The markers are scaled according to the reddening value obtained in the SED fit. The two distinct outliers in the SMC are binaries and are discussed in Section \ref{section:specialcasessmc}.}.
     \label{figure:PLCWI}
\end{figure}

The top and bottom panels of Figure \ref{figure:PLCWI} show a comparison between the luminosities derived from the dereddened SED models and the PLC. The top panel represents the LMC and bottom panel represents for the SMC.
We conclude from these two plots that the luminosities derived using the two methods correlate well within 1$\sigma$ except for the two distinct outliers in the SMC, which show ellipsoidal and eclipsing variations
in their light curves. This is further discussed in Section \ref{section:specialcasessmc}. We use the luminosities obtained from the SED in our analyses.

\begin{table*}
\tiny
\centering                        
\begin{tabular}{cllllllllll}
\hline 
Star               & P$_0$   & $L_\mathrm{PLC}$ & $\triangle\,L_\mathrm{PLC}$ & $L_\mathrm{SED}$ & $\triangle\,L_\mathrm{SED}$ &T$_{\rm eff}$ & $E(B-V)$ & $\log$\,$g$  & SED type     & Notes \\
                   & (days) & (L$_{\odot}$) & (L$_{\odot}$)            & (L$_{\odot}$) & (L$_{\odot}$)           & (K)          & (mag)    &              &              &           \\
\hline\hline                                                                                                                                            
OGLE-LMC-T2CEP-003 & 35.6     & 2256        & 270                      & 3236          &  129                    &  5700        &  0.01    &  2.49    & disc             &      \\
OGLE-LMC-T2CEP-014 & 38.5     & 3450        & 269                      & 2430          &  97                     &  5947        &  0.11    &  2.50    & disc             &      \\
OGLE-LMC-T2CEP-015 & 56.5     & 2879        & 524                      & 3141          &  125                    &  5013        &  0.09    &  0.45    & disc             &      \\    
OGLE-LMC-T2CEP-029 & 31.2     & 2629        & 188                      & 2728          &  109                    &  6143        &  0.14    &  0.55    & disc             &      \\
OGLE-LMC-T2CEP-032 & 44.5     & 2963        & 360                      & 3435          &  137                    &  5070        &  0.07    &  2.48    & disc             & Binary, P$_{\rm Orb}$ $\sim$ 916 d \\
OGLE-LMC-T2CEP-067 & 48.2     & 3434        & 344                      & 4163          &  166                    &  5687        &  0.01    &  0.55    & disc             &   \\
OGLE-LMC-T2CEP-091 & 35.7     & 2689        & 231                      & 2690          &  107                    &  5973        &  0.01    &  1.65    & disc             &    \\
OGLE-LMC-T2CEP-104 & 24.8     & 1165        & 165                      & 1352          &  54                     &  5182        &  0.01    &  0.96    & disc             &    \\
OGLE-LMC-T2CEP-119 & 33.8     & 5676        & 192                      & 5209          &  208                    &  7306        &  0.38    &  1.09    & disc             &     Period change \\
OGLE-LMC-T2CEP-129 & 62.5     & 7808        & 523                      & 4025          &  161                    &  6122        &  0.10    &  2.09    & disc             &  \\
OGLE-LMC-T2CEP-147 & 46.7     & 6989        & 349                      & 5936          &  237                    &  6599        &  0.20    &  0.55    & disc             &   \\
OGLE-LMC-T2CEP-149 & 42.4     & 4013        & 262                      & 3093          &  123                    &  6001        &  0.06    &  1.10    & disc             &   Period change     \\
OGLE-LMC-T2CEP-174 & 46.8     & 3560        & 346                      & 4920          &  106                    &  5733        &  0.01    &  1.64    & disc             &   \\
OGLE-LMC-T2CEP-180 & 30.9     & 2743        & 219                      & 2407          &  96                     &  5608        &  0.13    &  1.36    & disc             &  \\
OGLE-LMC-T2CEP-191 & 34.3     & 2652        & 203                      & 2996          &  119                    &  5805        &  0.08    &  1.65    & disc             & Period change \\
OGLE-LMC-T2CEP-200 & 34.9     & 2724        & 201                      & 1622          &  64                     &  4887        &  0.01    &  0.50    & disc             &  Binary, P$_{\rm Orb}$ $\sim$\,850 d   \\ \hline 

OGLE-LMC-T2CEP-011 & 39.2     & 4989         & 260                      & 3497          &  139                    &  6250       &  0.16    &  2.50    & uncertain        &                           \\
OGLE-LMC-T2CEP-016 & 20.2     & 2000         & 92                       & 967           &  38                     &  6752       &  0.00    &  2.50    & uncertain        &                            \\
OGLE-LMC-T2CEP-025 & 67.9     & 5164         & 530                      & 2963          &  118                    &  5129       &  0.11    &  1.28    & uncertain        &                            \\  
OGLE-LMC-T2CEP-045 & 63.3     & 3130         & 584                      & 3574          &  143                    &  5267       &  0.00    &  0.54    & uncertain        &                           \\  
OGLE-LMC-T2CEP-050 & 34.7     & 2440         & 195                      & 1908          &  76                     &  6518       &  0.00    &  2.48    & uncertain        &                           \\
OGLE-LMC-T2CEP-055 & 41.0     & 4617         & 269                      & 4056          &  162                    &  6738       &  0.21    &  2.37    & uncertain        &                           \\
OGLE-LMC-T2CEP-058 & 21.4     & 533          & 106                      & 848           &  33                     &  5630       &  0.00    &  1.77    & uncertain        &                          \\
OGLE-LMC-T2CEP-065 & 35.0     & 3176         & 224                      & 2281          &  91                     &  6045       &  0.22    &  2.47    & uncertain        &                          \\
OGLE-LMC-T2CEP-075 & 50.1     & 2869         & 344                      & 1803          &  72                     &  4814       &  0.10    &  1.09    & uncertain        &                          \\
OGLE-LMC-T2CEP-080 & 40.9     & 2534         & 286                      & 2249          &  90                     &  5591       &  0.00    &  2.41    & uncertain        &                          \\
OGLE-LMC-T2CEP-112 & 39.3     & 3953         & 264                      & 3436          &  137                    &  6100       &  0.07    &  1.94    & uncertain        &                          \\
OGLE-LMC-T2CEP-115 & 24.9     & 2137         & 143                      & 1377          &  55                     &  6031       &  0.36    &  2.50    & uncertain        &                         \\
OGLE-LMC-T2CEP-125 & 33.0     & 1886         & 251                      & 1660          &  66                     &  5750       &  0.13    &  2.49    & uncertain        &                         \\
OGLE-LMC-T2CEP-162 & 30.3     & 2348         & 187                      & 2114          &  84                     &  5808       &  0.31    &  2.50    & uncertain        &                           \\
OGLE-LMC-T2CEP-169 & 30.9     & 3216         & 195                      & 4973          &  199                    &  5473       &  0.28    &  1.09    & uncertain        &                               \\
OGLE-LMC-T2CEP-199 & 37.2     & 14073        & 206                      & 13049         &  522                    &  10624      &  0.00    &  1.00    & uncertain        &    Contamination                       \\
OGLE-LMC-T2CEP-202 & 38.1     & 1589         & 337                      & 1216          &  48                     &  4966       &  0.09    &  2.07    & uncertain        &                            \\ \hline 
OGLE-LMC-T2CEP-005 & 33.2     & 7586         & 243                      & 8372          &  335                    &  4412        &  0.44    &  2.50    & non-IR          &    Contamination                        \\ 
OGLE-LMC-T2CEP-051 & 40.6     & 2331         & 203                      & 1770          &  70                     &  5524        &  0.00    &  1.13    & non-IR          &                            \\      
OGLE-LMC-T2CEP-082 & 35.1     & 1806         & 223                      & 1635          &  65                     &  5059        &  0.10    &  2.50    & non-IR          &                            \\
OGLE-LMC-T2CEP-108 & 30.0     & 2567         & 166                      & 2061          &  82                     &  6201        &  0.10    &  2.50    & non-IR          &                                \\
OGLE-LMC-T2CEP-135 & 26.5     & 1128         & 152                      & 1066          &  42                     &  5045        &  0.00    &  0.50    & non-IR          &                            \\
OGLE-LMC-T2CEP-190 & 38.3     & 3049         & 246                      & 3890          &  155                    &  5746        &  0.10    &  2.47    & non-IR          &     Period change                \\
OGLE-LMC-T2CEP-192 & 26.1     & 2136         & 140                      & 1449          &  58                     &  6212        &  0.24    &  2.50    & non-IR          &                            \\
OGLE-LMC-T2CEP-198 & 38.2     & 2803         & 275                      & 1988          &  79                     &  4541        &  0.20    &  0.54    & non-IR          &                                \\ 
OGLE-LMC-T2CEP-203 & 37.1     & 1205         & 178                      & 972           &  38                     &  4677        &  0.07    &  1.79    & non-IR          &                            \\
 \hline \hline \\
OGLE-SMC-T2CEP-18  & 39.5     & 2687        & 325                      & 3181           &  194                    &  5699        & 0.19     &  0.79    & disc            &                             \\ \hline
OGLE-SMC-T2CEP-12  & 29.2     & 2448        & 296                      & 1691           & 103                     &  5665        & 0.20     &  2.50    & uncertain           &                             \\
OGLE-SMC-T2CEP-19  & 41.6     & 2489        & 301                      & 3642           & 222                     &  6283        & 0.06     &  2.50    & uncertain       &                             \\
OGLE-SMC-T2CEP-20  & 50.6     & 3486        & 421                      & 2039           & 124                     &  5540        & 0.01     &  2.14    & uncertain       &                             \\
OGLE-SMC-T2CEP-24  & 43.9     & 2600        & 314                      & 3182           & 194                     &  6056        & 0.01     &  2.40    & uncertain       &                             \\
OGLE-SMC-T2CEP-43  & 23.7     & 2210        & 267                      & 1546           & 94                      &  5306        & 0.50     &  2.45    & uncertain           &                             \\ \hline
OGLE-SMC-T2CEP-07  & 30.9     & 2258        & 273                      & 7011           & 429                     &  5743        & 0.01     &  2.45    & non-IR              &  Ellipsoidal variability     \\
OGLE-SMC-T2CEP-29  & 33.6     & 2427        & 293                      & 8456           & 517                     &  5789        & 0.16     &  2.49    & non-IR              &  Eclipsing                     \\
OGLE-SMC-T2CEP-41  & 29.1     & 2060        & 249                      & 1843           & 112                     &  6256        & 0.13     &  2.49    & non-IR              &                                \\
\hline
\end{tabular}
\caption{Luminosities of the LMC and SMC targets derived using the PLC relation and SED. The luminosity deduced from the PLC relation and the errors are
shown in columns 3 and 4, respectively. In column 5 we display the luminosities derived from the SED and the errors are shown in column 6. The effective temperature, reddening, and $\log\,g$ obtained from the SED model are shown in 
columns 7, 8, and 9, respectively.}   
\label{table:lumsLMC}   
\end{table*}                                
 
 \subsection{Over-luminous stars in the SMC and LMC} \label{section:specialcasessmc}
There are a few stars in our sample that do not follow the general luminosity trend of the rest of the stars.
OGLE-SMC-T2CEP-07 and OGLE-SMC-T2CEP-29 are two such examples, which have SED luminosities significantly higher than what is expected from the PLC. Both stars are the two distinct outliers in the Wesenheit plane  (see Figure \ref{figure:PLCsmc}) and in  Figure \ref{figure:PLCWI}.
As reported by \citet{soszynski10}, one of these stars (OGLE-SMC-T2CEP-07) also shows ellipsoidal variability in its light curve. Ellipsoidal variables are 
non-eclipsing close binaries in which one or both components is or are distorted \citep[e.g.][]{wood99,pawlak2014}.
On the other hand, OGLE-SMC-T2CEP-29 was shown to be an eclipsing binary by \citet{soszynski10}.
\cite{soszynski10} argued that for both objects, the luminous companion contributes significantly to the total observed luminosity. High-resolution data with a wide wavelength coverage is needed to investigate both stars in detail.

In the case of the over-luminous OGLE-LMC-T2CEP-005, we find that the carbon star [KDM2001]\,547 is within 1.4\,arcsec to the RV Tauri star. This is below the spatial resolution of the instruments used to generate the data in the photospheric
part of our SEDs; for example, the spatial resolution of the 2\,MASS survey is 4\,arcsec and the UBVI photometry by \citet{Zaritsky02} are at a spatial resolution of $\sim$ 3.6 arcsec.
The carbon star is around 1 magnitude brighter in the J band and almost 2 magnitudes brighter in the H and K bands. Thus, it is highly likely that the photospheric part of the SED of OGLE-LMC-T2CEP-005 is contaminated by this star, leading to a higher luminosity determination.

In a similar case, the SED photometry of OGLE-LMC-T2CEP-199 is likely contaminated by another long-period variable star (2MASS J05044388$-$6858371). 
This star is within 8\,arcsec of OGLE-LMC-T2CEP-199. It is as bright in the I band and is around 2 magnitudes brighter in the J- and H bands.

\subsection{Luminosity difference between dusty and non-dusty stars} \label{section:lumdiff}
We carried out a Kolmogorov-Smirnov (K-S) test on the luminosities of the LMC sample to investigate whether the stars showing three different SED characteristics are statistically different. 
We only performed the test for the LMC stars because the sample size was large enough for a statistical study. 
Assuming the null hypothesis that all three distributions (disc-type, non-IR, and uncertain) are drawn from the same sample, we tested the significance at which this hypothesis is rejected; the lower the $p$ value, the greater the statistical significance of rejecting the null hypothesis. We concluded the distributions are different at the $1 - p$ significance level.  We show in Table \ref{table:ks} the results of the K-S test.

\begin{table}
  \centering
  \begin{tabular}{@{} lcc @{}}
    \hline
    Test samples          & $p$-value & \% of rejecting the       \\
                          &           &  null hypothesis        \\
    \hline \hline
    Disc and non-IR       & 0.01      & 99                     \\ 
    non-IR and uncertain  & 0.33      & 67                     \\ 
    Disc and uncertain    & 0.04      & 96                      \\ 
    \hline
  \end{tabular}
  \caption{Results from the K-S test on the luminosity distribution, showing the significance level at which the null hypothesis is rejected.}
  \label{table:ks}
\end{table}

We obtained the average luminosity in each of the sample representing disc-type, non-IR and uncertain as 3336 L$_\odot$, 1854 L$_\odot$ and 2432 L$_\odot$, respectively.
Based on the results of the K-S test we concluded that on average, the dusty (disc-type) objects are significantly more luminous than the non-dusty stars (non-IRs). 
The implications of these results are discussed in Section \ref{section:discussion_lum_diff}.

\section{($O-C$)} \label{section:O_C}
A few of the RV Tauri stars showed significant period change in their phased light curves. This led us to investigate their pulsational 
behaviour via an ($O-C$) diagram. 

The ($O-C$) (observed - calculated) diagram is useful to investigate long-term period changes and hence the evolution of stars, 
RV Tauri stars in particular \citep[e.g.][]{percy97,percy98,percy2005,groenewegen2017a}. A preliminary study of the ($O-C$) diagrams of RV Tauri stars, carried out by \citet{percy91}, reported that 13 out of 16 RV Tauri stars
have decreasing periods. We apply the same method as an attempt to trace their period evolution.

If there are significant changes in the period over time, the ($O-C$) diagram will be a parabola with positive or negative curvature representing
an increase or decrease in period, respectively. In principle, this method can be used to trace the evolution of a 
star: if there is an increase in the radius of a star, the period increases and vice-versa. \citet{percy2005} applied this method 
to a sample of RV Tauri pulsators and SRd variables. Their results show that indeed the sample RV Tauri
stars are evolving towards lower radii. We note that their data coverage was of the order of a few tens of decades up to a century in some cases.
Their results predict the pulsational evolution of RV Tauri stars to be on the scale of a few decades to a century, as is the span of their data. 
This fast evolution was found to be in contradiction to the theoretical predictions, however, as theory predicts evolution at a much longer timescale of a few hundreds to a thousand years \citep{schonberner83,percy91}. 

Some of our stars do not show prominent RV Tauri characteristics, i.e. indiscernible deep and shallow minima. So, for the sake of homogeneity, we used the 
fundamental period ($P_{\rm 0}$) for all stars to construct the ($O-C$) diagram, following \citet{percy2005}.

The ($O-C$) analysis starts with knowing the average period and adopting a fixed epoch of a variable star.  The epoch ($T_{\rm 0}$) is chosen as the time of occurrence of the first deep minumum. The minima in the time series data, the pulsational fit represented by $T_{\rm i}$, and the fundamental period, $P_{\rm 0}$ were used in the following equation to obtain the ($O-C$) as follows:
\begin{equation}
 (O - C) = T_{\mathrm i} - (T_{\mathrm 0} + n_{\mathrm i} \times P_{\mathrm 0})
,\end{equation}
where $n_{\rm i}$ represents the number of cycles.

\begin{figure*}
\centering
  \subfloat[OGLE-LMC-T2CEP-119 ]{%
  \includegraphics[width=0.45\textwidth]{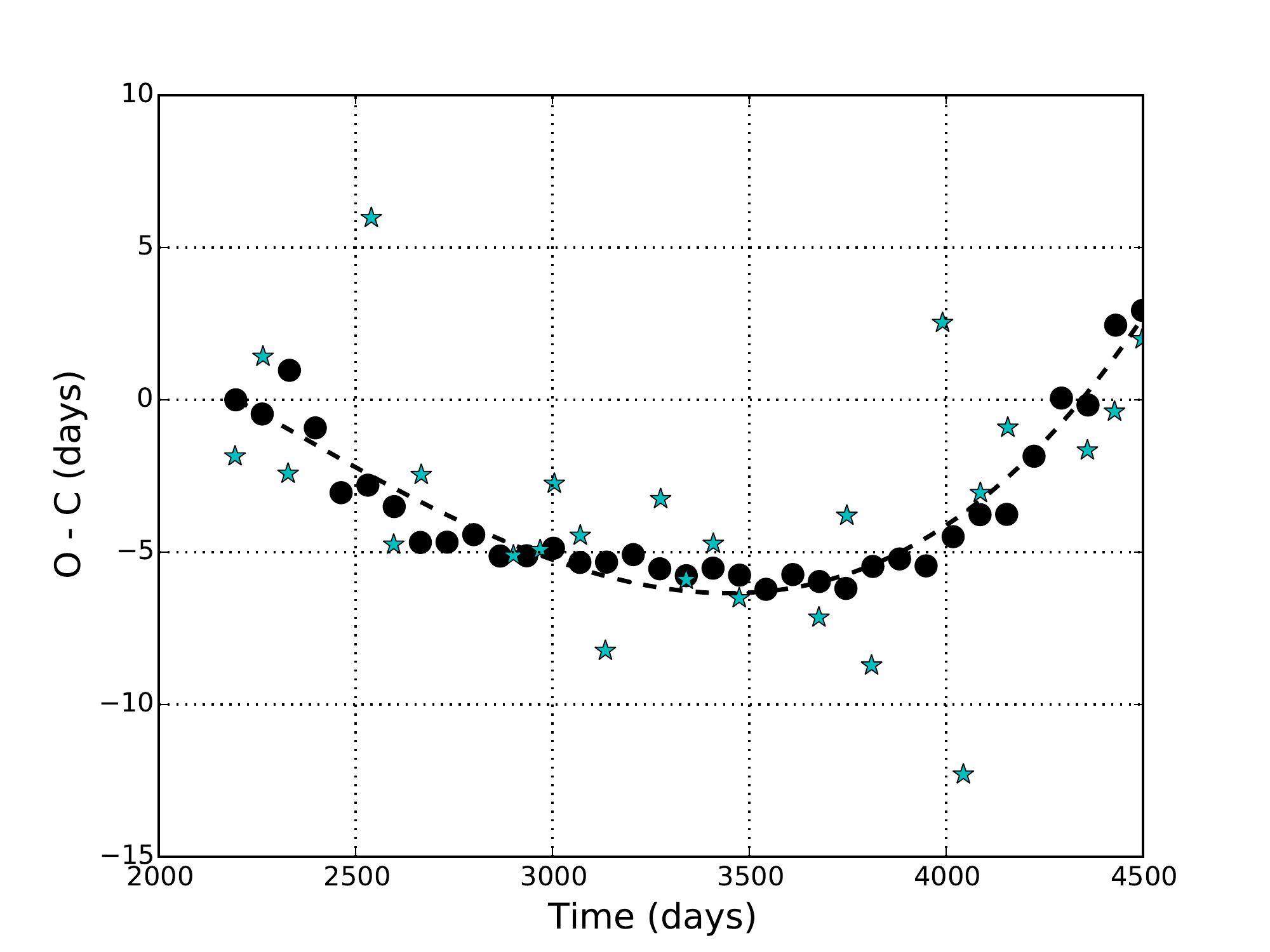}
  } 
  \subfloat[OGLE-LMC-T2CEP-149 ]{%
  \includegraphics[width=0.45\textwidth]{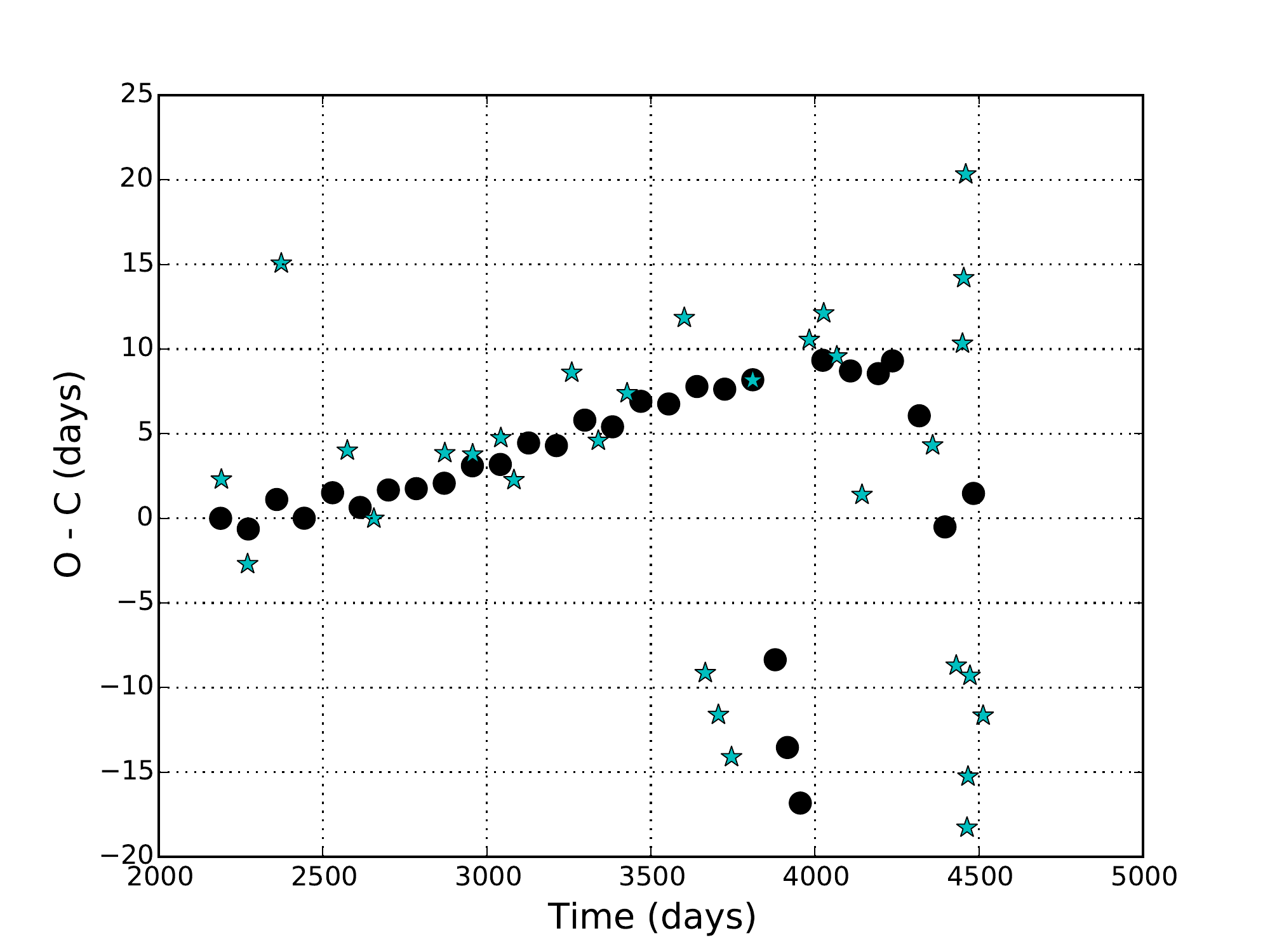}
  } \\
  \subfloat[OGLE-LMC-T2CEP-190 ]{%
  \includegraphics[width=0.45\textwidth]{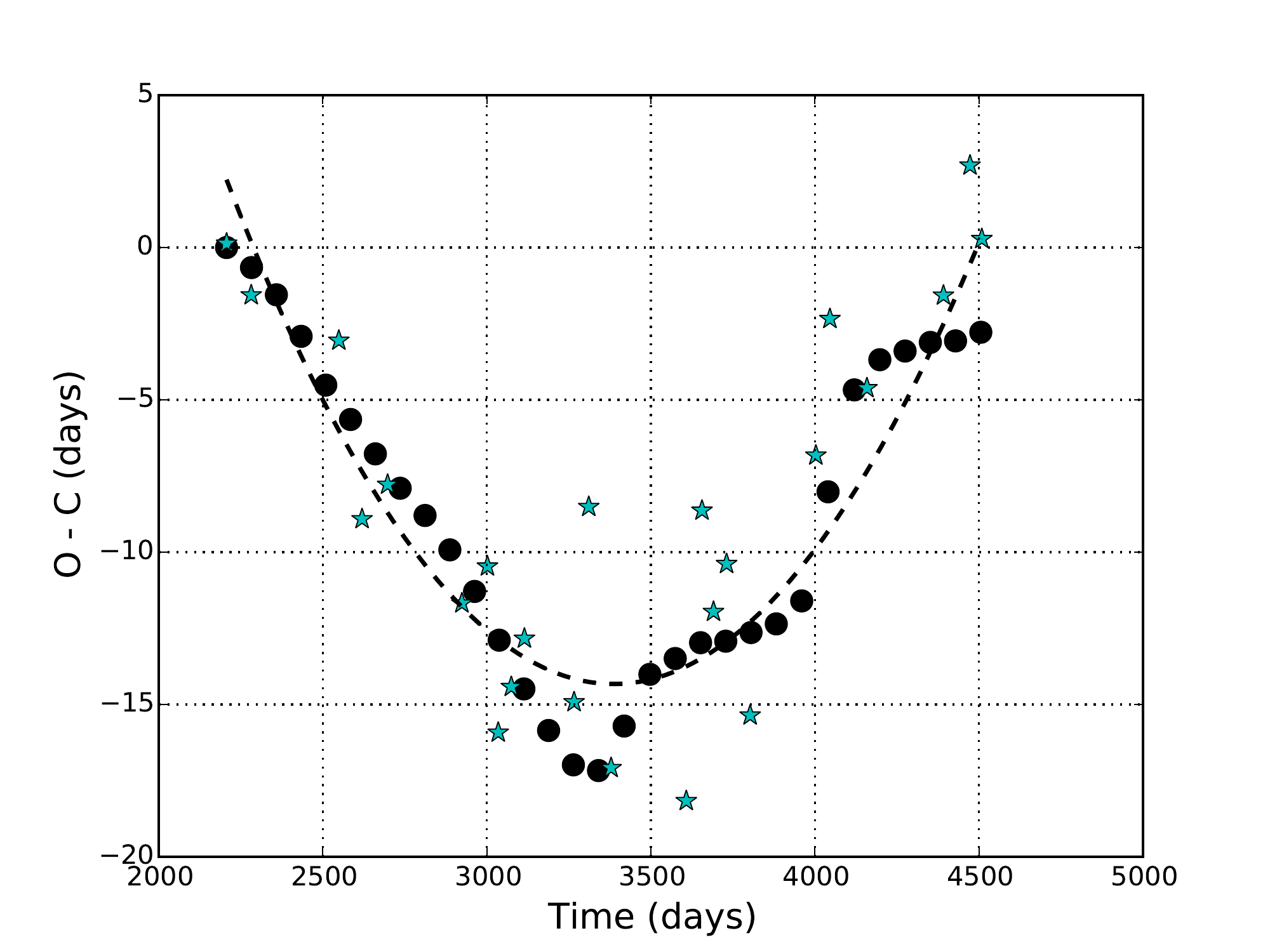}
  }
  \subfloat[OGLE-LMC-T2CEP-191 ]{%
  \includegraphics[width=0.45\textwidth]{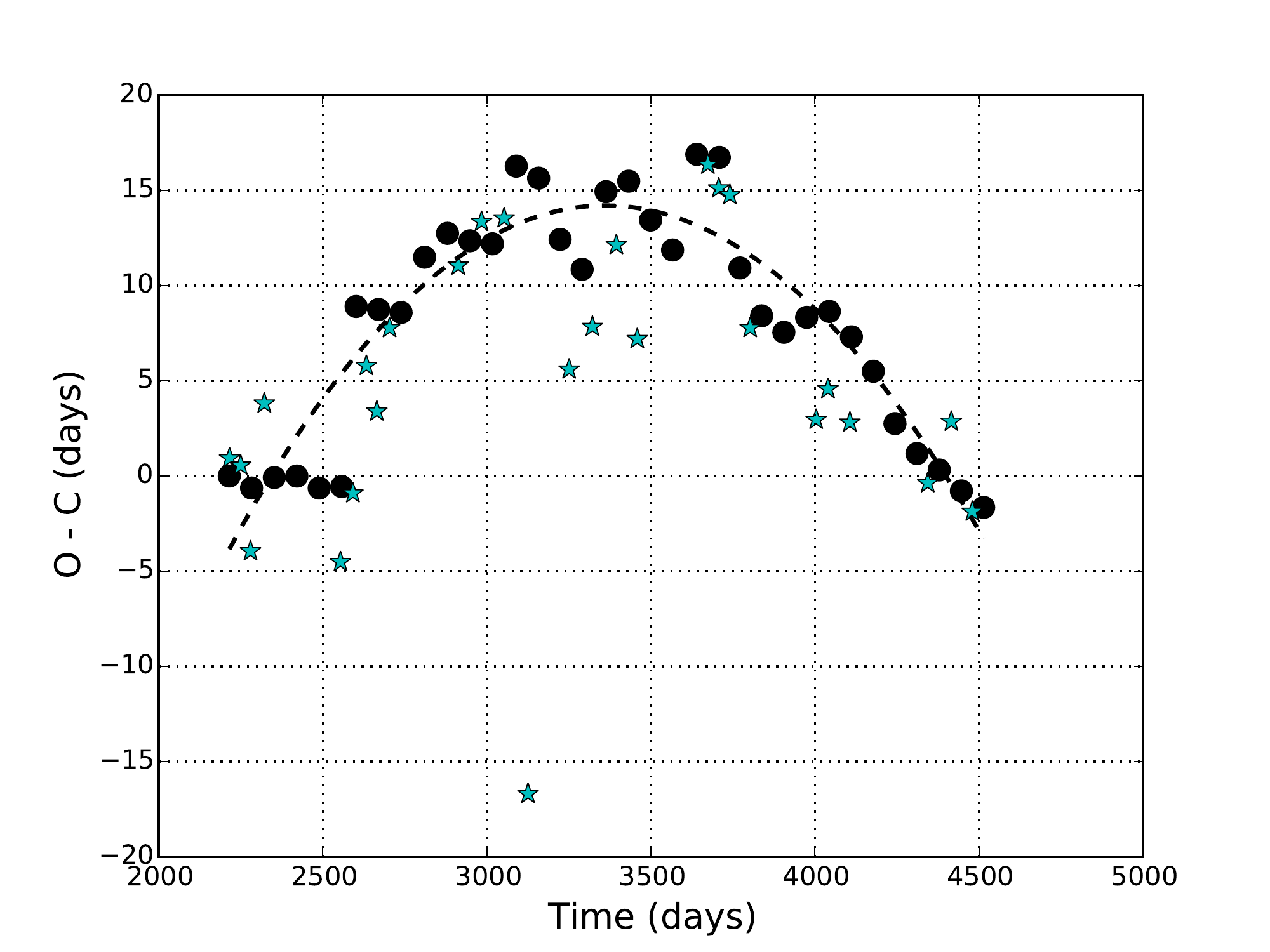}
  }
  \caption{($O-C$) diagrams of LMC targets. The black dots and cyan stars represent the ($O-C$) computed from the model fit and the data points, respectively. The dashed line indicates the parabolic fit to the ($O-C$).}
  \label{figure:O_C}
\end{figure*}

A pulsation model was fit with the significant periods found after iteratively pre-whitening the time series data. We added several overtones 
to the fit to obtain the best fitting model and thus accurate values for the subsequent minima. The black dots in Figure \ref{figure:O_C} represent the ($O-C$) computed from the minima of the pulsation model fit.
As a check, we compared the ($O-C$) from the model to that obtained from the data itself, plotted as cyan stars in Figure \ref{figure:O_C}. We traced the cycle-to-cycle 
variability by chunking the time series data and computing the mean period in each chunk. These are shown in Figure \ref{figure:p_dot} in Appendix \ref{appendix:AppendixE}. 

In the LMC sample, the following four stars show period change in the ($O-C$): OGLE-LMC-T2CEP-119, OGLE-LMC-T2CEP-149, OGLE-LMC-T2CEP-190, and OGLE-LMC-T2CEP-191.
The period change is more noticeable in the chunk time series (see Figure \ref{figure:p_dot} in Appendix \ref{appendix:AppendixE}), as increasing (OGLE-LMC-T2CEP-119 and OGLE-LMC-T2CEP-190) or decreasing periods (OGLE-LMC-T2CEP-149 and OGLE-LMC-T2CEP-191). 
These results are discussed in Section \ref{section:o_cdiscus}. We note that a few more stars in the LMC showed evidence of period changes in the ($O-C$), but the changes were not 
clear in the chunk time series.

In the SMC sample, OGLE-SMC-T2CEP-19, OGLE-SMC-T2CEP-20, and OGLE-SMC-T2CEP-24 show evidence of period changes in their phased light curves. Analysis
of their ($O-C$), however, did not reveal significant variability in their periods. Remarkably, OGLE-SMC-T2CEP-43 show variability in the ($O-C$) with 
a slightly negative curvature, which was also reported to have a decreasing period by \citet{groenewegen2017a}.

\section{Discussion} \label{section:discussion}
\subsection{SEDs}
Based on our classification of the SEDs, 
we identified 16 objects with dust in the form of a clear disc in the LMC (see Figures \ref{figure:seddisc1} to \ref{figure:seddisc16}) and 1 in the SMC (Figure \ref{figure:sedsmcdisc1}). 
In most cases, the IR excess already starts at $\sim$\,2 microns and provides a clear indication of the presence of hot dust at the dust sublimation radius \citep{hillen2016}.

\begin{figure}
   \centering
   \includegraphics[width=9cm,height=7cm]{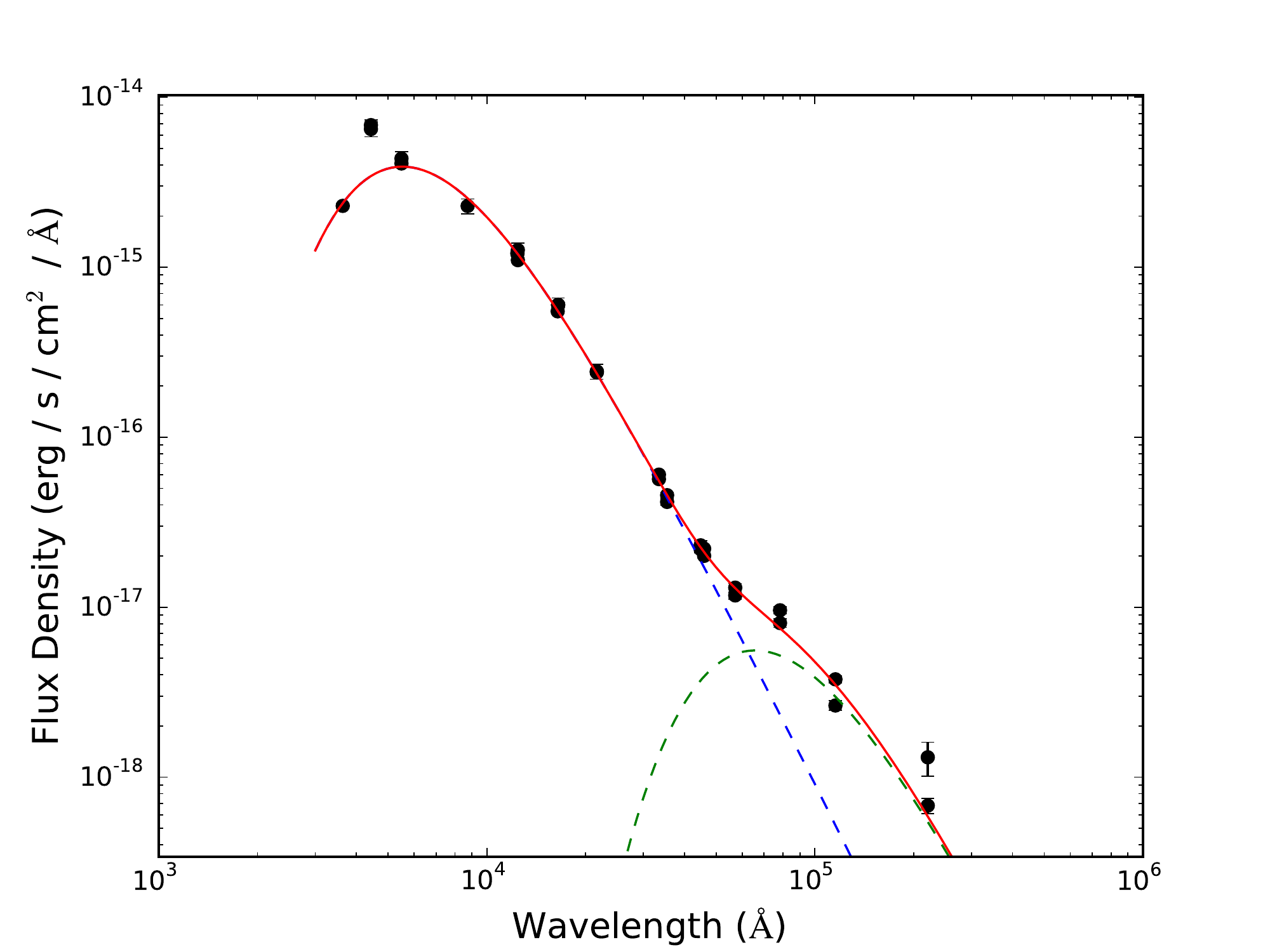}
   \caption{SED of OGLE-LMC-T2CEP-129 fitted using black bodies at temperatures of $\sim$ 5240 K (stellar component shown in blue dashed line) 
   and 440\,K (dust component shown in green dashed line). The red line shows the best fit to the data.}
     \label{figure:stardisc129}
\end{figure}

A few stars with a disc-type SED in the LMC display SEDs in which the IR excesses start at longer wavelengths, namely OGLE-LMC-T2CEP-014, 
OGLE-LMC-T2CEP-015, and OGLE-LMC-T2CEP-129. In the case of OGLE-LMC-T2CEP-015, the IR excess starts at $\sim$\,3 microns, while 
for OGLE-LMC-T2CEP-014 and OGLE-LMC-T2CEP-129, the excess starts at a much longer wavelength ($\sim$~4 microns see Figure \ref{figure:stardisc129}). 
The IR excesses in these systems appear similar to that of OGLE-LMC-T2CEP-149 (MACHO 81.8520.15) \citep{gielen09b} and represent the existence of 
dust further out from the dust sublimation radius. 

In Figure \ref{figure:stardisc129}, we show as an example, the 
SED of OGLE-LMC-T2CEP-129. Assuming the disc absorbs and emits as a black body we fit two black bodies (one for the photosphere and the other for the dust) to the SED data and estimate the dust temperature. 
We also assume that the inner rim of the disc dominates the IR luminosity.
The fitting was performed using a Markov Chain Monte Carlo (MCMC) method via the DREAM(ZS) package \citep{terBraak2008}. We estimate the dust temperature to be $\sim$ 440$\pm$9\,K.   
This temperature can be used to estimate the inner-rim radius ($R_{in}$) of the disc around OGLE-LMC-T2CEP-129, which scales with the bolometric luminosity of the star ($L_\mathrm{bol}$) and the dust temperature ($T_\mathrm{dust}$) through 
the equation \citep{dullemond01,kama2009} 
\begin{equation}
R_{in} = \frac{1}{2}\,(C_\mathrm{bw}/\epsilon)^{1/2}\,(L_\mathrm{bol}/4\pi\,\sigma\,T_\mathrm{dust}^4)^{1/2}
,\end{equation}
where $C_\mathrm{bw}$ is defined as the backwarming coefficient, 
ranging from 1 to 4 depending on the optical thickness of the disc \citep{kama2009}. The cooling efficiency ($\epsilon$) of the dust grains is a ratio of the Planck mean 
opacity of the dust species at their own temperature to the effective temperature of the star.
For an estimate of the inner-rim radius, these two coefficients can be assigned a value of 1 \citep[e.g.][]{lazareff2017}. 
Based on our assumptions, we compute the inner-rim radius of OGLE-LMC-T2CEP-129 to be $\sim$ 25\,au. This indicates that the disc around this system is clearly 
an evolved disc and is not at the dust sublimation radius \citep{hillen2016}. This is likely the case for the other stars we mentioned as displaying similar SED characteristics.

Interestingly, none of the RV Tauri stars in the Magellanic Clouds show clear shell type SEDs. 
This indicates that optically bright RV Tauri stars with dust might all exist as disc objects and evolve solely 
through a binary channel, given the strong link between the presence of a disc and binarity among post-AGB stars \citep{vanwinckel09} and RV Tauri stars \citep{manick2017}. 

In total, we found 12 RV Tauri stars in the Magellanic Clouds that displayed SEDs without IR excesses (see Figures \ref{figure:nonirsed1} 
to \ref{figure:nonirsed9} for the LMC and Figures \ref{figure:nonirsmc1} to \ref{figure:nonirsmc3} for the SMC). This represents $\sim$\,24\,\% of 
the RV Tauri population in the Magellanic Clouds, which is interestingly lower than the $\sim$\,50\,\% known non-IRs in the 
Galaxy \citep{gezer2015}. We note that the LMC is crowded, which makes it possible that the SEDs are contaminated by a nearby source 
(as in OGLE-LMC-T2CEP-005 and OGLE-LMC-T2CEP-199, see Section \ref{section:specialcasessmc}), 
thus making it hard to distinguish the correct SED characteristic of the star in question. It is therefore possible that some of the 
objects classified as ``uncertain'' are actually non-IRs, with the excess flux(es) caused by a potential background or nearby source.

The SEDs of the rest of the 22 stars in the Magellanic Clouds were classified as uncertain because the IR excesses were not clear either because of the lack of data or the fact that
only one or two flux points is in excess in the IR region without a clear disc or shell SED characteristic (see Figures \ref{figure:uncertsed1} to \ref{figure:uncertsed17} 
for the LMC and Figures \ref{figure:uncertsmc1} to \ref{figure:uncertsmc5} for the SMC).

\subsection{Disc-type objects}
In the following two subsections, we describe the evolutionary nature of the dusty (disc-type) RV Tauri stars based on their derived position on the HRD. 
We also report the discovery of two binaries via the RVb phenomenon among the LMC targets.

Figures \ref{figure:evoall_LMC} and \ref{figure:evoall_SMC} show evolutionary tracks from the zero age main sequence (ZAMS) until the AGB from \citet{bertelli09}. The 
post-AGB tracks were modelled by \citet{millerbertolami2016} and include evolution from the AGB 
to the white dwarf phase for stars with initial masses between 1.0 and 3.0 M$_{\odot}$. We identify the evolutionary nature of our sample stars 
based on these evolutionary tracks. However, we do not disregard the fact that these tracks are computed in the framework of single 
stellar evolution, while many stars in our sample (especially the disc-types) follow a binary evolution channel, for which models are unavailable at this stage and differ for different 
initial orbital periods, mass ratios, and eccentricities. 

\subsubsection{Evolutionary nature of disc objects}
\begin{figure*}
   \centering
   \includegraphics[width=12cm,height=12cm]{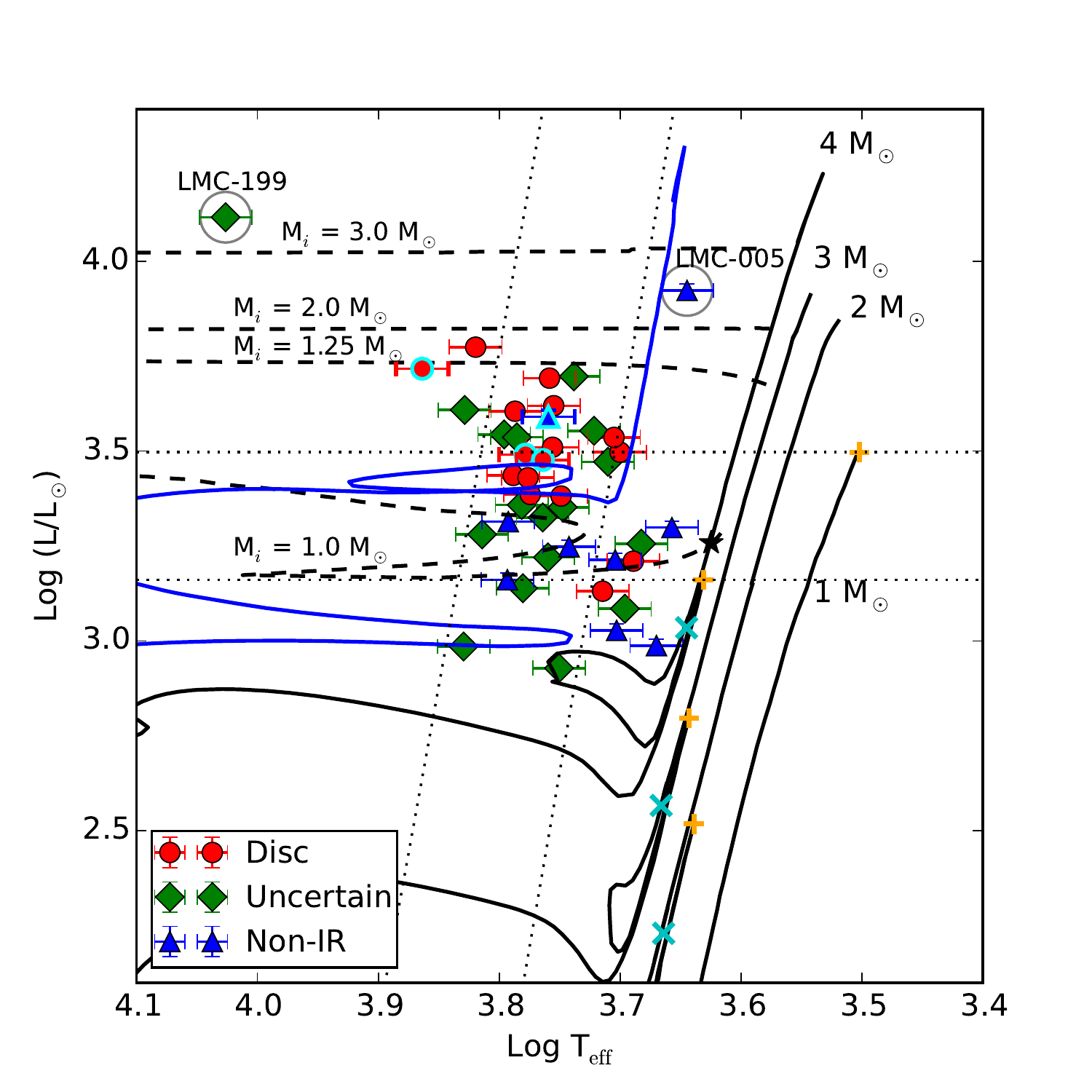} 
   \caption{LMC RV Tauri stars with different SED characteristic, plotted in the HRD.
   The black lines are evolutionary tracks for a 1 M$_\odot$, 2 M$_\odot$,  3 M$_\odot$, and 4 M$_\odot$ stars with a composition of Z = 0.008 and Y = 0.26 for the LMC \citep{bertelli09}. The dashed lines are post-AGB tracks from 
   \citet{millerbertolami2016} for 1 M$_\odot$, 1.25 M$_\odot$, 2 M$_\odot$, and 3 M$_\odot$ initial mass stars with an initial metallicity of Z = 0.001. The blue line represents the evolution of a 3 M$_\odot$ star, which goes through a blue-loop phase.
   The slanted dotted lines show the theoretical instability strip taken from \citet{kiss07}. 
   The four points with cyan marker edges indicate the position of OGLE-LMC-T2CEP-119, OGLE-LMC-T2CEP-149,OGLE-LMC-T2CEP-190, and OGLE-LMC-T2CEP-191, which display clear period change in their ($O-C$) diagram. 
   The stars in the grey circles denote the position of OGLE-LMC-T2CEP-005 and OGLE-LMC-T2CEP-199, which have higher-than-expected luminosities (see Section \ref{section:specialcasessmc}).
   The tip of the RGB for each track corresponding to a mean metallicity of the LMC are indicated with orange `$+$' signs, which are at L $\sim$\,3150, 330, 630, and 1450 L$_\odot$ for the 1, 2, 3, and 4 M$_\odot$ tracks, respectively. The upper and lower horizontal dotted lines
   indicate the tip of the RGB for a 1 and 4 M$_\odot$ star, respectively.
   The cyan `x' markers represent the start of the early AGB phase (E-AGB); these markers are at L $\sim$\,170, 370 and 1090 L$_\odot$ for the 2, 3 and 4 M$_\odot$ tracks, respectively.}
   \label{figure:evoall_LMC}
\end{figure*}

\begin{figure*}
   \centering
   \includegraphics[width=12cm,height=12cm]{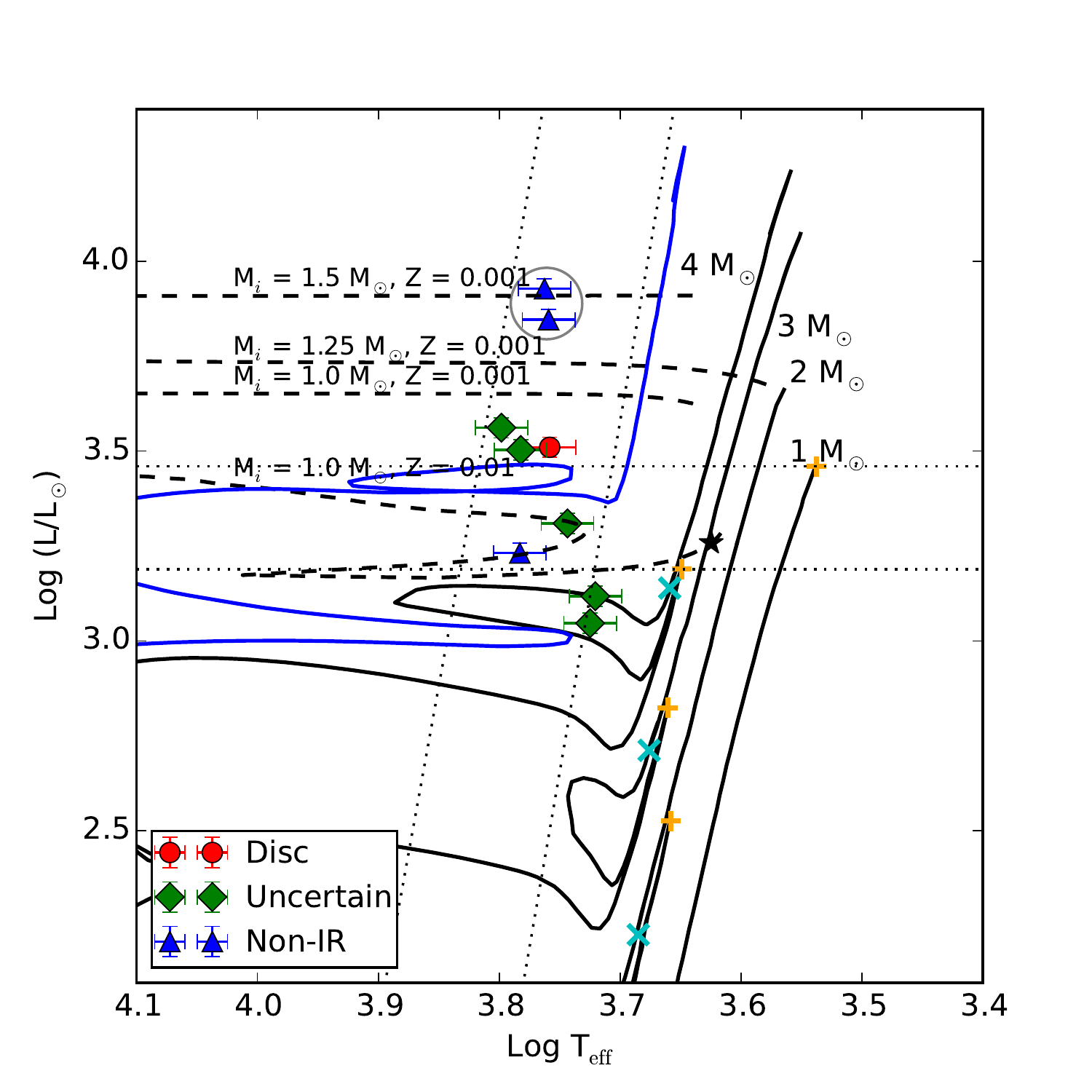} 
   \caption{SMC RV Tauri stars with different SED characteristic, plotted in the HRD.
   The black lines are evolutionary tracks for a 1 M$_\odot$, 2 M$_\odot$, 3 M$_\odot,$ and 4 M$_\odot$ stars with a composition of Z = 0.004 and Y = 0.26 for the SMC \citep{bertelli09}. The dashed lines are post-AGB tracks from 
   \citet{millerbertolami2016} for 1 M$_\odot$ and 1.5 M$_\odot$ stars and Z = 0.001, plotted as 2$^{nd}$ and 3$^{rd}$ from bottom, respectively. The bottom-most dashed track is for a 1 M$_\odot$ star with a metallicity of Z = 0.01, 
   which is at a lower luminosity than a post-AGB track of Z = 0.001. 
   The blue line represents the evolution of a 3 M$_\odot$ star, which goes through a blue-loop phase.
   The slanted dotted lines show the theoretical instability strip taken from \citet{kiss07}. The tip of the RGB for each track corresponding to a mean metallicity of the SMC are indicated with orange `$+$' signs , which are at L $\sim$\,2880, 330, 660, and 1560 L$_\odot$ for the 
   1, 2, 3, and 4 M$_\odot$ tracks, respectively. The cyan `x' markers represent the start of the early AGB phase (E-AGB), which are at L $\sim$ 170, 510, and 1380 L$_\odot$ for the 2, 3, and 4 M$_\odot$ tracks, respectively. The
   stars in the grey circle are the OGLE-SMC-T2CEP-07 and OGLE-SMC-T2CEP-29, which shows ellipsoidal variability and eclipses in their light curves, respectively. The upper and lower horizontal dotted lines
   indicate the tip of the RGB for a 1 and 4 M$_\odot$ star, respectively.}
     \label{figure:evoall_SMC}
\end{figure*}

The position of the dusty (disc) RV Tauri stars are indicated by the red circles in Figure \ref{figure:evoall_LMC} for the LMC and Figure \ref{figure:evoall_SMC} for the SMC. 
Our derived range of luminosities, effective temperatures, and their dusty (disc) nature suggest that these stars are either post-RGB or post-AGB objects. 
These stars could have evolve off into the post-RGB or post-AGB domain from different initial mass tracks. 
As an attempt to differentiate between their post-RGB or post-AGB nature and hence their progenitor-mass, we use as a criterion, the luminosity at the RGB-tip attained by stars with different initial masses.

The RGB-tip of a 1 M$_{\odot}$ star is denoted by a horizontal dotted line at L $\sim$ 3150 L$_{\odot}$ for the LMC (Figure \ref{figure:evoall_LMC}) and at L $\sim$ 2880 L$_{\odot}$ for the SMC (Figure \ref{figure:evoall_SMC}).
The 2, 3, and 4 M$_{\odot}$ stars reach the tip of the RGB at luminosities lower than that of a 1 M$_{\odot}$ star (see yellow `+' signs in Figures \ref{figure:evoall_LMC} and \ref{figure:evoall_SMC})
and start their early AGB phase at positions denoted by the cyan crosses.
The lower horizontal dotted line in Figure \ref{figure:evoall_LMC} indicates the RGB-tip of a 4 M$_{\odot}$ star at L $\sim$ 1450 L$_{\odot}$,
which is at a luminosity just below most of the disc stars. For the SMC (Figure \ref{figure:evoall_SMC}), the lower horizontal dotted line denotes the RGB-tip luminosity at L $\sim$ 1560 L$_{\odot}$ of a 4 M$_{\odot}$ star.

Using the RGB-tip-luminosity argument, if the disc-type stars lying between these two horizontal dotted lines are indeed progenies of stars with masses between 
$\sim$ 2 and 4 M$_{\odot}$, they would be post-AGB objects. Otherwise, they are likely post-RGB stars evolved from a lower mass progenitor.

The only objects that we are relatively certain are post-AGBs among the RV Tauri sample are those that lie above the RGB-tip of a 1\,M$_{\odot}$ star. 
These are stars plotted as red markers above the upper dotted horizontal line in Figures \ref{figure:evoall_SMC} and \ref{figure:evoall_LMC}. 
The fact that these stars lie above the RGB-tip of all four stellar model tracks, their high luminosities in combination with their dusty nature,
indicates that they likely post-AGB progenies of stars with initial masses higher than $\sim$ 1 M$_{\odot}$. 
This viewpoint is reinforced by the fact that two of these stars (OGLE-SMC-T2CEP-18 and OGLE-LMC-T2CEP-015) have previously been proven to be genuine post-AGB stars chemically; 
their photospheres were found to be C and s-process enriched by \cite{reyniers07a} and \citet{kamath2015}, respectively.

\begin{figure*}
   \centering
   \includegraphics[width=12cm,height=12cm]{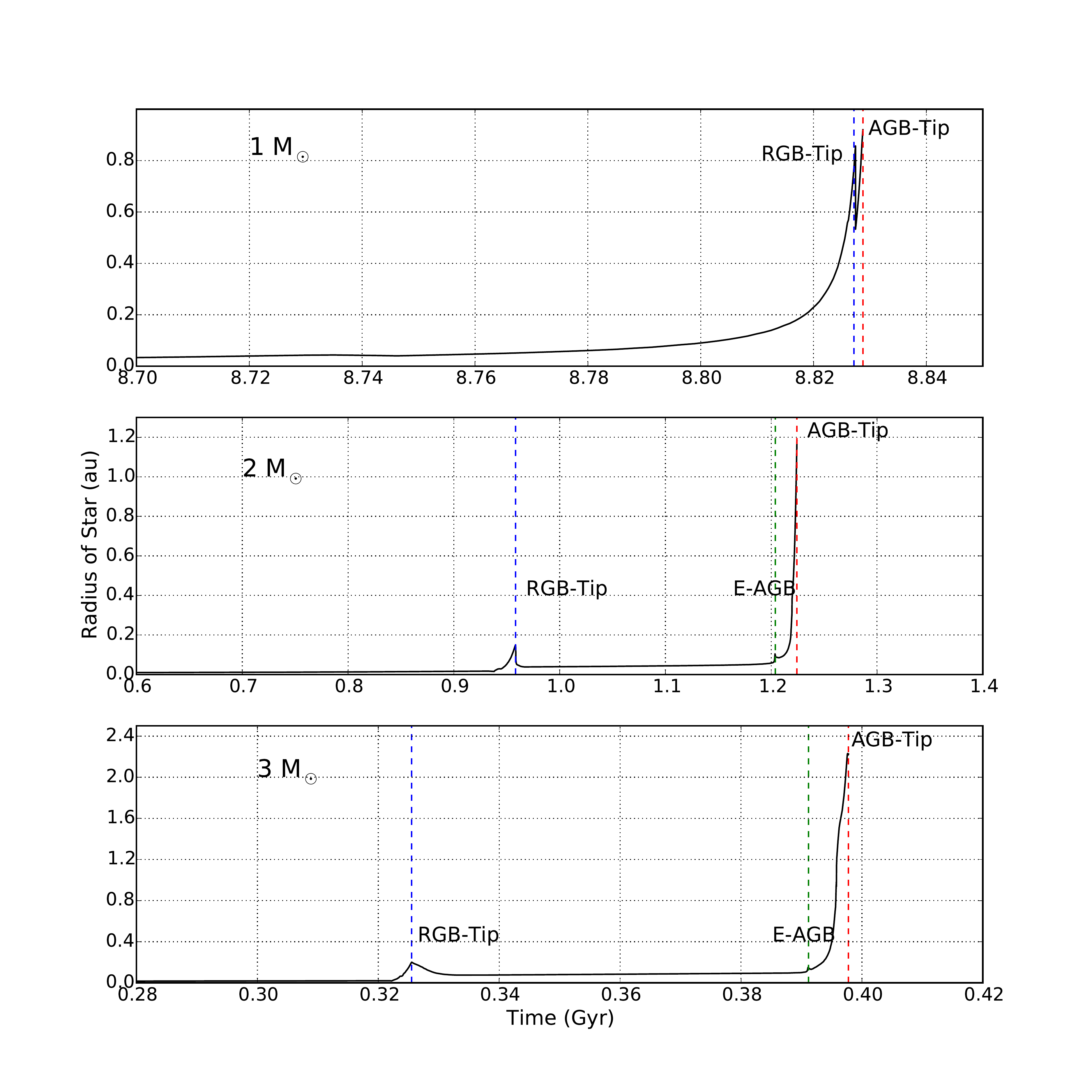}
   \caption{Evolution of the stellar radius of a 1, 2, and 3 M$_\odot$ star, from the models of \citet{bertelli09}. This figure shows the late stages of the stars' evolution, i.e. from the onset of a noticeable change
   in radius through the evolution.}
     \label{figure:radiusevo}
\end{figure*}

The stars with a disc-type SED have evolved off the RGB or AGB likely from binary interaction, 
making these objects post-RGB or post-AGB binaries. Their disc SED nature and the strong correlation between disc-types and binarity, justifies their likely binary nature \citep{vanwinckel99,manick2017}. 
Assuming they are indeed binaries, to differentiate between their post-RGB or post-AGB nature, we analysed the probability of 
interaction between the primary and the companion at various stages of the evolution of the primary. In Figure \ref{figure:radiusevo}, we show the radius evolution of the primary as a 
function of the initial mass. The radii are computed using the luminosities and T$_{\rm eff}$ output from the models of \citet{bertelli09}. 

If the progenitor mass of the disc-type objects is $\sim$\,1 M$_{\odot}$ 
then the star would have swelled considerably (up to $\sim$\,0.8\,au) upon reaching the RGB stage
(see top panel of Figure \ref{figure:radiusevo}) and hence with a higher probability of interaction
with the companion already on the RGB. This evolutionary scenario would make most of the disc stars below a luminosity of $\sim$\,3150, 
which is the tip of the RGB for a 1 M$_{\odot}$ star, post-RGB binaries.
On the other hand, if the progenitor is more massive, for example higher than $\sim$ 2 M$_{\odot}$, we can deduce from the middle and bottom panel of 
Figure \ref{figure:radiusevo} that the radius of the primary is much larger on the AGB than on the RGB. Thus, the probability of interaction with 
the companion would be much higher on the AGB, consequently making the disc objects more likely to be post-AGB binaries.

If the disc RV Tauri stars are indeed post-RGB or post-AGB binaries, they are likely in wide orbits. 
We already have indications of the existence of long period binaries among the LMC and SMC RV Tauri 
sample with the long orbital periods of 850 and 916 days we found in OGLE-LMC-T2CEP-200 and OGLE-LMC-T2CEP-032, respectively.
Addtionally, \citet{groenewegen2017a} found three more binary candidates: OGLE-LMC-T2CEP-011, OGLE-SMC-T2CEP-018, and OGLE-SMC-T2CEP-029, which have periods between $\sim$\,609 and 1657 days. 
It is worth mentioning that their Galactic counterparts presented in \citet{vanwinckel99} and \citet{manick2017}, have wide orbits as well.

Interestingly, OGLE-SMC-T2CEP-07 and OGLE-SMC-T2CEP-29 do not show dust excess in their SEDs (see Figures \ref{figure:nonirsmc1} and \ref{figure:nonirsmc2}), 
but are strong binary candidates with orbital of periods of 394 days and 609 days, respectively (see Section \ref{section:timeseriessmc}) \citep{soszynski10,groenewegen2017a}. 
If binarity is indeed confirmed via spectroscopy in these two stars, they would be the first known RV Tauri stars in a binary system that display no IR excess.

\subsubsection{Binary nature of OGLE-LMC-T2CEP-200 and OGLE-LMC-T2CEP-032} \label{section:RVb_bin}
The link between the RVb phenomenon and binarity was first predicted by 
\citet{waelkens91a}, based on the long-term variability seen in the post-AGB star HR\,4049. These authors hypothesised that it was likely due to variable obscuration 
by the circumstellar dust of the order of the orbital period of the binary system. Since then, this concept has been discussed and explored in the literature 
owing to the discovery of the same phenomenon in many more binary stars with a dusty circumbinary disc. \citet{vanwinckel99} and \citet{manick2017} showed that 
the long-term period in the photometric time series due to the RVb nature, corresponds to the periods found in the radial velocities. These results are thus best 
explained by variable extinction in our line of sight through the dust that is caused by the orbital motion of the binary. 

\begin{figure}
   \centering
   \includegraphics[width=9.5cm,height=5cm]{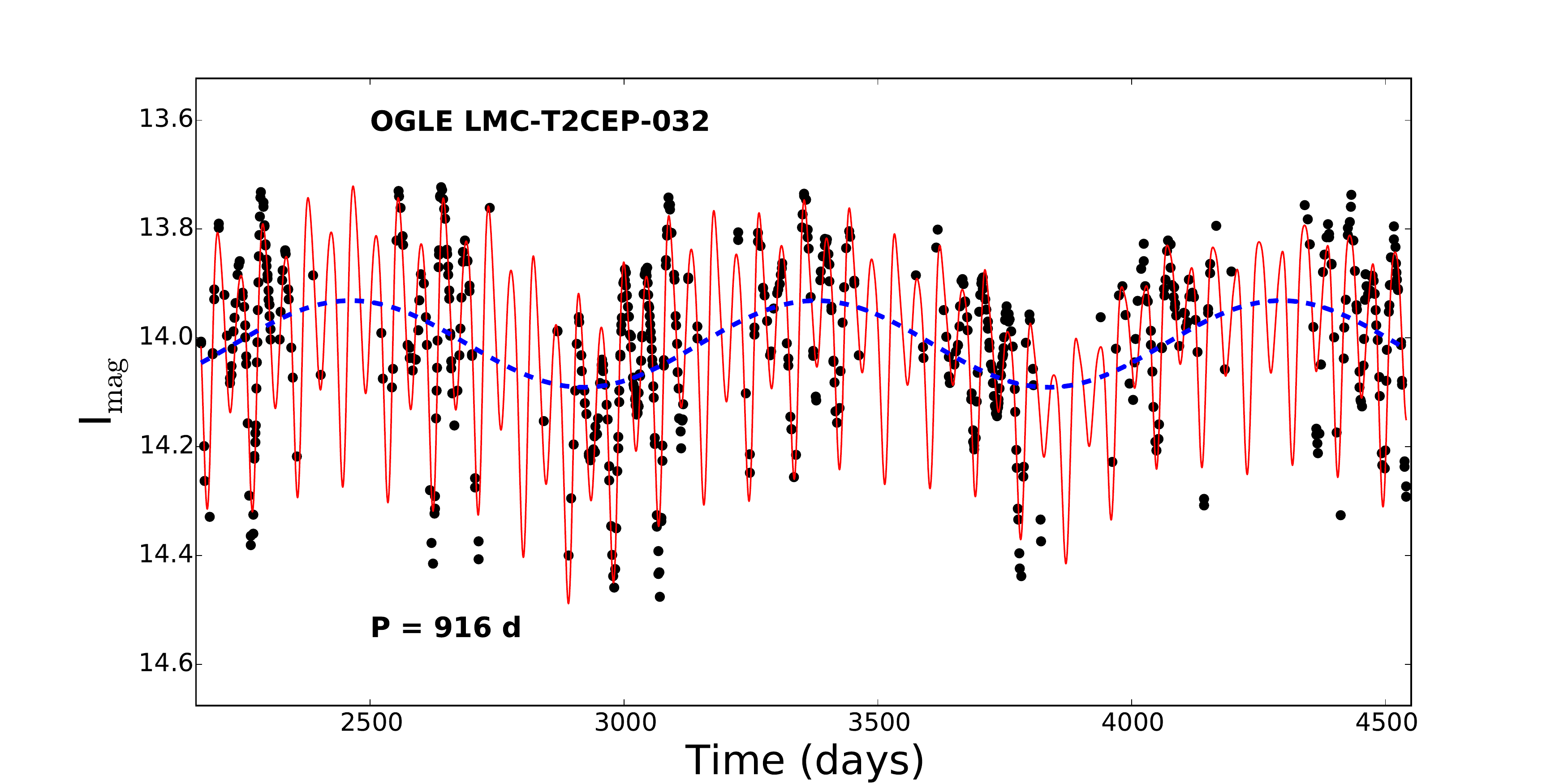} \\
   \includegraphics[width=9.5cm,height=5cm]{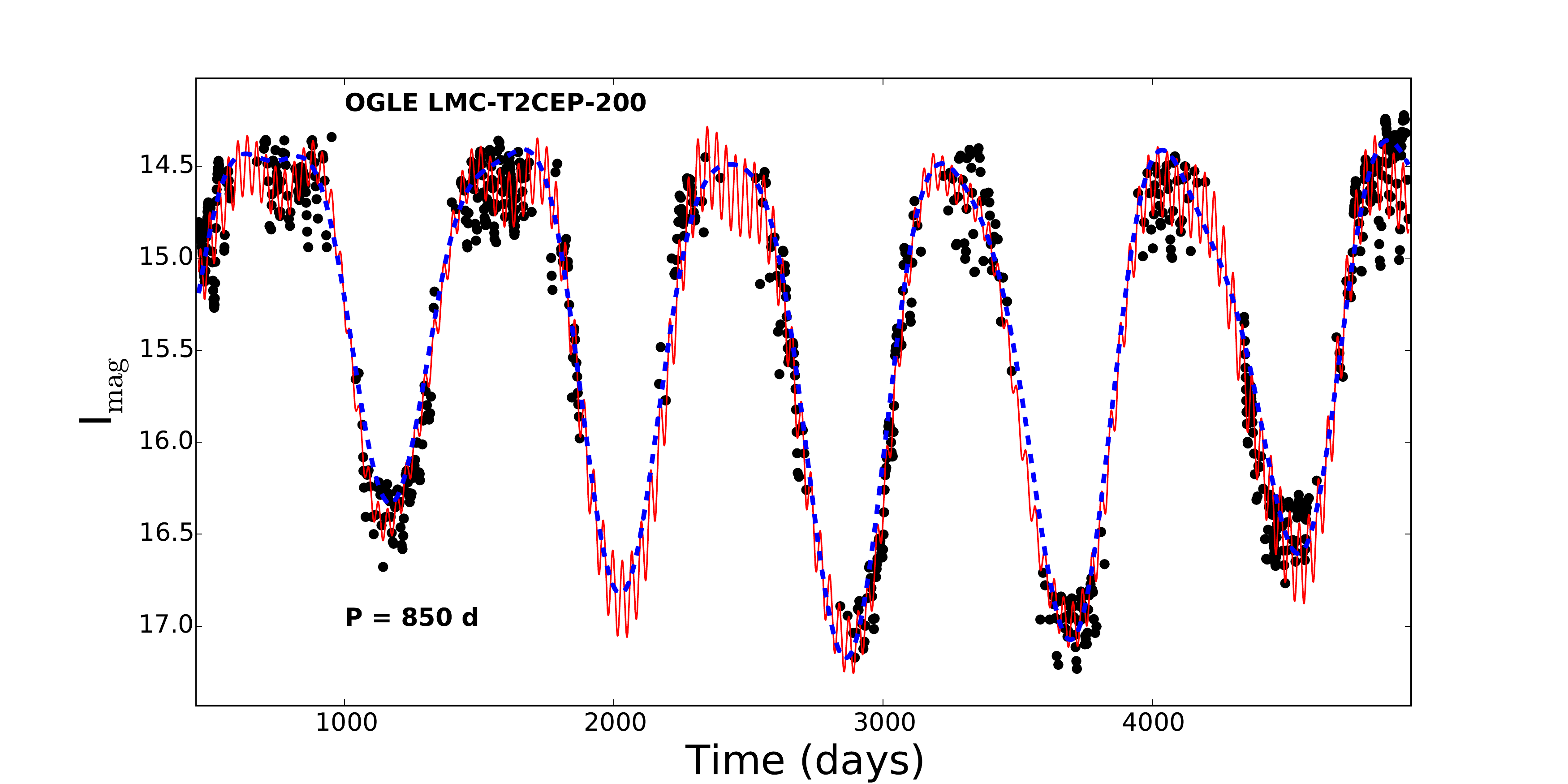} 
   \caption{Stars in the LMC that display RVb type light curves. The red line shows the pulsation model combined with the long-term variable period of the orbit. The blue dashed line shows the long-term variability.}
     \label{figure:RVb_binaries}
\end{figure} 

We therefore claim that the RV Tauri stars OGLE-LMC-T2CEP-200 and OGLE-LMC-T2CEP-032, which display the RVb phenomenon in their light curves, are
binaries with orbital periods of $\sim$ 850 and 916 days (see Figure \ref{figure:RVb_binaries}). The RVb nature of these two stars was already pointed out by \citet{soszynski2008a}, 
who mentioned a long-term period of 854 days in only one of this pair, OGLE-LMC-T2CEP-200. In the case of OGLE-LMC-T2CEP-032, the RVb phenomenon is not 
as pronounced as in OGLE-LMC-T2CEP-200, most probably because of an inclination effect.

These two stars are perfect candidates within our LMC RV Tauri sample for a spectroscopic follow-up to confirm their binary nature. However, tracing orbital motions in RV Tauri stars using radial velocity shifts 
could be challenging for the Magellanic Cloud targets: they are faint and are in wide orbital configurations \citep{vanwinckel99, manick2017}. 
In addition, their high amplitude pulsations and line splitting due to shocks in their atmospheres, subdue the orbital motions and make binary detection challenging \citep{baird82,gillet90,pollard97}. 

\subsection{Evolutionary nature of the non-IR objects}
Since the discovery of RV Tauri stars in the early 1900s, their post-AGB nature has been extensively hypothesised in the literature \cite[e.g.][]{jura86,alcock98}. 
Their link to post-AGB stars has been only considered the dusty RV Tauri stars without much consideration to the non-IR stars, which represent a sizeable portion 
($\sim$\,42\%) of all the known RV Tauri stars in the Milky Way and Magellanic clouds \citep{gezer2015, soszynski2008a, soszynski10}. 

Their derived position on the HRD raises interesting questions regarding their evolutionary nature. 
Given their derived luminosities and their apparent evolved nature, we expect to see dust in 
the form of IR excess at longer wavelengths. If these are single post-AGB stars, the dust should be in the form 
of an expanding shell (double peaked SED) in the infrared region as seen in many single post-AGB stars \citep{volk89, vanwinckel03}, whereas in the case of binaries, we expect a dusty circumbinary disc \citep{vanwinckel09, hillen2017}.
However, their non-dusty SED characteristics contradict their post-AGB or post-RGB nature.
One possibility is that they left the AGB or RGB long enough ago for the dust to have dispersed. 

The dust (disc or shell) dissipation lifetime is not well constrained in the literature, but can be estimated in the case of a shell.
Our WISE SED data range is wavelength-limited up to around 22 microns. We can calculate roughly how long it takes for an expanding shell of dust to create an IR excess at $\sim$ 22 microns.
This gives us an idea of the age of the dust shell (if indeed present around these stars) to create a detectable IR excess around 22 microns.

\begin{figure}
   \centering
   \includegraphics[width=9cm,height=7cm]{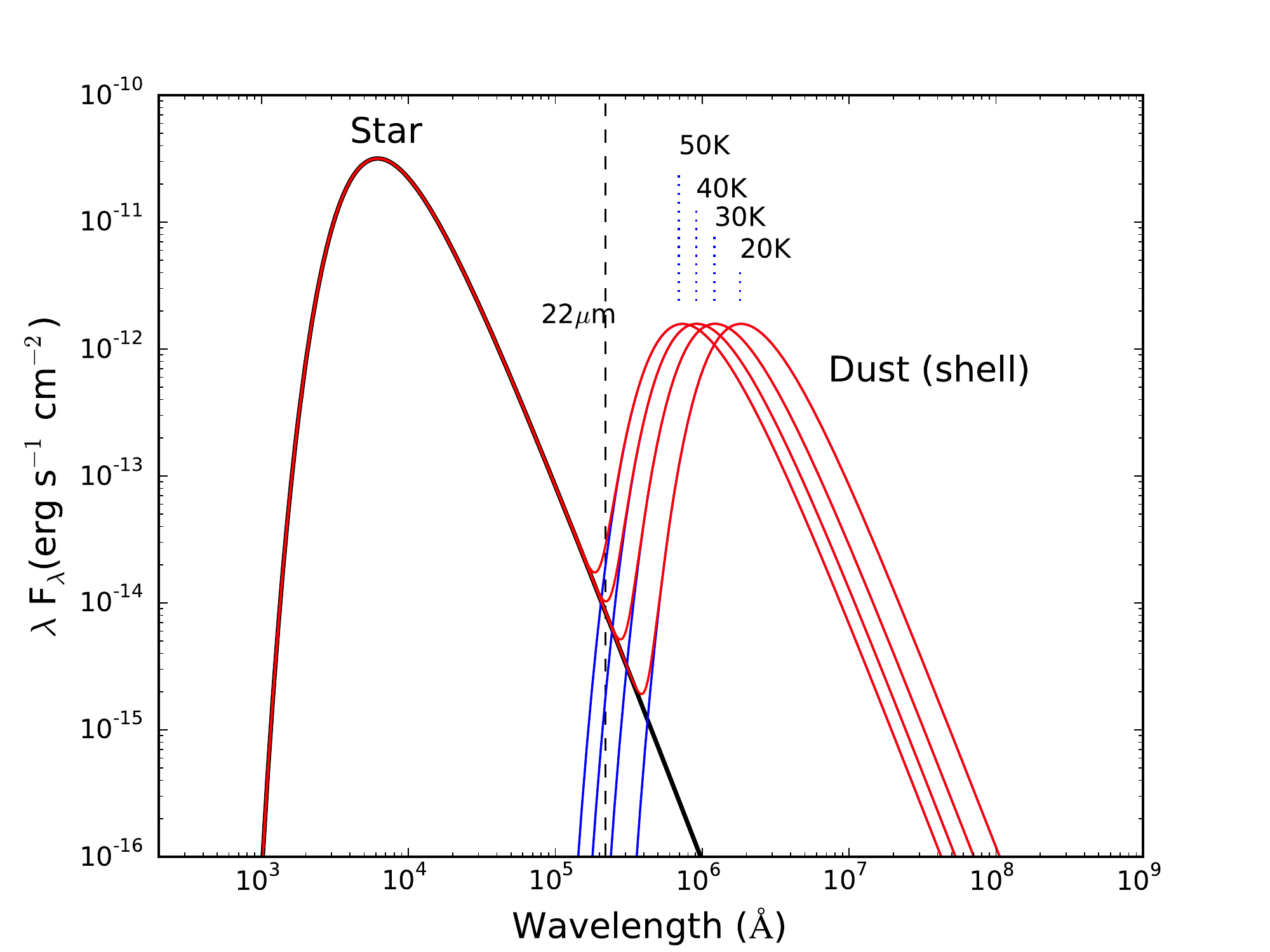}
   \caption{Star with luminosity, L $\sim$ 3300 L$_{\odot}$ represented by a black body with T$_{\rm eff}$ $\sim$ 5900 K. The dust shells, assuming they absorb and emit as black bodies represented by the blue curves with T = 50 K, 40 K, 30 K, and 20 K. The red curves represent, in each case, the sum of the black-body spectra of the star and  shell. The dashed vertical line is indicated at 22 microns.}
     \label{figure:blackbodystardust}
\end{figure} 

In Figure \ref{figure:blackbodystardust} we show the evolution of the SED produced by an expanding dust shell around a post-AGB star assuming T$_{\rm eff}$ of 5900 K and L = 3300 L$_\odot$.
We assume that the dust shell absorbs and emits as a black body and that 5\% of the luminosity of the star is in the dust.
According to this plot, the highest temperature that the dust can have without creating an IR excess at 22 microns is around 40 Kelvin.
This temperature can be used to compute the shell radius, '$a$', around the star using the equation T$_{\rm dust}$ = T$_{\rm eff}$\,$\times$\,$(R/2\,a)^{0.5}$, where T$_{\rm dust}$ is the dust temperature, 
T$_{\rm eff}$ is the effective temperature of the star, and $R$ is the stellar radius.
We find the dust-shell radius to be of the order of 4.7\,$\times$10$^{11}$ $km$. 
Assuming a 15 km s$^{-1}$ expansion rate of the shell \citep[e.g.][]{likkel91,loup1993}, the age is estimated to be $\sim$ 1000 years. 

The positions of most of the non-dusty stars in the SMC and LMC coincide with the post-AGB track of a 1.0 M$_\odot$ initial mass star.
Our estimated dust-shell lifetime is shown by a black star marker 
on this track (see Figures \ref{figure:evoall_LMC} and \ref{figure:evoall_SMC}). This is shortly after the start of the post-AGB phase of such a star. 
The implication is that if these stars are indeed evolving through this track, they would be low-luminous single post AGB stars in which the dust has dissipated $\sim$ 1000 years after the start of the post-AGB phase.
It is very unlikely that these are single post-AGB stars with initial masses of $\sim$ 1.25 M$_{\odot}$ or higher because the post-AGB models predict that single stars with such initial masses would have 
luminosities much higher than those of our non-dusty RV Tauri sample (see dashed post-AGB track of a 1.25 M$_{\odot}$ in Figures \ref{figure:evoall_LMC} and \ref{figure:evoall_SMC}).

On the other hand, considering the possibility that these stars have evolved off  the track due to binary interaction, it is more likely that they are post-RGB binaries with the primary star having an initial mass of 1 M $_{\odot}$ or less. There are two main explanations for this possibility:
First, a 1 M $_{\odot}$ star would become large enough on the RGB-tip already to interact with the companion (see Figure \ref{figure:radiusevo}) and, second, a low mass primary would evolve slowly enough for the dust to have dissipated 
at this stage.

It is also possible that the dust around these stars is in the form of a disc, but is too evolved for the IR excess to be detectable. 
The best example of an evolved circumbinary disc is probably that around the Galactic post-AGB binary BD$+$39 4926 \citep{kodaira70,deruyter06}. 
This is a strongly depleted binary that evolved off the AGB, but  shows only a very small infrared excess at 22 microns \citep{gezer2015}. 
If such a small excess is present in our non-dusty RV Tauri sample, given they are Magellanic cloud objects, the excess would be too faint to be detected by the WISE instrument.

Interestingly, the region occupied by the non-IRs on the HRD corresponds to that of blue-loop stars as well.
Blue-loop evolution in intermediate-mass stars ( $\sim$ 3 to 9 M$_{\odot}$) occurs when they start burning helium in their cores \citep{walmswell2015}.  
In the first part of the loop, the stars move leftward of the HRD as a result of contraction and heating up. They eventually attain a maximum effective 
temperature and return to the Hayashi track via the second part of the loop.

The occurrence of blue loops is very metallicity-dependent in the models of \citet{bertelli09}  and occurs mostly at low metallicities.
The blue line in Figures \ref{figure:evoall_LMC} and \ref{figure:evoall_SMC}, shows the blue-loop track of a 3 M$_{\odot}$ star 
with the lowest metallicity (Z = 0.001) and highest helium content (Y = 0.4). 
According to the models, low-metallicity stars with a range of initial masses between $\sim$ 3 M$_{\odot}$ and 4 M$_{\odot}$,
evolve along blue loops around the same region too, but for the sake of convenience we show only 
the blue-loop track of the 3 M$_{\odot}$ star.  

We argue that it is highly unlikely that these are blue-loop stars mainly because
blue loops occur in stars with masses above 3 M$_{\odot}$ and this mass regime is more representative of population I Cepheids, 
which follow a completely different PL relation \citep[e.g.][]{tammann2003}. In addition, the light-curve characteristics of our non-IR sample 
clearly contradicts the hypothesis that these could be population I Cepheids.

\subsection{Uncertain objects}
The uncertain objects display an SED with only one or a few points in excess in the IR region, but none of these objects display a clear disc- or shell-type characteristic. 
If these excesses are indeed due to dust around the stars, they might be coming from either an expanding shell or an evolved disc. As discussed by
\citet{gezer2015}, an expanding shell of cold gas would imply that these stars likely evolve in the same way as single post-AGB stars. 
On the other hand, an evolved disc would mean that the central source is likely a binary. 

The results of our statistics on the luminosities, however, show that the objects with an uncertain SED are not significantly different from the non-dusty objects in terms of luminosities (see Table \ref{table:ks}).
It is therefore plausible that, similar to the non-dusty objects, the dust is dispersed in these systems and the few IR fluxes in excess are arising because of the remnant dust. 
More photometry at longer wavelengths is needed to better classify their SEDs and be conclusive about their evolution.

\subsection{Most luminous dusty stars} \label{section:discussion_lum_diff}
Interestingly, we find, with high statistical significance, that the luminosities of the dusty (disc) stars are on average highest among 
the RV Tauri stars (See section \ref{section:lumdiff} and Figure \ref{figure:evoall_LMC}).
Assuming that these are binaries with orbits similar to the Galactic disc-type sources (see introduction), they evolved as interacting binaries on the RGB or AGB.
We therefore expect their evolution to be interrupted by a phase of binary interaction on the giant branch. As a consequence, they should, in principle, 
appear at lower luminosities than the single stars on the HRD. The results presented in this work, however, contradict this hypothesis.

We believe that our derived luminosities are not a result of an observational or methodological bias because the higher luminosities of the dusty objects are already 
seen in the Wesenheit plane in the form of their pulsation periods being higher on average (see the bottom panel of Figure \ref{figure:PLClmc}). 

The apparent high luminosities among the disc-type objects was also reported by \citet{groenewegen2017a}, which was attributed to a 
potential contribution in flux from the companion. Given the high luminosity of the primary star, this seems to be highly improbable.

\subsection{($O-C$) variability} \label{section:o_cdiscus}
\citet{percy2005} used time series data of the order of several decades to trace the period evolution and 
hence the evolutionary stage of RV Tauri stars across the HRD. The downward bending parabola in the ($O-C$) variation of the two RV Tauri stars they studied provides  a
clear indication of an overall decrease in period and hence, a decrease in overall radius. This indicates that the stars are evolving towards higher temperatures at
constant luminosity, thereby complying with post-AGB evolution.

Despite our $\sim$ 2500 days time series data span, we computed the ($O-C$) as an attempt to trace the period evolution. Indeed, we notice a change in period 
for four stars in the LMC, namely OGLE-LMC-T2CEP-119, OGLE-LMC-T2CEP-149, OGLE-LMC-T2CEP-190, and OGLE-LMC-T2CEP-191. Closer inspection of the ($O-C$) curvature showed that the periods of
OGLE-LMC-T2CEP-119 and OGLE-LMC-T2CEP-190 are increasing and those of OGLE-LMC-T2CEP-149 and OGLE-LMC-T2CEP-191 are decreasing (see Figure \ref{figure:p_dot} in Appendix \ref{appendix:AppendixE}). Our results are in accordance with 
those reported by \citet{groenewegen2017a} for these stars. Moreover, we note an interesting feature in the ($O-C$) of OGLE-LMC-T2CEP-149: there is a sudden 
period jump at JD $\sim$ 3800 days (also reported by \citet{groenewegen2017a}) and a steep decrease in the curvature as from JD $\sim$\,4200 days.
The reason behind this is not clear.

We speculate that the period change that we observe is a likely consequence of random cycle-to-cycle variations \citep{percy2005}, since the time span of our data 
is not enough to derive any strong evolutionary conclusion. Random cycle-to-cycle variations results in wave-like patterns in the ($O-C$). 
One possibility could be that the cyclic changes are 
the result of large-scale convectional turbulence in the stellar envelope or periodic thermal pulses in the stellar interiors \citep{percy91}. Moreover, \citet{zsoldos88} 
and \citet{zsoldos91b} found essentially cyclic variations in the ($O-C$) diagrams of AC Her and SS Gem, respectively. These works suggest the long-term modulation might be due to either internal structural changes or external effects, for example binarity.

The timescale at which we can readily predict a period change in RV Tauri stars and thus their period evolution seems to be of the order of $\sim$\,25000 days (TX Per) 
to $\sim$\,30000 days (UZ Oph) \citep{percy2005}. Thus, time series data of the order of many decades is required to derive any strong conclusions about the period evolution 
and hence their evolutionary nature across the HRD, from a pulsational point of view.

\section{Conclusions}   \label{section:conclusion}  
The  late stages of stellar evolution from the AGB to the WD phase are still poorly understood. The study of RV Tauri stars could potentially increase this 
understanding mainly because their luminosities can be constrained accurately via the PLC relation followed by the type II Cepheids.

We have shown that the SEDs of RV Tauri stars in the Magellanic clouds can be classified in three groups: dusty, non-dusty, and uncertain.
The SEDs of the dusty stars indicate that when dust is present in RV Tauri stars, it exists only in the form of a stable disc.
Given the strong link between the presence of a disc and binarity, these disc-type RV Tauri stars are likely all binaries and follow a binary evolution channel. 
The RVb phenomenon has enabled us to detect two binaries among the disc-types (OGLE-LMC-T2CEP-032 and OGLE-LMC-T2CEP-200) that have orbital periods of 916 and 850 days.
The derived luminosities and position on the HRD of the dusty RV Tauri stars suggest that while the luminous RV Tauri stars with dust indeed comply with post-AGB evolution, 
the low-luminosity stars are post-RGB objects.

The derived position of the non-dusty RV Tauri stars on the HRD suggests that they should have undergone an episode of mass loss after which the 
remnant mass should be seen in the form of an IR excess in their SEDs at this stage of their evolution. However, no dust is detected. We speculate that if these are single stars, they are
low-initial mass (< 1.25 M$_{\odot}$) and low-luminous post-AGB stars in which the dust has dissipated shortly after the start of their post-AGB phase. However, if they are binaries, it is most likely that they are in a post-RGB stage of evolution as a result of binary interaction on the RGB. In this evolutionary scenario, the low-mass primary evolved slow enough
for the dusty disc to have dispersed.

Based on our K-S statistics, the objects with uncertain SEDs are not a too different population from non-dusty objects in terms of luminosities. 
We speculate that the objects with uncertain SEDs could be stars in which the dust has dispersed and the few IR points in excess are arising because of remnant dust.
More SED data at longer wavelengths would be required to better probe their SED characteristics, thereby putting these objects in an evolutionary context.

One of the most striking results of this paper is the statistical difference showing the high significance that the dusty RV Tauri stars are 
on average more luminous than the non-dusty RV Tauri stars. In principle, we would expect the dusty RV Tauri stars to appear at lower luminosities 
than the non-dusty as a consequence of their evolution being interrupted by a phase of binary interaction. The dusty stars being most luminous may indicate 
that these come from a more massive progenitor population than the rest of the RV Tauri stars. Spectroscopic studies are needed to investigate this.                                                                                                                                                                                                                                                                                                                   

Four stars in the LMC sample display variability in their ($O-C$) diagram. We speculate that the variability is most likely due to random cycle-to-cycle variations in the time series or due to binarity,
although pulsation evolution on this timescale cannot be completely ruled out. Time series of the order of many decades are a prerequisite to probe their evolutionary status, pulsation-wise. Whether RV Tauri stars have a dusty disc or do not have proven circumstellar dust, we conclude that these stars cannot simply be seen as post-AGB stars evolving on tracks predicted by single star evolutionary models. 

\begin{acknowledgements} 
This research has been carried out based on the publicly available data from the OGLE-III survey.
RM, DK, and HVW acknowledge support from the Research Council of K.U. Leuven under contract GOA/13/012 and the Belgian Science Policy Office under contract BR/143/A2/STARLAB. HVW acknowledges additional support from the Research Council of K.U. Leuven under contract C14/17/082.
The research leading to these results has (partially) received funding from the Fonds Wetenschappelijk Onderzoek - Vlaanderen (FWO) under the grant agreement G0H5416N (ERC Opvangproject).
We used the following internet-based resources: NASA Astrophysics Data System for bibliographic services; Simbad;
the VizieR on-line catalogue operated by CDS; and the ASAS and AAVSO photometric databases. RM thanks Brent Miszalski, Bram Buysschaert, Jacques Kluska, Mike Laverick, and Jack Gallimore for their helpful comments and inputs. We thank the anonymous referee for very helpful
comments and input, which helped improve the manuscript.
\end{acknowledgements} 

 \bibliographystyle{aa}          
 \bibliography{allreferences}       

  \appendix
  \section{SEDs LMC} \label{appendix:AppendixA}
 \captionsetup[sub]{font=small,labelfont={bf,sf}}
 \begin{figure*}
   \subfloat[OGLE-LMC-T2CEP-003 ]{%
     \includegraphics[width=0.33\textwidth]{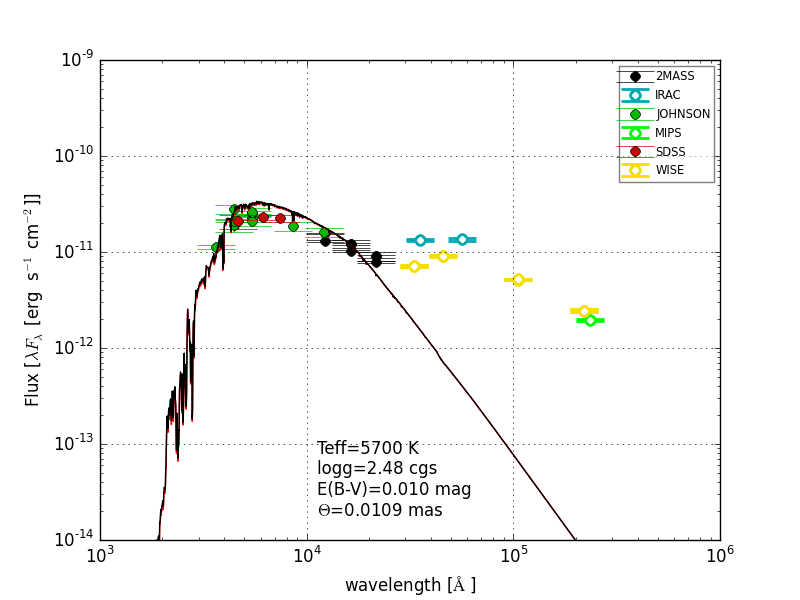}          \label{figure:seddisc1}
   }
   \subfloat[OGLE-LMC-T2CEP-014]{%
     \includegraphics[width=0.33\textwidth]{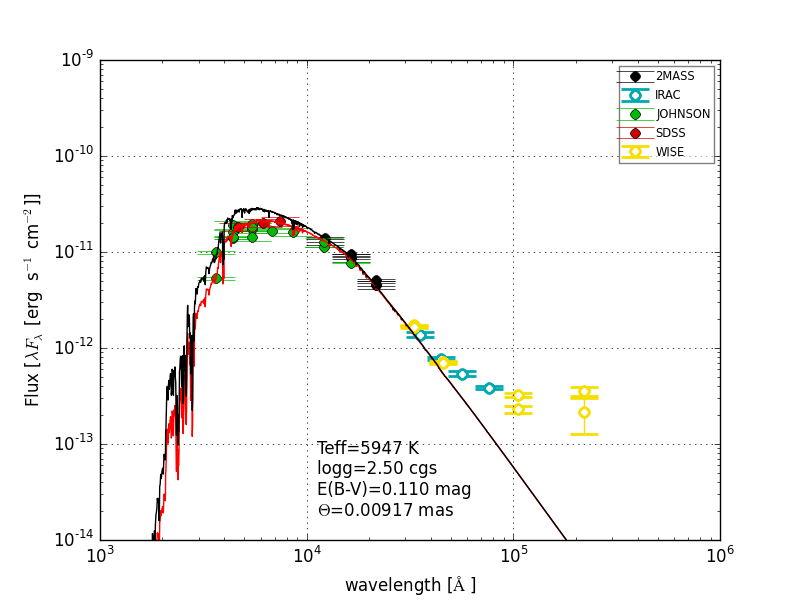}        \label{figure:seddisc2}
   }
   \subfloat[OGLE-LMC-T2CEP-015]{%
     \includegraphics[width=0.33\textwidth]{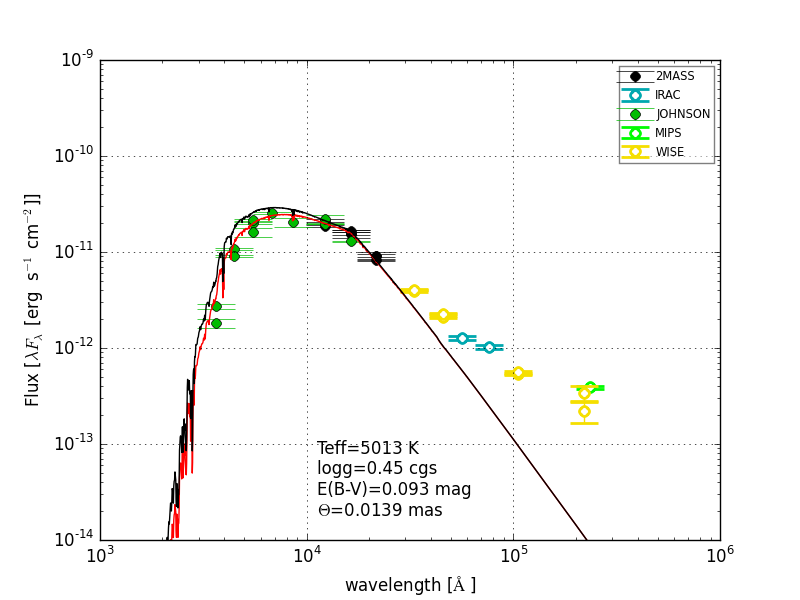}       \label{figure:seddisc3}
   }\\

   \subfloat[OGLE-LMC-T2CEP-029]{%
     \includegraphics[width=0.33\textwidth]{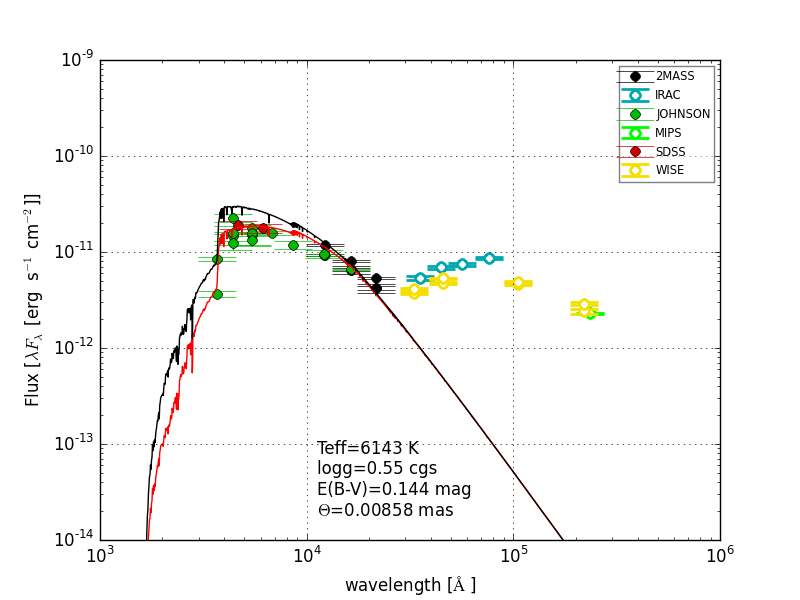}        \label{figure:seddisc4}
   }
   \subfloat[OGLE-LMC-T2CEP-032]{%
     \includegraphics[width=0.33\textwidth]{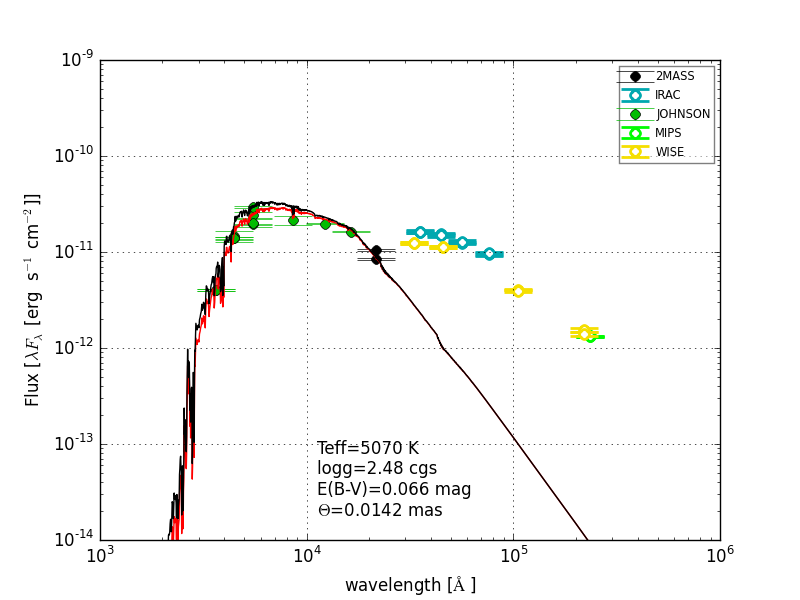}       \label{figure:seddisc5}
   }
   \subfloat[OGLE-LMC-T2CEP-067]{%
     \includegraphics[width=0.33\textwidth]{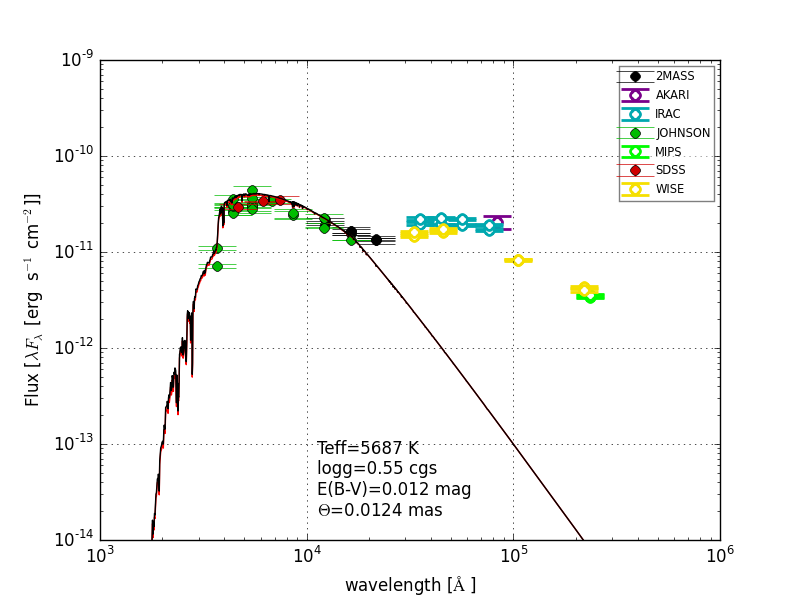}        \label{figure:seddisc6}
   }\\

   \subfloat[OGLE-LMC-T2CEP-091]{%
     \includegraphics[width=0.33\textwidth]{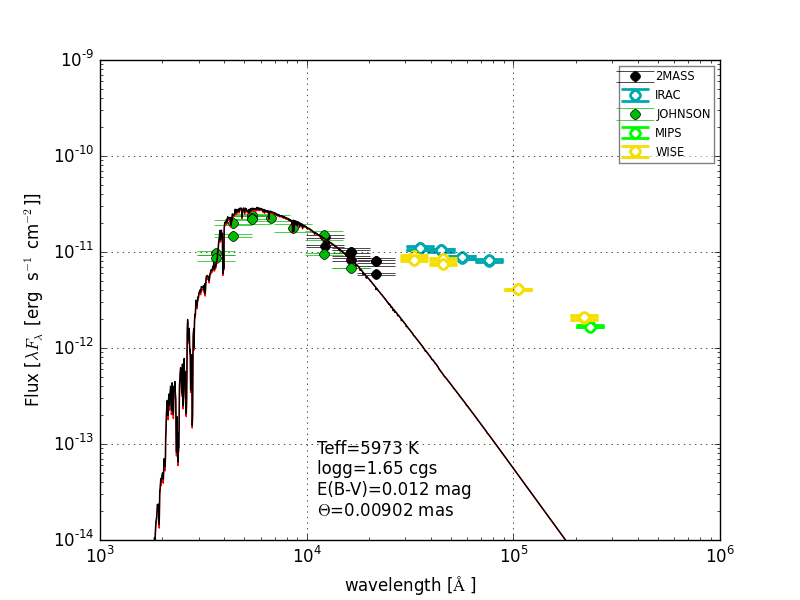}      \label{figure:seddisc7}
   }
   \subfloat[OGLE-LMC-T2CEP-104]{%
     \includegraphics[width=0.33\textwidth]{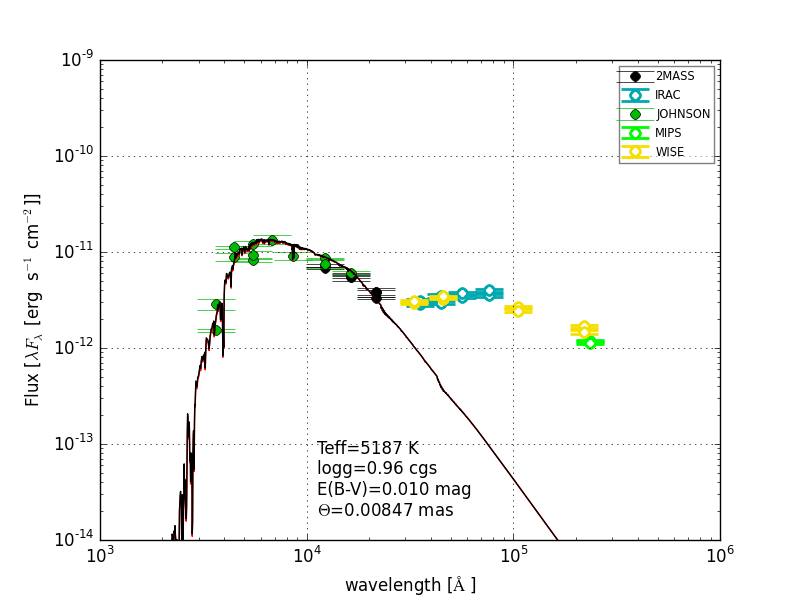}      \label{figure:seddisc8}
   }
      \subfloat[OGLE-LMC-T2CEP-119]{%
     \includegraphics[width=0.33\textwidth]{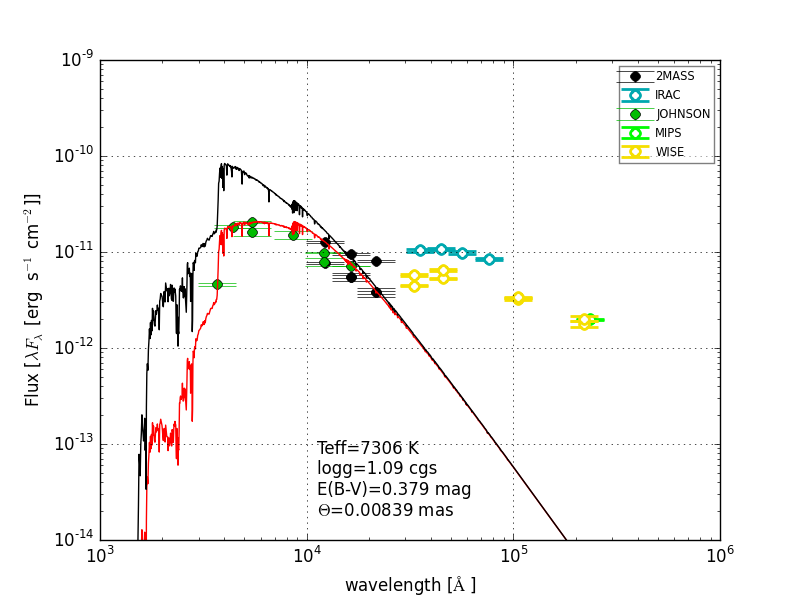}     \label{figure:seddisc9}
   }\\
   \caption{LMC disc SEDs fitted with Kurucz model (Red). The black line shows the de-reddened SED model.}
 \end{figure*} 
 
 \begin{figure*} \label{figure:modulation}

   \subfloat[OGLE-LMC-T2CEP-129]{%
     \includegraphics[width=0.33\textwidth]{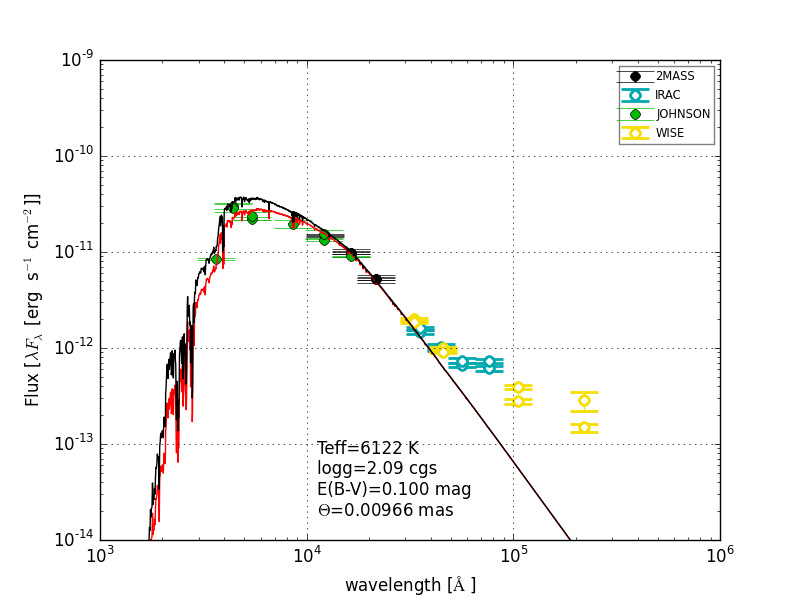}     \label{figure:seddisc10}
   }
   \subfloat[OGLE-LMC-T2CEP-147]{%
    \includegraphics[width=0.33\textwidth]{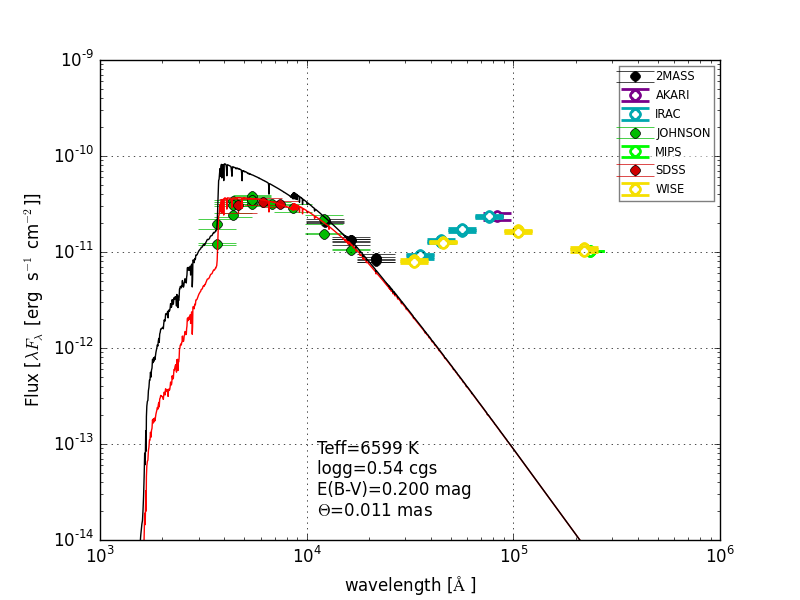}     \label{figure:seddisc11}
   }
   \subfloat[OGLE-LMC-T2CEP-149]{%
     \includegraphics[width=0.33\textwidth]{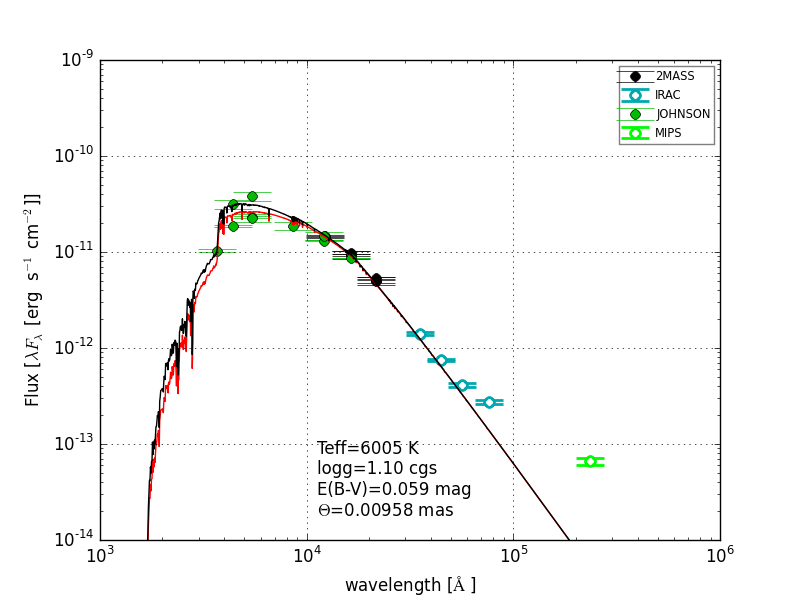}    \label{figure:seddisc12}
   }\\

   \subfloat[OGLE-LMC-T2CEP-174]{%
     \includegraphics[width=0.33\textwidth]{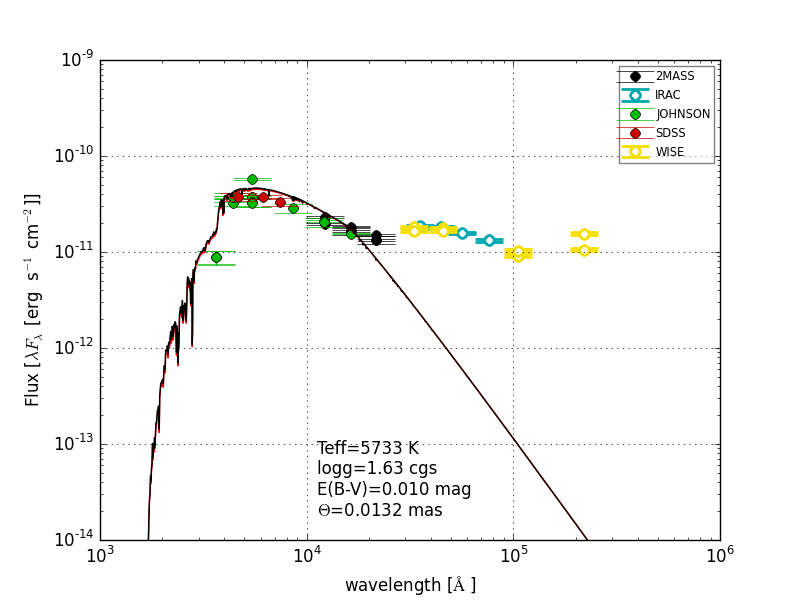}    \label{figure:seddisc13}
   }
   \subfloat[OGLE-LMC-T2CEP-180]{%
     \includegraphics[width=0.33\textwidth]{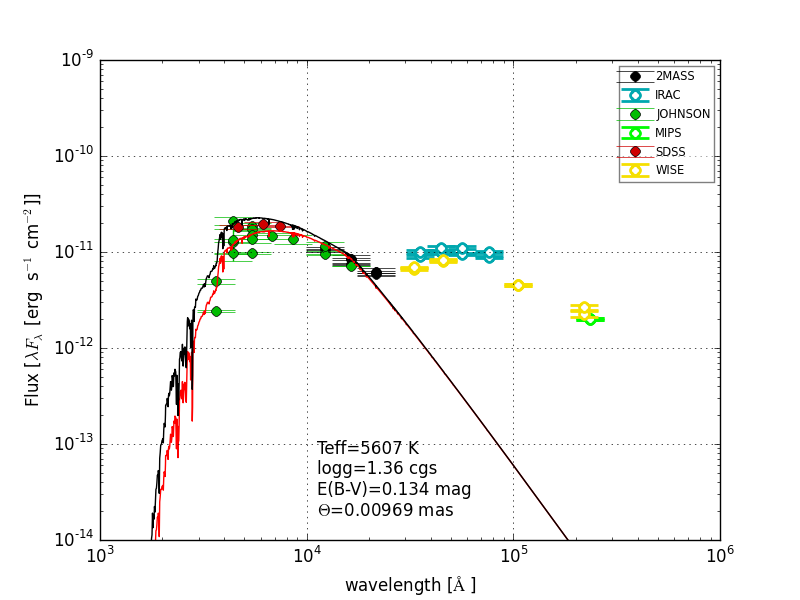}   \label{figure:seddisc14}
   }
   \subfloat[OGLE-LMC-T2CEP-191\label{}]{%
     \includegraphics[width=0.33\textwidth]{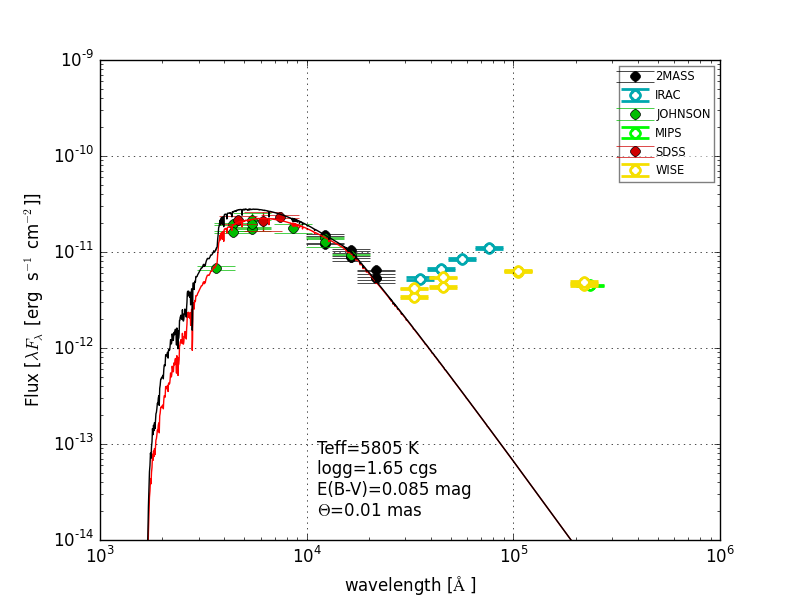}     \label{figure:seddisc15}
   } \\
   \subfloat[OGLE-LMC-T2CEP-200\label{subfig:nowise2}]{%
     \includegraphics[width=0.33\textwidth]{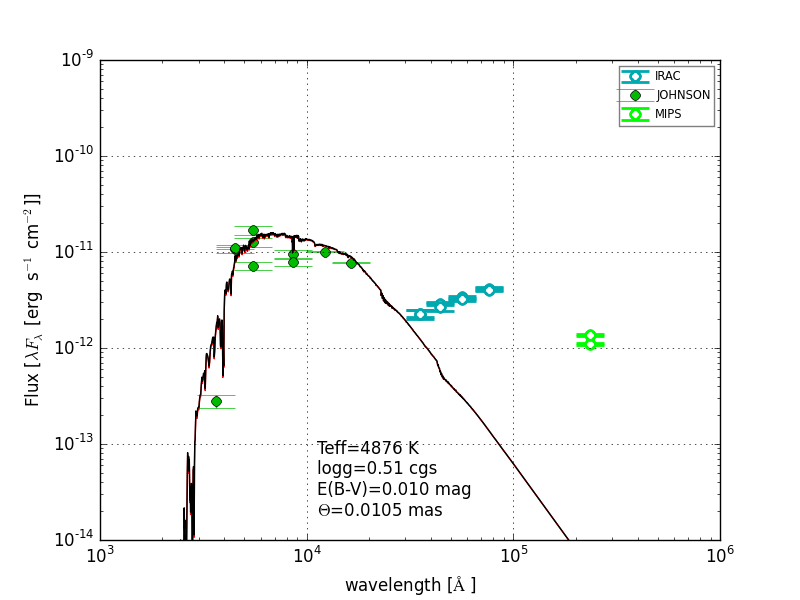} \label{figure:seddisc16}
   }
   \caption{LMC disc SEDs fitted with Kurucz model (red) and de-reddened model (black). - continued.}
 \end{figure*}   
 %
 %
 
 \begin{figure*}
   \subfloat[OGLE-LMC-T2CEP-011]{%
     \includegraphics[width=0.33\textwidth]{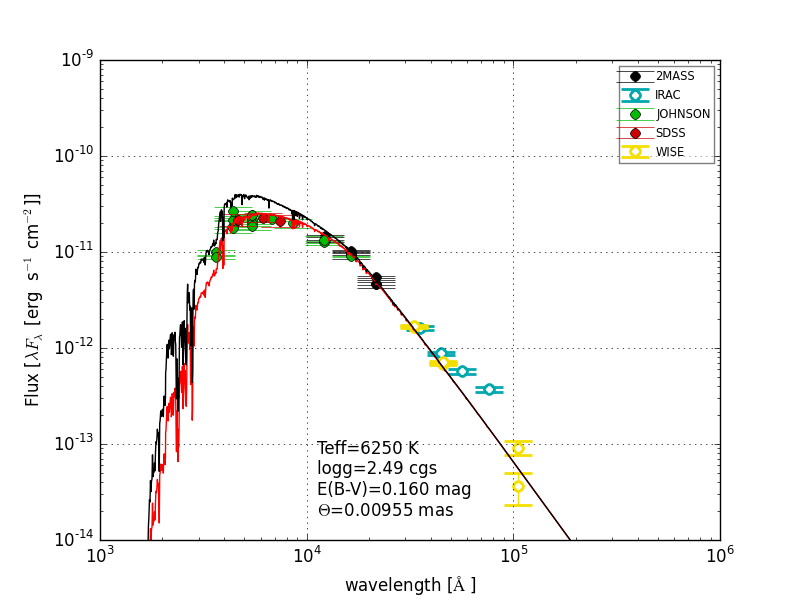} \label{figure:uncertsed1}
   }
   \subfloat[OGLE-LMC-T2CEP-016 ]{%
     \includegraphics[width=0.33\textwidth]{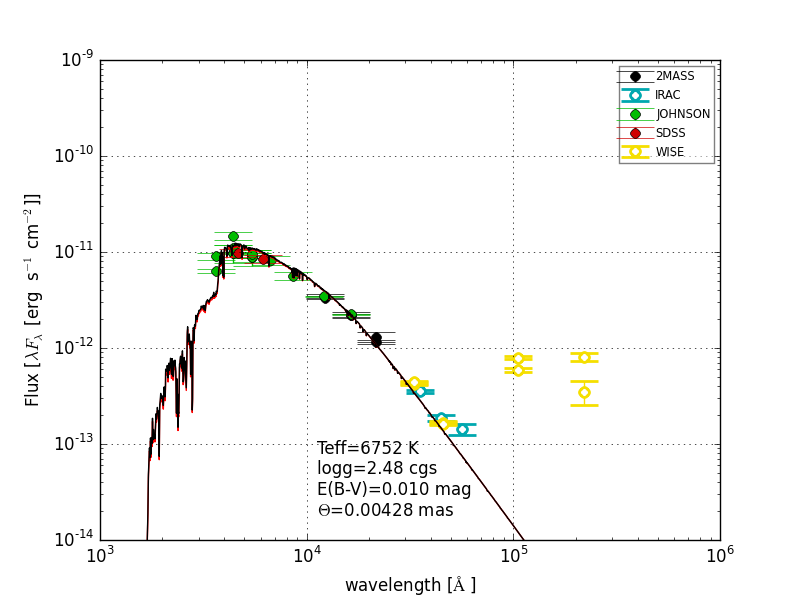} \label{figure:uncertsed2}
   }
    \subfloat[OGLE-LMC-T2CEP-025]{%
     \includegraphics[width=0.33\textwidth]{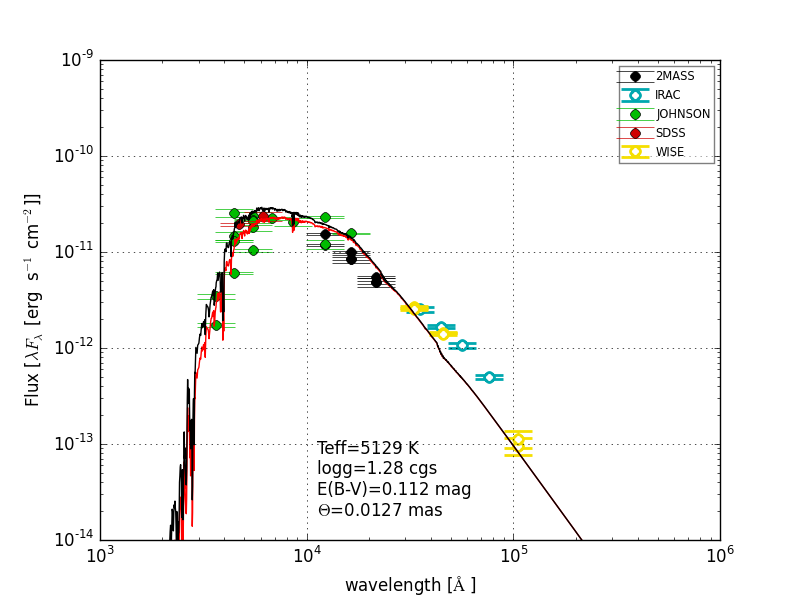} \label{figure:uncertsed3}
   }\\
   
    \subfloat[OGLE-LMC-T2CEP-045]{%
    \includegraphics[width=0.33\textwidth]{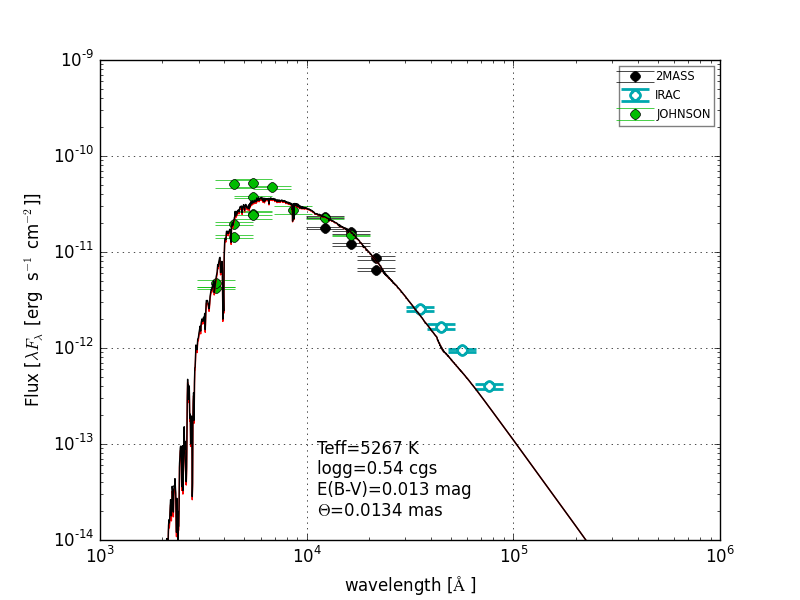} \label{figure:uncertsed4}
   }
    \subfloat[OGLE-LMC-T2CEP-050]{%
    \includegraphics[width=0.33\textwidth]{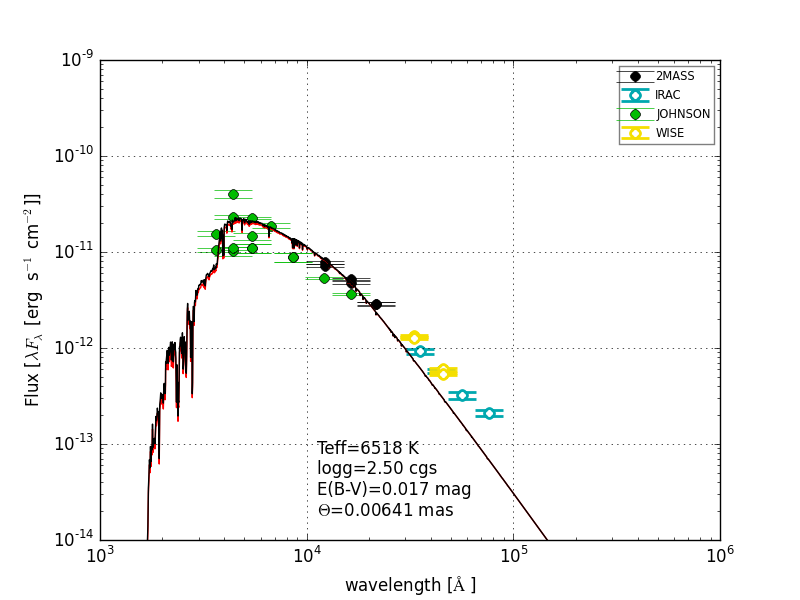} \label{figure:uncertsed5}
   }
   \subfloat[OGLE-LMC-T2CEP-055]{%
     \includegraphics[width=0.33\textwidth]{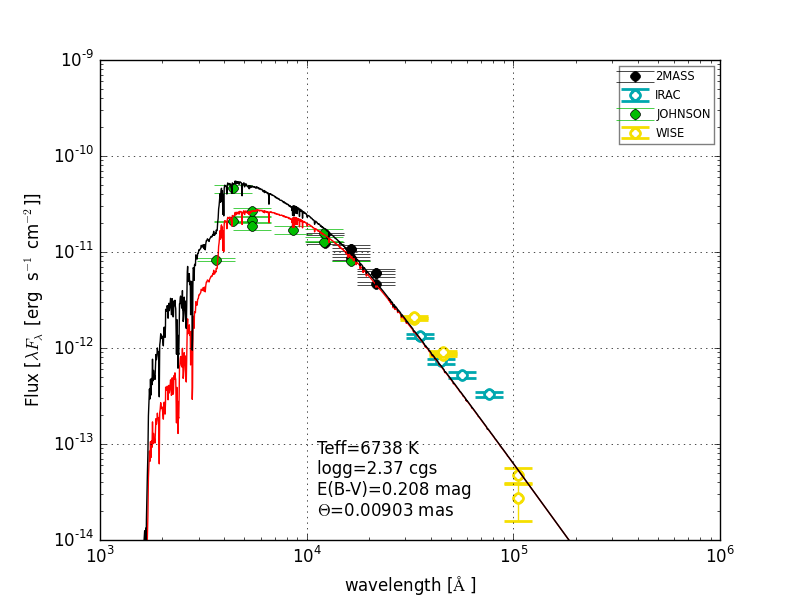} \label{figure:uncertsed6}
   }\\
   
    \subfloat[OGLE-LMC-T2CEP-058]{%
    \includegraphics[width=0.33\textwidth]{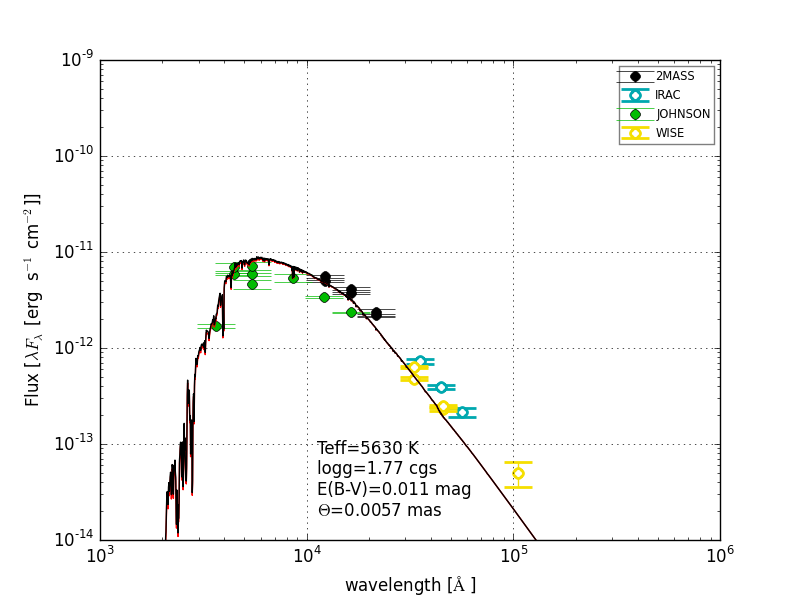} \label{figure:uncertsed7}
   }
   \subfloat[OGLE-LMC-T2CEP-065]{%
     \includegraphics[width=0.33\textwidth]{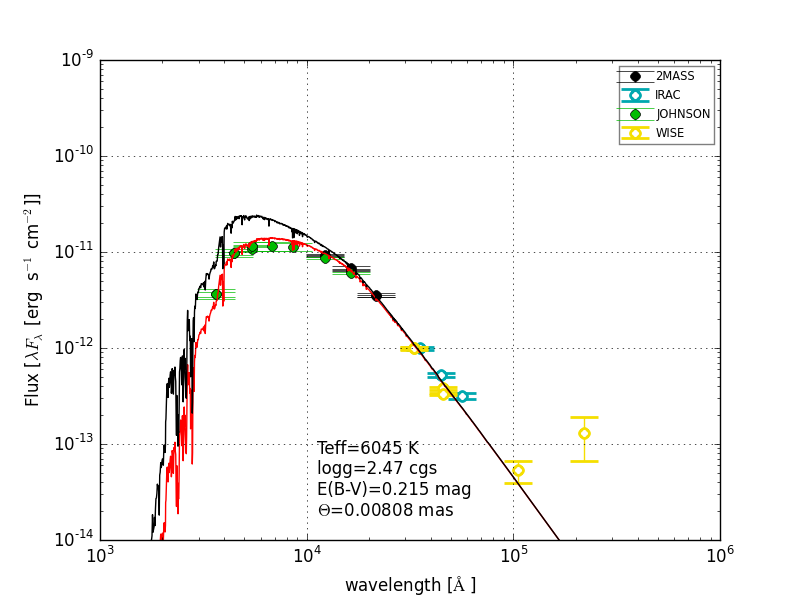} \label{figure:uncertsed8}
   }
   \subfloat[OGLE-LMC-T2CEP-075]{%
     \includegraphics[width=0.33\textwidth]{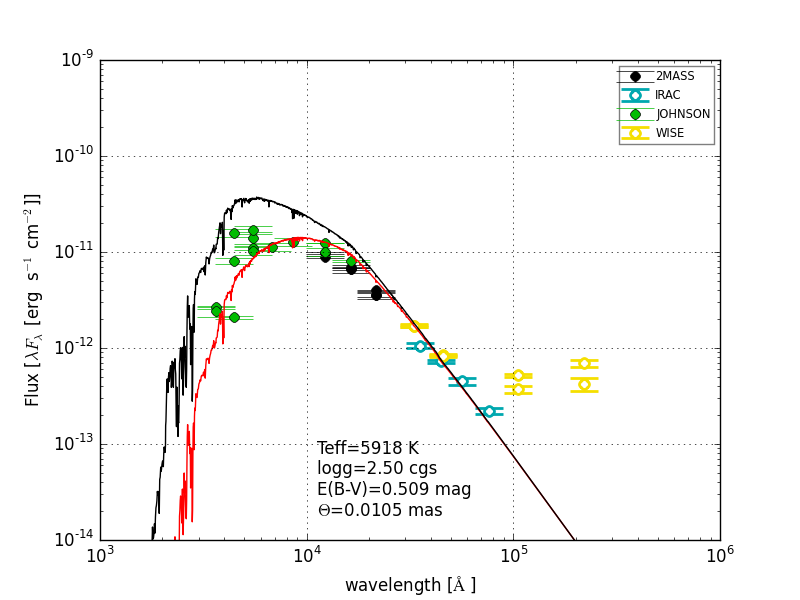} \label{figure:uncertsed9}
   }\\
   \caption{LMC uncertain SEDs fitted with Kurucz model (red) and de-reddened model (black).}
 \end{figure*} 
 
  \begin{figure*} 
    \subfloat[OGLE-LMC-T2CEP-080]{%
     \includegraphics[width=0.33\textwidth]{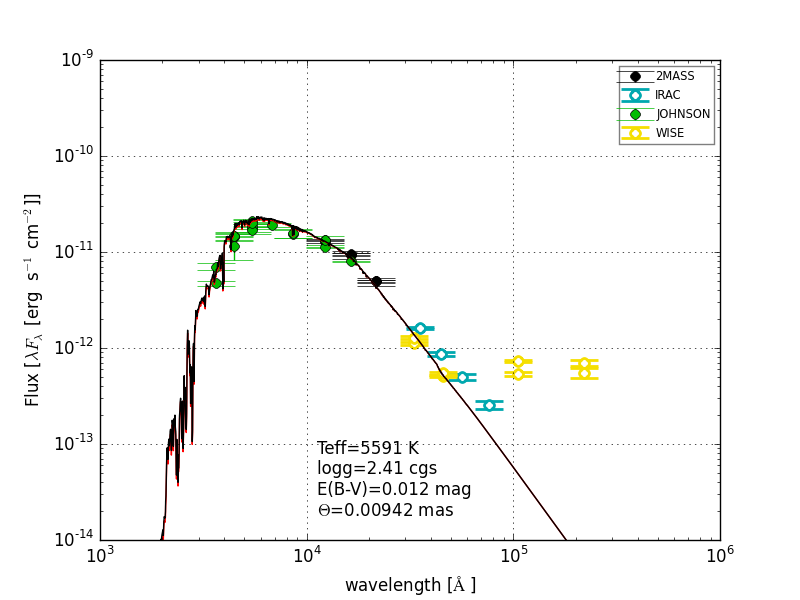} \label{figure:uncertsed10}
   }
   \subfloat[OGLE-LMC-T2CEP-112]{%
    \includegraphics[width=0.33\textwidth]{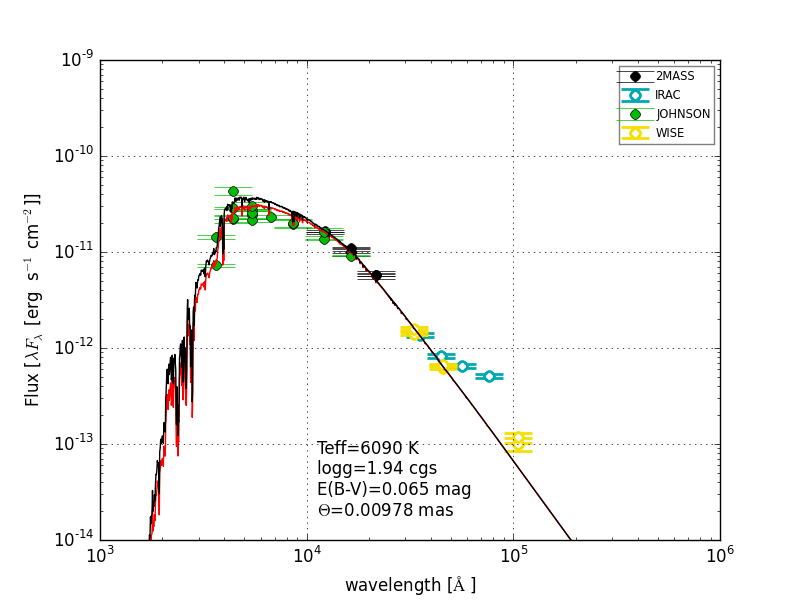} \label{figure:uncertsed11}
   }
     \subfloat[OGLE-LMC-T2CEP-115]{%
    \includegraphics[width=0.33\textwidth]{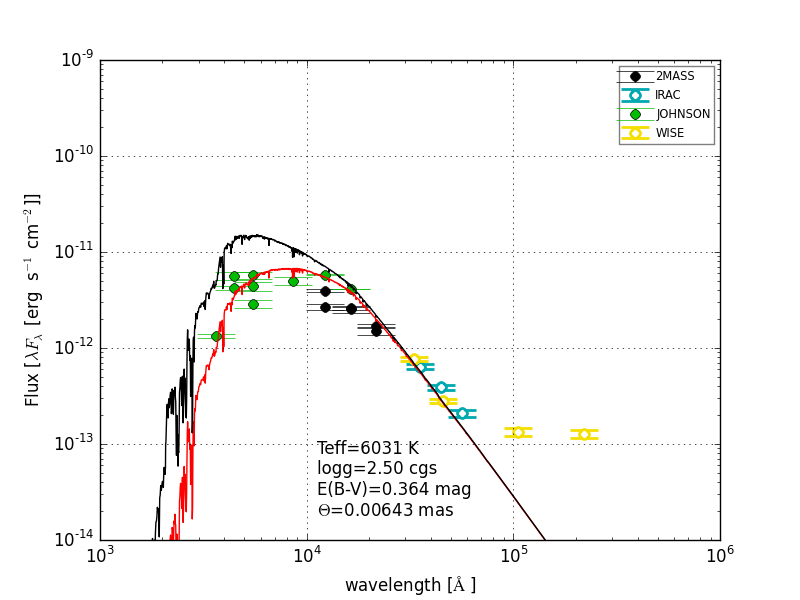} \label{figure:uncertsed12}
   }\\
   
    \subfloat[OGLE-LMC-T2CEP-125\label{subfig:dummy}]{%
    \includegraphics[width=0.33\textwidth]{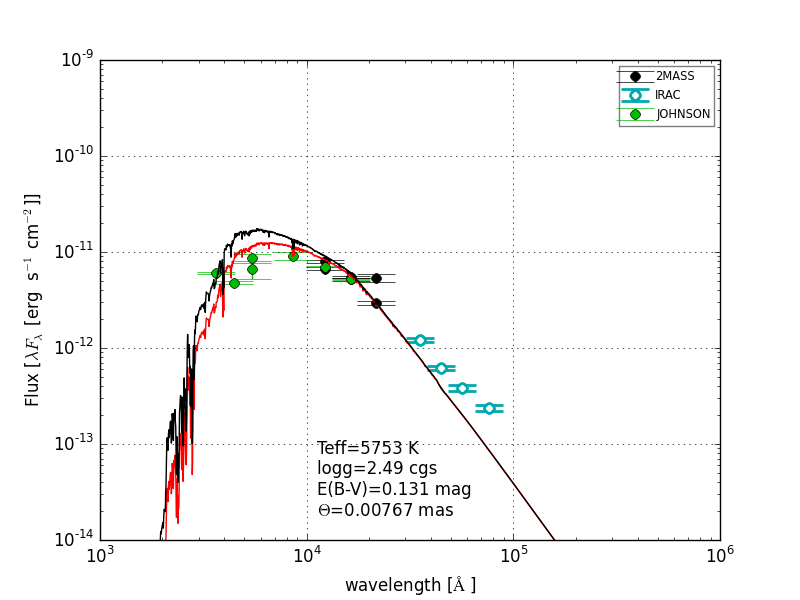} \label{figure:uncertsed13}
   }
   \subfloat[OGLE-LMC-T2CEP-162]{%
     \includegraphics[width=0.33\textwidth]{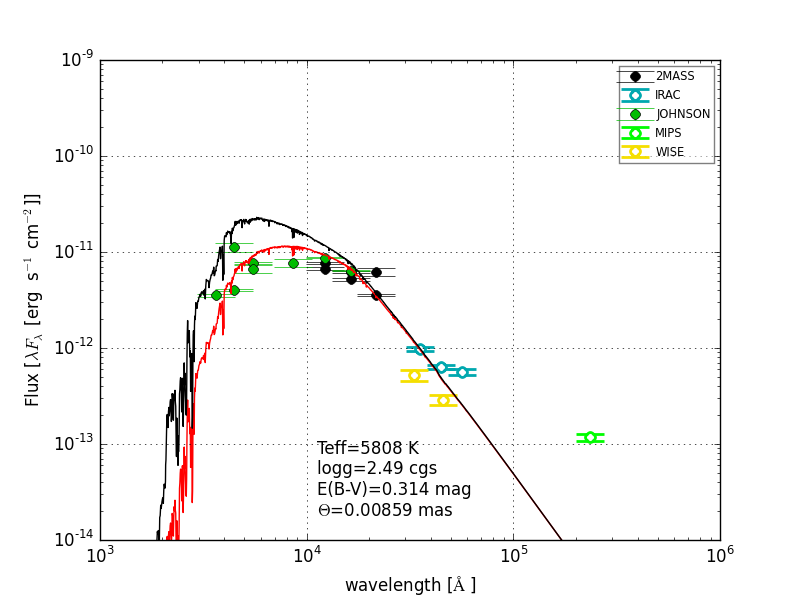} \label{figure:uncertsed14}
   } 
   \subfloat[OGLE-LMC-T2CEP-169]{%
     \includegraphics[width=0.33\textwidth]{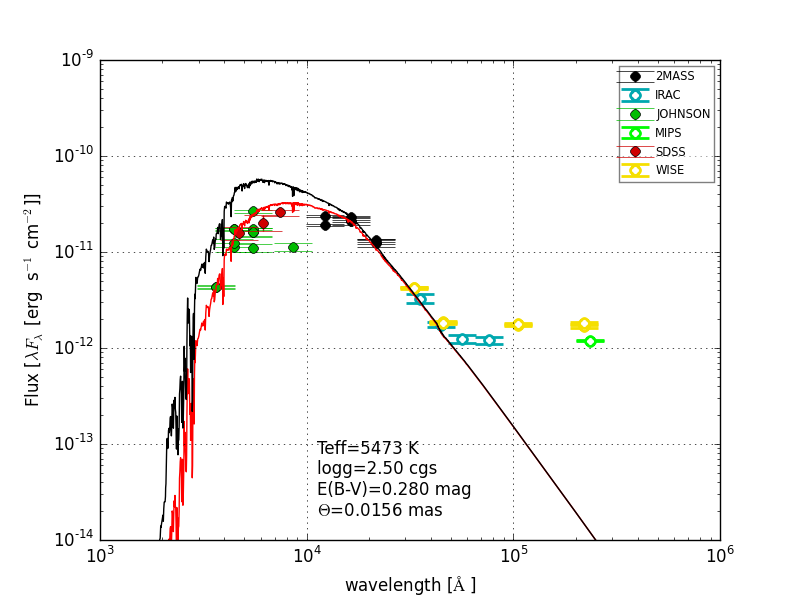} \label{figure:uncertsed15}
   }\\
      \subfloat[OGLE-LMC-T2CEP-199]{%
     \includegraphics[width=0.33\textwidth]{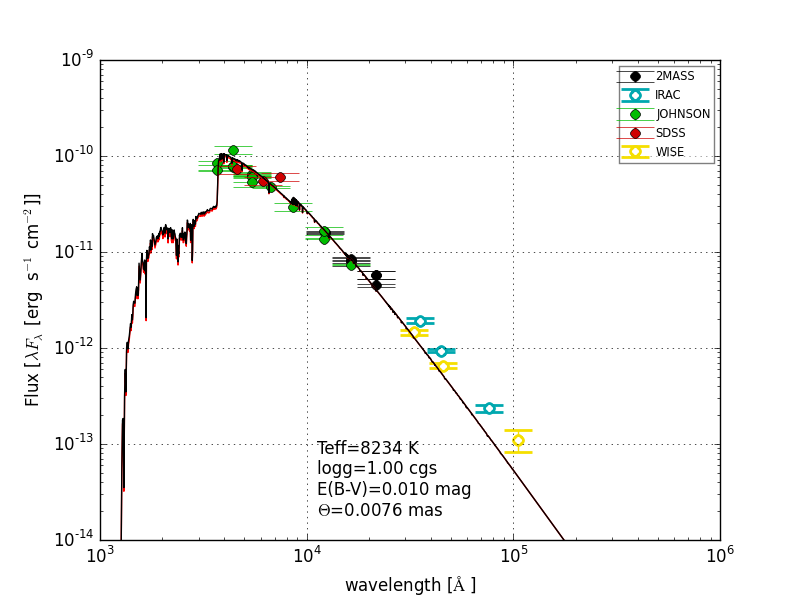} \label{figure:uncertsed16}
   }
   \subfloat[OGLE-LMC-T2CEP-202]{%
     \includegraphics[width=0.33\textwidth]{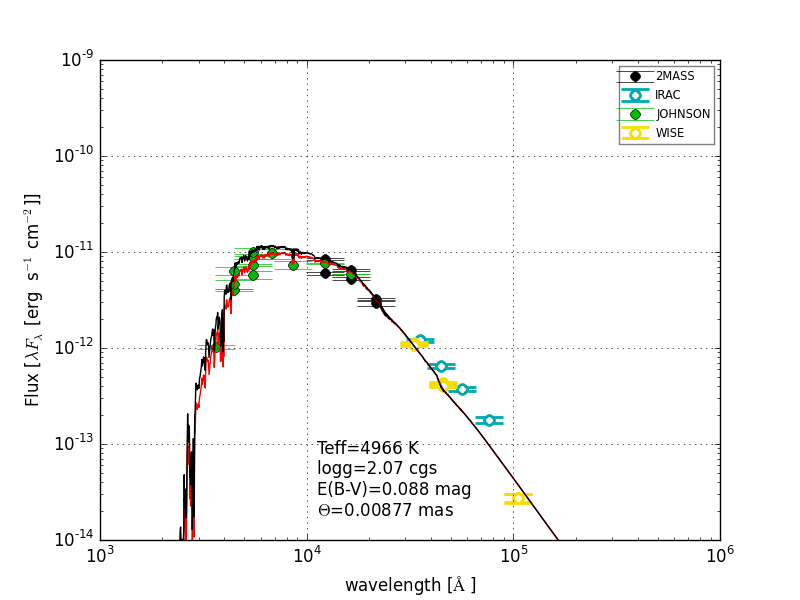} \label{figure:uncertsed17}
   }
   \caption{LMC uncertain SEDs - continued.}
 \end{figure*}

 
 \begin{figure*} \label{figure:non-IR}
   \subfloat[OGLE-LMC-T2CEP-005]{%
    \includegraphics[width=0.33\textwidth]{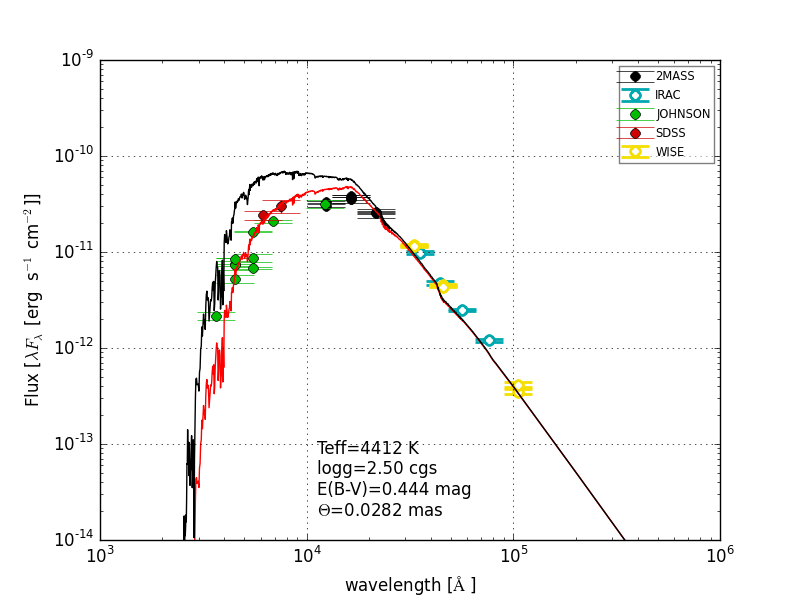} \label{figure:nonirsed1}
   }
   \subfloat[OGLE-LMC-T2CEP-051\label{subfig:ogle051}]{%
    \includegraphics[width=0.33\textwidth]{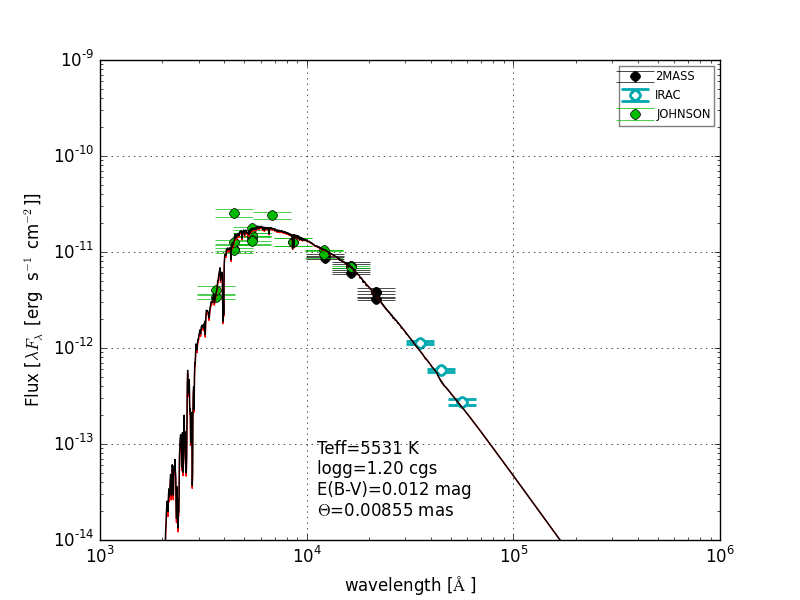} \label{figure:nonirsed2}
   } 
   \subfloat[OGLE-LMC-T2CEP-082]{%
    \includegraphics[width=0.33\textwidth]{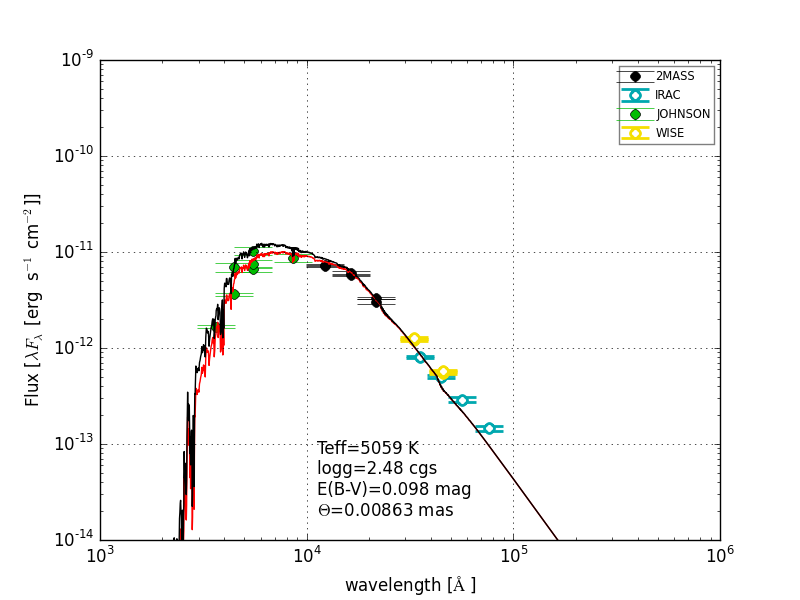} \label{figure:nonirsed3}
   }\\
   
   \subfloat[OGLE-LMC-T2CEP-108]{%
    \includegraphics[width=0.33\textwidth]{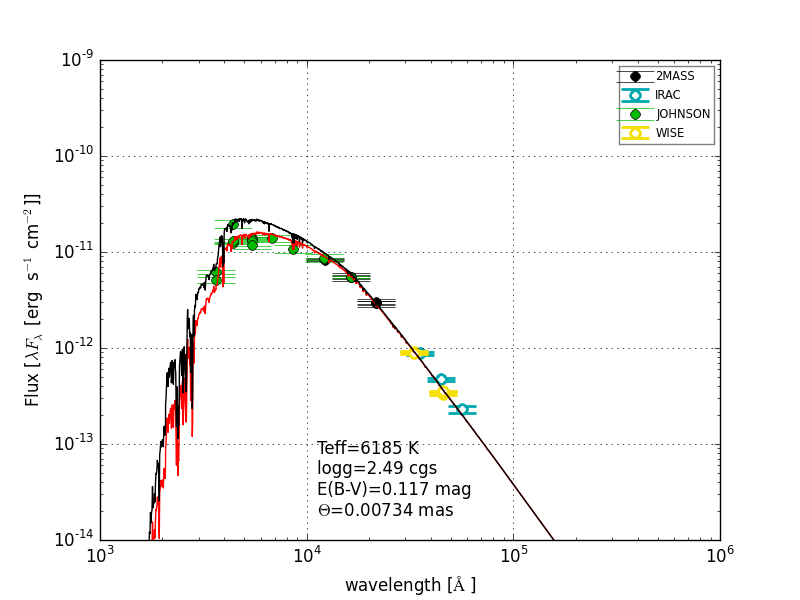} \label{figure:nonirsed4}
   }
     \subfloat[OGLE-LMC-T2CEP-135]{%
     \includegraphics[width=0.33\textwidth]{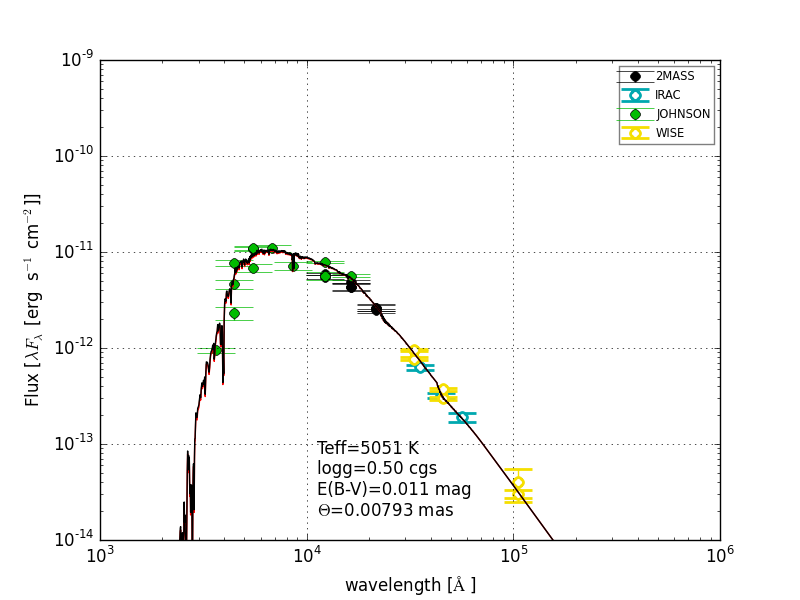} \label{figure:nonirsed5}
   } 
   \subfloat[OGLE-LMC-T2CEP-190]{%
    \includegraphics[width=0.33\textwidth]{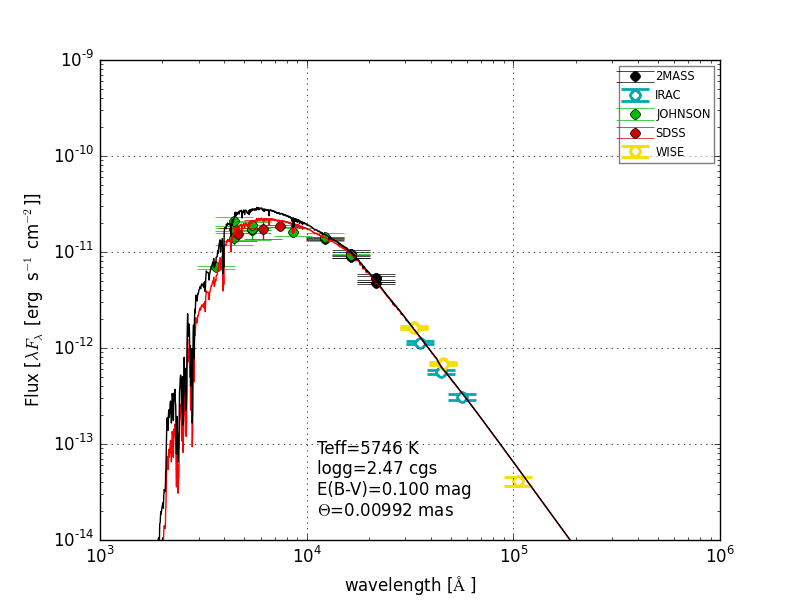} \label{figure:nonirsed6}
   }\\
   
   \subfloat[OGLE-LMC-T2CEP-192]{%
     \includegraphics[width=0.33\textwidth]{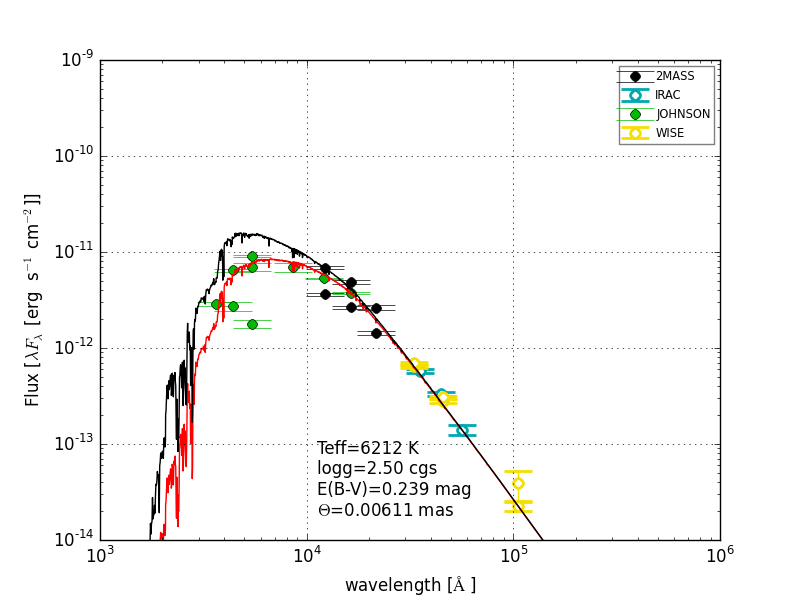} \label{figure:nonirsed7}
   } 
   \subfloat[OGLE-LMC-T2CEP-198]{%
     \includegraphics[width=0.33\textwidth]{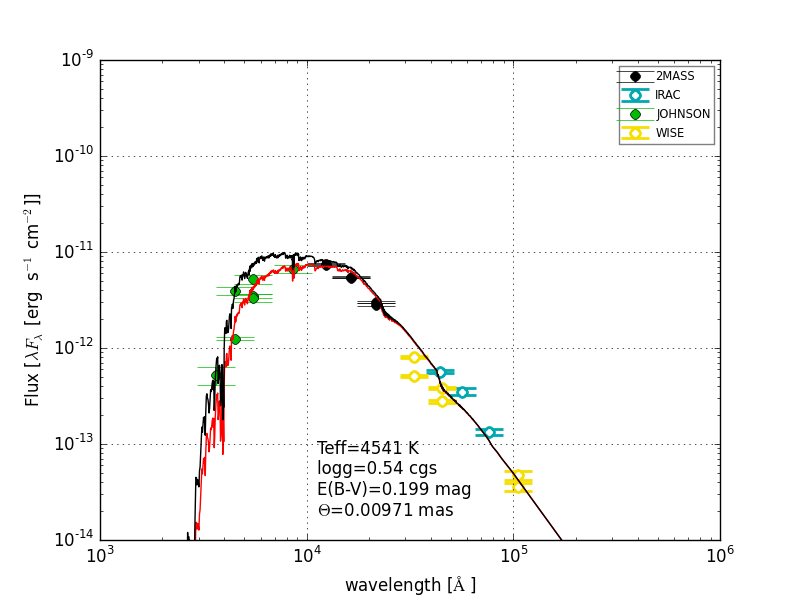} \label{figure:nonirsed8}
   } 
   \subfloat[OGLE-LMC-T2CEP-203 \label{subfig:ogle203}]{%
     \includegraphics[width=0.33\textwidth]{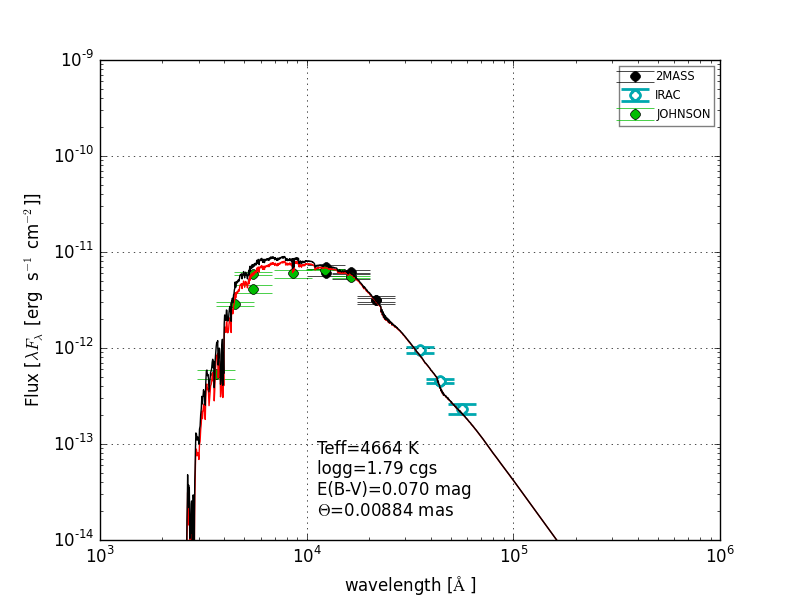} \label{figure:nonirsed9}
   }
   \caption{LMC non-IR SEDs fitted with Kurucz model (red). The black line shows the de-reddened SED model.}
   \label{fig:nonirsedsLMC}
 \end{figure*}    
 
 \section{SEDs SMC} \label{appendix:AppendixB}
 \begin{figure*}
   \subfloat[OGLE-SMC-T2CEP-018 ]{%
     \includegraphics[width=0.33\textwidth]{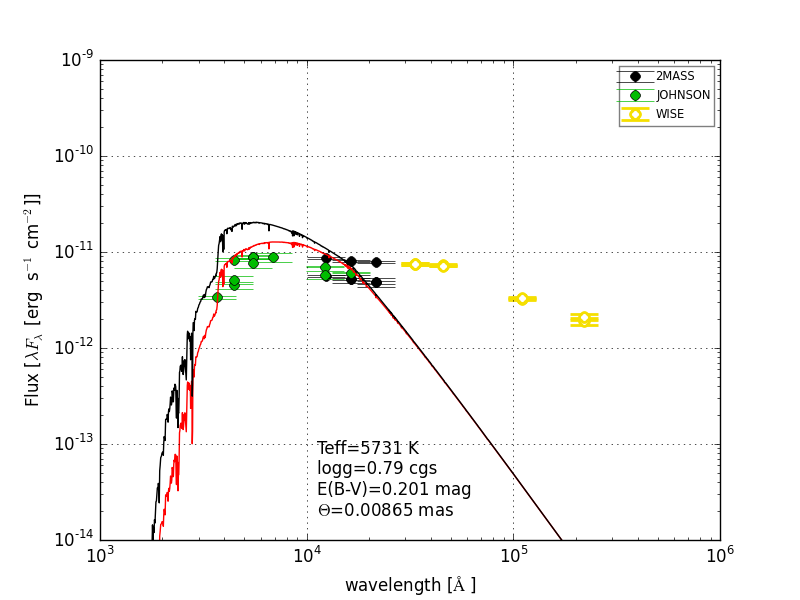}     \label{figure:sedsmcdisc1}
   }
   \caption{SMC disc SED fitted with Kurucz model (red). The black line shows the de-reddened SED model.}
 \end{figure*}  
 
 
 \begin{figure*}
   \subfloat[OGLE-SMC-T2CEP-012 ]{%
     \includegraphics[width=0.33\textwidth]{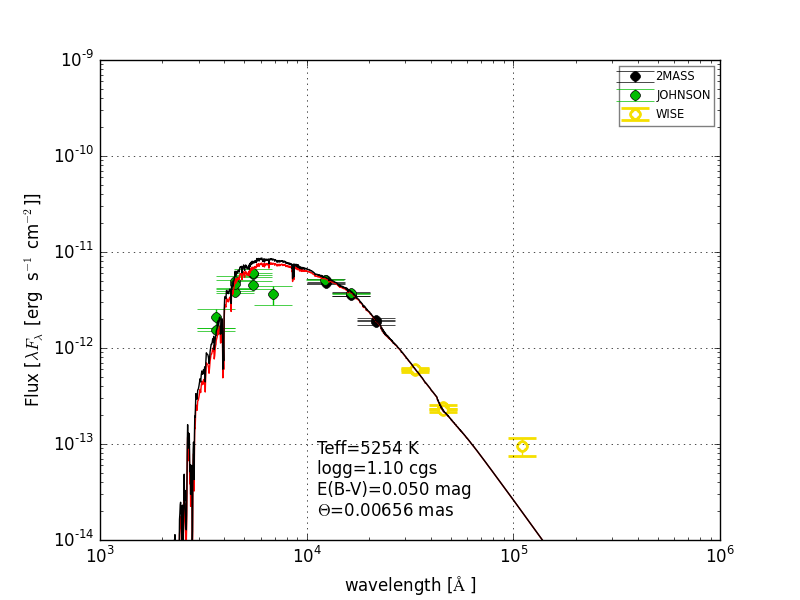} \label{figure:uncertsmc1}
   }
   \subfloat[OGLE-SMC-T2CEP-019 ]{%
     \includegraphics[width=0.33\textwidth]{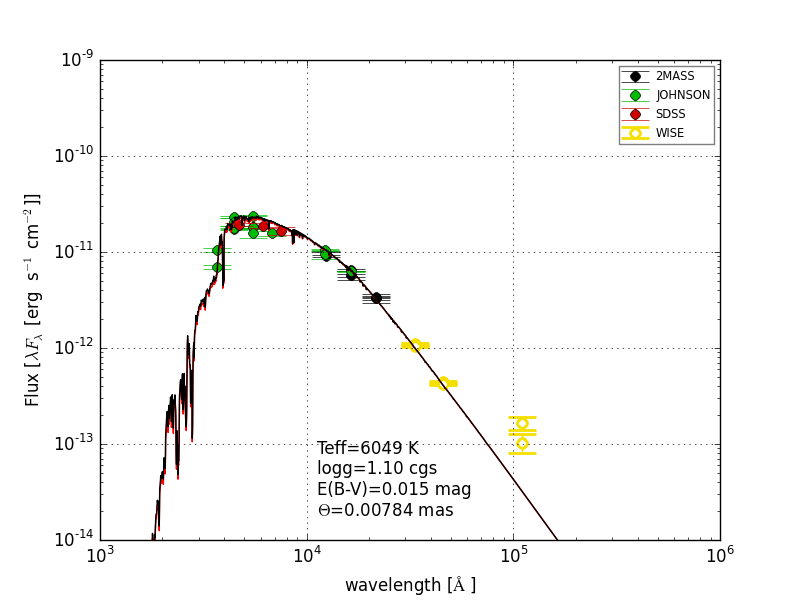} \label{figure:uncertsmc2}
   }
   \subfloat[OGLE-SMC-T2CEP-020 ]{%
     \includegraphics[width=0.33\textwidth]{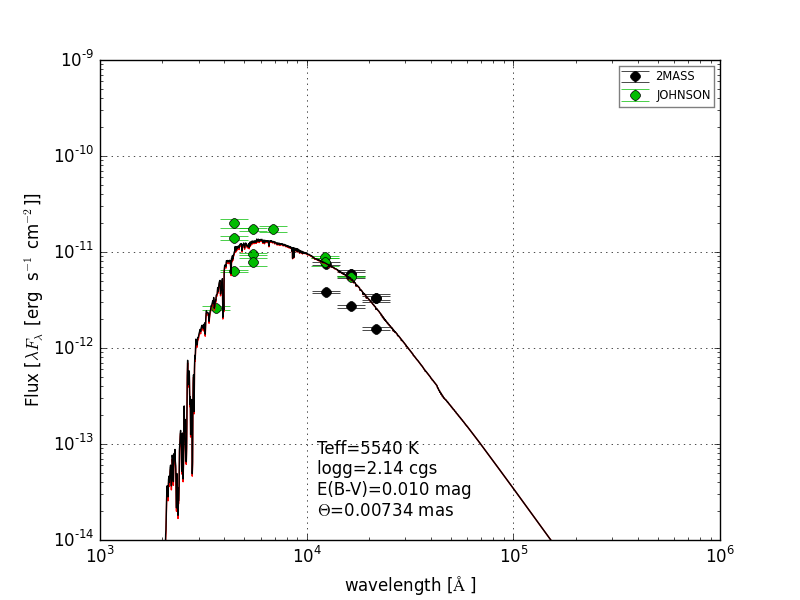} \label{figure:uncertsmc3}
   }\\

   \subfloat[OGLE-SMC-T2CEP-024 ]{%
     \includegraphics[width=0.33\textwidth]{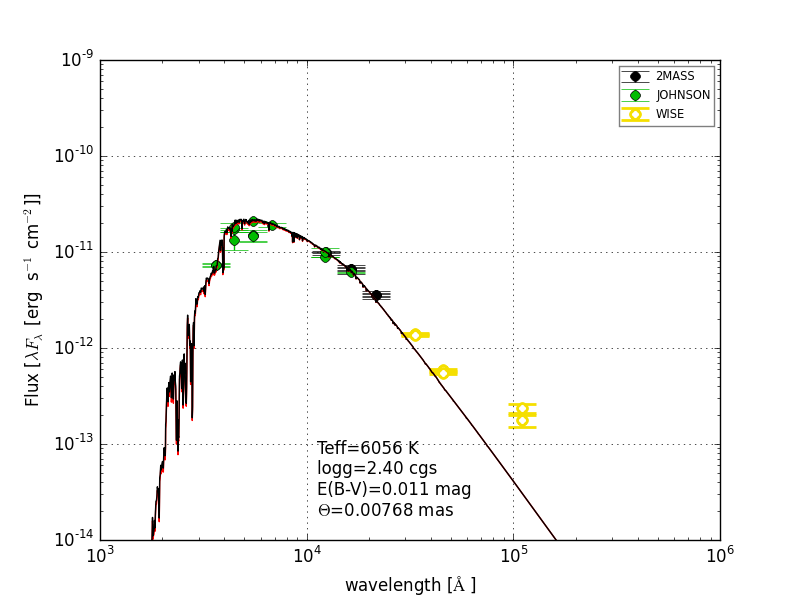} \label{figure:uncertsmc4}
   }  
   \subfloat[OGLE-SMC-T2CEP-043 ]{%
     \includegraphics[width=0.33\textwidth]{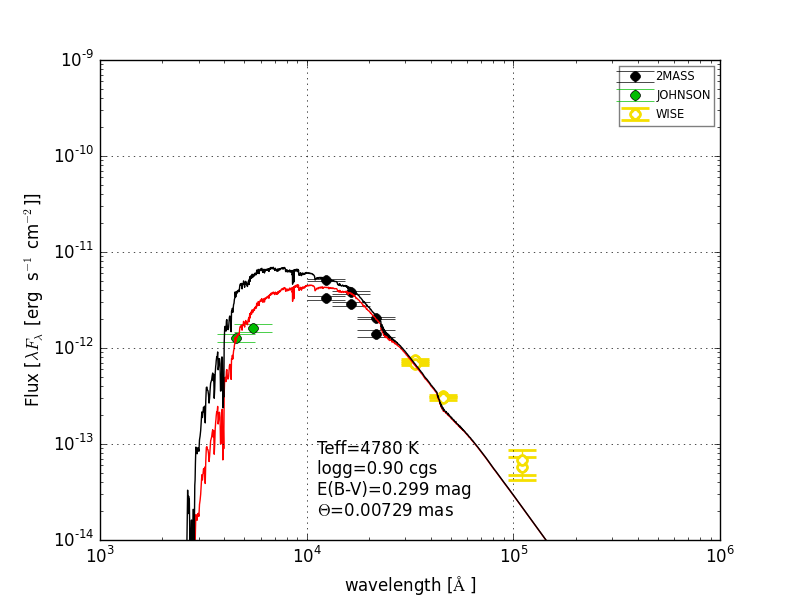} \label{figure:uncertsmc5}
   }
   \caption{SMC uncertain SEDs fitted with Kurucz model (red). The black line shows the de-reddened SED model.}
 \end{figure*}   
  \begin{figure*}
    \subfloat[OGLE-SMC-T2CEP-007 \label{figure:ellip1}]{%
      \includegraphics[width=0.33\textwidth]{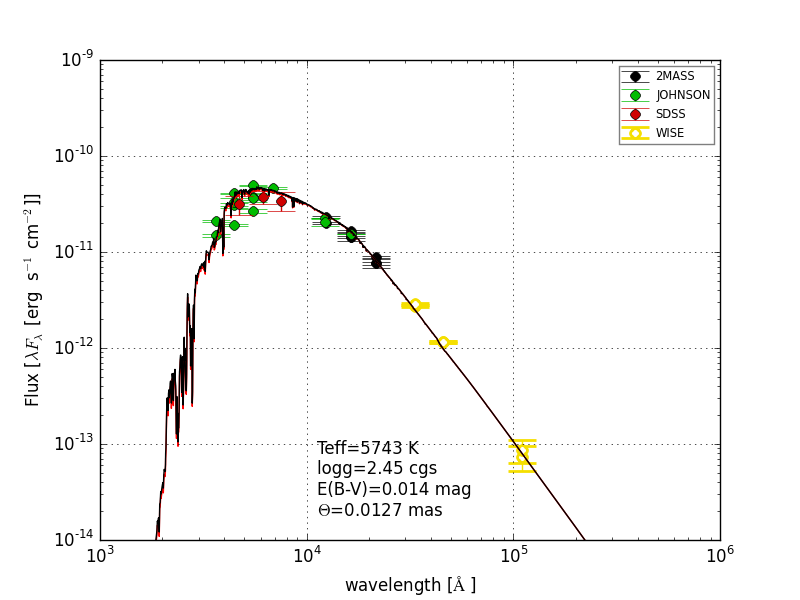} \label{figure:nonirsmc1}
    }
    \subfloat[OGLE-SMC-T2CEP-029 \label{figure:ellip2}]{%
      \includegraphics[width=0.33\textwidth]{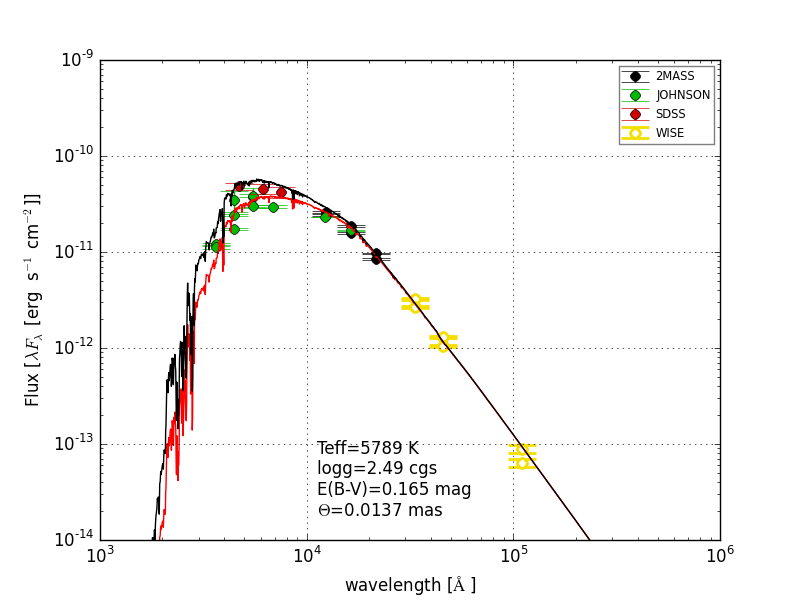} \label{figure:nonirsmc2}
    } 
    \subfloat[OGLE-SMC-T2CEP-041 ]{%
      \includegraphics[width=0.33\textwidth]{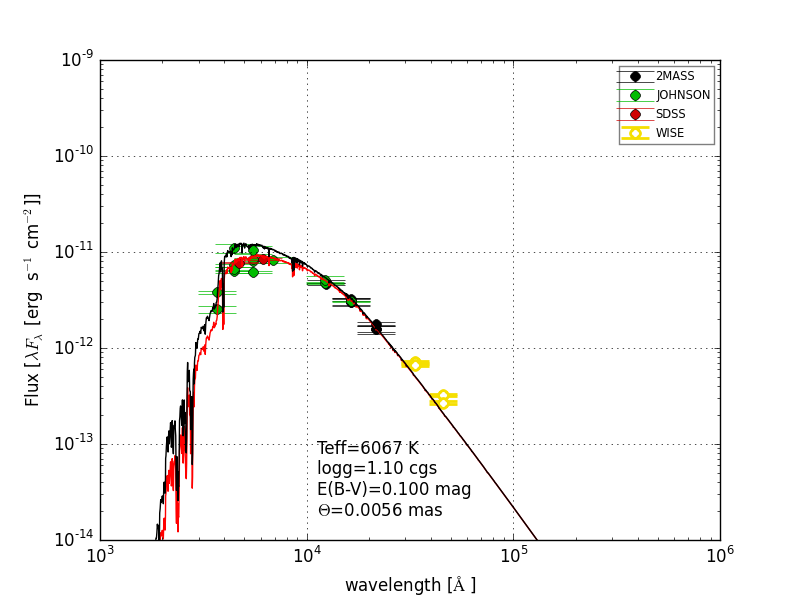} \label{figure:nonirsmc3}
    } 
    \caption{SMC non-IR SEDs fitted with Kurucz model (red). The black line shows the de-reddened SED model.}
  \end{figure*} 
 %
 \section{Pulsations LMC} \label{appendix:AppendixC}
 \begin{figure*}
  \subfloat[]{\includegraphics[width=0.33\textwidth]{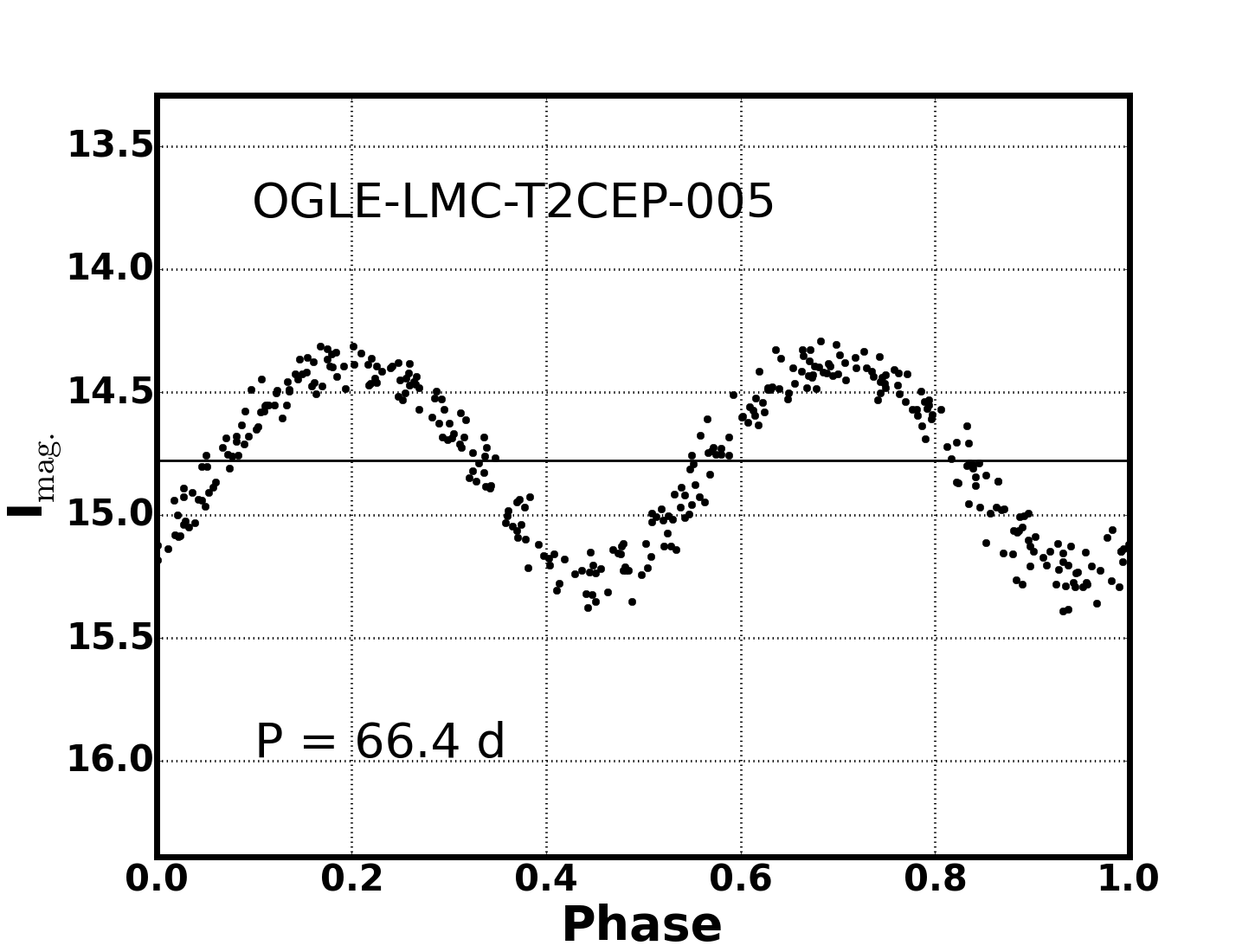}
  }
  \subfloat[]{\includegraphics[width=0.33\textwidth]{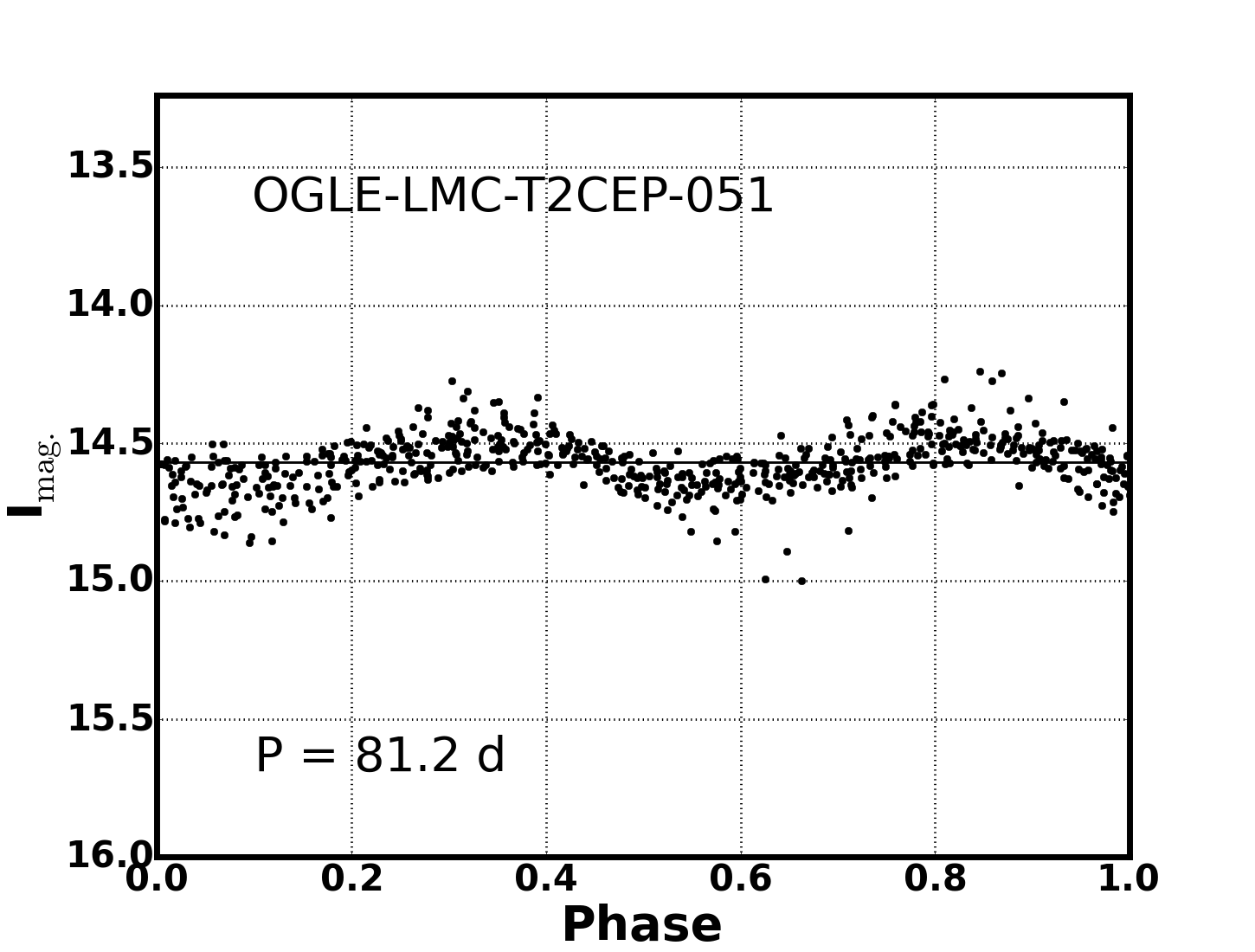} 
  }
  \subfloat[]{\includegraphics[width=0.33\textwidth]{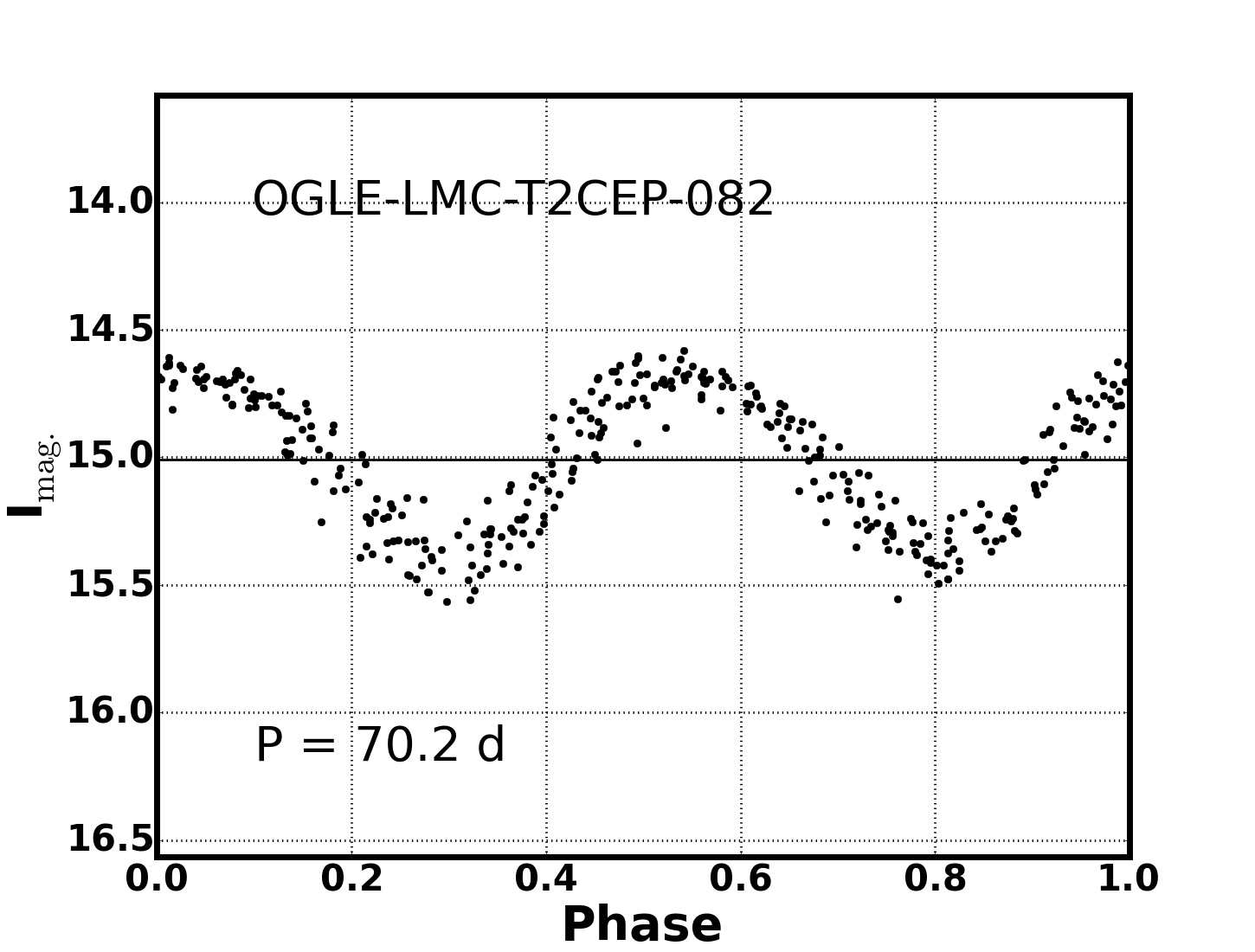}
  } \\
  
  \subfloat[]{\includegraphics[width=0.33\textwidth]{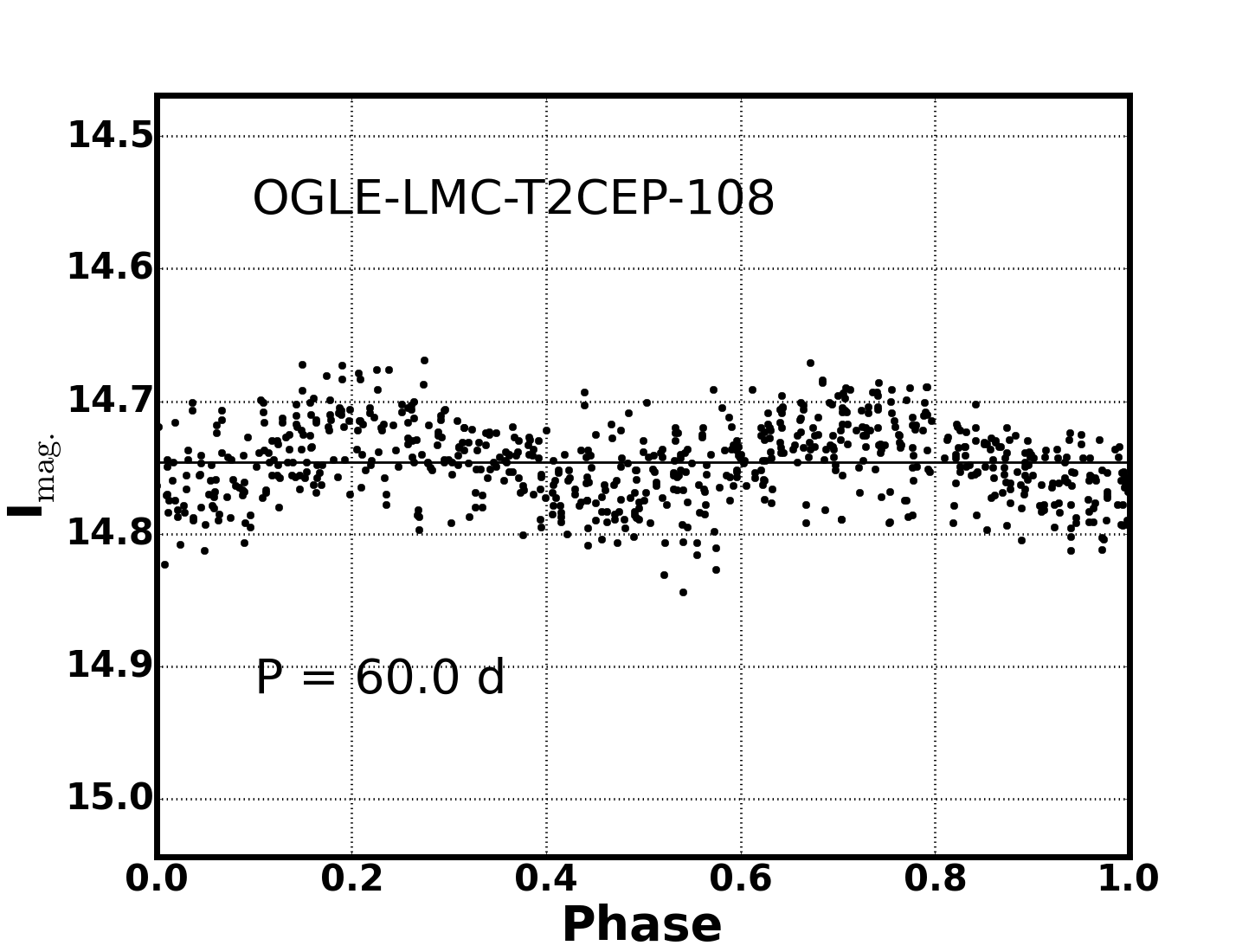} \label{figure:LMC108pulsation}
  } 
   \subfloat[]{\includegraphics[width=0.33\textwidth]{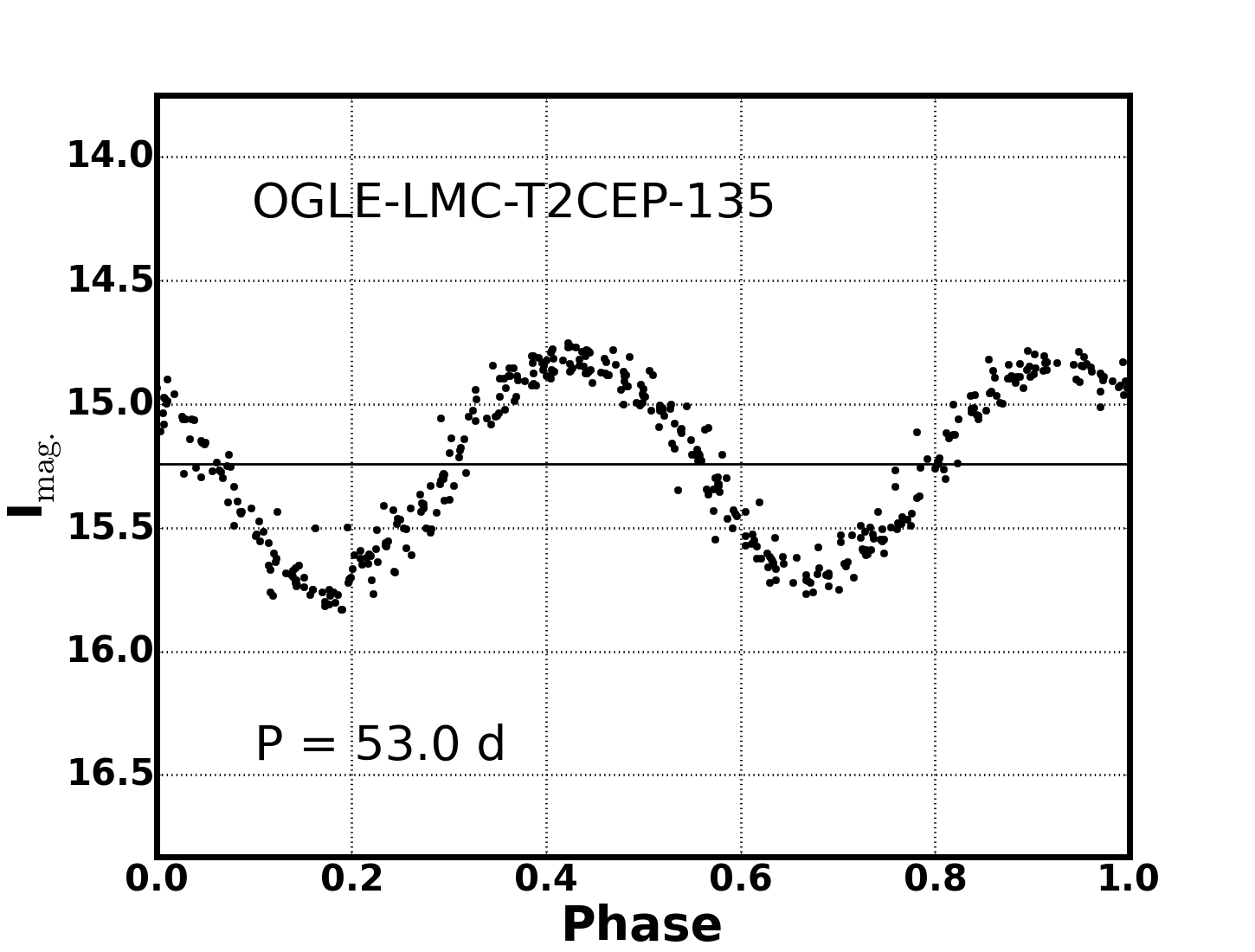}
  }
  \subfloat[]{\includegraphics[width=0.33\textwidth]{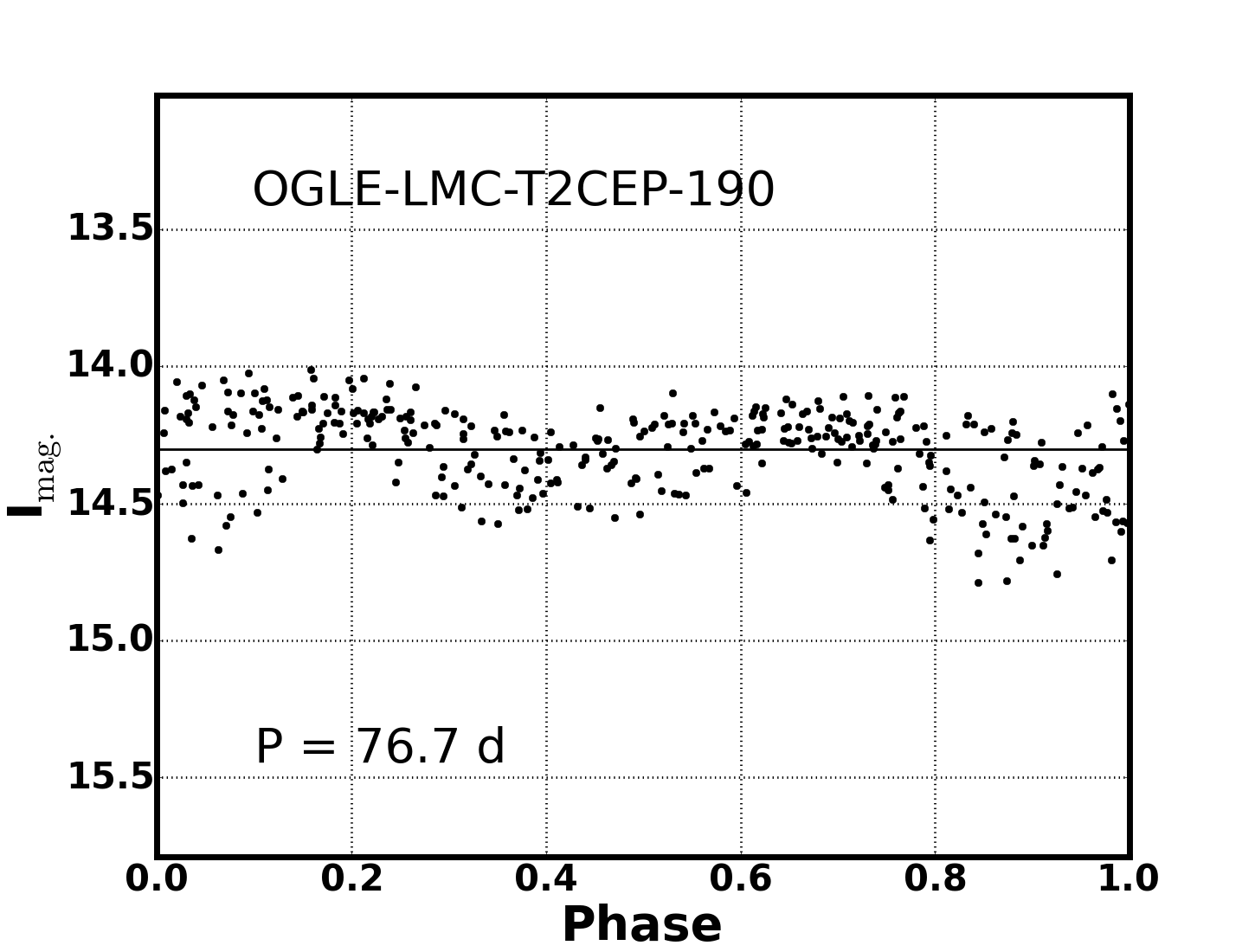}
  } \\
  
  \subfloat[]{\includegraphics[width=0.33\textwidth]{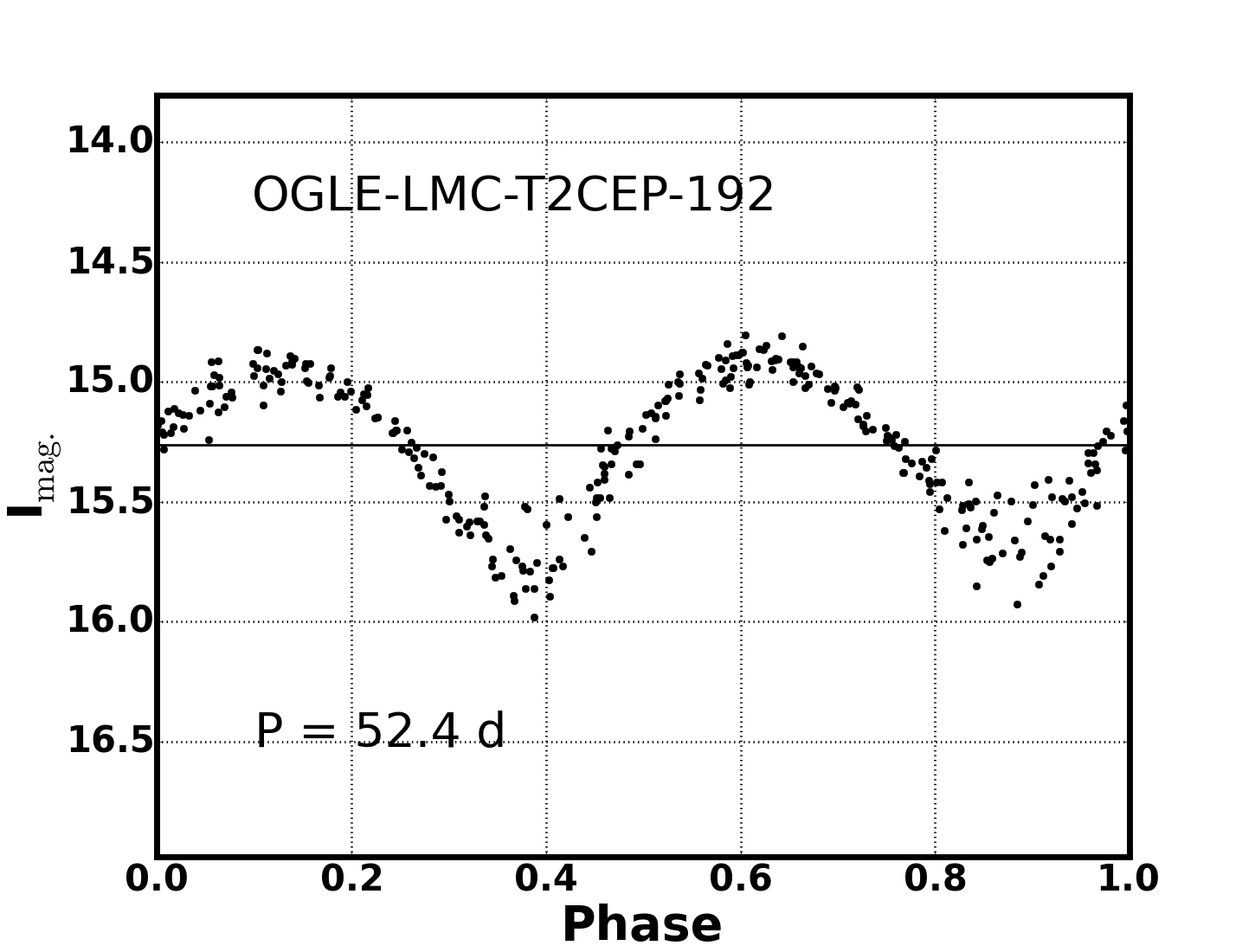}
  }
  \subfloat[]{\includegraphics[width=0.33\textwidth]{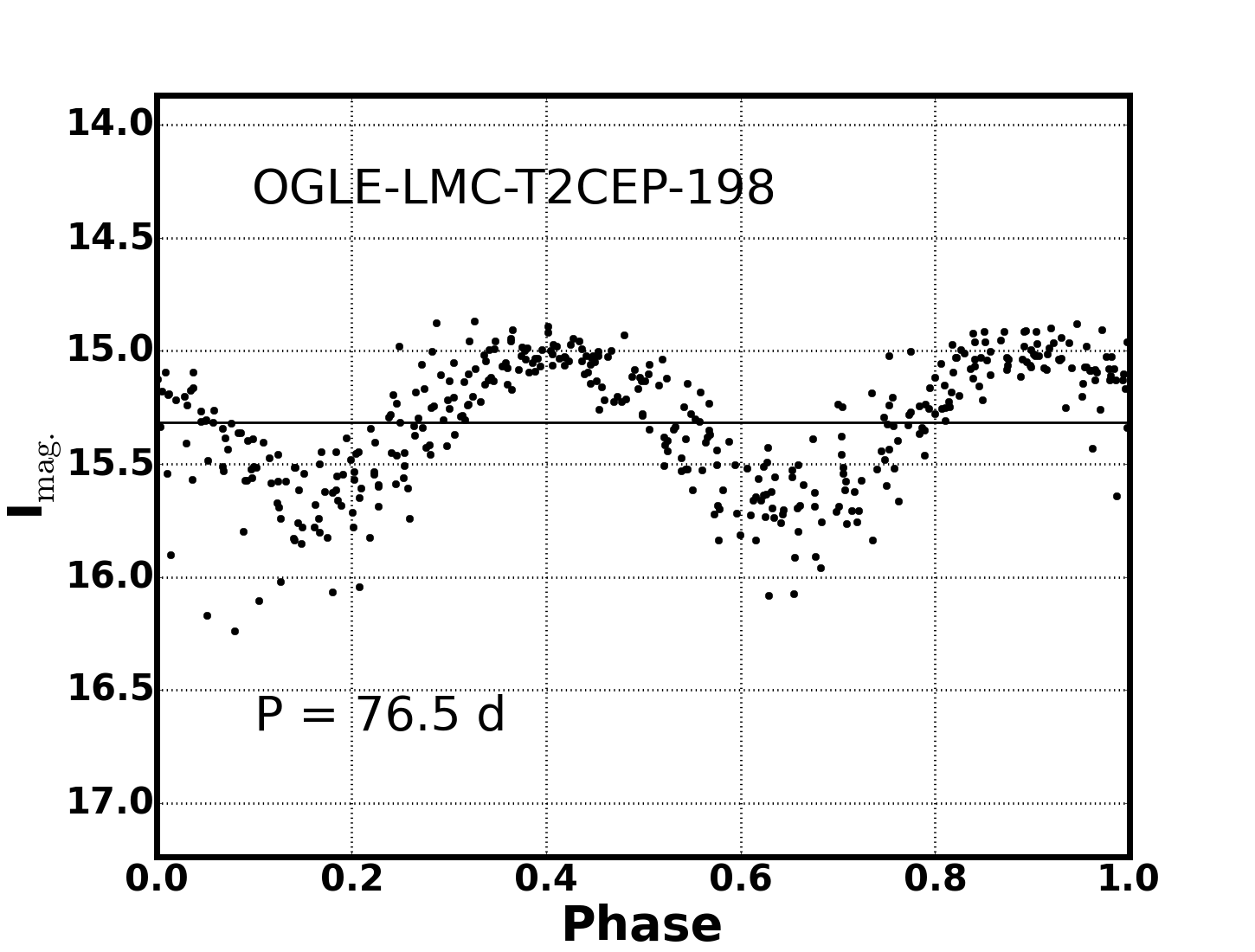}
  } 
  \subfloat[]{\includegraphics[width=0.33\textwidth]{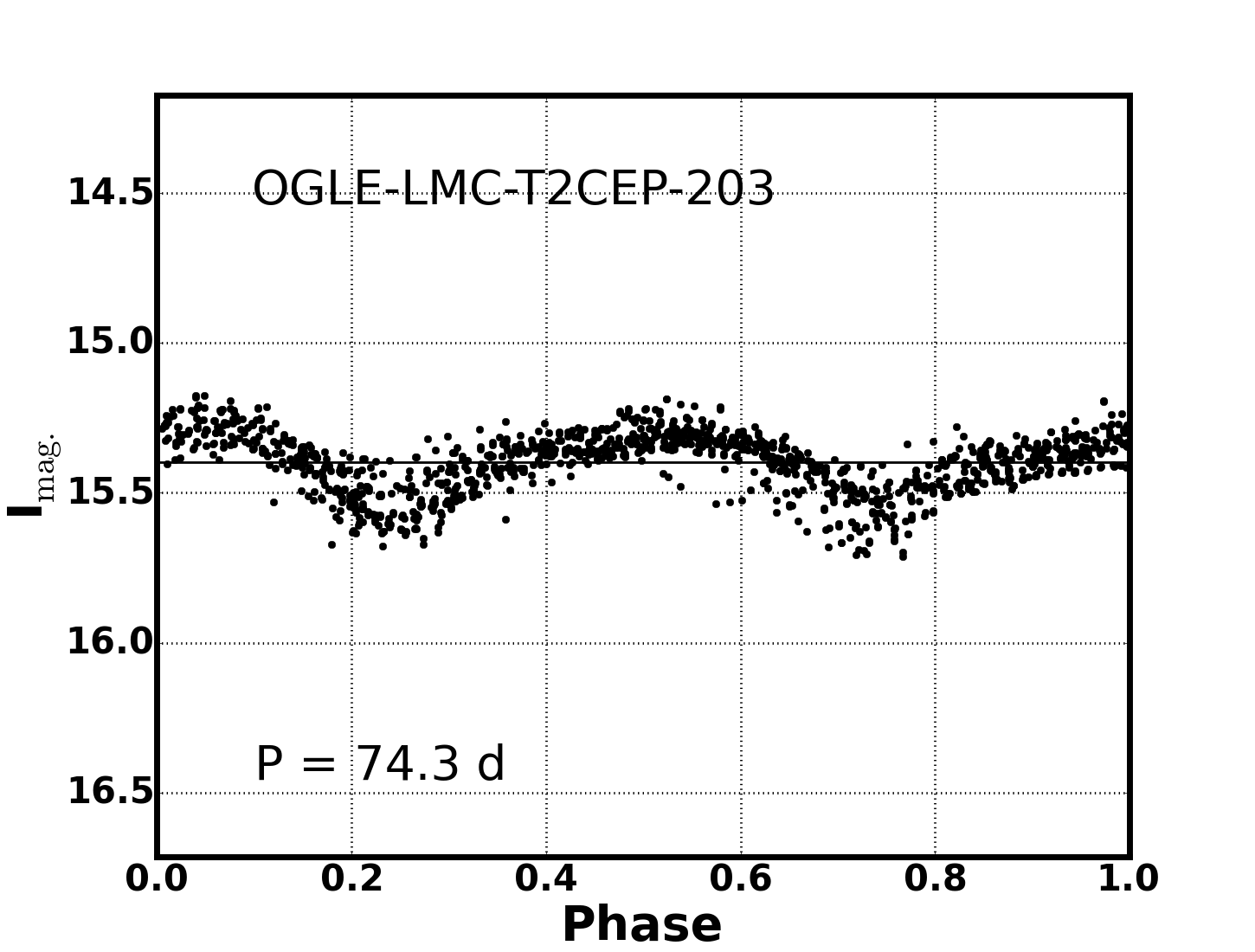}
  } 
  \caption{Photometric time series of the LMC RV Tauri stars that have SEDs showing no IR excess, phase-folded on their formal periods. 
  The horizontal line indicates the mean I$_{\rm mag}$}.
 \end{figure*}  
 
 \begin{figure*}
  \subfloat[]{\includegraphics[width=0.33\textwidth]{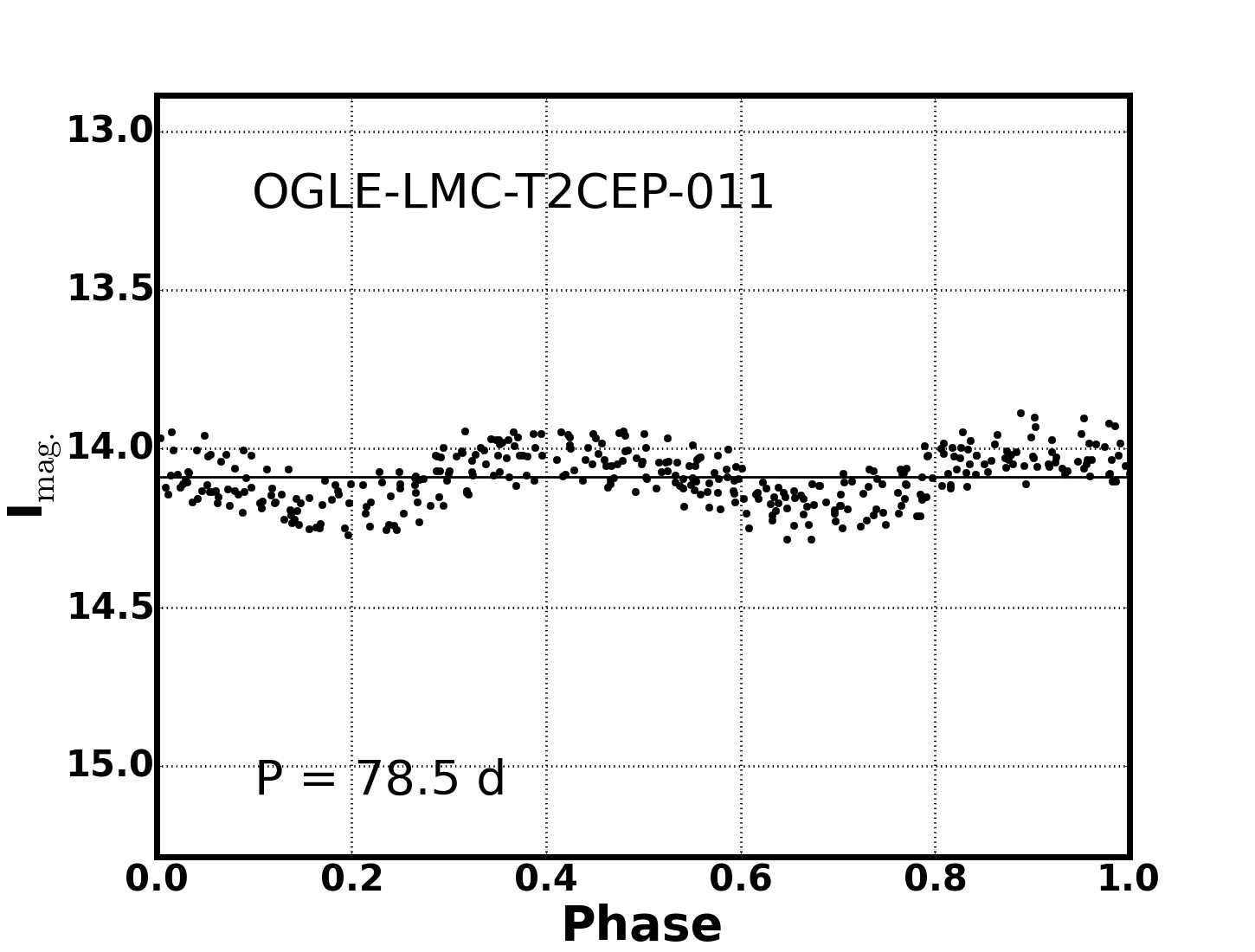}
  }
  \subfloat[]{\includegraphics[width=0.33\textwidth]{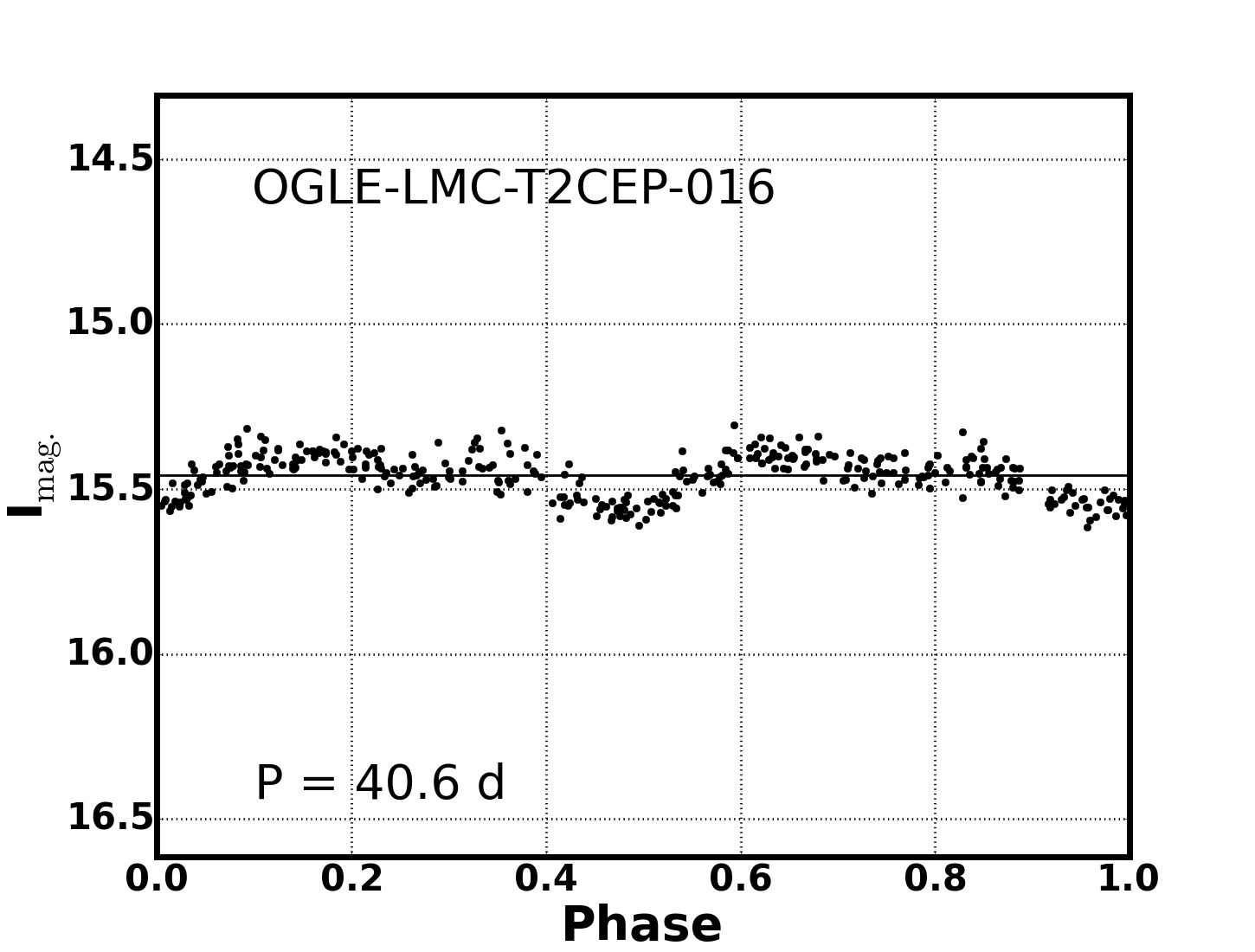}
  }
  \subfloat[]{\includegraphics[width=0.33\textwidth]{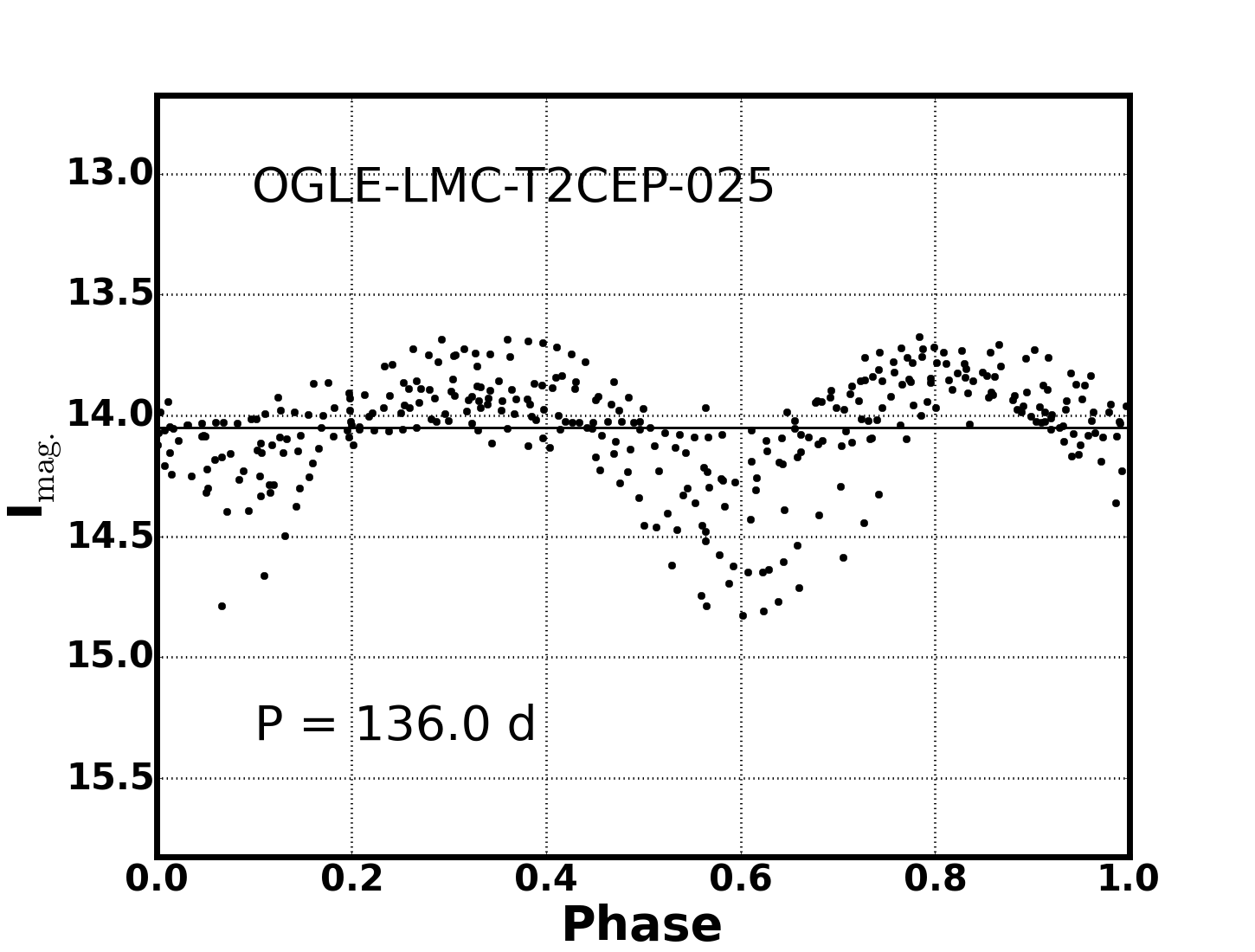}
  } \\
  
  \subfloat[]{\includegraphics[width=0.33\textwidth]{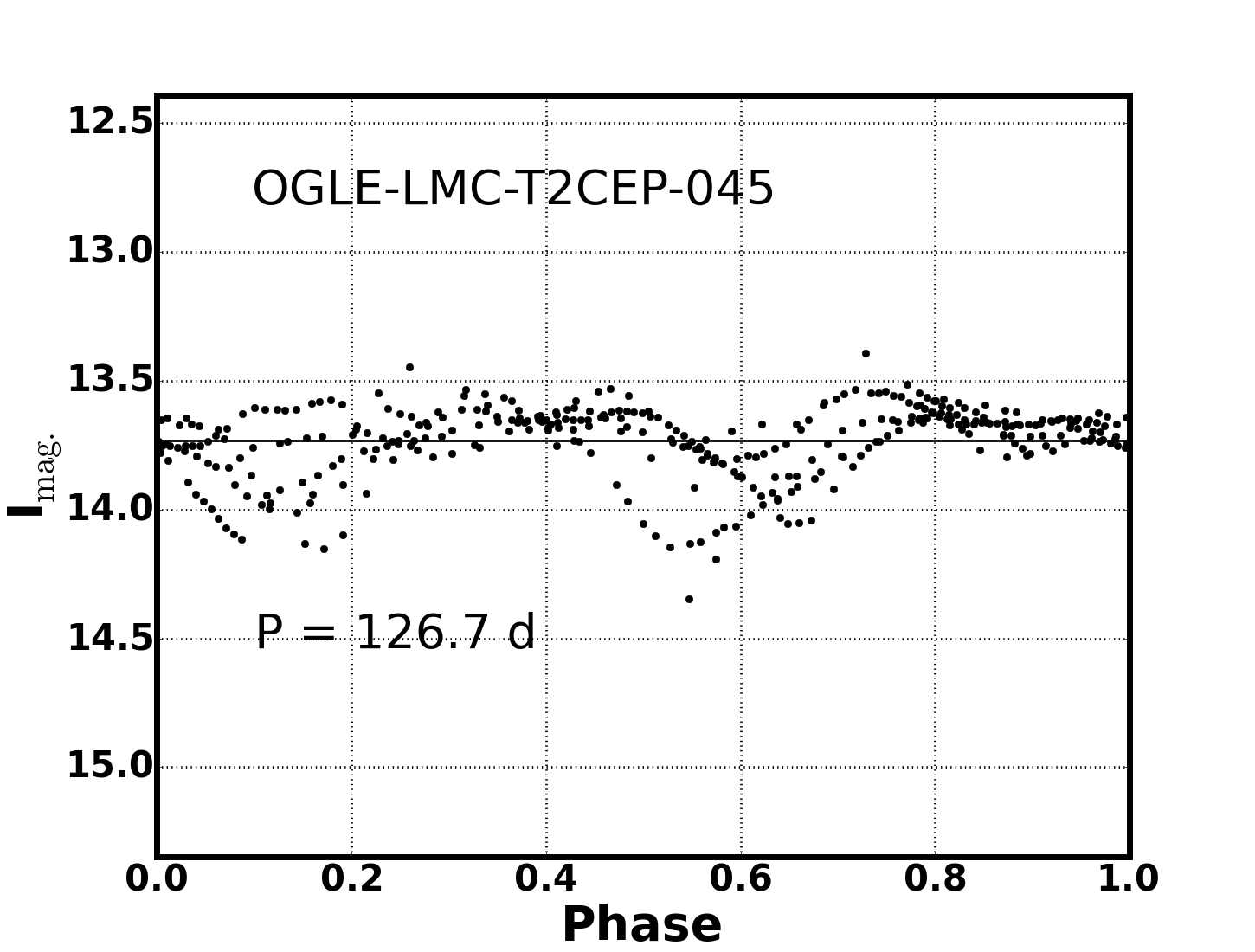}
  }
  \subfloat[]{\includegraphics[width=0.33\textwidth]{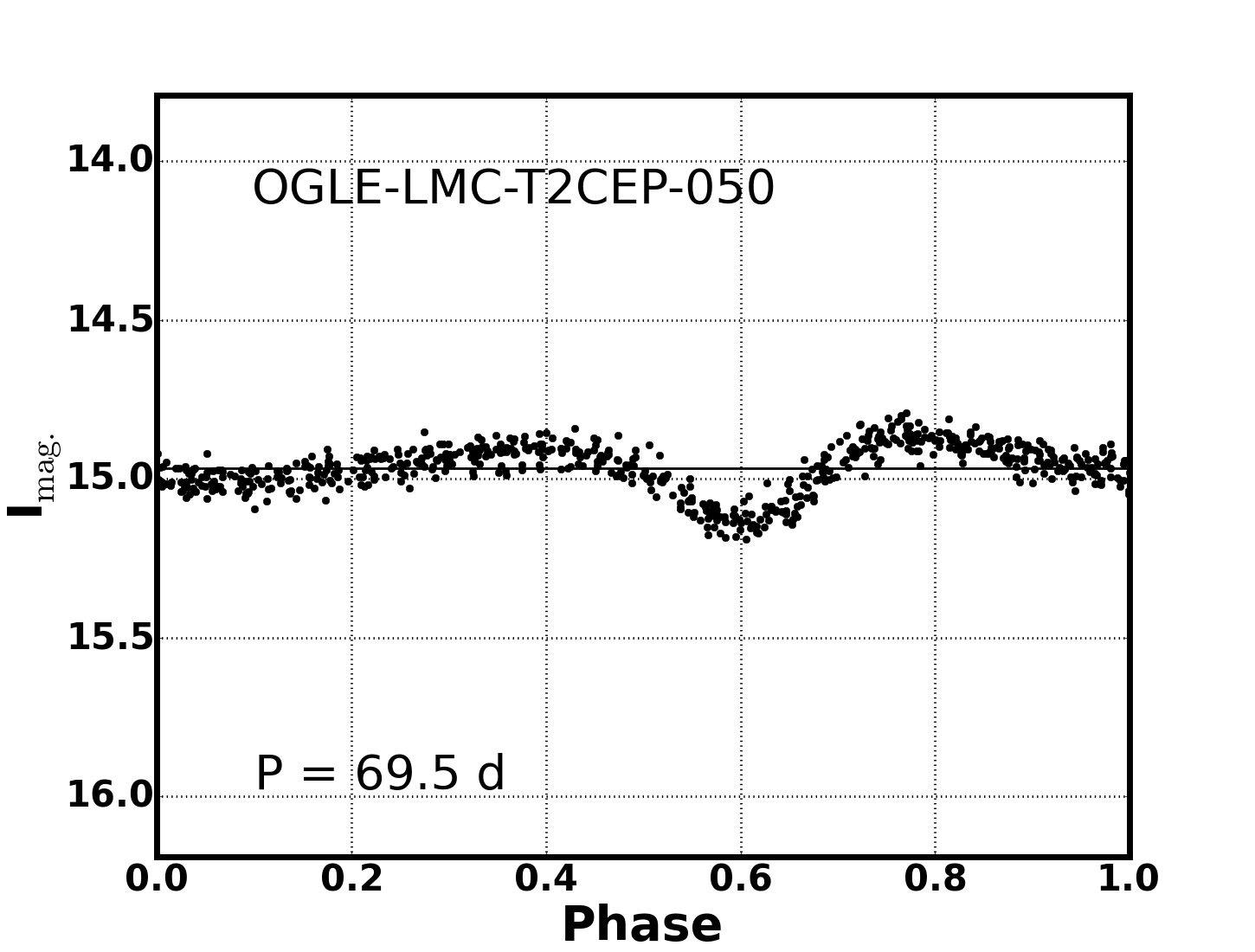}
  }
  \subfloat[]{\includegraphics[width=0.33\textwidth]{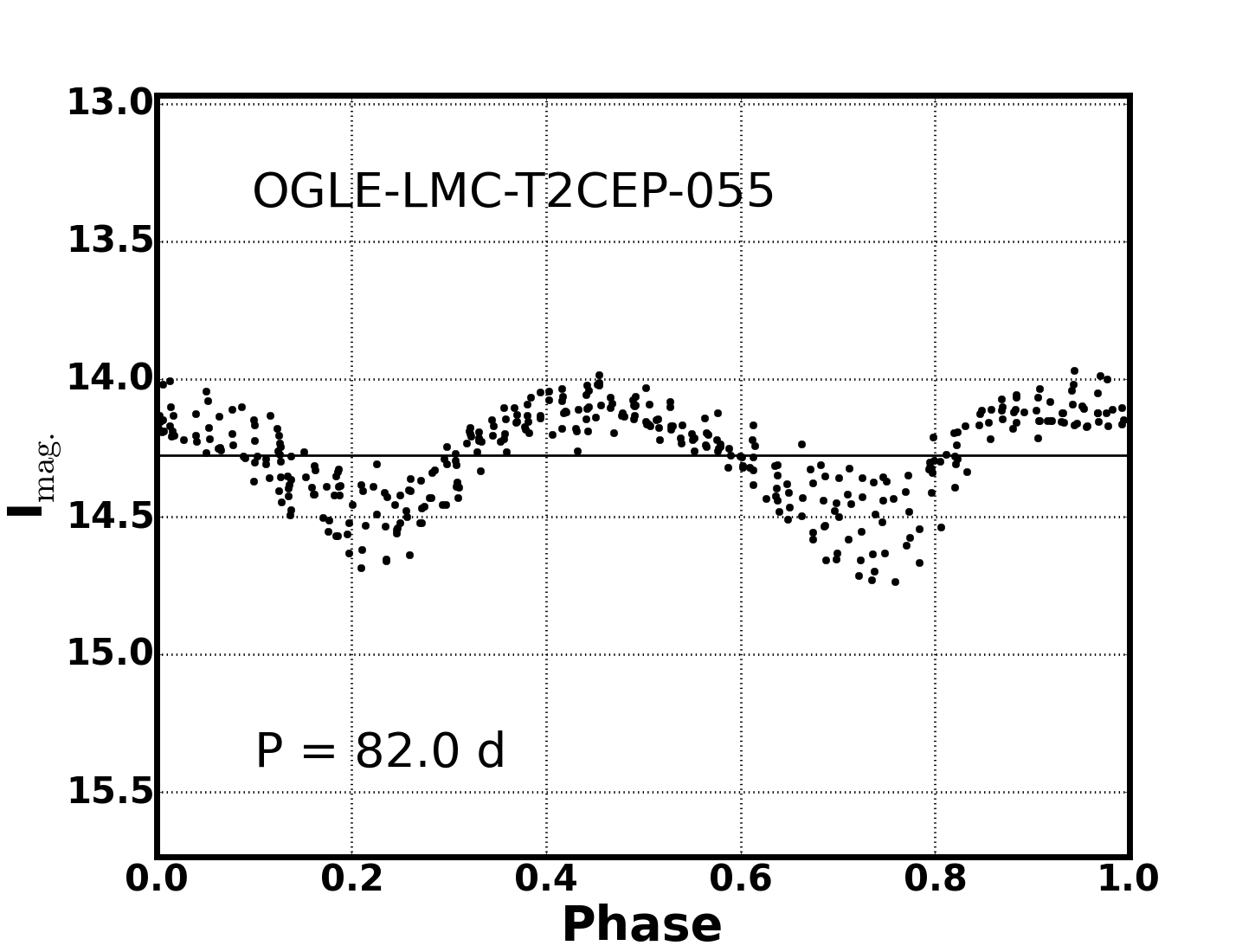}
  } \\
  
  \subfloat[]{\includegraphics[width=0.33\textwidth]{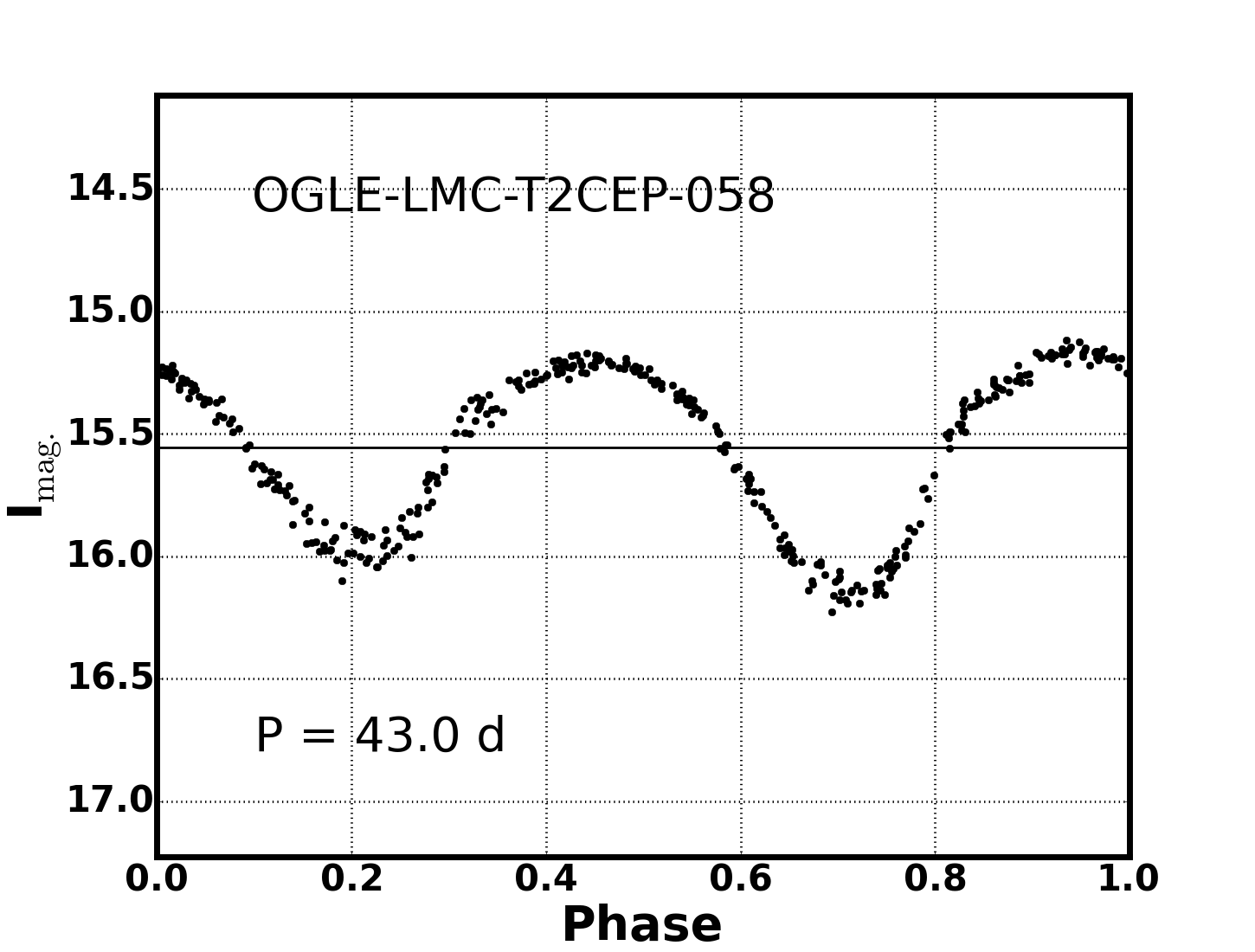} 
  }
  \subfloat[]{\includegraphics[width=0.33\textwidth]{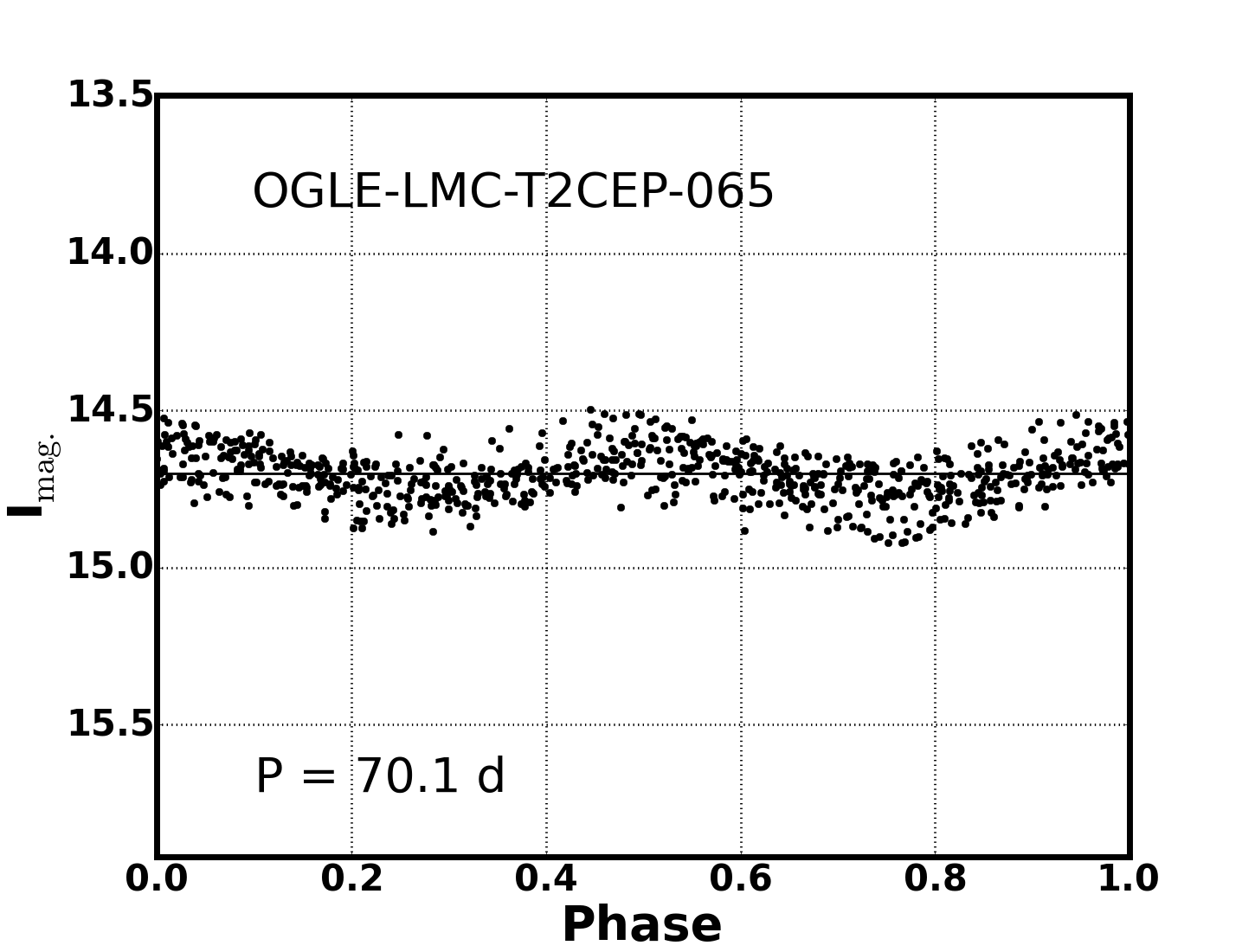}
  } 
  \subfloat[]{\includegraphics[width=0.33\textwidth]{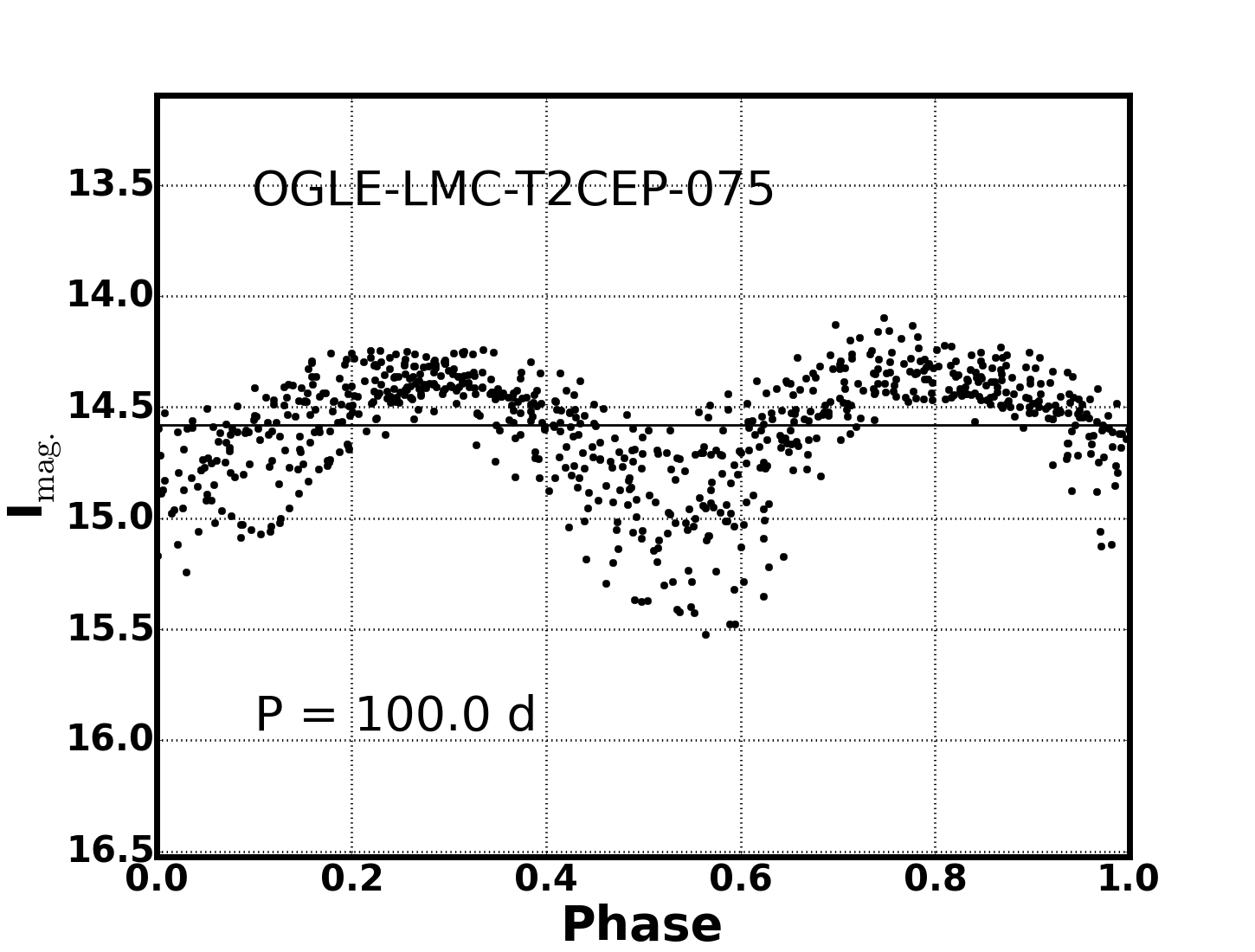}
  }\\

  \subfloat[]{\includegraphics[width=0.33\textwidth]{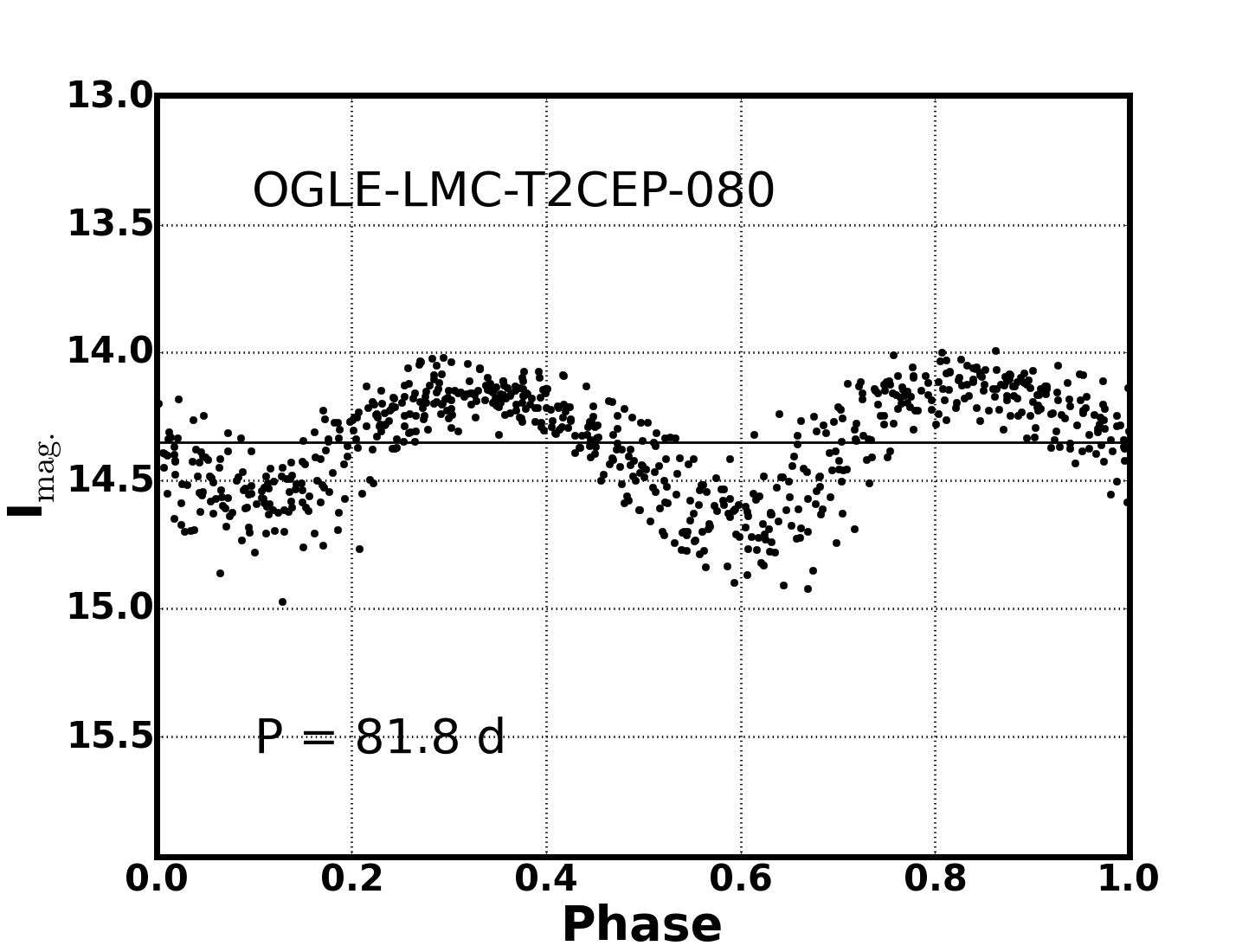}
  }
  \subfloat[]{\includegraphics[width=0.33\textwidth]{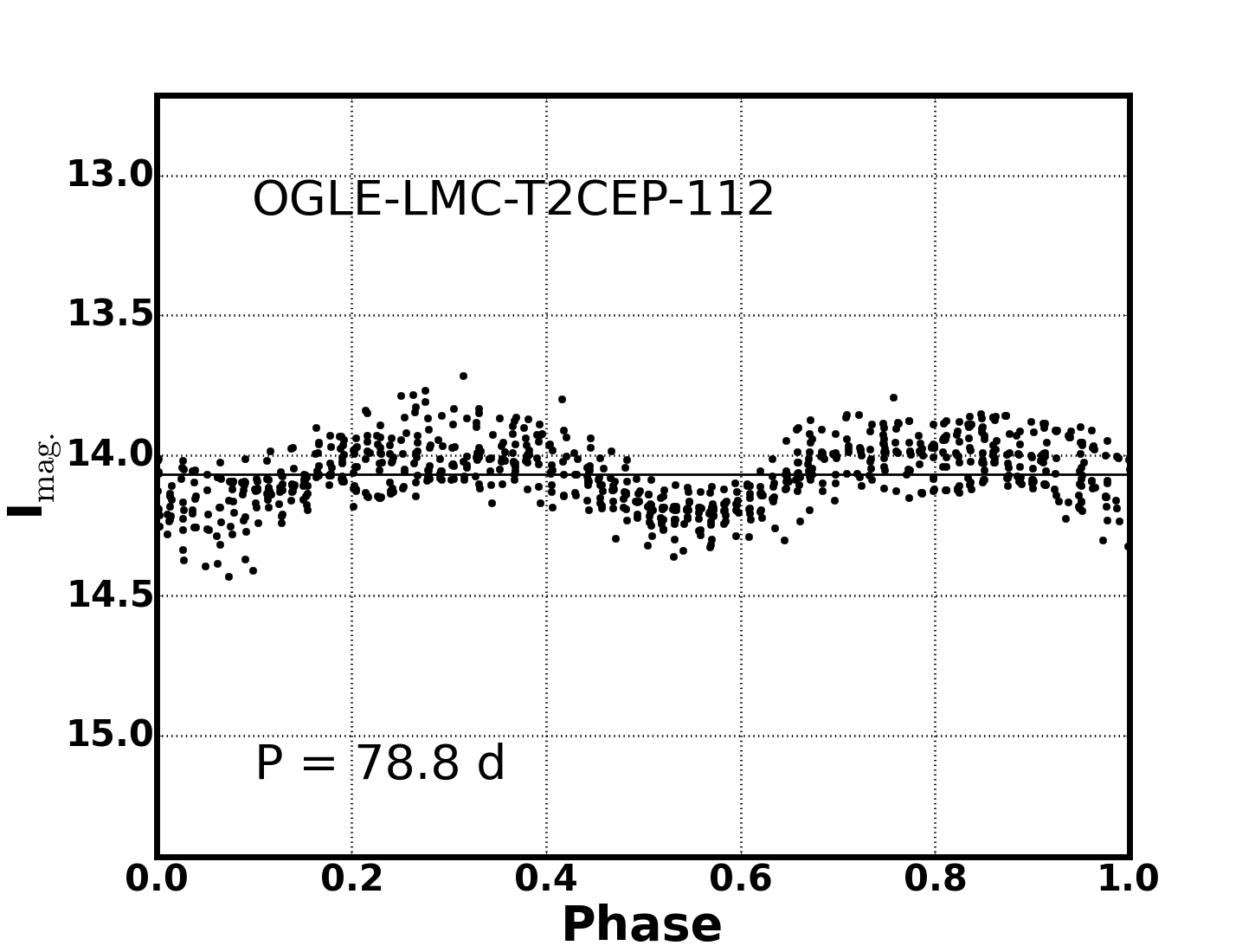}
  }
  \subfloat[]{\includegraphics[width=0.33\textwidth]{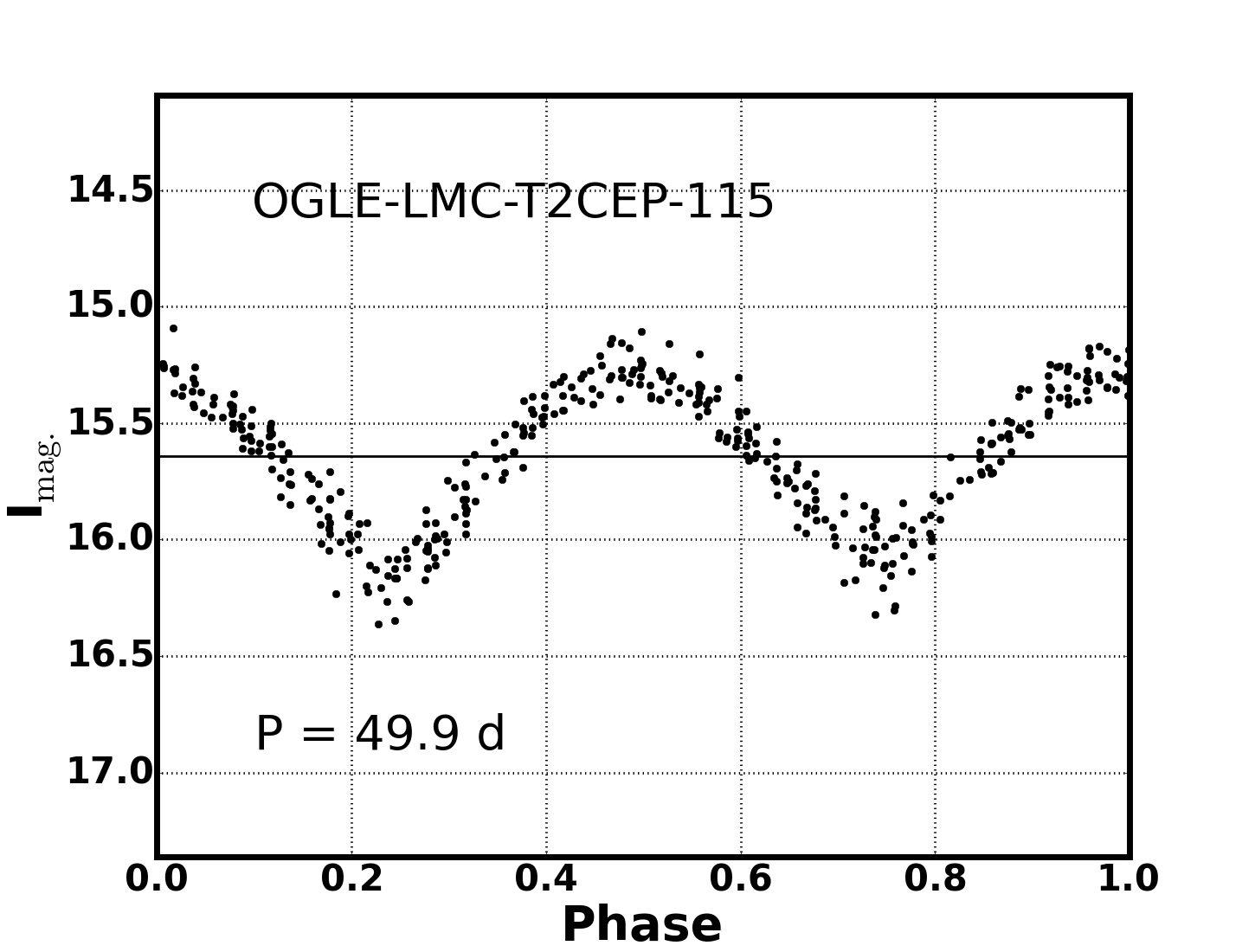}
  } \\
  
  \caption{Photometric timeseries of the LMC RV Tauri stars that have uncertain SEDs, phase-folded on their formal periods. The horizontal line shows the mean I$_{\rm mag}$}.
 \end{figure*}

 \begin{figure*}
  \subfloat[]{\includegraphics[width=0.33\textwidth]{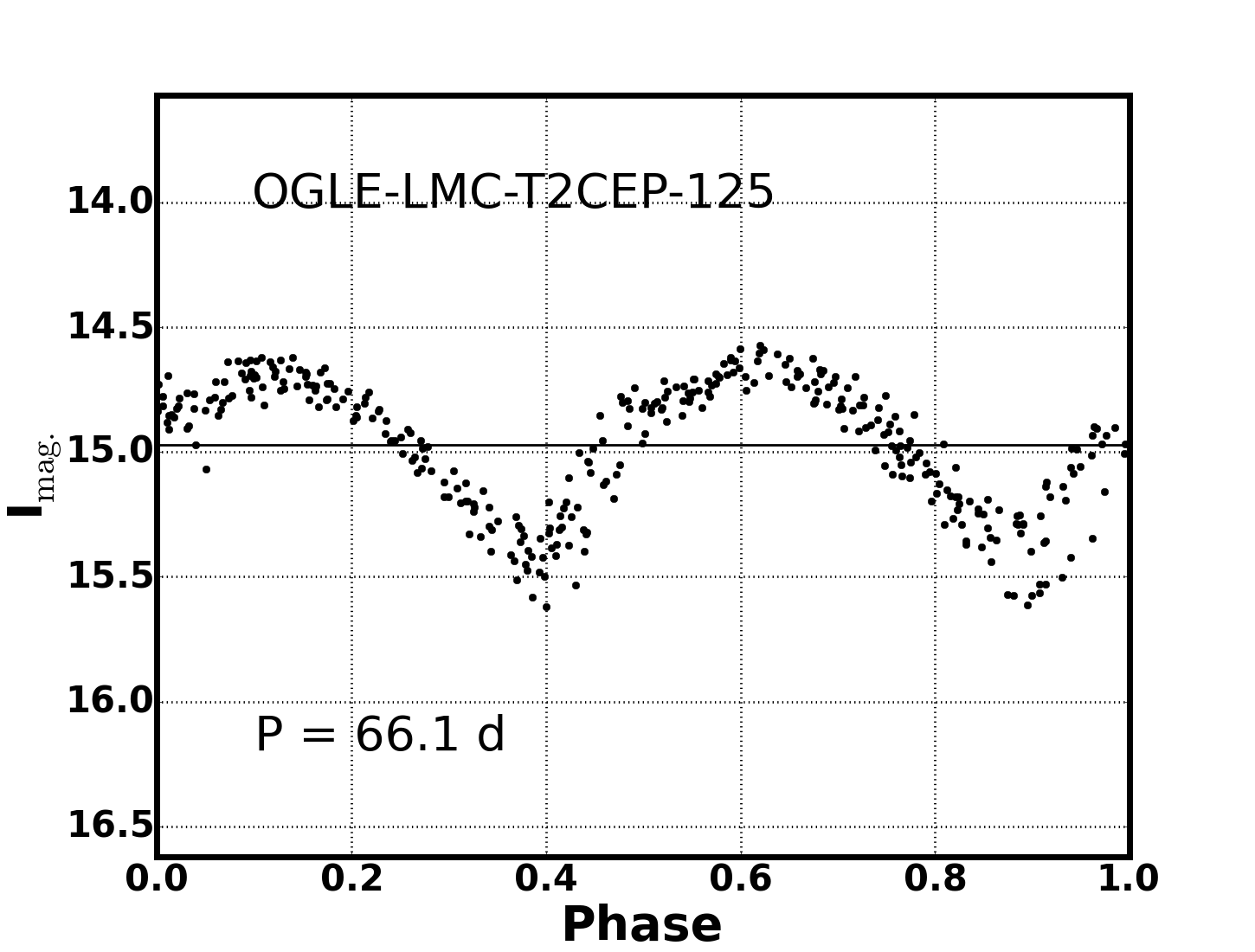}
  }
  \subfloat[]{\includegraphics[width=0.33\textwidth]{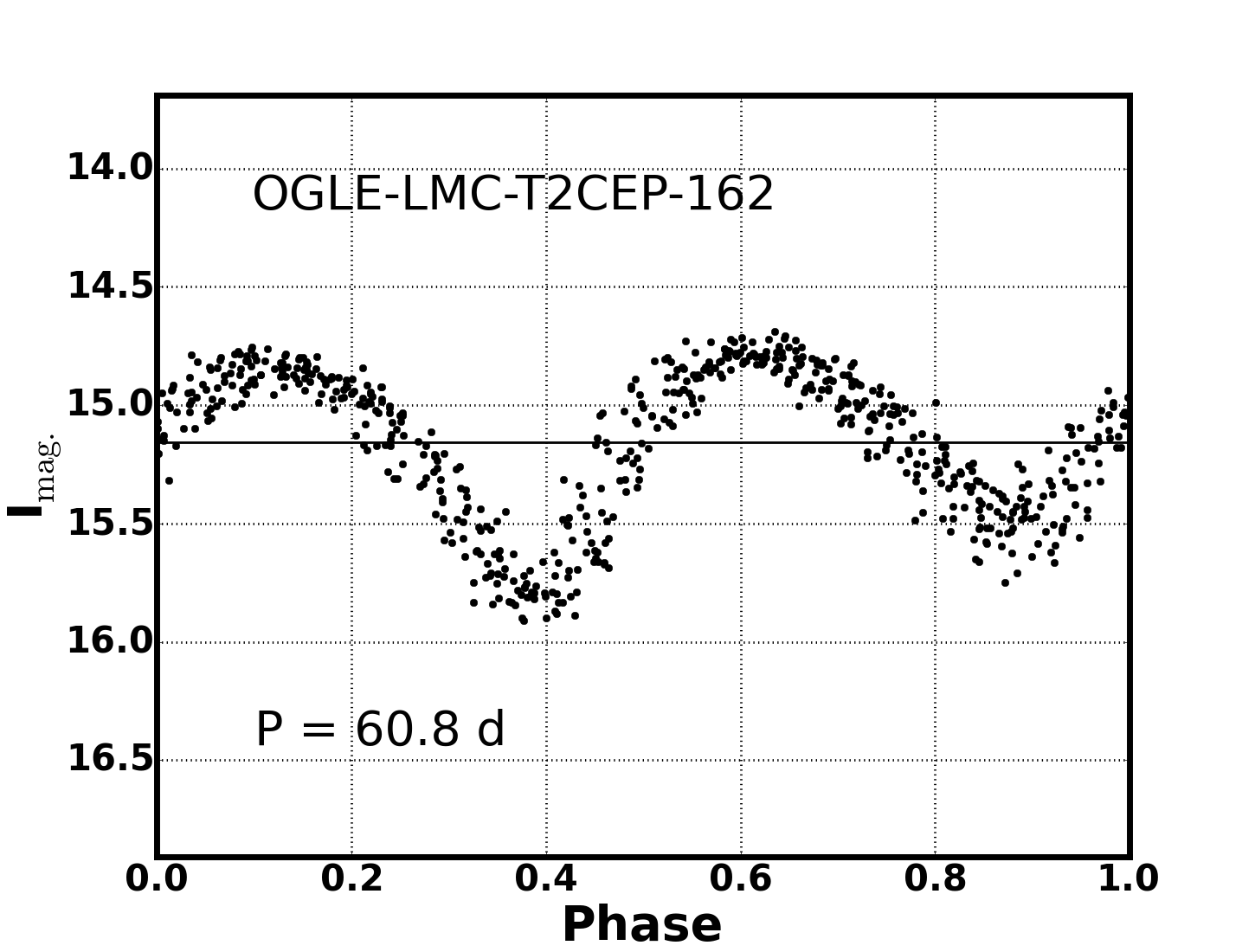}
  }
  \subfloat[]{\includegraphics[width=0.33\textwidth]{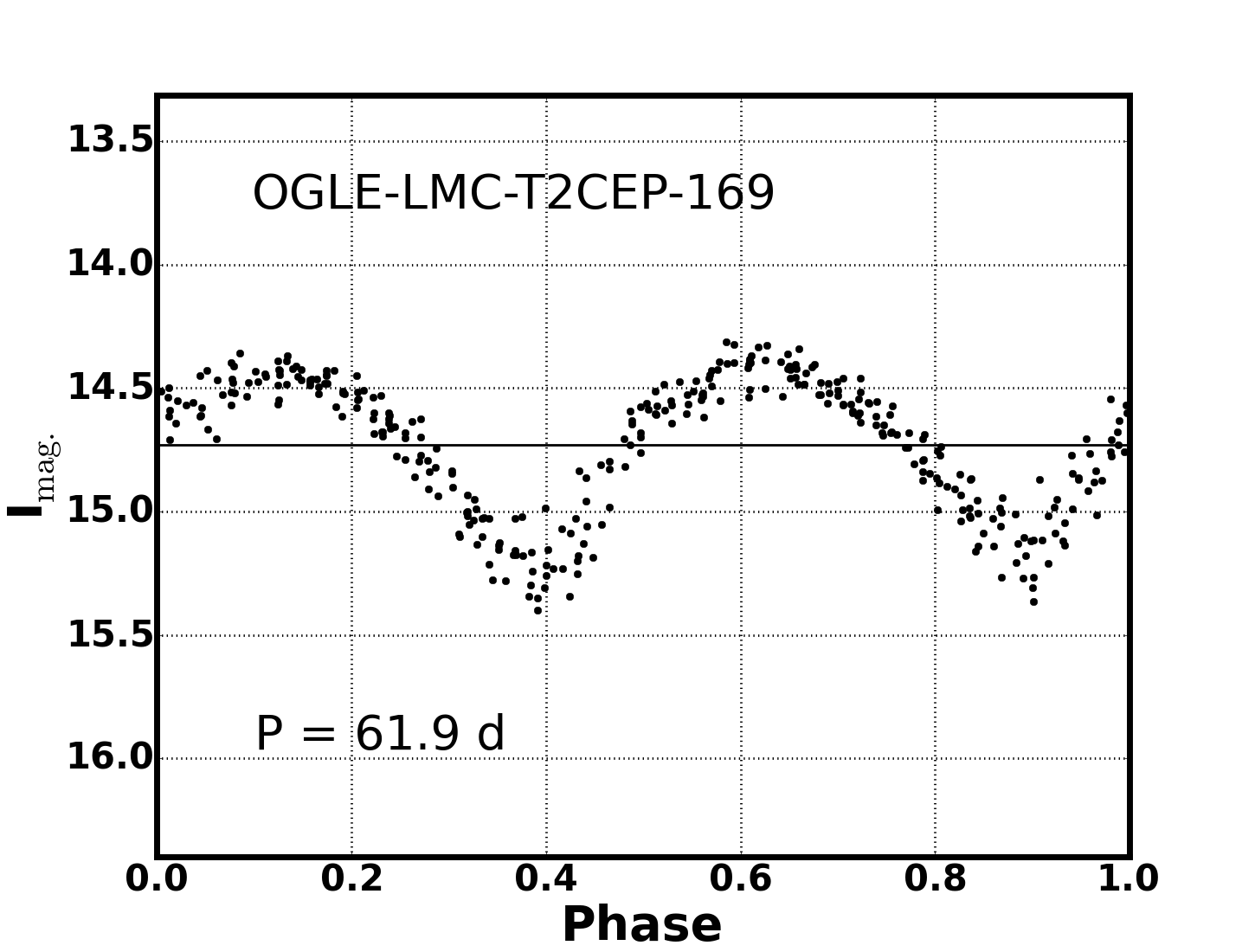}
  }\\
  
    \subfloat[]{\includegraphics[width=0.33\textwidth]{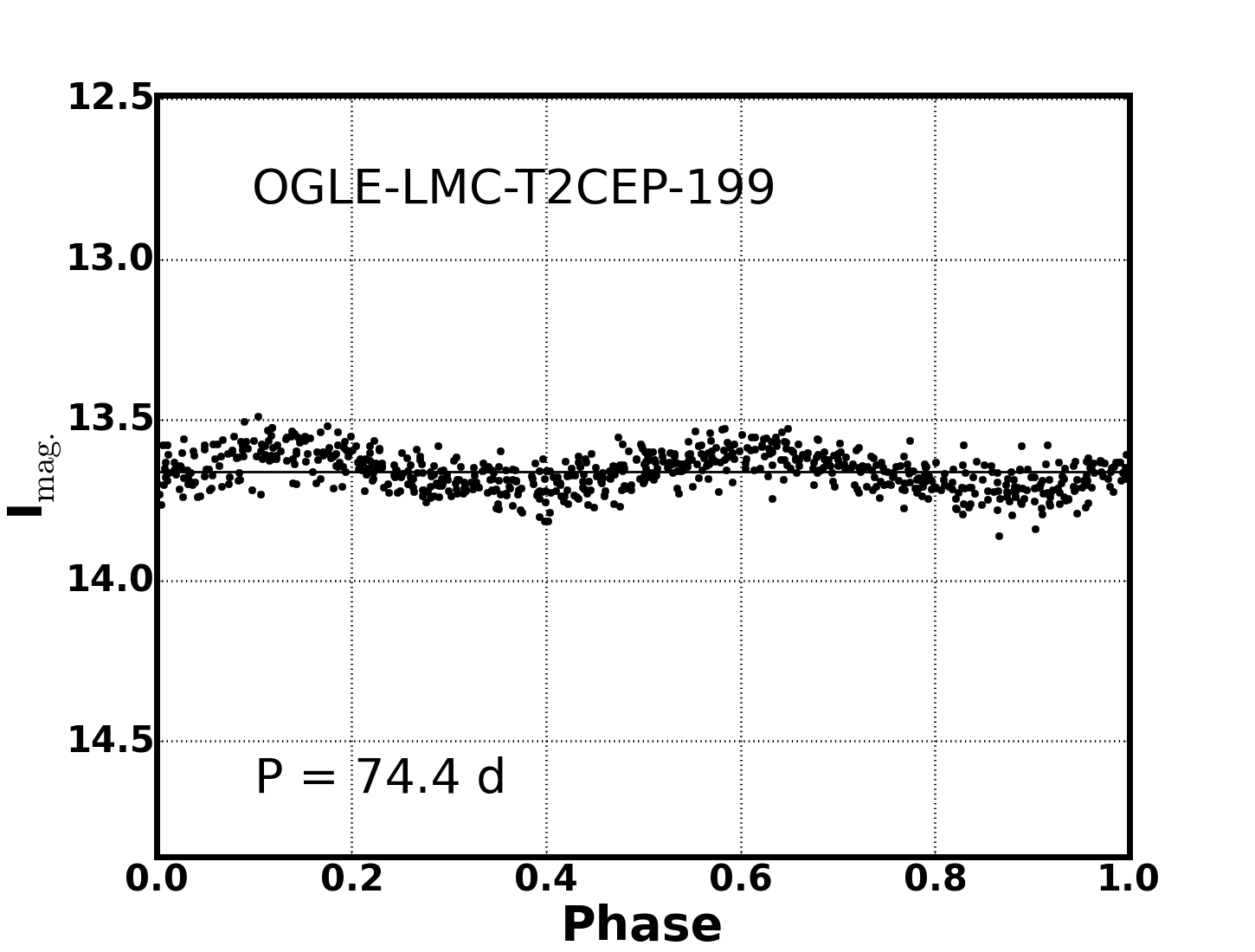}
  } 
  \subfloat[]{\includegraphics[width=0.33\textwidth]{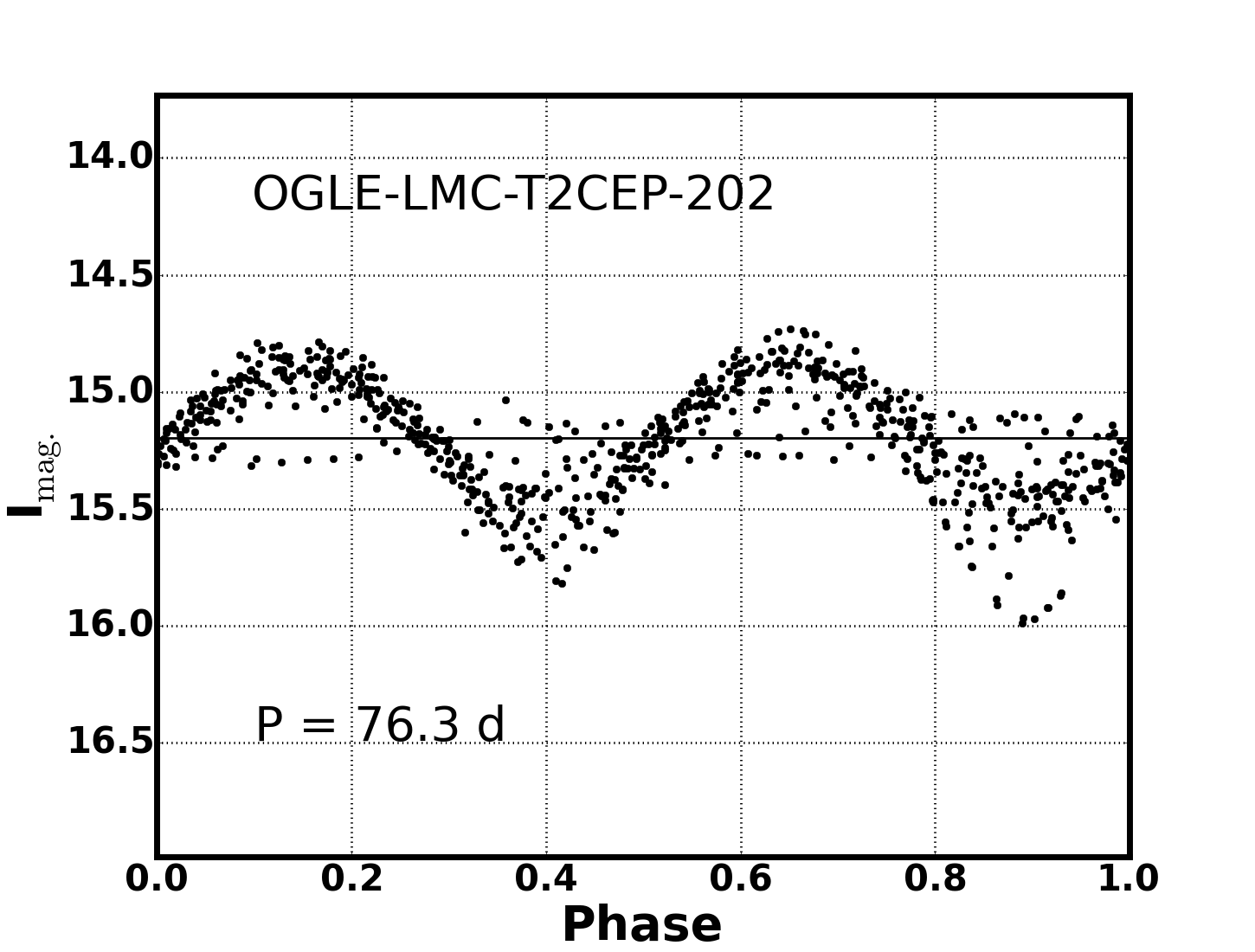}
  } 
  \caption{Photometric time series of the LMC RV Tauri stars that have uncertain SEDs, continued.}
 \end{figure*}  
 %
 \begin{figure*}
  \subfloat[]{\includegraphics[width=0.33\textwidth]{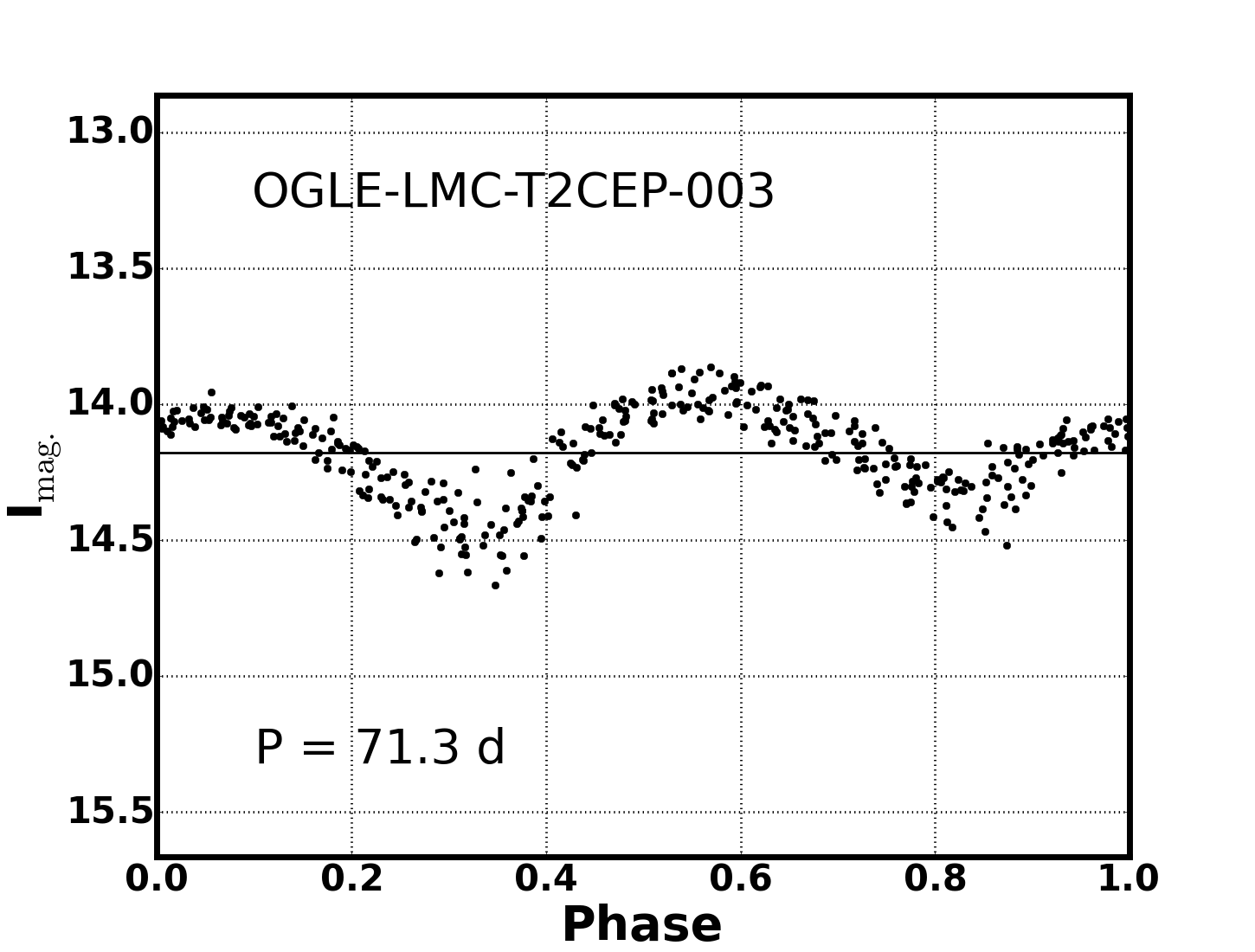}
  }
  \subfloat[]{\includegraphics[width=0.33\textwidth]{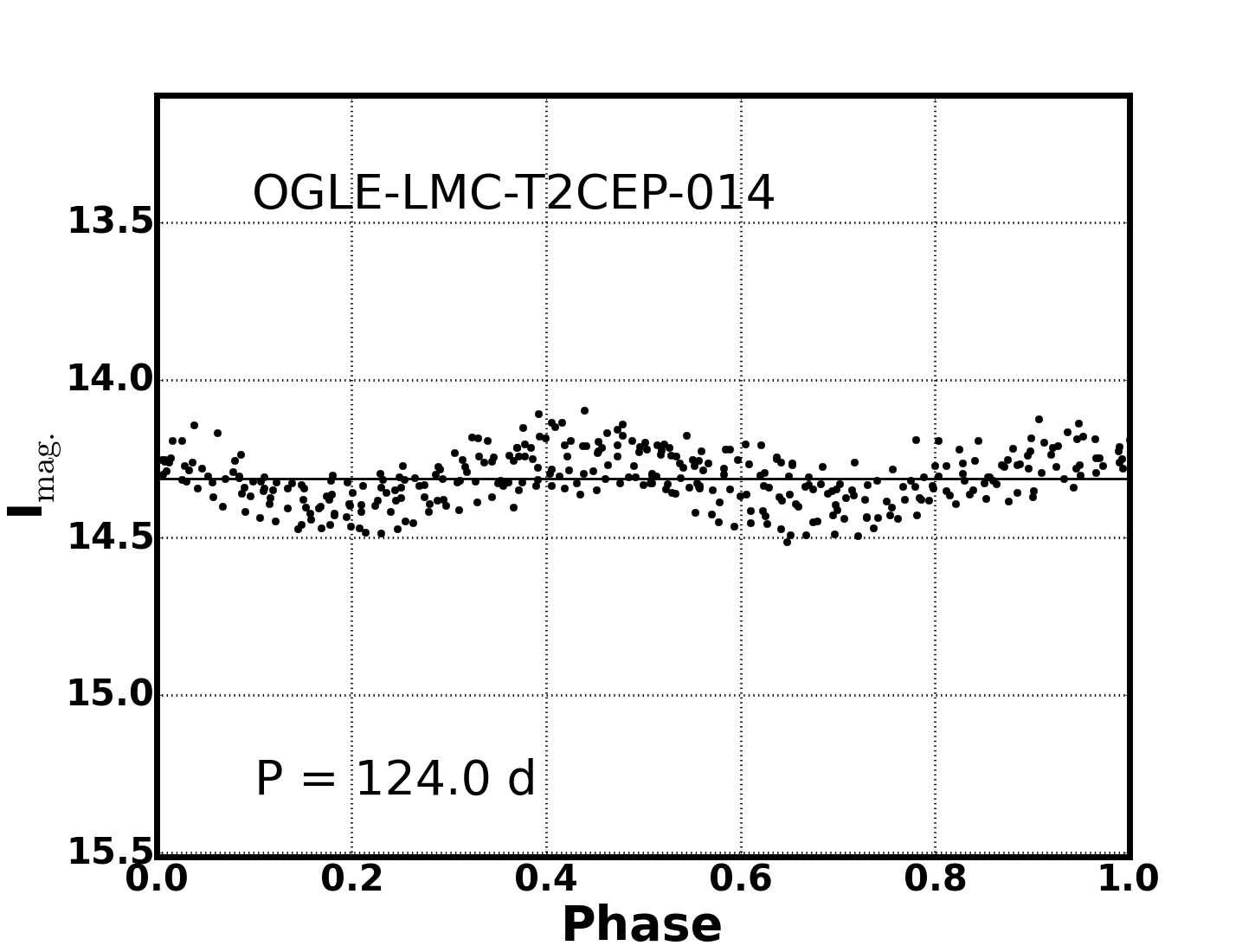}
  }
  \subfloat[]{\includegraphics[width=0.33\textwidth]{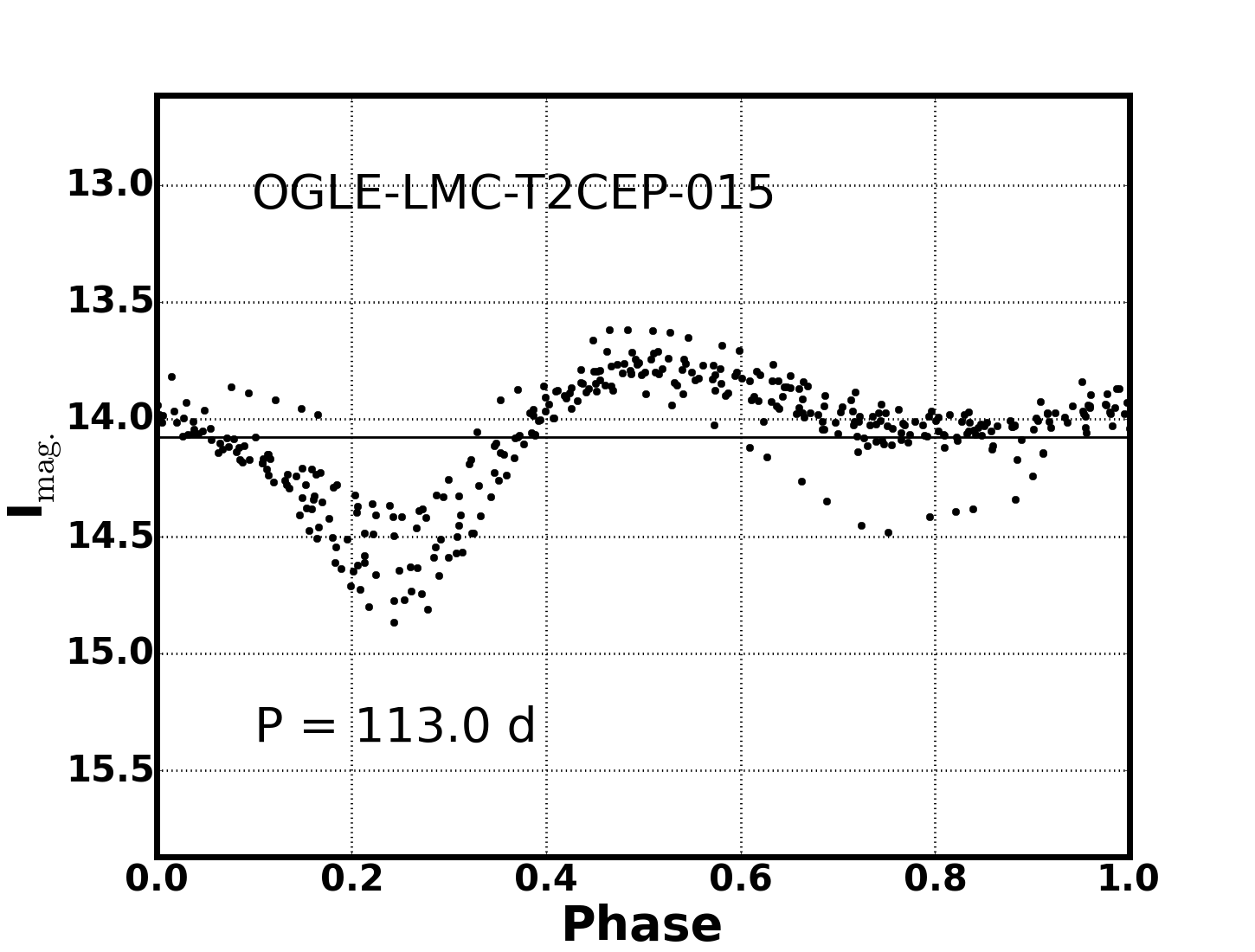}
  }\\
  
  \subfloat[]{\includegraphics[width=0.33\textwidth]{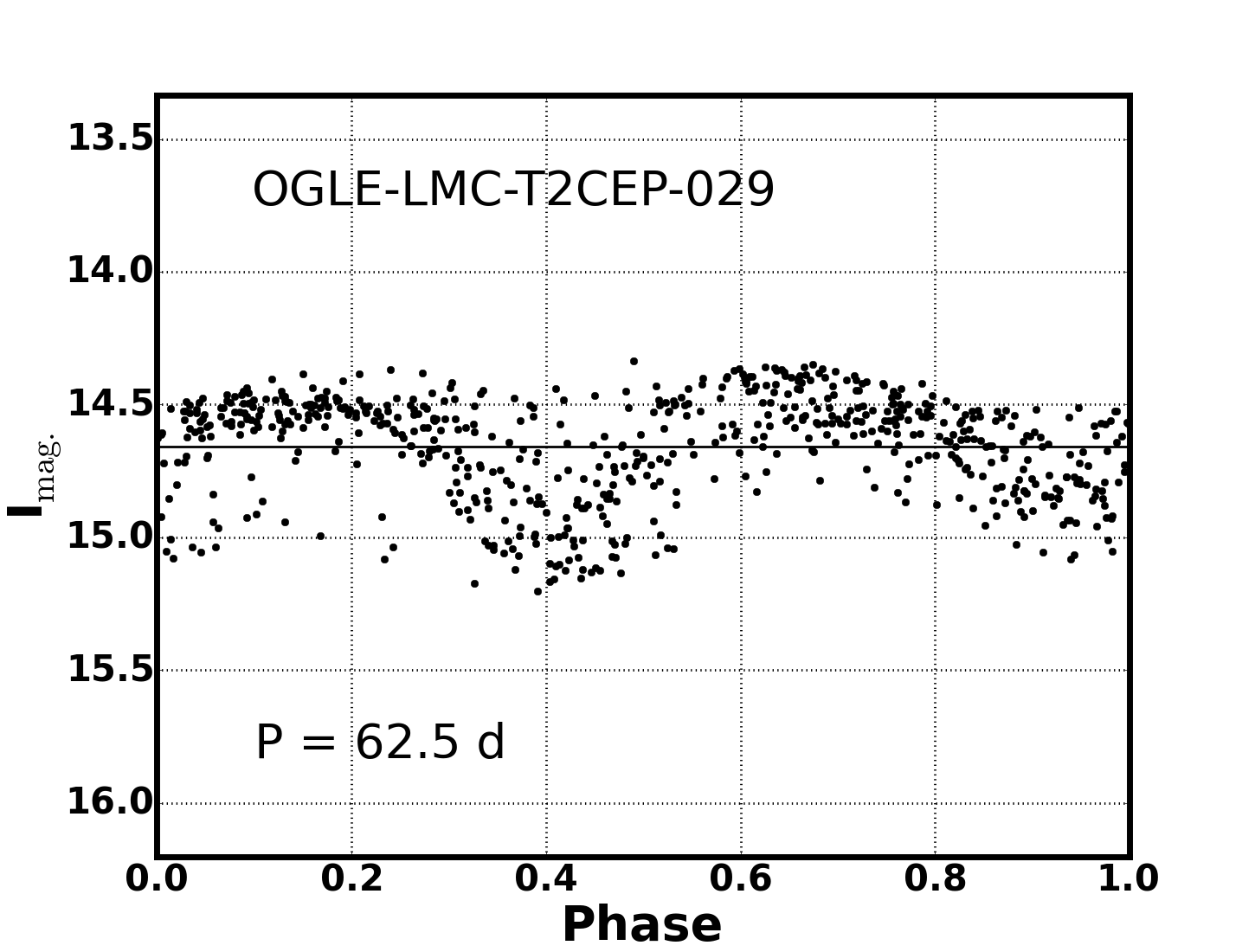}
  } 
   \subfloat[]{\includegraphics[width=0.33\textwidth]{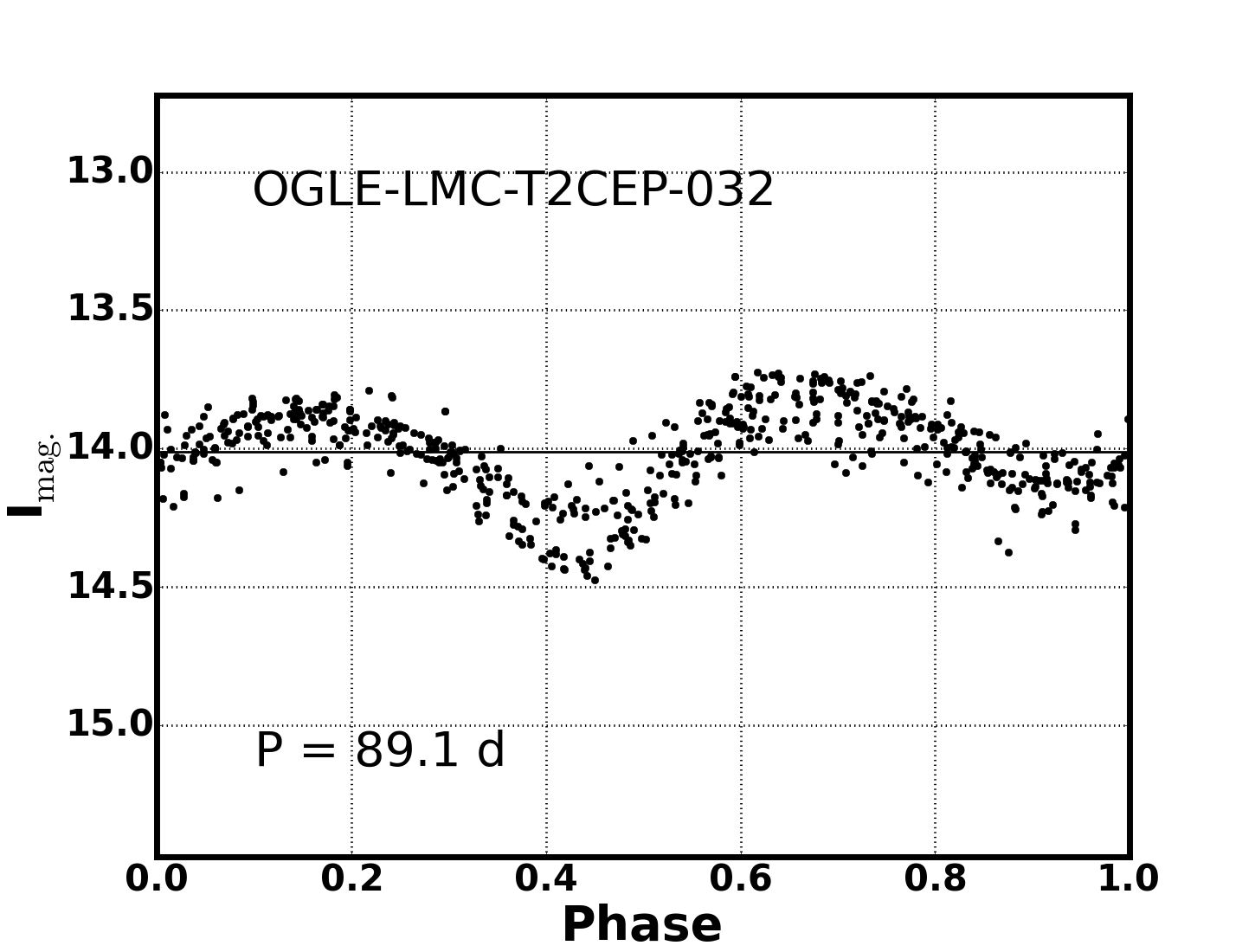}
  }
  \subfloat[]{\includegraphics[width=0.33\textwidth]{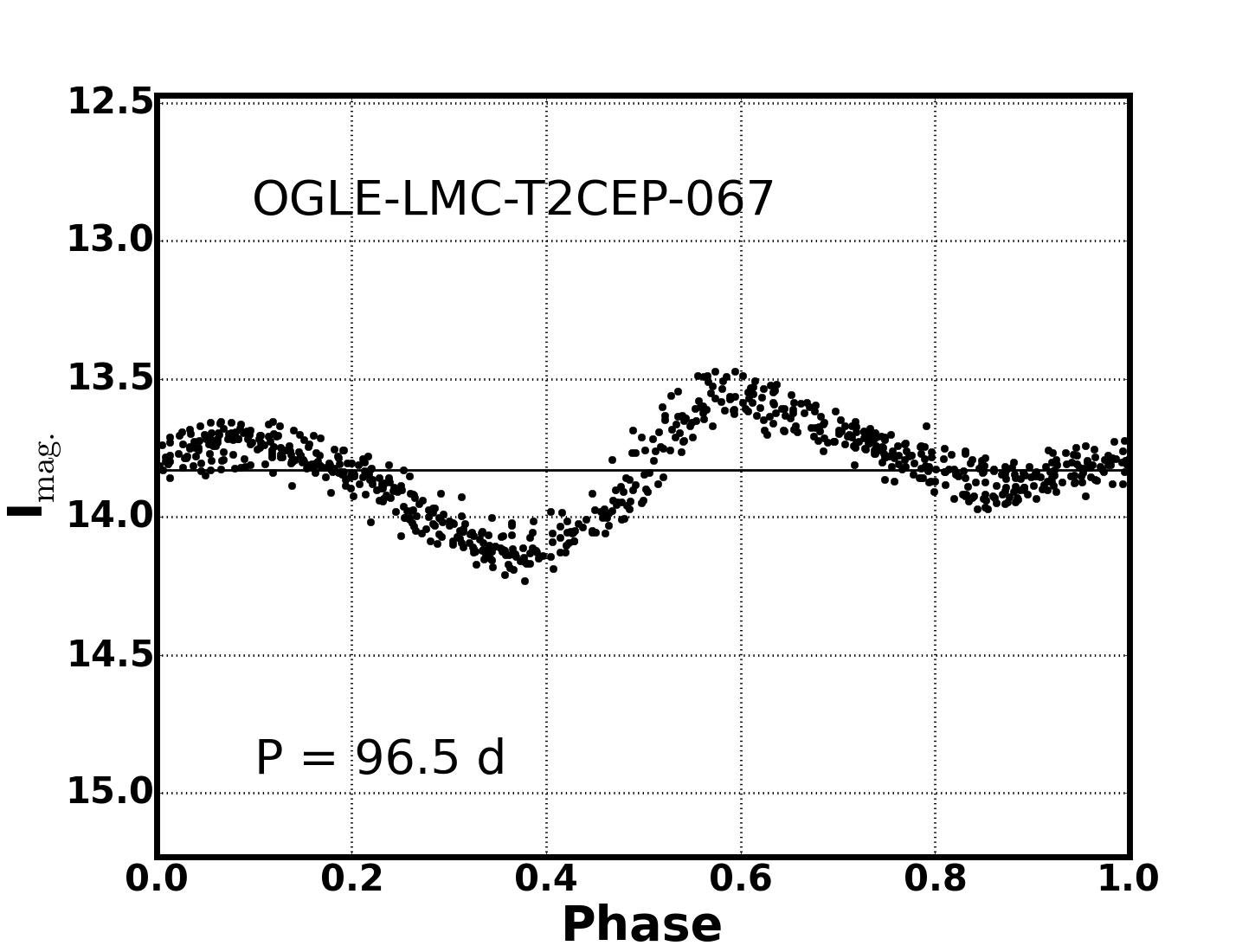}
  }\\
  
  \subfloat[]{\includegraphics[width=0.33\textwidth]{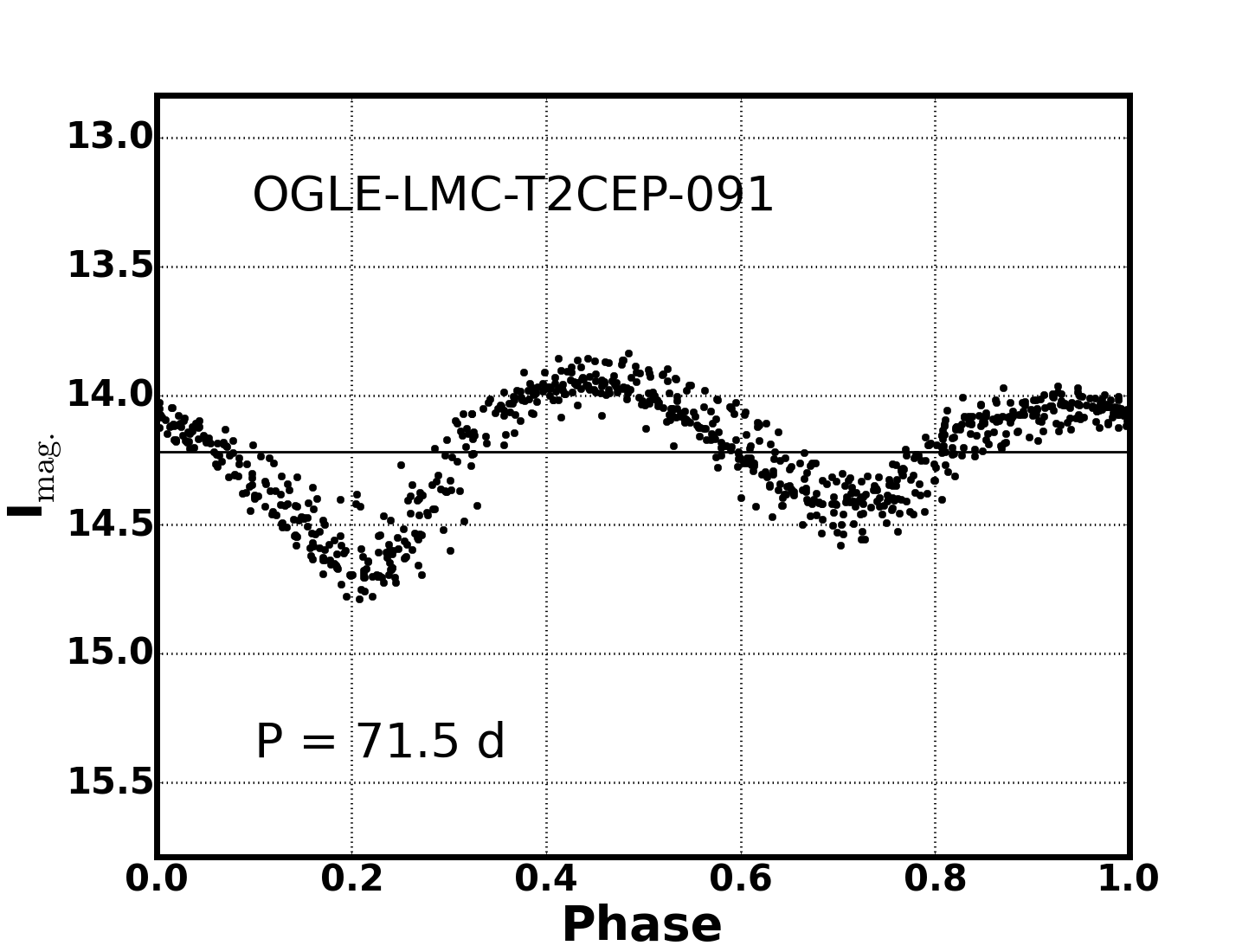}
  }
  \subfloat[]{\includegraphics[width=0.33\textwidth]{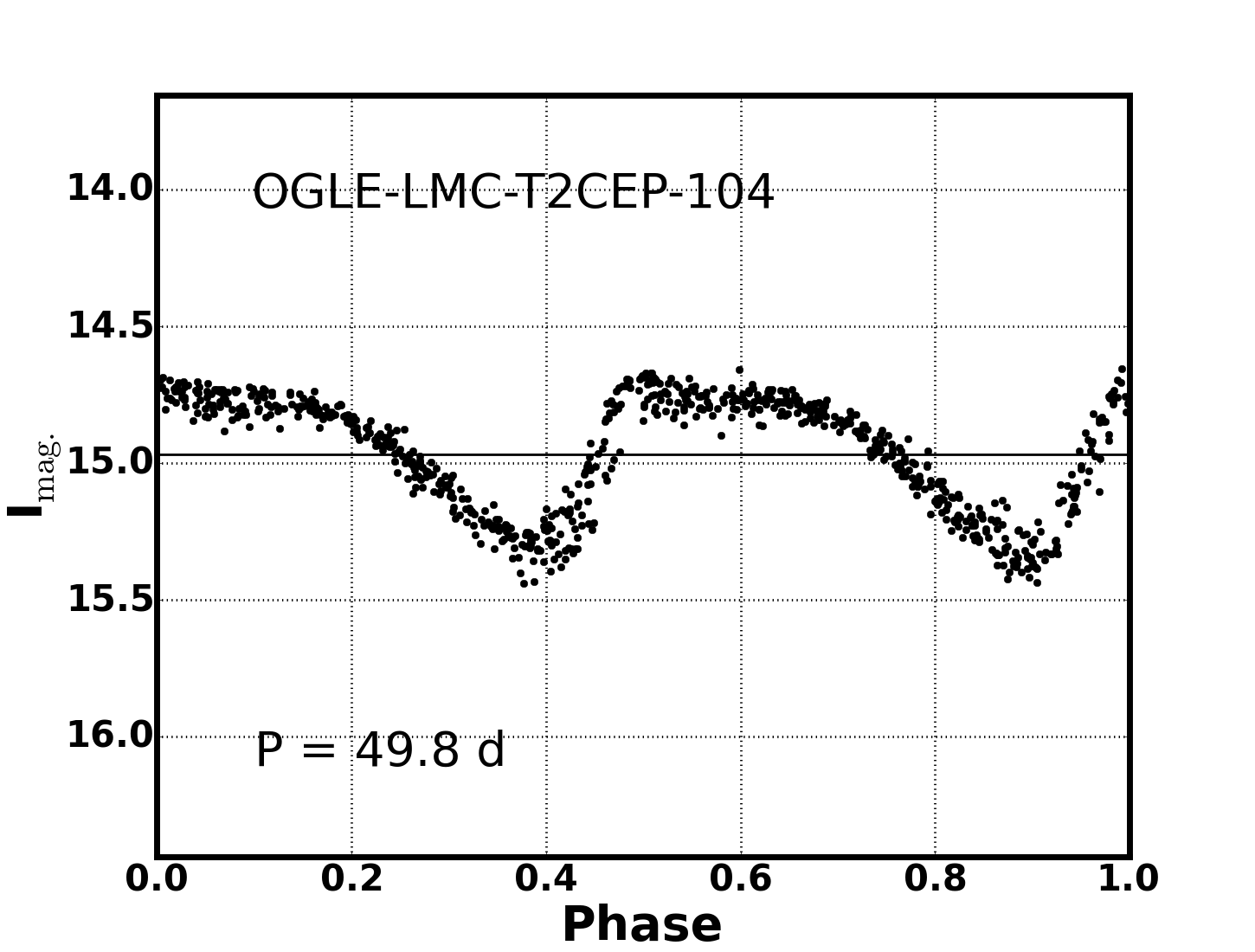}
  } 
   \subfloat[]{\includegraphics[width=0.33\textwidth]{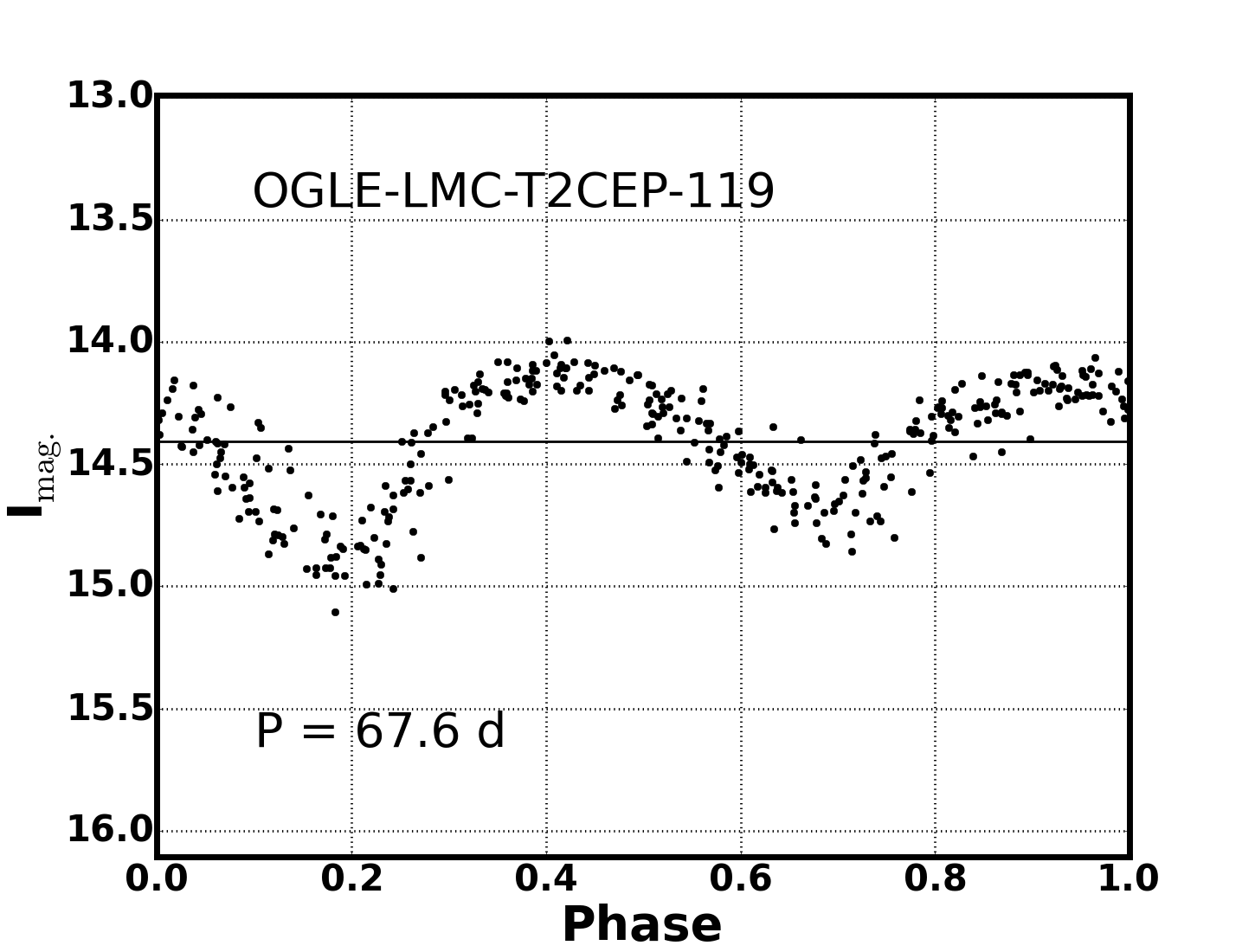}
  }\\
    \caption{Photometric time series of the LMC RV Tauri stars that have SEDs showing a broad IR excess (disc-type), phase folded on their formal periods. The horizontal line shows the mean I$_{\rm mag}$}.
 \end{figure*}

 \begin{figure*}
  \subfloat[]{\includegraphics[width=0.33\textwidth]{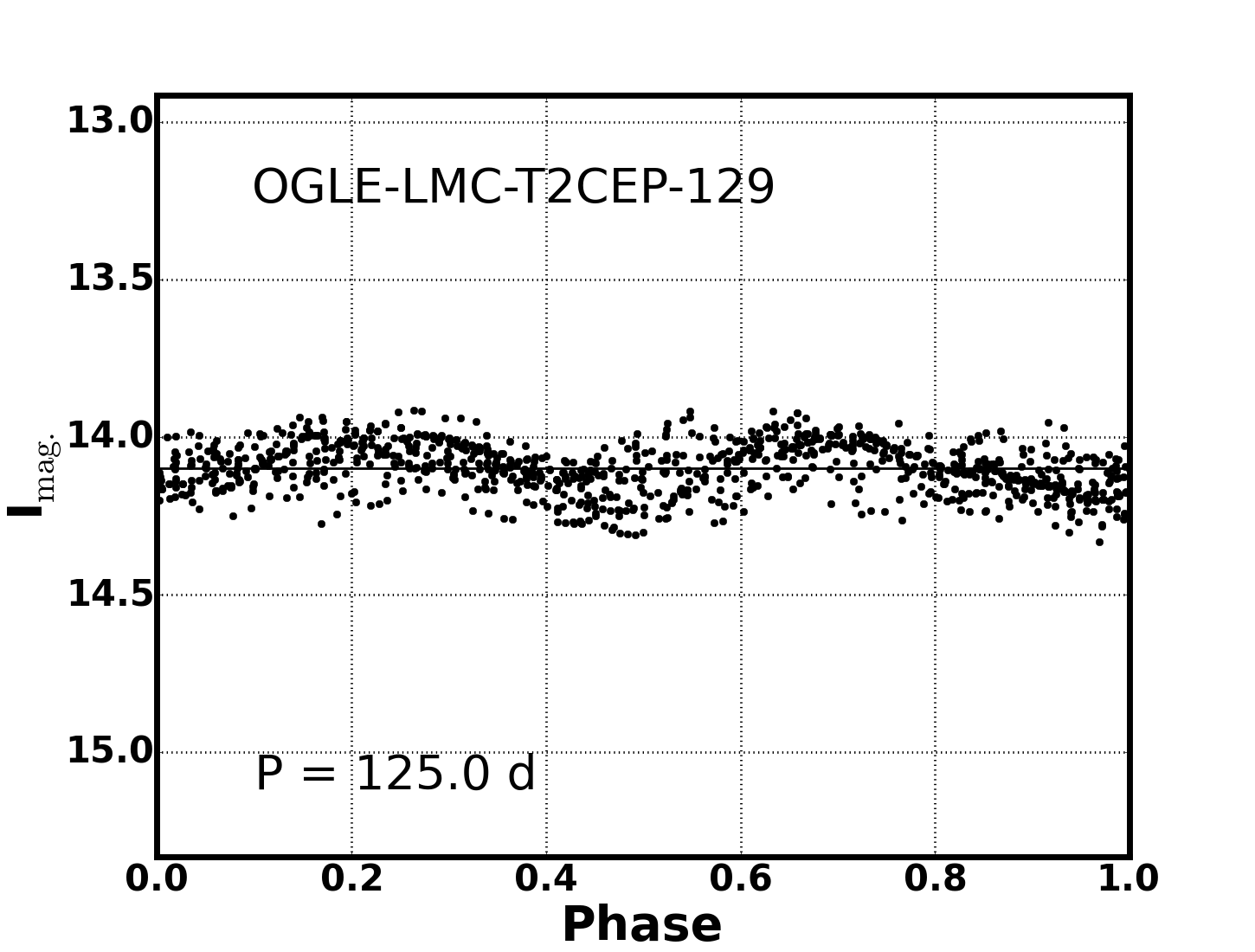}
  }
  \subfloat[]{\includegraphics[width=0.33\textwidth]{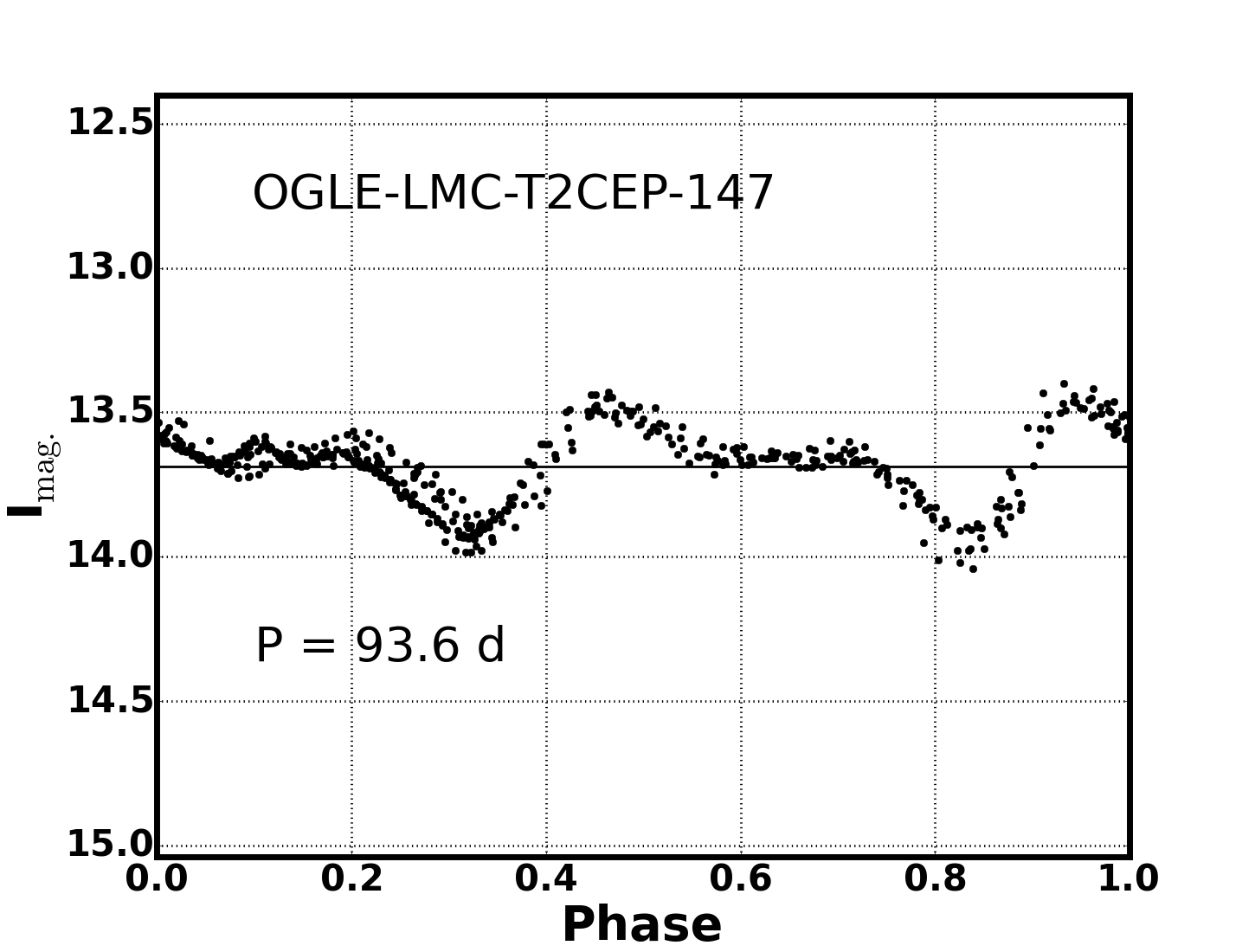}
  }
  \subfloat[]{\includegraphics[width=0.33\textwidth]{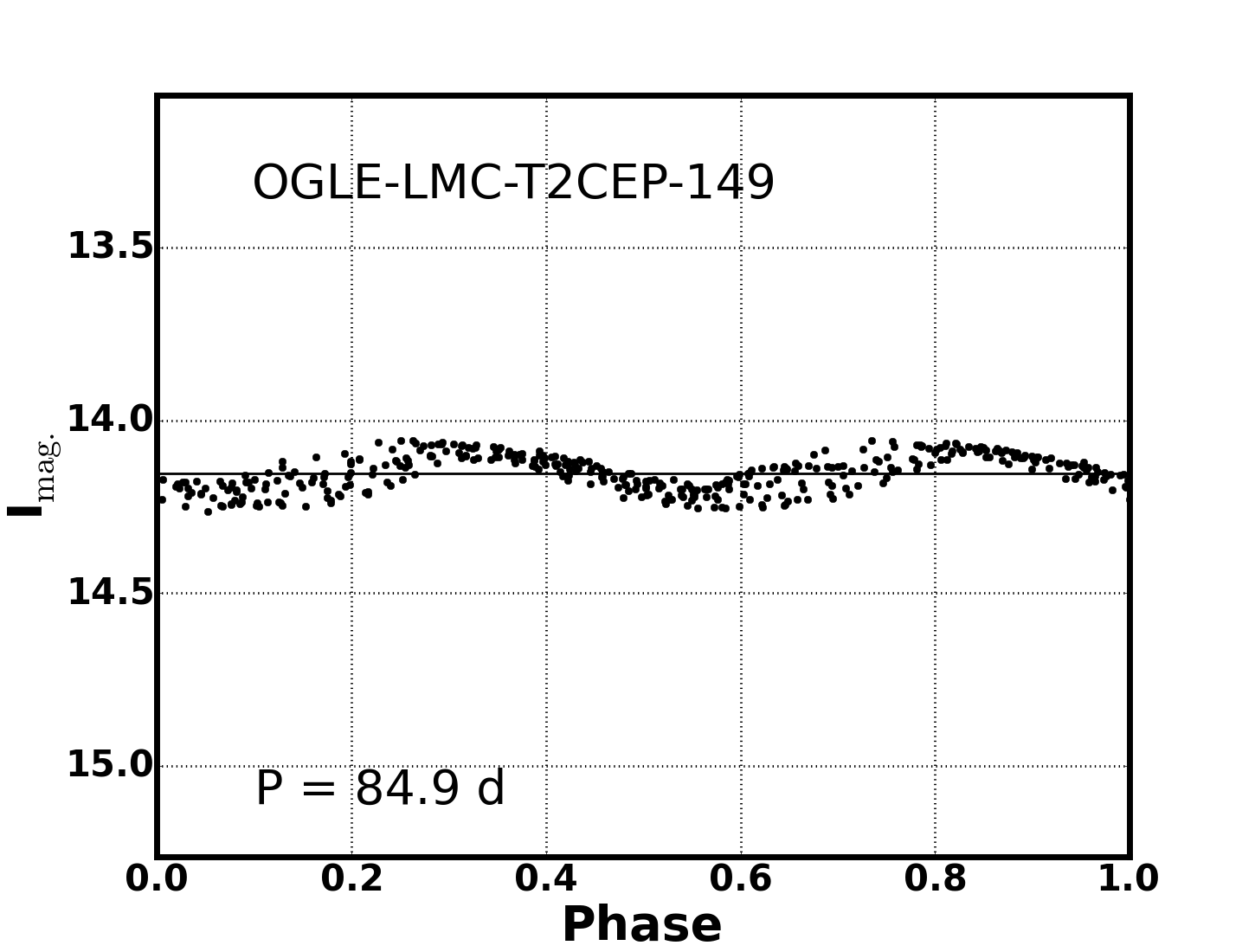}
  }\\
  
  \subfloat[]{\includegraphics[width=0.33\textwidth]{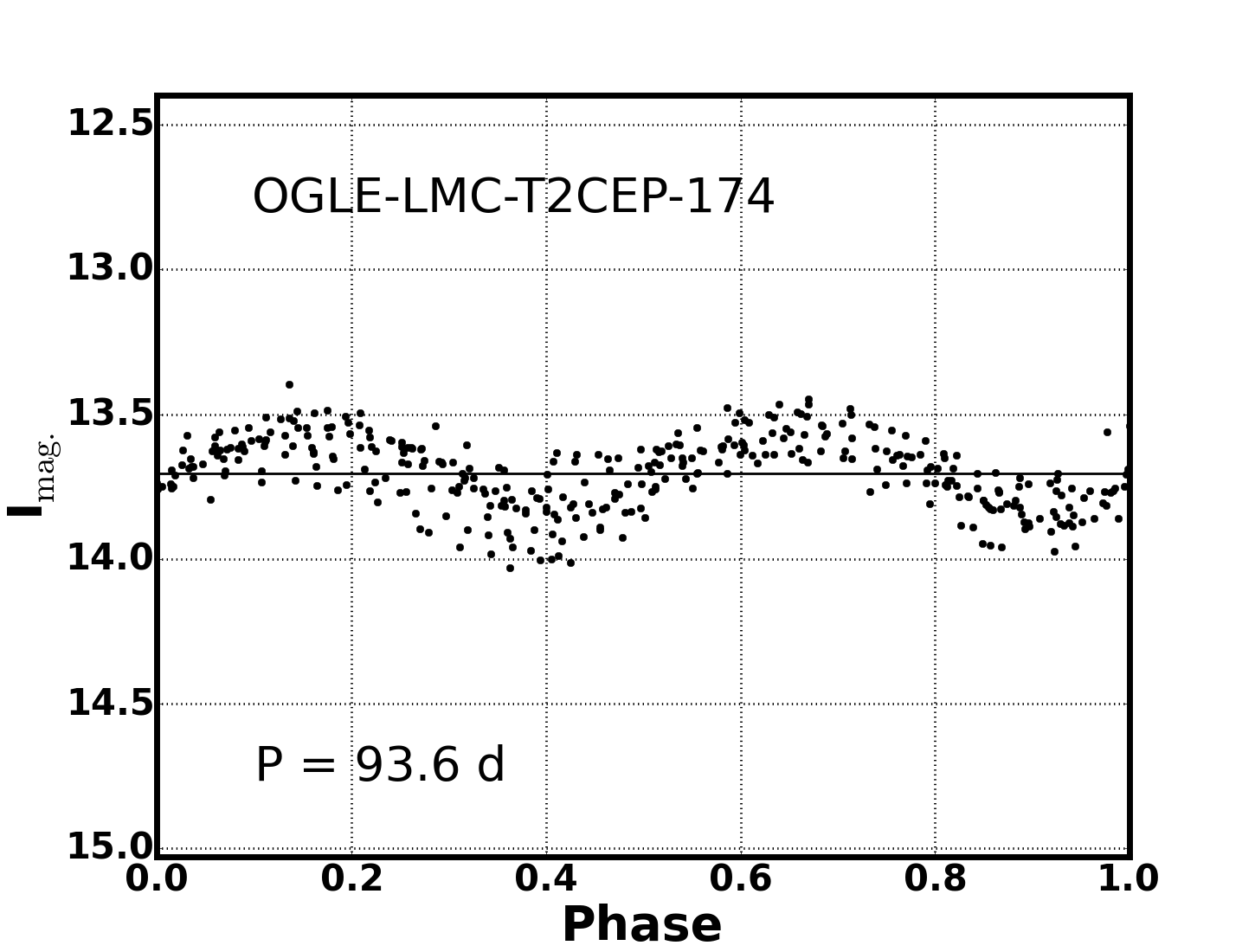}
  }
  \subfloat[]{\includegraphics[width=0.33\textwidth]{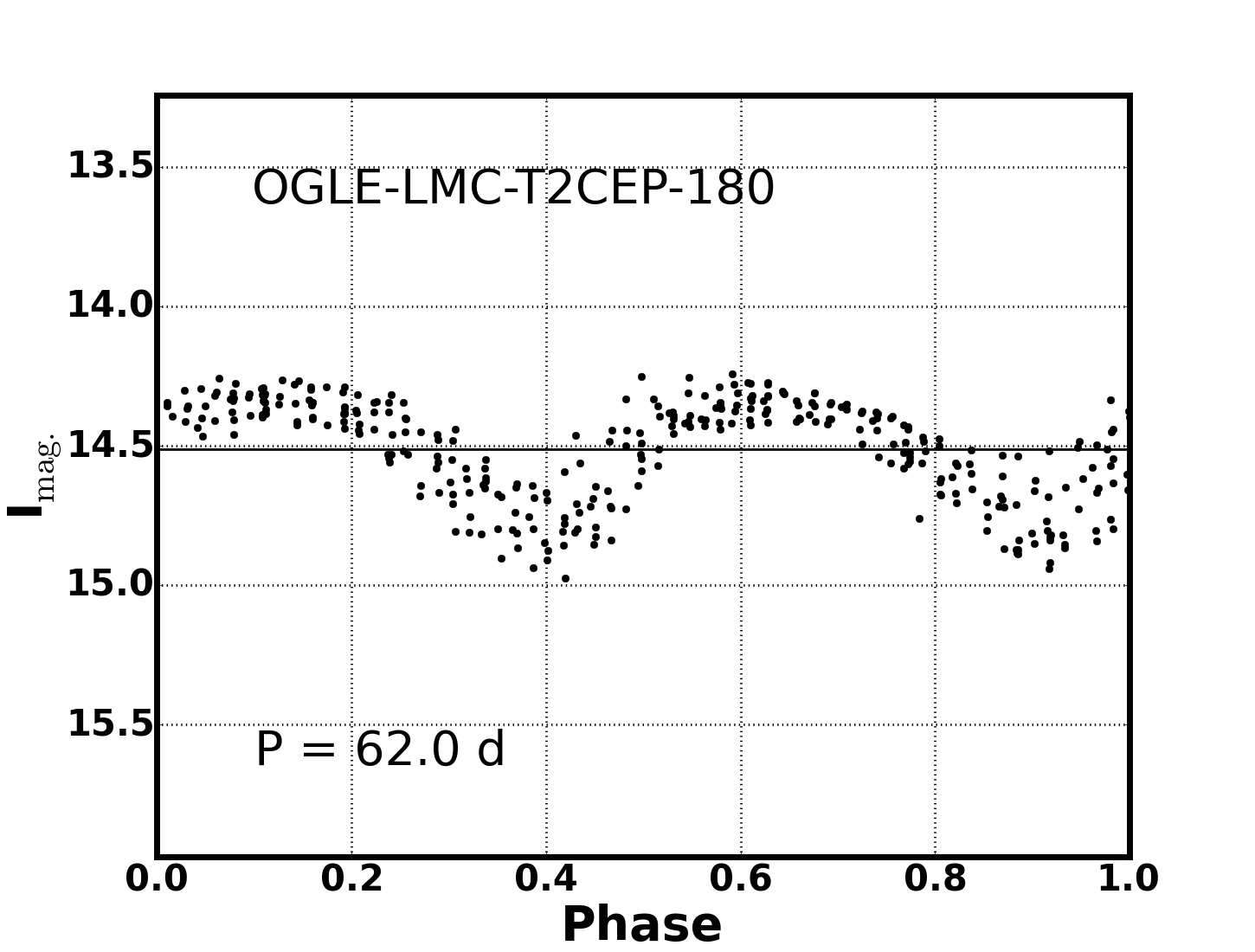}
  }
  \subfloat[]{\includegraphics[width=0.33\textwidth ]{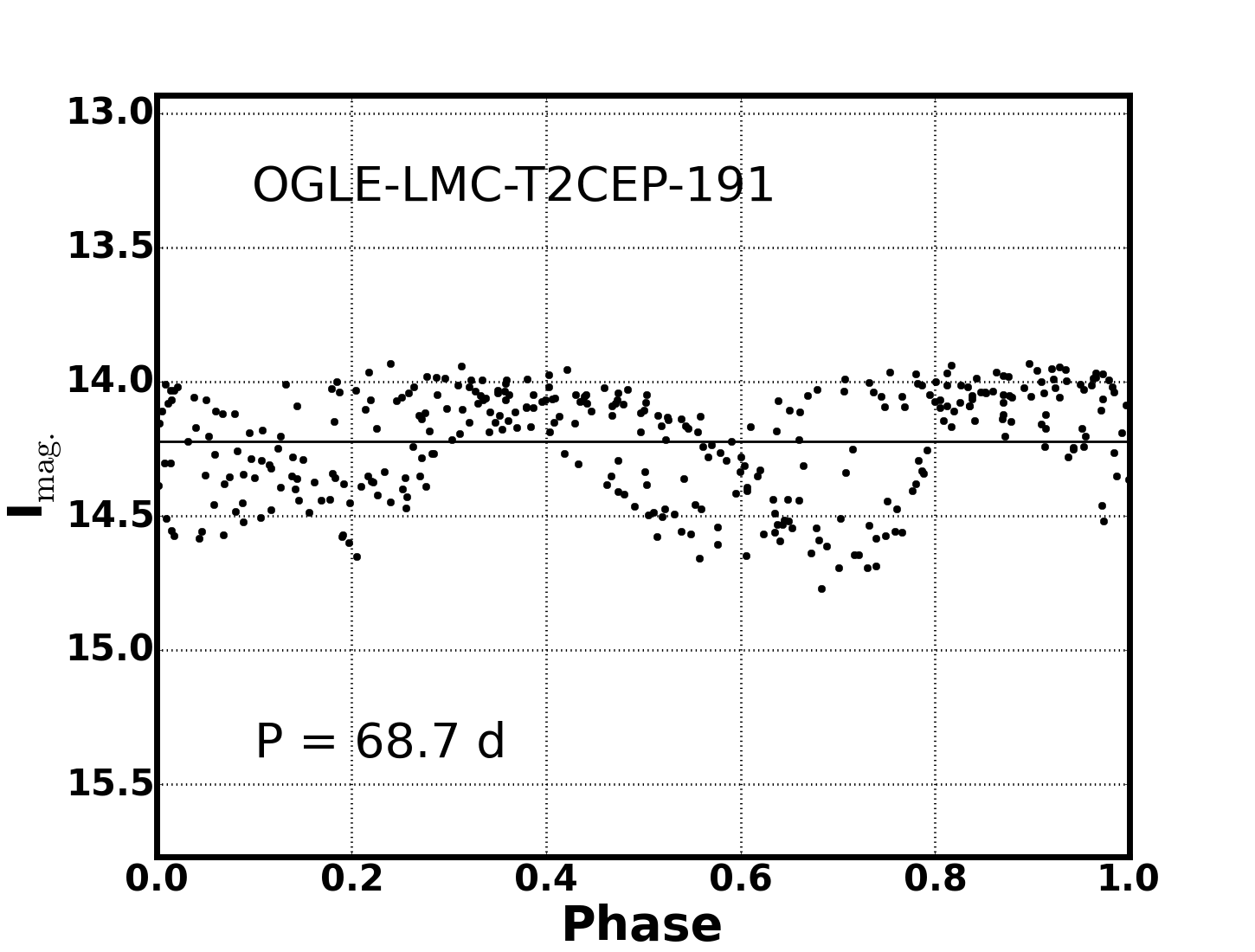} \label{subfig:modulation}
  }\\
  
  \subfloat[]{\includegraphics[width=0.33\textwidth]{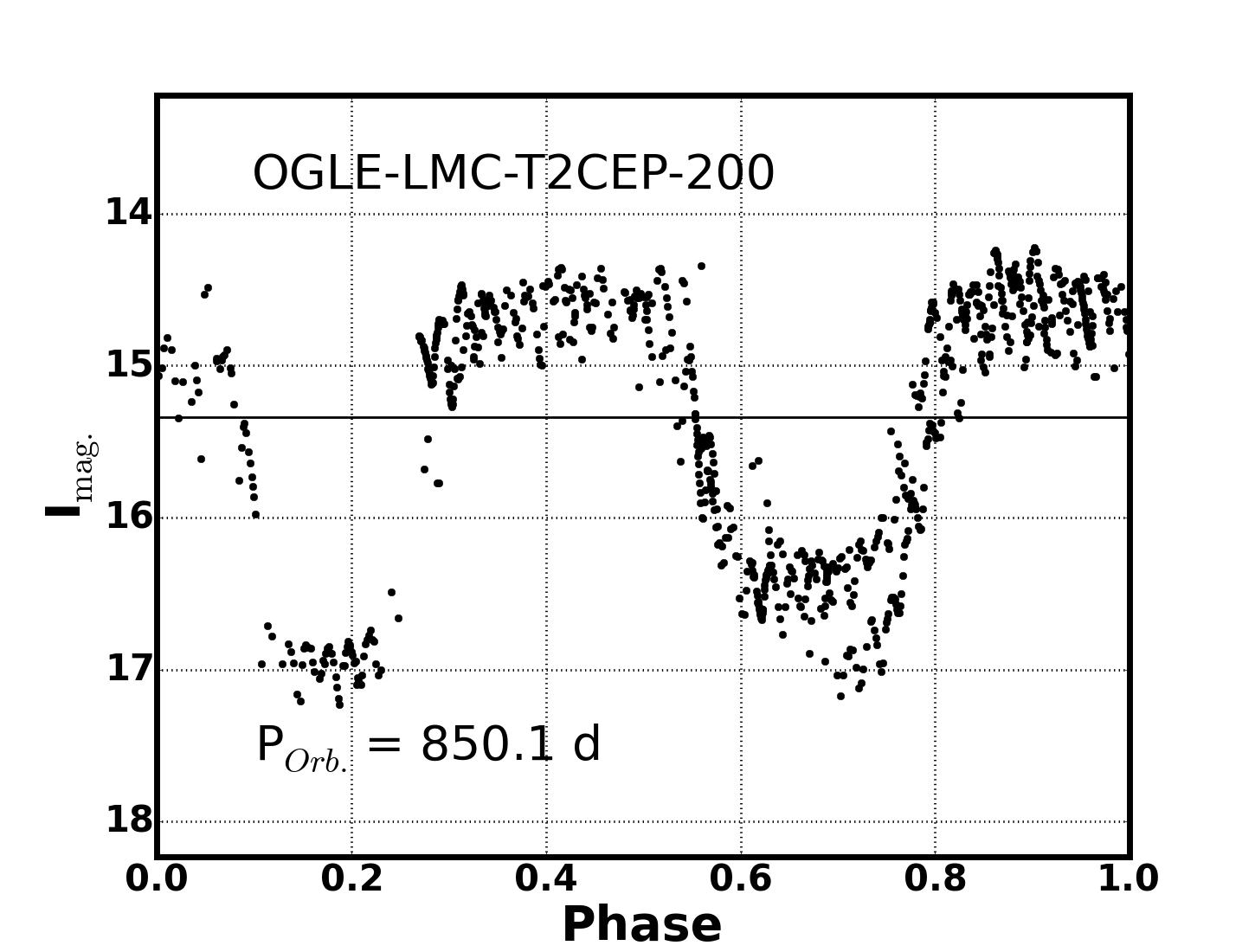}
  }
  \caption{continued. For OGLE-LMC-T2CEP-200, the light curve is folded on the 850 d orbital period of the binary.}
 \end{figure*}

 \section{Pulsations SMC} \label{appendix:AppendixD}
 \begin{figure*}
  \subfloat[]{\includegraphics[width=0.33\textwidth]{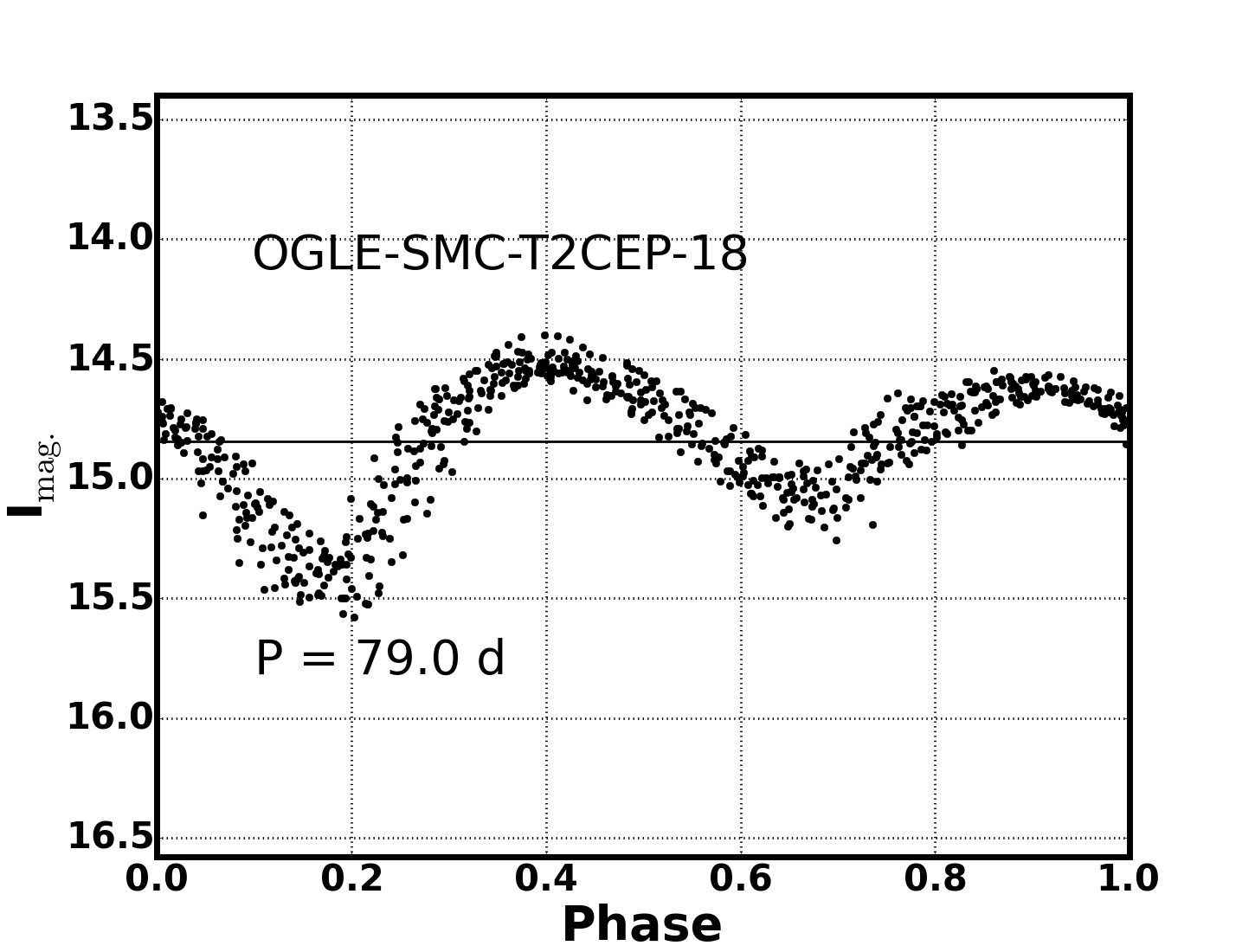}
  }
  \caption{Photometric time series of the only disc object in the SMC, phase folded on its formal period. The horizontal line shows the mean I$_{\rm mag.}$}
 \end{figure*}

 \begin{figure*}
  \subfloat[]{\includegraphics[width=0.33\textwidth]{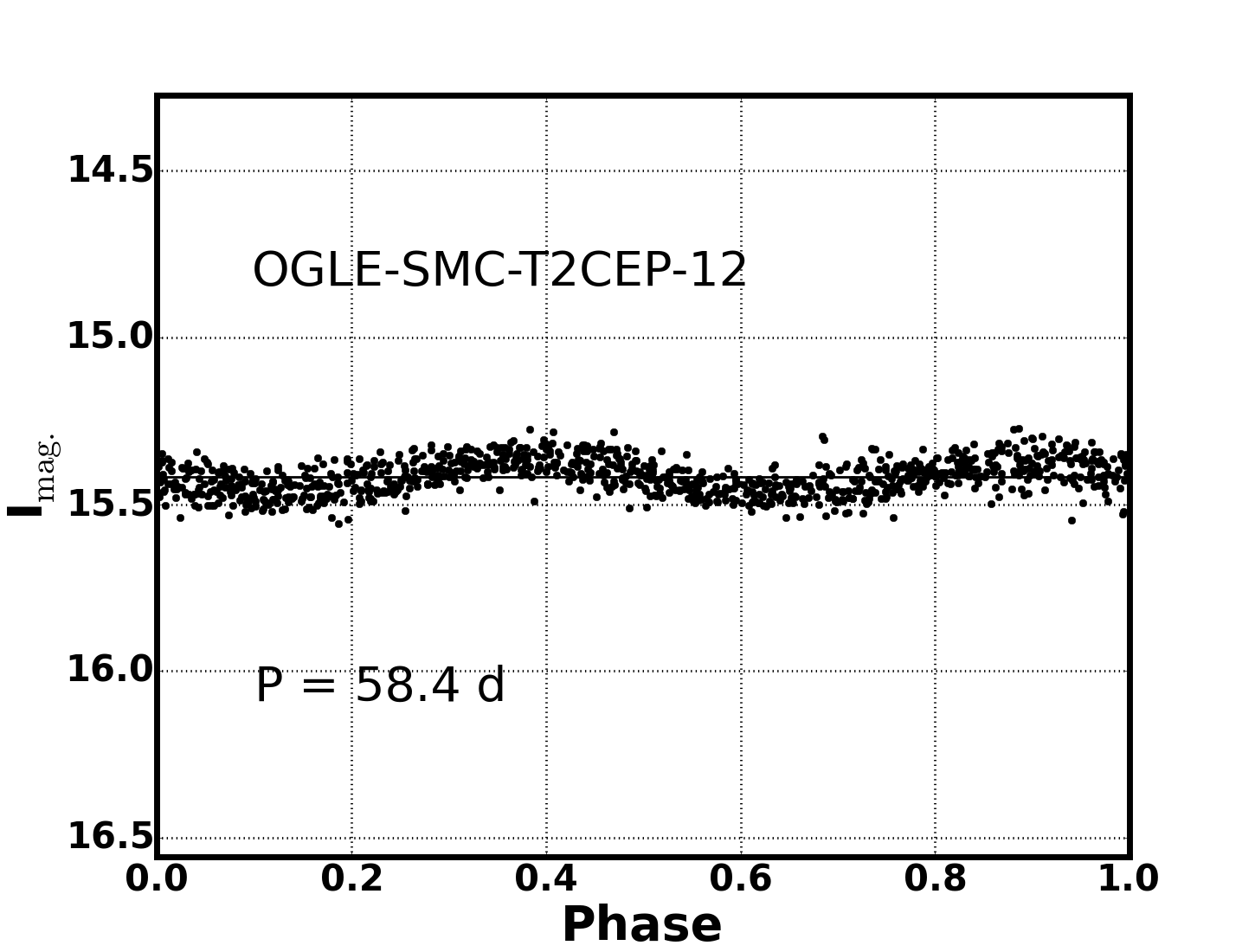}
  }
  \subfloat[]{\includegraphics[width=0.33\textwidth]{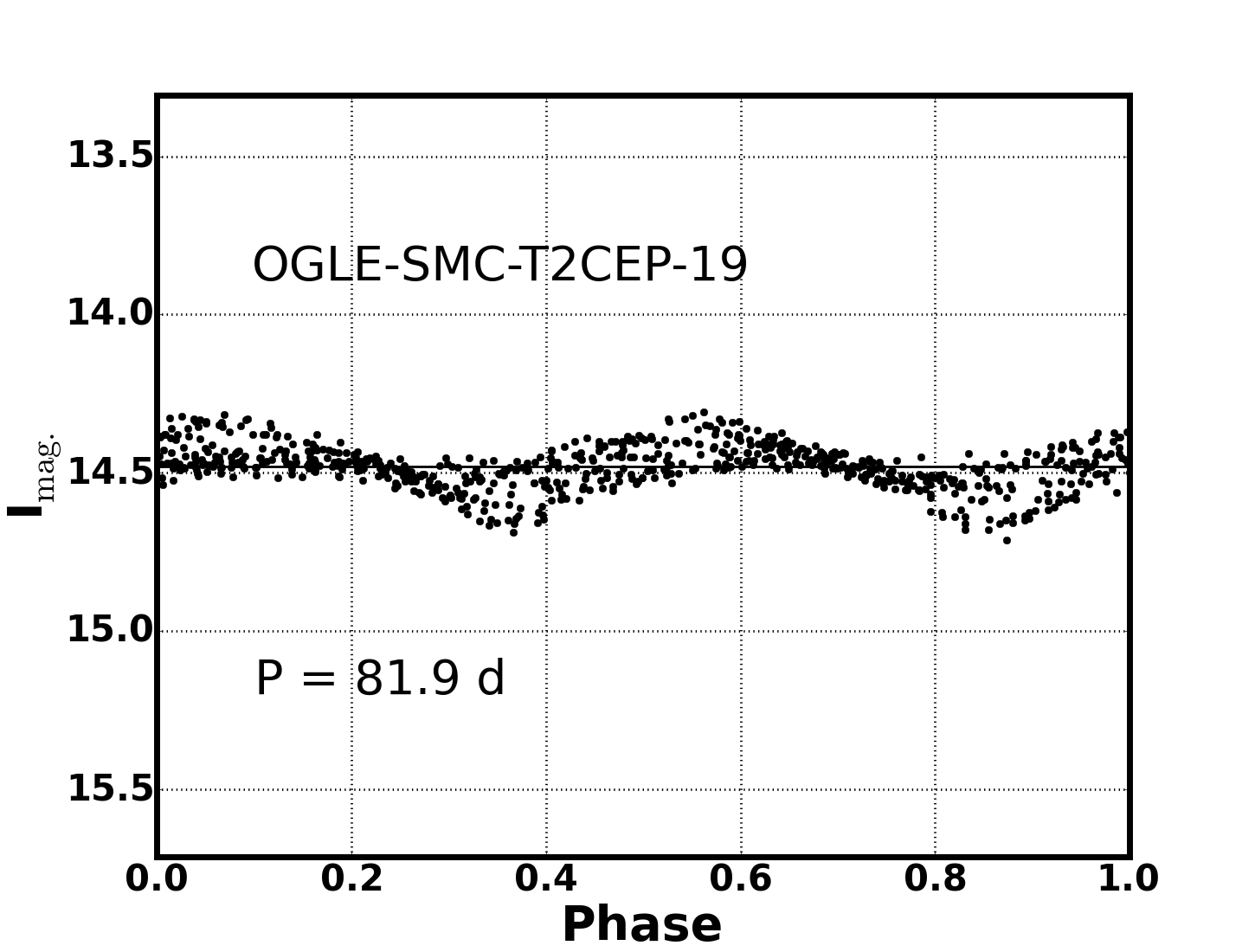}
  }
  \subfloat[]{\includegraphics[width=0.33\textwidth]{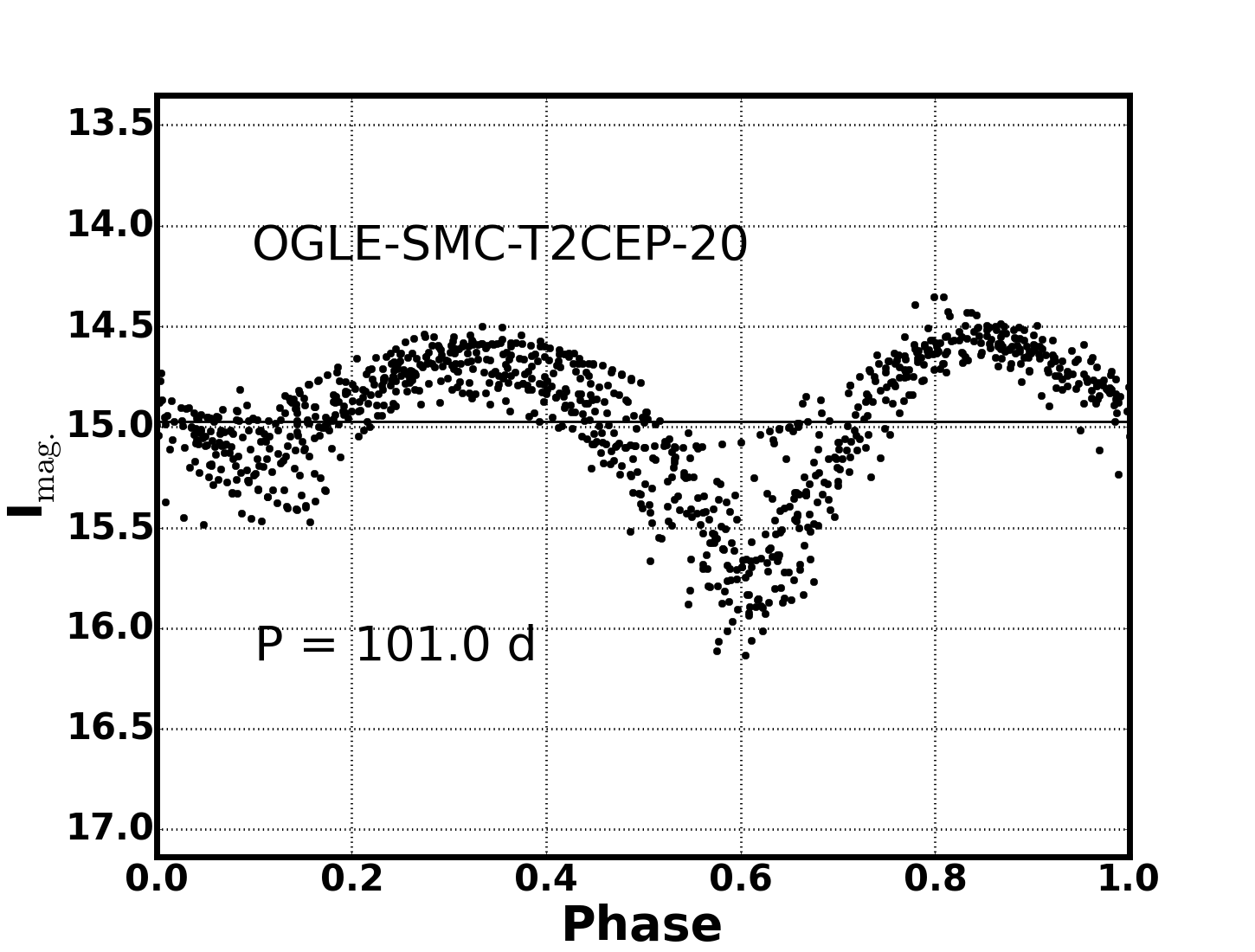}
  }\\
  
  \subfloat[]{\includegraphics[width=0.33\textwidth]{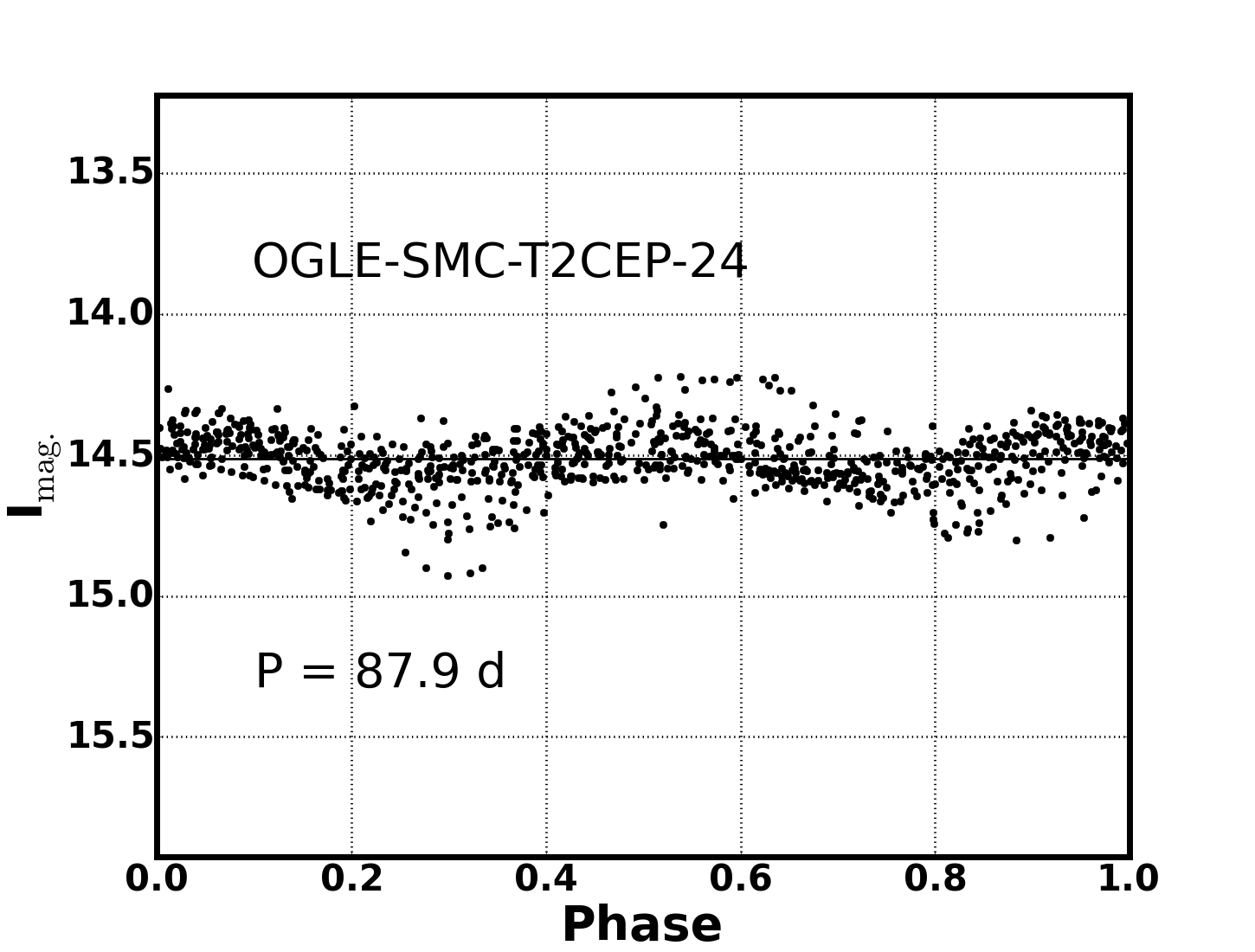}
  }
  \subfloat[]{\includegraphics[width=0.33\textwidth]{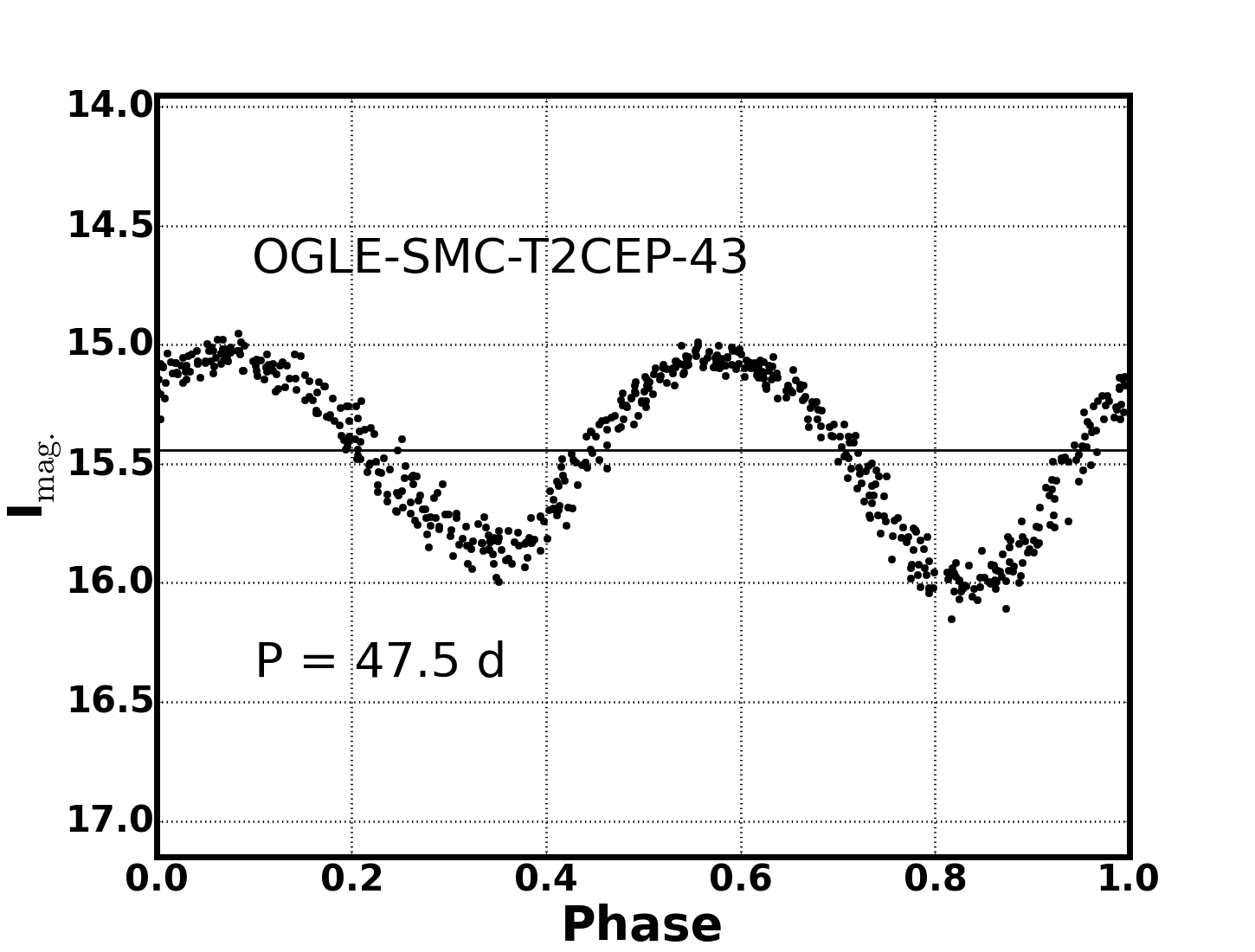}
  }
  \caption{Photometric time series of the SMC RV Tauri stars that have uncertain SEDs, phase folded on their formal period, shown in each panel. The horizontal line shows the mean I$_{\rm mag.}$}
 \end{figure*}  
  \begin{figure*}
   \subfloat[]{\includegraphics[width=0.33\textwidth]{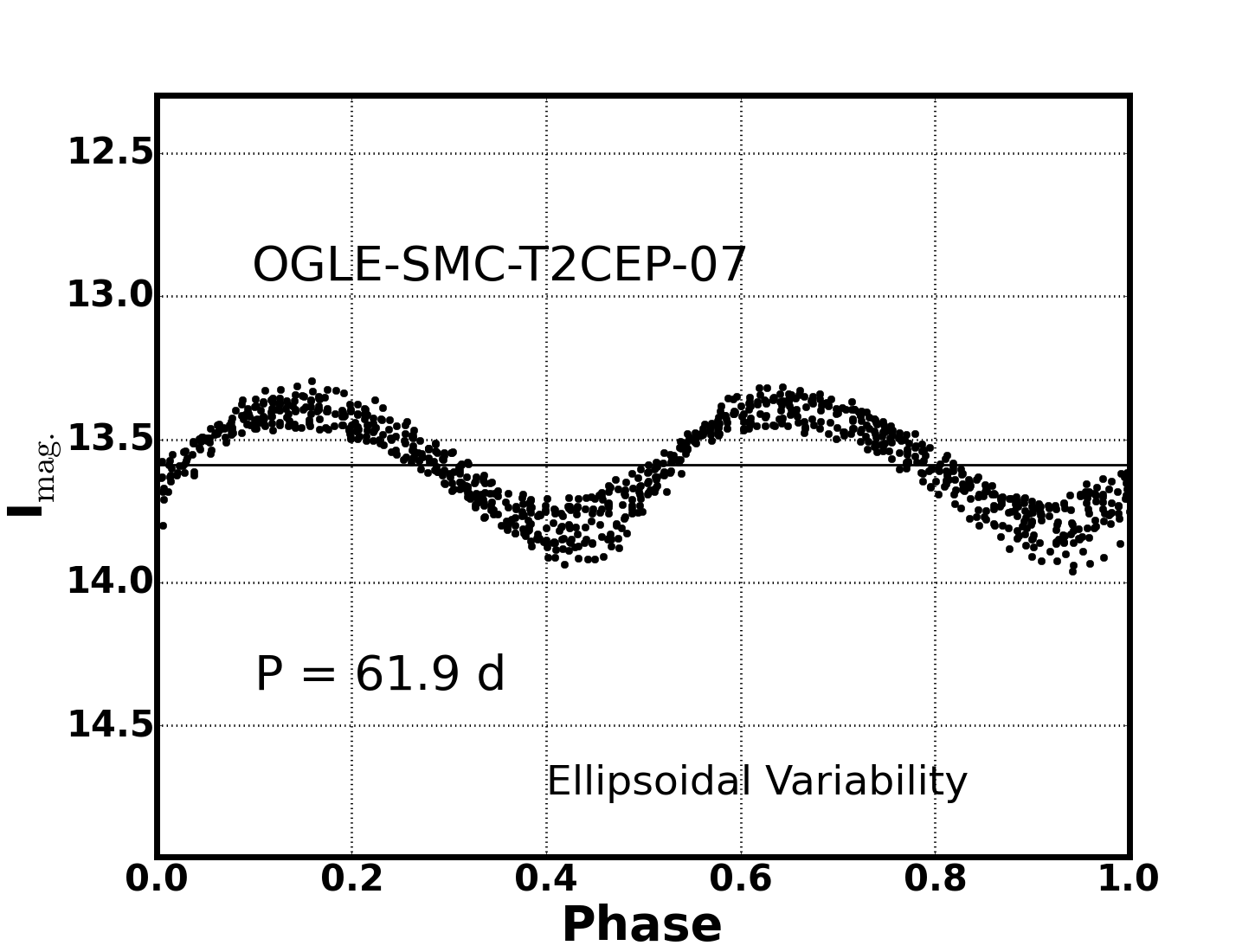}
   }
   \subfloat[]{\includegraphics[width=0.33\textwidth]{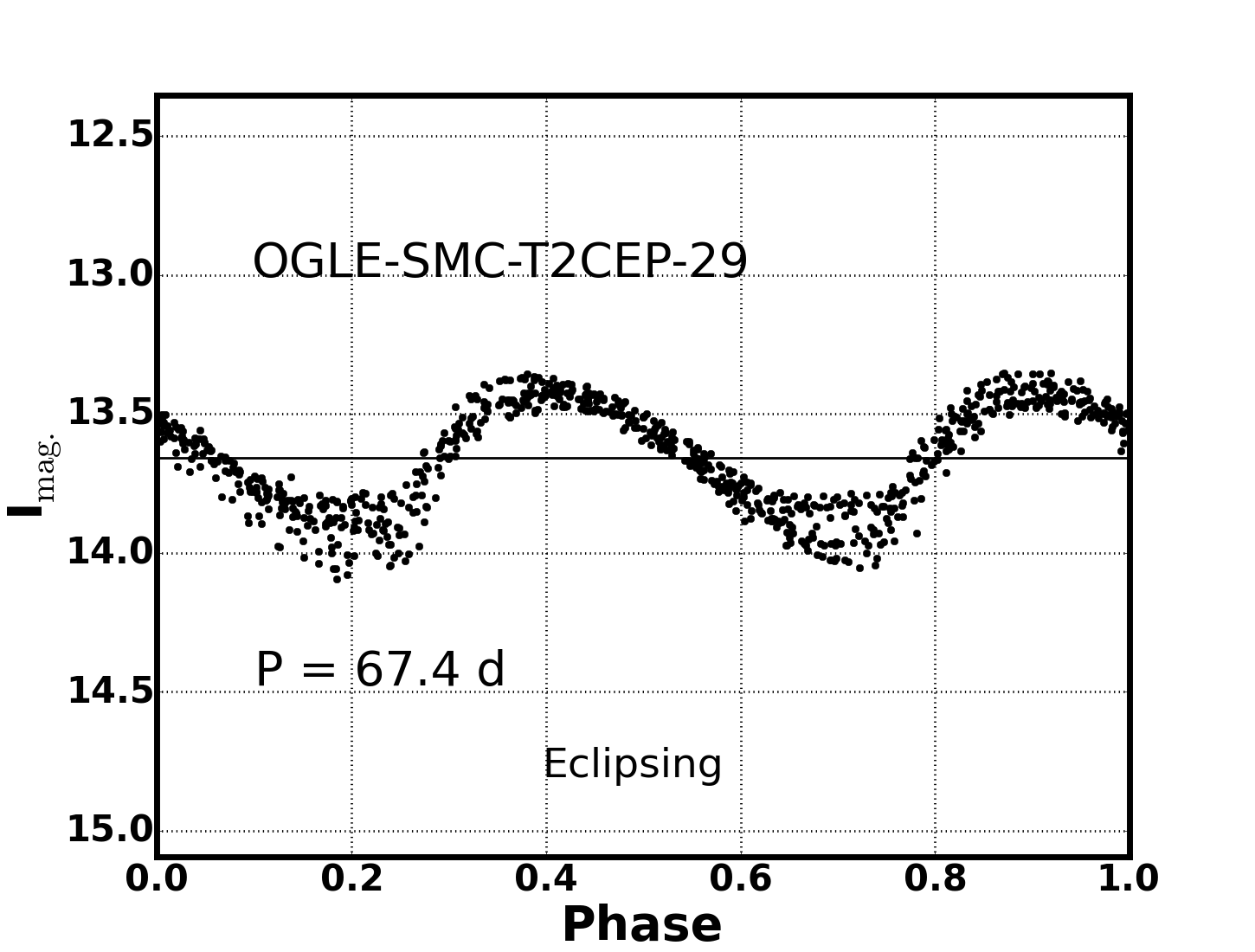}
   }
   \subfloat[]{\includegraphics[width=0.33\textwidth]{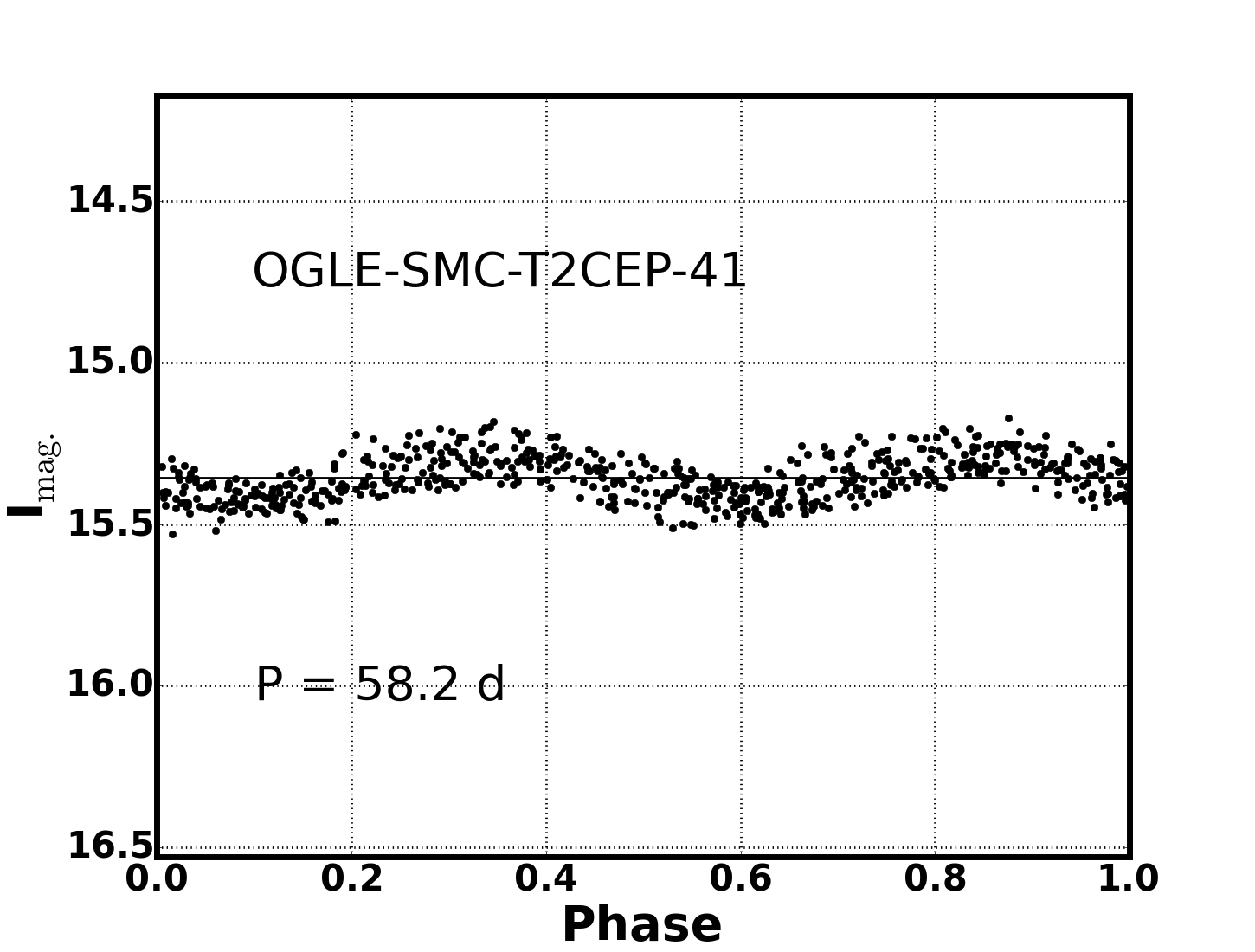}
   }
   \caption{Photometric time series of the SMC RV Tauri stars that have non-IR SEDs, phase folded on their formal period. The horizontal line shows the mean I$_{\rm mag.}$}
  \end{figure*}                          

 \section{Period evolution in the time series} \label{appendix:AppendixE}  
\begin{figure*}
  \subfloat[OGLE-LMC-T2CEP-119 ]{%
  \includegraphics[width=0.5\textwidth]{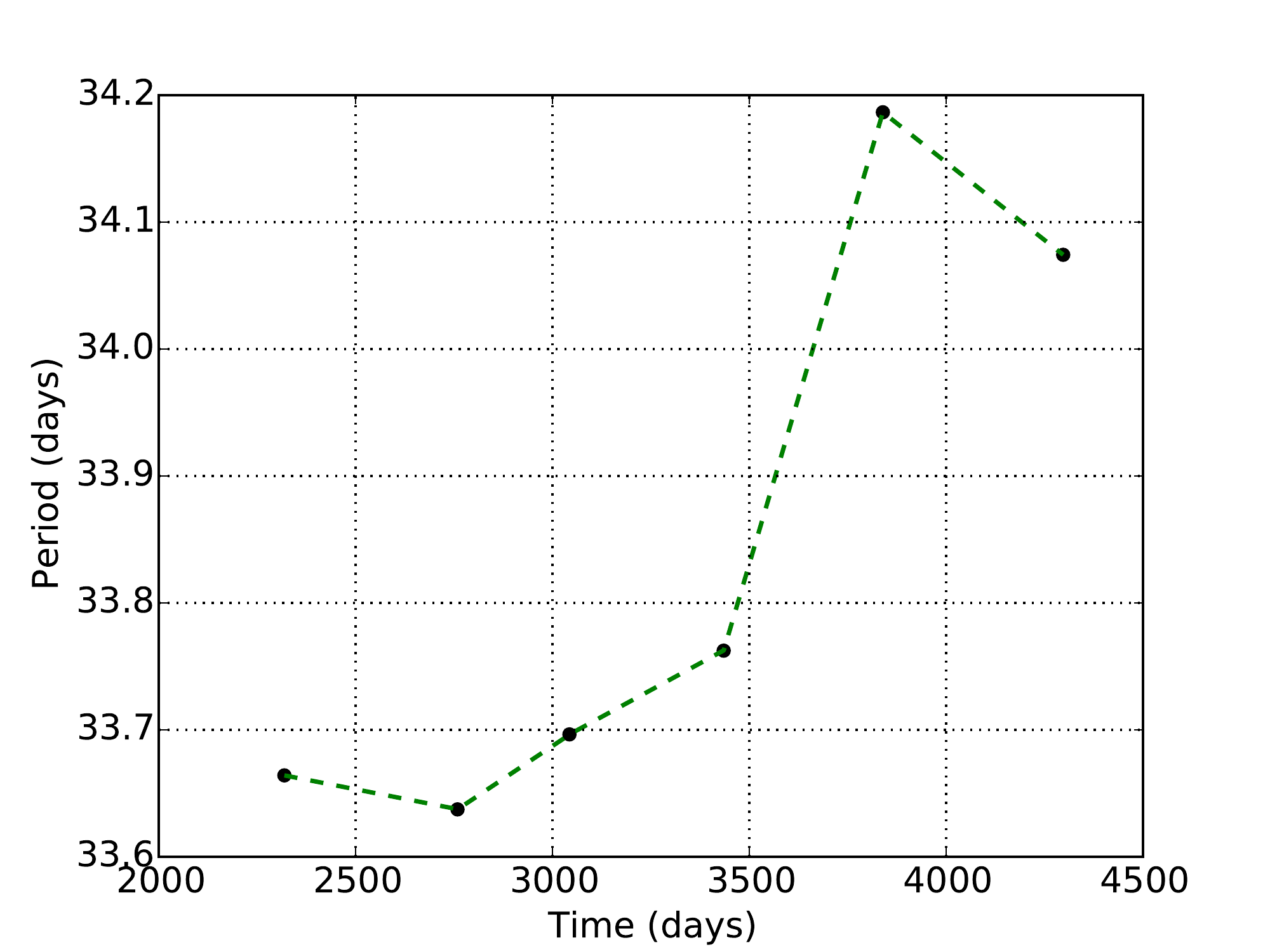}
  } 
  \subfloat[OGLE-LMC-T2CEP-149 ]{%
  \includegraphics[width=0.5\textwidth]{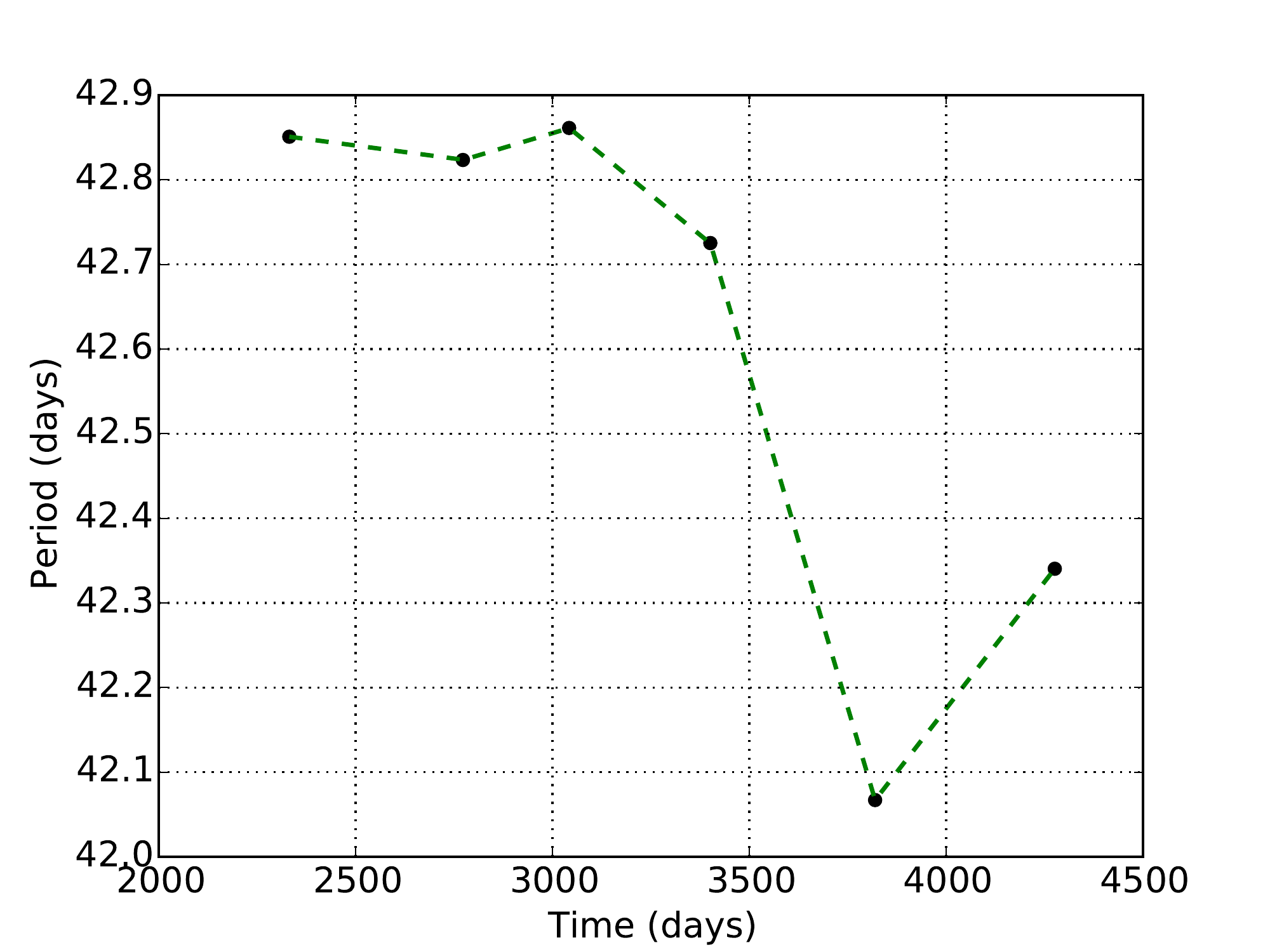}
  } \\
  \subfloat[OGLE-LMC-T2CEP-190 ]{%
  \includegraphics[width=0.5\textwidth]{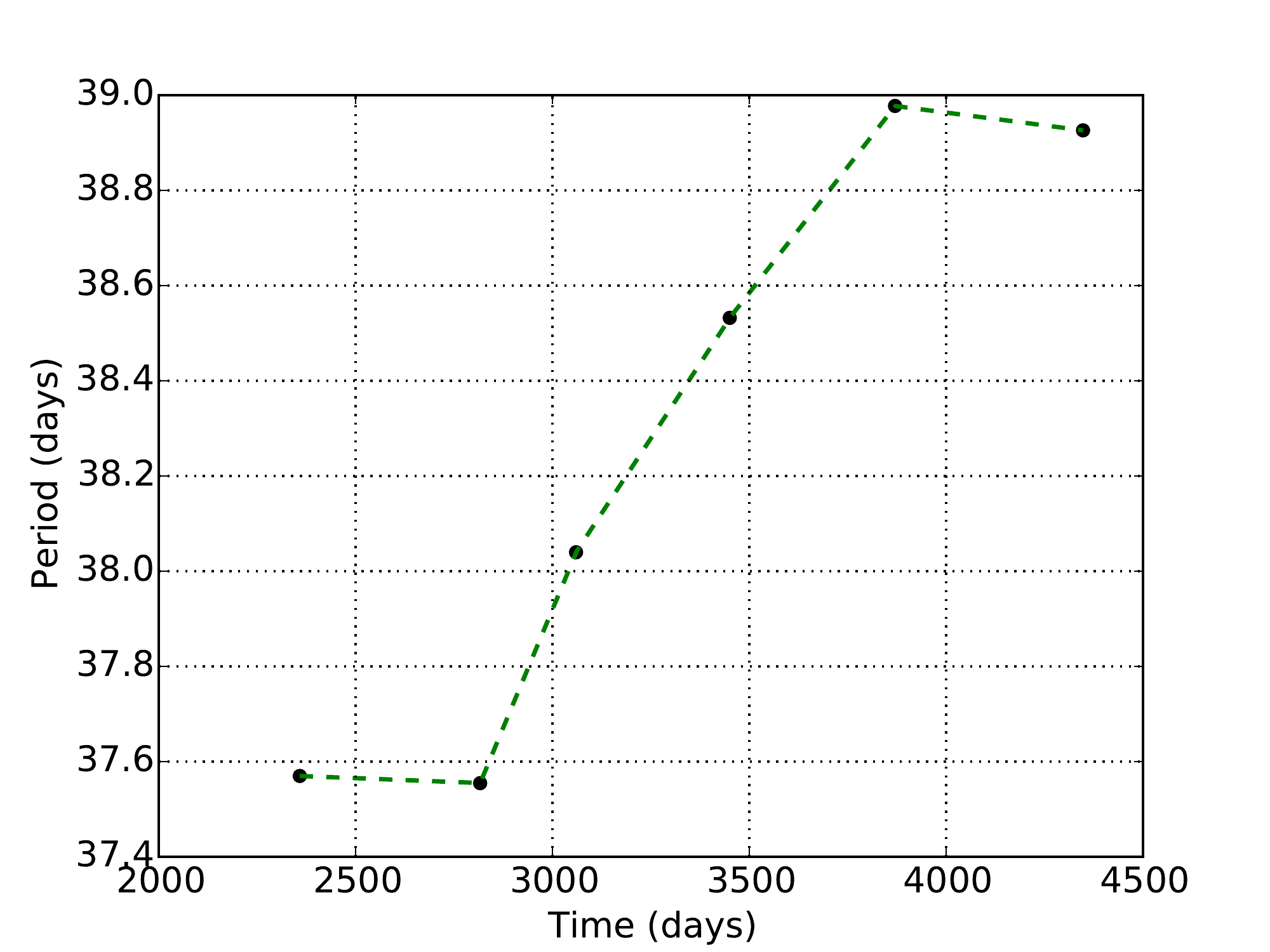}
  }
  \subfloat[OGLE-LMC-T2CEP-191 ]{%
  \includegraphics[width=0.5\textwidth]{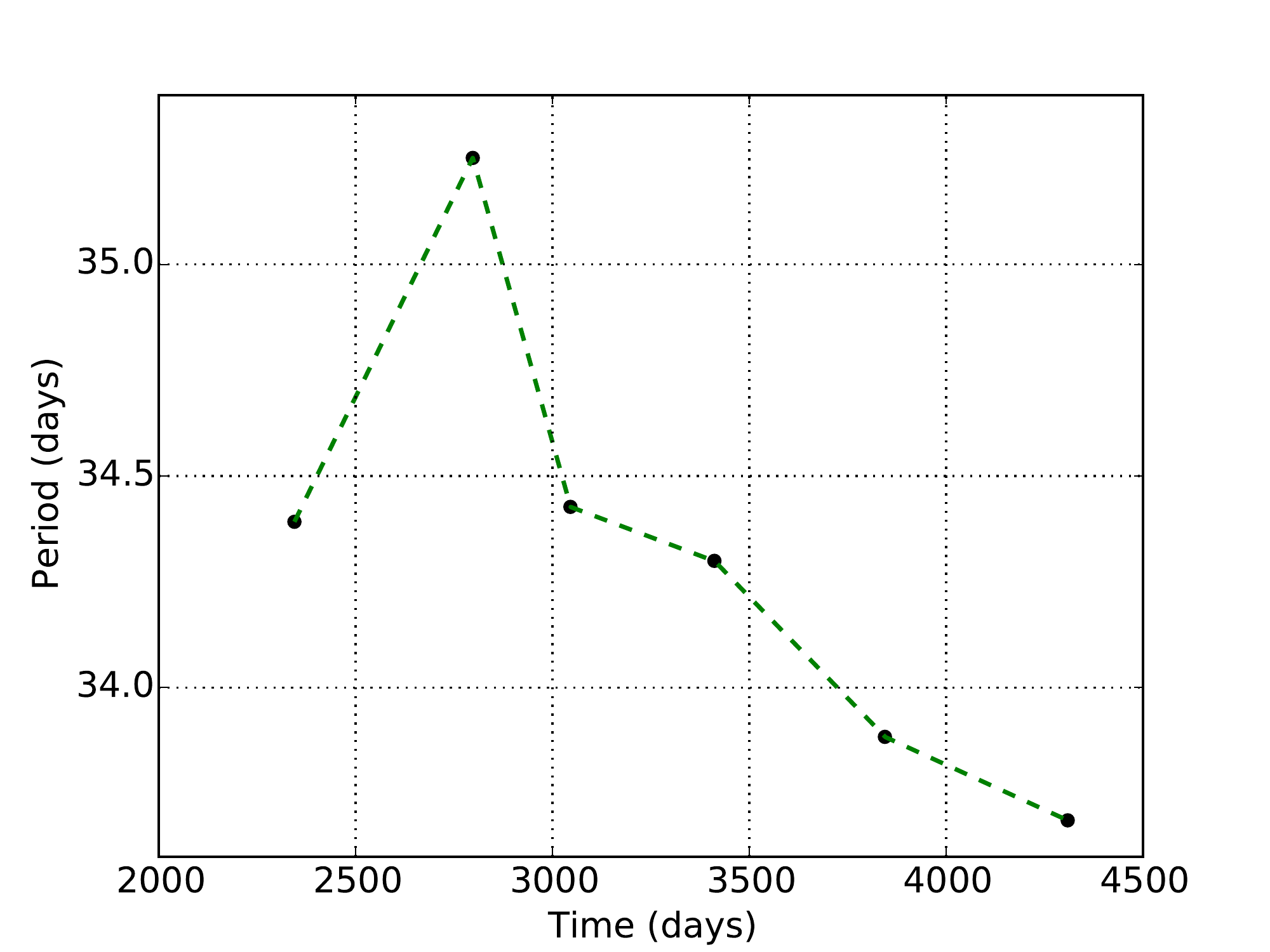}
  }
  \caption{Period evolution of LMC targets. The black dots represent the mean period in each time series chunk. The green dashed line indicates an interpolation through the points.}
  \label{figure:p_dot}
\end{figure*}    
  
\end{document}